\documentclass[1p]{elsarticle}

\makeatletter
\def\Hline{%
\noalign{\ifnum0=`}\fi\hrule \@height 1.5pt \futurelet
\reserved@a\@xhline}
\makeatother
\usepackage[bookmarks=false]{hyperref}
\usepackage{graphicx}
\usepackage{float}
\usepackage{subfigure}
\usepackage{subfigmat}
\usepackage{amsmath}
\usepackage{enumitem}
\usepackage{lipsum}

\setlist[itemize]{leftmargin=*}
\usepackage{amssymb}
\usepackage{xcolor}
\usepackage{rotating} 
\usepackage{pdflscape} 
\usepackage{silence}
\usepackage{ulem,cancel}

 \usepackage{lineno}

\begin{document}

\begin{frontmatter}



\title{{\color{black}High-order upwind and non-oscillatory approach for {\color{black}steady state diffusion}, advection-diffusion and application to magnetized electrons}}

\author[AA_address]{Amareshwara Sainadh Ch.}
\cortext[cor1]{Corresponding author. \\ 
E-mail address: s.chamarthi@al.t.u-tokyo.ac.jp, skywayman9@gmail.com (Amareshwara Sainadh Ch.).}
\author[AA_address]{Kimiya Komurasaki}
\author[AA_address]{Rei Kawashima}
\address[AA_address]{Department of Aeronautics and Astronautics, The University of Tokyo, 7-3-1 Hongo, Bunkyo, Tokyo 113-8656, Japan}
\begin{abstract}
{\color{black} Steady state simulations} of magnetized electron fluid equations with strong anisotropic diffusion based on the first-order hyperbolic approach is carried out using cell-centered higher order upwind schemes, linear and weighted essentially non-oscillatory (WENO). Along with the magnetized electrons, the diffusion equation is also simulated to demonstrate the implementation and design order of the accuracy of the approach due to their similar upwind structure. We show the adequacy of linear upwind schemes for diffusion equation and the use of shock-capturing scheme like WENO does not have any adverse effect on the solution, unlike the total-variation diminishing (TVD) methods. We further extended the approach to advection-diffusion equation, and appropriate boundary conditions have obtained a consistent design accuracy of the third and fifth order. We implemented the WENO approach to advection-diffusion equation by using the split hyperbolic method to demonstrate the advantage of non-oscillatory schemes to capture sharp gradients in boundary layer type problems without spurious oscillations. Finally, numerical results for magnetized electrons simulations indicate that with increasing strength of magnetic confinement it is possible to capture sharp gradients without oscillations by WENO scheme.
\end{abstract}

\end{frontmatter}



\section{Introduction}
An upwind hyperbolic-equation-system approach for two-dimensional electron fluid equations has been first proposed by Kawashima et al. \cite{Kawashima2015} for robust calculation on a vertical-horizontal uniform mesh. While this work is a novel attempt for such problems, there are few drawbacks. The underlying scheme for the original paper was first order upwind scheme which was highly diffusive and might require extremely fine mesh which can be computationally expensive. Also, the original work was focused on improving diagonal dominance by avoiding cross-diffusion terms and speed of the computation but did not specifically address the issue of spurious oscillations due to misalignment of the grid and magnetic field that may occur with highly anisotropic diffusion problems on non-aligned meshes. However, mesh convergence carried out by evaluating the transverse electron flux for both the hyperbolic approach and field-aligned approach indicated that the numerical diffusion, which can be inferred as spurious oscillations, is reduced on very fine meshes. It has been proven that the hyperbolic approach attains good efficiency and accuracy in the presence of strong magnetic confinement. Kawashima et al. \cite{kawashima2016} had later extended their approach to {\color{black}{\color{black}non-isothermal system by including energy equation}. They considered $3^{rd}$ order TVD scheme to reduce the numerical diffusion, but it will be shown in this paper that the TVD scheme will reduce to third order linear scheme and can still have considerable numerical errors. TVD scheme may not be able to capture the sharp gradients without oscillations. The issues that are addressed in this paper are:
\begin{itemize}
{\color{black}\item spurious oscillations due to misalignment of the grid and magnetic field}
\item implementation of essentially non-oscillatory schemes to reduce numerical diffusion
{\color{black}\item implementation of boundary conditions}
\end{itemize}}

The primary influence for the magnetized electron fluid simulations is the upwind formulation for diffusion equation introduced by Nishikawa \cite{Nishikawa2007} based on residual-distribution (RD) method. The mathematical strategy of this approach is to split the second order partial differential equation into a set of first-order differential equations by adding new variables and pseudo-time advancement terms the diffusion equation is computed as a hyperbolic system. This radical approach has been shown to offer several advantages over conventional methods, like accelerated convergence for steady state solution, higher order of accuracy for both primary and gradient variables. The original approach of Nishikawa has been further extended to edge-based finite volume schemes \cite{Nishikawa2014b} and Lee et al. have introduced cell-centered finite volume approach \cite{Hong2018}. This approach was also extended to advection-diffusion equation by Nishikawa \cite{Nishikawa2010b, Nishikawa2014}, to time-dependent problems by Mazaheri and Nishikawa \cite{mazaheri2014N} and Navier-Stokes equation \cite{Nishikawa2011a}. In addition to the magnetized electron simulations, in this paper, we extend the upwind formulation introduced for the diffusion and advection-diffusion equations to higher order and high-resolution cell centered schemes on uniform meshes. The similar upwind structure of diffusion and magnetized electron fluid equations and the ability of WENO scheme to capture the sharp gradients in advection-diffusion without oscillations provide essential insights for the plasma simulations. 

The accuracy of the finite volume method (FVM) depends on the accuracy of the numerical fluxes computed at the cell interfaces. These numerical fluxes are evaluated based on the flux Jacobian matrices and variables defined at the cell-centers. For a hyperbolic equation the upwind fluxes at the cell interfaces, say $\hat E_{i+\frac{1}{2}}$, are constructed either by a Godunov type approximate Riemann solver(Roe \cite{roe1981approximate}, HLLC \cite{Harten1983}, etc.) or Boltzmann type solver, also known as flux vector splitting method (AUSM \cite{liou2006sequel}, Steger-Warming \cite{Steger1981} etc.). According to the Godunov's theorem \cite{godunov1959}: ``For simulations of flows involving discontinuities and sharp gradients, Gibbs phenomenon or spurious oscillations appear in the solutions near the discontinuities if the computations are carried out by linear numerical schemes that are greater than first-order accurate".

 In the last three decades, several high-order accurate schemes were developed to satisfy this criterion, and among these methods, TVD and WENO are most popular. The total-variation diminishing monotone upstream scheme for conservation laws (TVD-MUSCL) \cite{van1977towards} is one of the popular high-order space accuracy schemes to compute the fluxes because of its simplicity and robustness. A well-known drawback of the slope limiters that are associated with these methods is that they tend to `clip' smooth extrema of the flow and the accuracy necessarily degenerates to first order. On the other end, WENO schemes first introduced by Jiang and Shu \cite{jiang1995} to capture discontinuities without spurious oscillations and are also able to achieve an arbitrarily high formal order of accuracy in smooth flows. Various versions, like WENO-5M \cite{Henrick2005}, WENO-3YC \cite{Yamaleev2009} and WENO-5Z \cite{Borges2008}, of the WENO scheme, are proposed over the years to improve the accuracy and reduce their dissipative nature. Tan and Shu \cite{Tan2011} had also proposed high-order boundary conditions based on the WENO idea to prevent numerical oscillations contaminating the solution due to shockwaves near the boundary. Another well-known high-order method is the family of central compact schemes developed by Lele \cite{lele1992compact}. Unfortunately, the compact central schemes cannot be used in the current `upwind' formulation for diffusion and also cannot capture the discontinuities without oscillations. Pirozzoli \cite{Pirozzoli2002} had proposed a conservative compact upwind method in combination with WENO scheme in-order to capture discontinuities. {\color{black} Recently, Ghosh and Baeder \cite{ghosh2012compact} has developed a class of upwind biased compact-reconstruction finite difference WENO schemes called CRWENO. This concept was extended in \cite{guo2014positivity} where a positivity-preserving fifth-order finite volume compact-WENO (FVCW) scheme was developed. On the other side, an alternative approach is developed by Deng et al. \cite{Deng2000} known as weighted compact non-linear schemes (WCNS) with similar discontinuity capturing abilities of WENO. These schemes are more flexible that we can interpolate not only fluxes \cite{Zhang2008}, but also conservative variables \cite{nonomura2012}, primitive variables or variables that are projected to the characteristic fields \cite{Wong2017}. The other advantages of WCNS are that they can be used with flux difference splitting methods like Roe and HLLC and still maintain high-order of accuracy and also have good freestream and vortex preservation capabilities on curvilinear grids \cite{Nonomura2010}.}

In this paper, we present the adequacy of linear upwind schemes for diffusion equation, and the use of shock-capturing scheme like WENO does not have any adverse effect on the solution, unlike TVD methods. While the third order explicit and fifth order explicit, compact and WENO reconstructions are presented in this paper, the schemes can be extended to arbitrarily high-order \cite{Balsara2000a,shukla2005derivation,deng2015family}. We can obtain a significant increase in efficiency because the high-order method can attain better results than the low order method on a coarse mesh. We further extended the approach to advection-diffusion equation, and appropriate boundary conditions have obtained a consistent design accuracy of third and fifth order. We implemented the WENO approach to advection-diffusion equation by using the split hyperbolic method to demonstrate the advantage of non-oscillatory schemes to capture sharp gradients in boundary layer type problems without spurious oscillations. Through detailed analysis, it is shown that the WENO scheme can also provide a robust approach in capturing sharp gradients in strongly anisotropic diffusion problems on a non-magnetic field aligned mesh. The objective is to incorporate the advantages of hyperbolic approach, higher order schemes and WENO methodology to develop a robust and efficient method that can also be extended to other equations like incompressible and compressible Navier - Stokes, non-neutral plasma simulations, multi-phase flows, etc. in future. 

The rest of the paper is organized as follows. Hyperbolic approach and upwind formulation of electron fluid equations, diffusion equation, and advection-diffusion equation are described in Section \ref{sec-2}. The cell-centered explicit and compact upwind schemes along with WENO and the implementation of boundary conditions are presented in Section \ref{sec-3}. Several test cases in one and two-dimensional problems for diffusion equation and advection-diffusion equation are presented in Section \ref{sec-4}. These numerical experiments validate our numerical schemes and corroborate the high-order accuracy, oscillation-free performance and implementation of boundary conditions. Finally, the simulations magnetized electron fluids are discussed in Section \ref{sec-4} and Section \ref{sec-5} summarizes our conclusions and provide suggestions for future work.

									\section{Governing equations}\label{sec-2}
\subsection{Magnetized electron fluid equations}
The system of equations, electron mass and momentum equations, for the magnetized electron fluid in quasi-neutral flow described by Kawashima et al. \cite{Kawashima2015} are given by
   \begin{equation}
      \nabla \cdot \left(n_{\rm e} \vec{u}_{\rm e}\right) = n_{\rm e}\nu_{\rm {ion}},   
      \label{eq:mass}
   \end{equation}
     \begin{equation}
     n_{\rm e}\left[\mu \right]\nabla \phi-\left[\mu\right]\nabla\left(n_{\rm e}T_{\rm e}\right)  =n_{\rm e}\vec{u}_{\rm e}, 
     \label{eq:momentum}
   \end{equation}
where  \(n_{\rm e}\), \(\vec{u}_{\rm e}\), \(\phi\), \(\nu_{\rm {ion}}\), and \(T_{\rm e}\), are the electron number density, electron velocity, space potential, {\color{black} ionization collision frequency, and electron temperature respectively}. The electron mobility tensor $\left[\mu\right]$ can be expressed as follows, 
   \begin{equation}
   \begin{aligned}
      \left[\mu\right]=\left[
   \begin{array}{cccc}
   \mu_x  & \mu_{\rm c}    \\
   \mu_{\rm c}  & \mu_y
   \end{array}
   \right]&=\Theta^{-1}\left[
   \begin{array}{cccc}
   \mu_{||}  & \    \\
   \  & \mu_{\perp}
   \end{array}
   \right]\Theta,\hspace{20pt}
   \Theta=\left[
   \begin{array}{cccc}
    \cos\theta  & -\sin\theta \\
    \sin\theta & \cos\theta
   \end{array}
   \right] \\ 
   \\
    \mu_{||}&= \frac{e}{m_{\rm e}\nu_{\rm {col}}}, \quad \mu_{\perp}=\frac{\mu_{||}}{1+\left(\mu_{||}B\right)^2}
       \end{aligned}
          \label{eqn:mobility}
   \end{equation} 
{\color{black}where \(e\), $B$, and \(\nu_{\rm {col}}\) are the elemental charge, magnetic flux density, and electron-neutral total collision frequency, respectively. Also, \(\Theta\) is the rotation matrix, and \(\theta\) is the angle between the magnetic lines of force and the grid, shown in Fig. \ref{fig:theta}. The electron flux in parallel (\(||\)) and perpendicular (\(\perp\)) directions of the magnetic lines of force are described by using the electron mobility \(\mu_{||}\) and \(\mu_{\perp}\) respectively. 
    \begin{figure}[H]
\centering
{\includegraphics[width=0.40\textwidth]{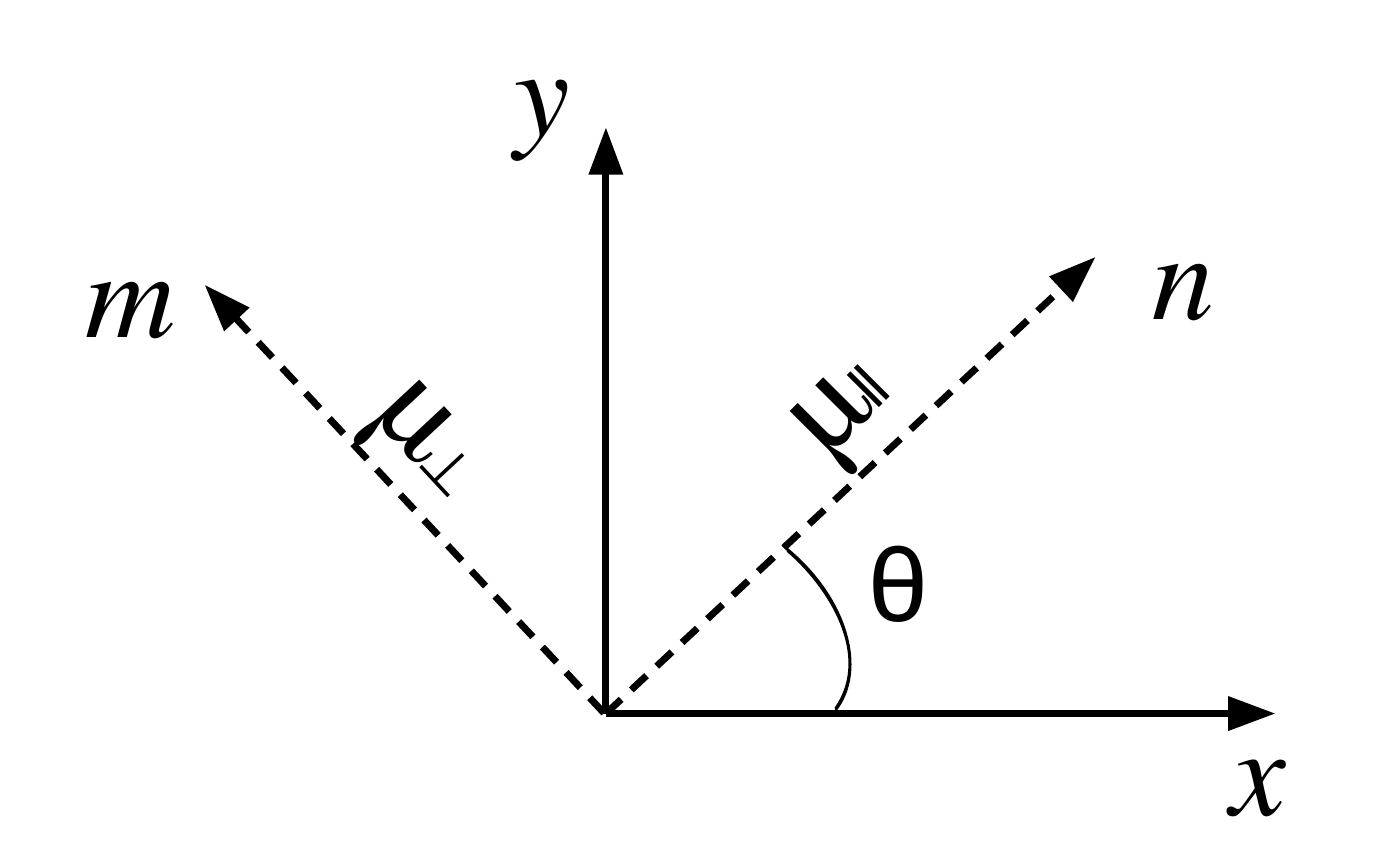}}
\caption{{\color{black}Illustration of symbols, \(\mu_{||}\), \(\mu_{\perp}\) and $\theta$.}}
\label{fig:theta}
\end{figure}}
{\color{black}For steady state condition, the following plasma diffusion equation is obtained by substituting momentum equation \ref{eq:momentum} in to continuity equation \ref{eq:mass}
   \begin{equation}
      \nabla\cdot\left(n_{\rm e}\left[\mu \right]\nabla \phi-\left[\mu\right]\nabla\left(n_{\rm e}T_{\rm e}\right)\right)=n_{\rm e}\nu_{\rm {ion}}.
     	\label{eq:diffusion-cross}
   \end{equation}
   For a given electron number density and temperature distribution, this equation can then be solved for plasma potential, $\phi$, using a general Poisson's equation solver \cite{komurasaki1995two,koo2006modeling}. Equation (\ref{eq:diffusion-cross}) is usually solved by a central scheme \cite{dragnea2017development,geng2013} and it is a well known fact that the central scheme can lead to numerical oscillations if sharp gradients are present in the flow. This approach also suffers from poor iterative convergence due to the large disparity between \(\mu_{\rm ||}\) and \(\mu_{\rm \perp}\). To avoid these difficulties Kawashima et al. \cite{Kawashima2015} have proposed hyperbolic approach by introducing pseudo-time terms in the continuity and momentum equations. This approach is similar to the artificial compressibility method utilized in solving incompressible Navier-Stokes \cite{chorin1967numerical}, where a pseudo-time derivative of the pressure is added to the continuity equation. Similarly, a pseudo-time derivative of plasma potential is added in the continuity equation and the corresponding equations are as follows,
   \begin{equation}
   \begin{aligned}
      \frac{n_{\rm e}}{T_{\rm e}}\frac{\partial \phi}{\partial t}-\nabla \cdot \left(n_{\rm e} \vec{u}_{\rm e}\right)&=-n_{\rm e}\nu_{\rm {ion}},\\
      \frac{1}{\nu_{\rm {col}}}
      \frac{\partial}{\partial t}\left(n_{\rm e}\vec{u}_{\rm e}\right)
      - n_{\rm e}\left[\mu \right]\nabla \phi+\left[\mu\right]\nabla\left(n_{\rm e}T_{\rm e}\right)&=-n_{\rm e}\vec{u}_{\rm e}.
   	\label{eq:mass-momentum}
	\end{aligned}
   \end{equation}}


For the simulations, the equations are expressed in non-dimensional form. The non-dimensionalization procedure of the equations \ref{eq:mass-momentum} and the definitions of dimensionless quantities are same as that of Ref. \cite{Kawashima2015} which are as follows:
   \begin{equation}
      \tilde{n}_{\rm e}=\frac{n_{\rm e}}{n_{\rm e}^*},\hspace{20pt}
      \tilde{T}_{\rm e}=\frac{T_{\rm e}}{T_{\rm e}^*},\hspace{20pt}
      \tilde{\phi}=\frac{\phi}{\phi_{\rm a}},\hspace{20pt}
      \left(\tilde{x}, \tilde{y}\right)^T=\frac{1}{\lambda_{\rm m}^*}\left(x, y\right)^T,
   \end{equation}
   \begin{equation}
      \tilde{t}=\frac{1}{\lambda_{\rm m}^*}\sqrt{\frac{2eT_{\rm e}^*}{m_{\rm e}}}t,\hspace{20pt}
      \tilde{\vec{u}}_{\rm e}=\frac{\vec{u}_{\rm e}}{c_{\rm s}^*}=\frac{\vec{u}_{\rm e}}{\sqrt{\frac{\gamma eT_{\rm e}^*}{m_{\rm e}}}}, \hspace{20pt} \tilde{\nu}_{\rm {col}}=\tau_{\rm m}^*\nu_{\rm {col}}, \hspace{10pt} \tilde{\nu}_{\rm {ion}}=\tau_{\rm m}^*\nu_{\rm {ion}}.
   \end{equation}
     
\noindent where \(\tau_{\rm m}\), \(v_{\rm {e,th}}\), \(c_{\rm s}\) and $\gamma$ are the mean free time, electron thermal velocity, electron acoustic velocity, {\color{red}and specific heat ratio} respectively.  Also, the electron number density \(n_{\rm e}^*\), electron temperature \({T_{\rm e}^*}\), plasma potential ${\phi_a}$, and mean free path \(\lambda_{\rm m}^*\) are the reference values. For simplified analysis the following values are assumed for all the test calculations,
   \begin{equation}
      \tilde{n}_{\rm e}=1,\hspace{20pt}\tilde{T}_{\rm e}=1,\hspace{20pt}\tilde{\nu}_{{\rm col}}=1,\hspace{20pt}\tilde{\nu}_{\rm {ion}}=0.
   \end{equation}
 Finally, the simplified nondimensional system of electron fluids for the Eqs. (\ref{eq:mass-momentum}) can be expressed as follows:  
   \begin{equation}      \label{vector:non}
   \begin{aligned}
\frac{\partial \tilde{\phi}}{\partial \tilde{t}}-\sqrt{\frac{\gamma}{2}}\tilde{\nabla} \cdot \vec{\tilde{u}}_{\rm e}&=0, \\ 
   \frac{\partial \vec{\tilde{u}}_{\rm e}}{\partial \tilde{t}}
   - \frac{1}{\sqrt{2\gamma}}\left[\tilde{\mu} \right]\tilde{\nabla} \tilde{\phi}&=-\vec{\tilde{u}}_{\rm e}.
      \end{aligned}
   \end{equation}

%
%
%
\subsubsection{Preconditioned system}
Due to the significant discrepancy between parallel, \(\mu_{||}= 10^3\), and orthogonal, \(\mu_{\perp}=1\), mobilities, the condition number of the system increases significantly and degrades the convergence performance. Convergence can be made independent of mobility by altering the eigenvalues of the system such that all of them are of the same order. A popular approach is to multiply the system of equations by a preconditioning matrix to normalize the eigenvalues \cite{turkel1993}. The system can be written in the preconditioned form as,
 \begin{equation}
           \mathbf{\hat P^{-1}}\frac{\mathbf{\partial Q}}{\partial \tilde{t}}+\frac{\mathbf{\partial E_x}}{\partial \tilde{x}}+\frac{\mathbf{\partial E_y}}{\partial \tilde{y}}=S,
      \label{vector:non_precon}
   \end{equation} 
  \noindent where $\mathbf{\hat P}$  = $\Theta^{-1}P \Theta$ is a preconditioning matrix which is derived based on the idea of electron mobility tensor rotation 
   \begin{equation}
      P=\left(
         \begin{array}{ccc}
         \sqrt{\frac{2}{\gamma}} & 0 & 0   \\
         0 & \frac{\sqrt{2\gamma}}{\tilde{\mu}_{||}} & 0 \\
         0 & 0 & \frac{\sqrt{2\gamma}}{\tilde{\mu}_{\perp}}
         \end{array}
      \right),\hspace{20pt}
      \Theta=\left(
   		\begin{array}{ccc}
   		1 & 0 & 0   \\
   		0 & \cos\theta & \sin\theta \\
   		0 & -\sin\theta & \cos\theta
   		\end{array}
   	\right).
      \label{eqn:precon_rotation}
   \end{equation}
  \noindent Finally, the preconditioned system of equations can be rewritten as follows:
   \begin{equation}\label{final_preconeqn}
   \begin{aligned}
      \frac{\partial \tilde{\phi}}{\partial \tilde{t}}- \nabla \cdot {\vec{\tilde u}_{\rm e}} &=0,\\
\frac{\partial {\vec{\tilde u}_{\rm e}}}{\partial \tilde{t}}- \nabla{ \tilde{\phi}} &=-[g] {\vec{\tilde u}_{\rm e}},
   \end{aligned}
   \end{equation}
where, $\mathbf{[g]}$ is a tensor which is written as follows:
   \begin{equation} 
	\mathbf{[g]}=\frac{\sqrt{2\gamma}}{{\tilde{\mu}_{||}}}\left[
   	\begin{array}{cc}
   	   	\cos^2\theta+ \frac{{\tilde{\mu}_{||}}}{{\tilde{\mu}_{\perp}}} \sin^2\theta &
     (\frac{{\tilde{\mu}_{||}}}{{\tilde{\mu}_{\perp}}} -1)\cos\theta\sin\theta\\
		
     (\frac{{\tilde{\mu}_{||}}}{{\tilde{\mu}_{\perp}}} -1)\cos\theta\sin\theta&		
   	   	\sin^2\theta+ \frac{{\tilde{\mu}_{||}}}{{\tilde{\mu}_{\perp}}} \cos^2\theta
   	\end{array}
   	\right].
      \label{cha3-eq:source_conservative}
   \end{equation}
 The flux Jacobian matrices and the corresponding eigenvalues for the preconditioned system of equations can be expressed as follows,
   \begin{equation}
      {\mathbf {\hat {PJ}} }_{x}=\left(
         \begin{array}{ccc}
         0 &-1 & 0   \\
         -1 & 0 & 0 \\
         0 & 0 & 0
         \end{array}
         \right),\hspace{20pt}
         {\mathbf {\hat {PJ}}}_{y}=\left(
         \begin{array}{ccc}
         0 & 0 &-1   \\
         0 & 0 & 0 \\
         -1 & 0 & 0
         \end{array}
         \right), \quad \lambda =\pm 1,0 .
   	\label{eqn:Jacobian-precon}
   \end{equation}
Therefore, the Jacobian matrices are not affected by the magnetic field as they are included in the source terms and also the eigenvalues are significantly simplified which improves the convergence speed. The construction of upwind fluxes will be explained in the next subsection along with the diffusion equation.
															\subsection{Diffusion equation and construction of hyperbolic scheme}
 In this subsection, the hyperbolic approach for diffusion equation, first described by Nishikawa \cite{Nishikawa2007}, is briefly explained here. Consider the following diffusion equation in two dimensions,
\begin{equation}\label{diffusion}
\begin{aligned}
\frac{\partial u}{\partial t} - \nu \left(\frac{\partial ^2 u}{\partial x^2} + \frac{\partial ^2 u}{\partial y^2}\right) &= -S ,
\end{aligned}
\end{equation}
\noindent where $\nu$ is the diffusion coefficient and S is the source term. In the original idea of Nishikawa \cite{Nishikawa2007}, new variables representing the gradients of the primary variable are introduced, and the original diffusion equation has been converted into a system of three coupled first-order equations. By defining new variables $p$ = $\frac{\partial u}{\partial x}$ and $q$ = $\frac{\partial u}{\partial y}$, {\color{black} and introducing pseudo-time terms} we can obtain the following first order hyperbolic system:

\begin{equation}
\begin{aligned}
{ \color{black}\frac{\partial u}{\partial \tau}} - \nu \frac{\partial p}{\partial x} - \nu \frac{\partial q}{\partial y} & = -S ,\\
{ \color{black}\frac{\partial p}{\partial \tau}} - \frac{\partial u}{\partial x} & = -p ,\\
{ \color{black}\frac{\partial q}{\partial \tau}} - \frac{\partial u}{\partial y} & = -q.
\label{eqn:hyper-diff}
\end{aligned}
\end{equation}

{\color{black}It is important to state the fact that the first-order system reduces to the diffusion equation at the steady state, i.e. the pseudo-time terms will be zero at steady state,
\begin{equation}
\begin{aligned}
&\begin{cases}
\begin{aligned}
\cancelto{}{\frac{\partial u}{\partial \tau}}  - \nu \frac{\partial p}{\partial x} - \nu \frac{\partial q}{\partial y} & = -S ,\\
\cancelto{}{\frac{\partial p}{\partial \tau}}  - \frac{\partial u}{\partial x} & = -p ,\\
\cancelto{}{\frac{\partial q}{\partial \tau}}  - \frac{\partial u}{\partial y} & = -q.
\end{aligned}
\end{cases}\rightarrow
&\begin{cases}
\begin{aligned}
&0 =  \nu \left(\frac{\partial p}{\partial x} + \frac{\partial q}{\partial y}\right) -S ,\\
&\frac{\partial u}{\partial x}=p,\\
&\frac{\partial u}{\partial y}=q.
\end{aligned}
\end{cases}\rightarrow
0 = \nu \left(\frac{\partial ^2 u}{\partial x^2} + \frac{\partial ^2 u}{\partial y^2}\right) -S.
\end{aligned}
\end{equation}}
\noindent The Eqs. (\ref{eqn:hyper-diff}) can be represented in vector form as,
 \begin{equation}\label{eqn:vector-form}
{ \color{black}\frac{\mathbf {\partial Q}}{\partial {\tau}}}+\frac{\mathbf{\partial E_x}}{\partial {x}}+\frac{\mathbf{\partial E_y}}{\partial {y}}=\mathbf{S},
\end{equation}
\noindent where the conservative variables, fluxes in x and y-direction and source terms are,
\begin{equation}\label{eq:diff-fluxes}
\mathbf{Q}= \left[ \begin{array}{ccc} u  \\ p \\ q \end{array} \right], 
\quad
\mathbf{E_x} = \left[ \begin{array}{ccc} -\nu p  \\  -u \\ 0 \end{array} \right], 
\quad 
\mathbf{E_y} = \left[ \begin{array}{ccc}-\nu q  \\  0 \\ -u \end{array} \right],
\quad 
\mathbf{S} = \left[ \begin{array}{ccc} -S  \\  -p \\ -q \end{array} \right],
\end{equation}
 respectively. The Jacobian matrices, $\mathbf{J_x}$ and $\mathbf{J_y}$, in the x- and y-directions and the corresponding eigenvalues, $\lambda$, are given by
 \begin{equation}\label{eqn:jacobians-hyper}
{\bf J}_x = \nu \left( \begin{array}{ccc}
0 & -1 &0 \\
-1  & 0 &0\\
0 & 0&0\\
\end{array} \right) , \quad {\bf J}_y = \nu \left( \begin{array}{ccc}
0 & 0 &-1 \\
0  & 0 &0\\
-1 & 0&0\\
\end{array} \right) , \quad \lambda =\pm \sqrt \nu,0 .
\end{equation}
We can notice that the above Jacobian matrices are the same as that of the hyperbolic form of magnetized electron equations derived in the earlier section, Eq. (\ref{eqn:Jacobian-precon}). The upwind fluxes are the same for both set of equations, and by changing the source terms and by using appropriate boundary conditions the magnetized electron fluid equations and diffusion can be simulated by the same code. In steady state, the diffusion equation will reduce to Laplace's or Poisson's equation. By the replacing the primary variable $u$ with $\phi$, space potential, the hyperbolic formulation can be used to model many physics problems that are time independent. By using the right, $\mathbf{R}$, and left, $\mathbf{L}$, eigenvectors of Jacobian matrices, and the diagonal eigenvalue matrix, $\mathbf{|\Lambda|}$, the absolute Jacobian or the Roe matrix in x-direction, $\mathbf{|J_{dx}|}$, can be written as,
\begin{equation}\label{eqn:absol}
|{\bf J}_{dx}|= \bf R_x|\Lambda|L_x = \sqrt \nu \left( \begin{array}{ccc}
1 & 0 &0 \\
0  & 1 & 0\\
0 & 0  & 0\\
\end{array} \right).
\end{equation}
Finally, the upwind flux in the x-direction can be expressed as Eq. (\ref{eqn:upwind}),
\begin{equation}\label{eqn:upwind}
\begin{aligned}
\hat E_{i+\frac{1}{2},j} &= \frac{1}{2}(\mathbf{E_L} + \mathbf{E_R}) - \frac{1}{2} |\mathbf {J_{dx}}|(\mathbf{Q_R}-\mathbf{Q_L})\\
 &=  \frac{1}{2} \left( \begin{array}{ccc}
-(p_R+p_L) \\
-(u_R+u_L)\\
0\\
\end{array} \right) - \frac{\sqrt\nu}{2} \left( \begin{array}{ccc}
1 & 0 &0 \\
0  & 1 & 0\\
0 & 0  & 0\\
\end{array} \right)  \left( \begin{array}{ccc}
u_R-u_L \\
p_R-p_L\\
q_R-q_L\\
\end{array} \right) .
\end{aligned}
\end{equation}
where the left and right fluxes, $\mathbf{E_L}$ and $\mathbf{E_R}$, are defined at the cell interfaces which are evaluated by the interpolation polynomials that are discussed later. Similarly, the upwind flux for the magnetized electrons will be constructed by following the standard procedure of local preconditioning method \cite{nishikawa2003general}, i.e., the preconditioned Jacobian is multiplied by $\mathbf P^{-1}$ to cancel the effect of $\mathbf P$,
\begin{equation}
\hat E_{i+\frac{1}{2},j} = \frac{1}{2}(\mathbf{E_L} + \mathbf{E_R}) - \frac{1}{2} \mathbf{\hat P^{-1}}|\mathbf{\hat P}\mathbf {\hat J_{x}}|(\mathbf{Q_R}-\mathbf{Q_L}).
\label{eqn:upwind-mag}
\end{equation}

\subsubsection{Homogeneity and flux vector splitting}

Interestingly, the hyperbolic form of the diffusion equation \ref{eqn:hyper-diff} satisfy the homogeneity property,
\begin{equation}
\mathbf{E} (Q) = \frac{\partial {\mathbf{E}}}{\partial {\mathbf{Q}}} Q = J({\mathbf{Q}})Q.
\end{equation}
The proof of this property is easy to notice. By multiplying the Jacobian matrices shown in Eq. (\ref{eqn:jacobians-hyper}) by the vector Q we can reproduce the flux vector E(Q). This remarkable property of the hyperbolic form can be used to solve the diffusion equation by employing Flux vector splitting schemes. For example, in the original work of Kawashima et al. \cite{Kawashima2015} the upwind fluxes are computed by using Steger-Warming \cite{Steger1981} flux vector splitting scheme, shown in Eq. (\ref{eqn:flux})
\begin{equation}\label{eqn:flux}
\begin{aligned}
\mathbf{E} &= \mathbf{E^+} + \mathbf{E^-} ,
\end{aligned}
\end{equation}
with,
 \begin{equation}
 \begin{aligned} 
\mathbf{E^+}&= {\bf J}^+Q=(R\Lambda^+ L) Q, \\ 
\mathbf{E^-}&= {\bf J}^-Q=(R\Lambda^- L) Q.
\end{aligned}
\end{equation}
where ${\bf J}^+$ and ${\bf J}^-$ in x-direction are expressed as follows,
 \begin{equation}
{\bf J}^+_x = \left( \begin{array}{ccc}
\frac{1}{2} & \frac{1}{2} &0 \\
\frac{1}{2}  & \frac{1}{2} &0\\
0 & 0&0\\
\end{array} \right) , \quad {\bf J}^-_x = \left( \begin{array}{ccc}
\frac{-1}{2} & \frac{1}{2} &0 \\
\frac{1}{2}  & \frac{-1}{2} &0\\
0 & 0&0\\
\end{array} \right).
\end{equation}
%
%
and $\Lambda^+$ is the diagonal matrix of positive eigenvalues, and $\Lambda^-$ is the diagonal matrix of negative eigenvalues. The Jacobian matrices satisfy, ${\bf J}^+ + {\bf J}^- = {\bf J} $ and the reconstruction for $\mathbf{E^+}$ uses a biased stencil with one more point to the left, and that for $\mathbf{E^-}$ uses a biased stencil with one more point to the right, to obey correct upwinding. In the current upwind formulation for diffusion equation, there is no difference between flux vector splitting and Roe solver.  For all the simulations in this paper, Roe fluxes are used unless otherwise stated.
										\subsection{Advection-Diffusion equation and construction of hyperbolic scheme}
\noindent In this section, we consider the advection-diffusion equation, 
 \begin{equation}
   \frac{\partial  u}{\partial t} +    a\frac{\partial  u}{\partial x} + b\frac{\partial  u}{\partial y} = \nu  (\frac{\partial ^2 u}{\partial x^2} +\frac{\partial ^2 u}{\partial y^2}),
   \label{eq:advec}
   \end{equation}
where where $u$ is the solution variable, (a, b) is a constant advection vector, and $\nu$ is a constant positive diffusion coefficient. Similar to the diffusion equation new variables $p$ = $\frac{\partial u}{\partial x}$ and $q$ = $\frac{\partial u}{\partial y}$ are defined {\color {black}and pseudo-time derivatives} are introduced to obtain the following first order hyperbolic system:
\begin{equation}\label{eqn:advec-hyp}
\begin{aligned}
{\color {black}\frac{\partial u}{\partial \tau}} +    a\frac{\partial  u}{\partial x} + b\frac{\partial  u}{\partial y} &= \nu (\frac{\partial p}{\partial x} +  \frac{\partial q}{\partial y}) ,\\
{\color {black}\frac{\partial p}{\partial \tau}} - \frac{\partial u}{\partial x} & = -p ,\\
{\color {black}\frac{\partial q}{\partial \tau}} - \frac{\partial u}{\partial y} & = -q.
\end{aligned}
\end{equation}
 The Eqs. (\ref{eqn:advec-hyp}) can be represented in vector form as,
 \begin{equation}\label{eqn:vector-form-ad}
{\color {black} \frac{\bf \partial Q}{\partial {\tau}}}+\frac{\bf \partial E_x}{\partial {x}}+\frac{\bf \partial E_y}{\partial {y}}=\bf S,
\end{equation}
where the conservative variables, fluxes in x and y-direction and source terms are,
\begin{equation}\label{eq:split}
\bf Q= \left[ \begin{array}{ccc} u  \\ p \\ q \end{array} \right], 
 \bf E_x = \left[ \begin{array}{ccc} au-\nu p  \\  -u \\ 0 \end{array} \right],  
\bf E_y = \left[ \begin{array}{ccc} bu-\nu q  \\  0 \\ -u \end{array} \right],
\quad 
\bf S = \left[ \begin{array}{ccc} 0  \\  -p \\ -q \end{array} \right],
\end{equation}
respectively. The split hyperbolic formulation introduced by Nishikawa in Ref. \cite{Nishikawa2014} is considered in this paper. The advection, $E^a_x$, and diffusion, $E^d_x$, fluxes in x-direction are separated as 
\begin{equation}\label{eq:splitting}
\bf E_x = \bf E^a_x + E^d_x = \left[ \begin{array}{ccc} au  \\  0 \\ 0 \end{array} \right] + \left[ \begin{array}{ccc} -\nu p  \\  -u \\ 0 \end{array} \right].
\end{equation}
The absolute Jacobian matrix for the diffusive fluxes would be the same as in Eq. (\ref{eqn:absol}) and the corresponding absolute Jacobian matrix for the advection flux is straightforward
\begin{equation}\label{eq:split-jac}
|\bf J_{ax}|= \left[ \begin{array}{ccc} |a| &0&0 \\ 0&0&0 \\ 0&0&0 \end{array} \right].
\end{equation}
and finally, the upwind flux can be constructed as, 
\begin{equation}\label{eqn:upwind-split}
\hat E_{i+\frac{1}{2},j} = \frac{1}{2}(\mathbf{E_L} + \mathbf{E_R}) - \frac{1}{2} |\mathbf {J_{dx}}+\mathbf {J_{ax}}|(\mathbf{Q_R}-\mathbf{Q_L}).
\end{equation}
The upwind fluxes can also be constructed by the unified advection-diffusion approach proposed by Nishikawa in Ref. \cite{Nishikawa2010b} and the numerical schemes presented here can be used in a straight forward manner and is currently beyond the scope of the current paper.

										\section{ Numerical method}\label{sec-3}
\begin{figure}[H]
\centering
 \subfigure{\includegraphics[width = 4in]{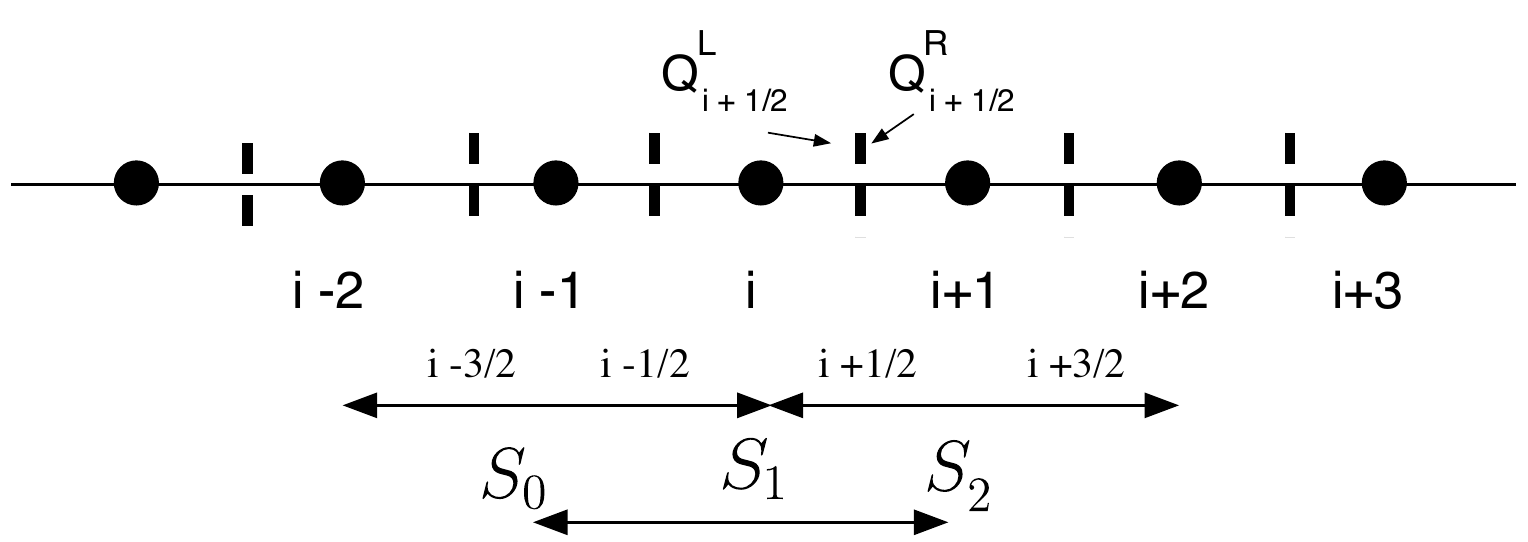}}     
  \caption{Left and right states at the cell interfaces and interpolation stencils for WENO.}
  \label{fig:WENO}
\end{figure}

In this section, the procedures of WENO and linear upwind schemes are briefly explained.  From the Eq. (\ref{eqn:upwind-split}) we can see that the computation of $E_{i+\frac{1}{2}}$ requires knowledge of the values  $q_R$ and $q_L$ at the cell interface. The values of $q_R$ and $q_L$ can be obtained by upwind interpolation to the same order of accuracy.  The ${Q}_{i+\frac{1}{2}}^{(L)}$ should be biased to left and similarly ${Q}_{i+\frac{1}{2}}^{(R)}$ has to be biased to the right for correct upwinding, shown in Fig. \ref{fig:WENO}. 

%
\subsection{Weighted Essentially Non-Oscillatory Schemes}
In the WENO scheme, the fifth order upwind-biased interpolation is nonlinearly weighted from three different third order interpolations on sub-stencils, $S_0$, $S_1$ and $S_2$, shown in Fig. \ref{fig:WENO}. For simplicity, the interpolation polynomials to the left side of the cell interface at $x_{i+\frac{1}{2}}$ are only presented here. The three third order interpolation formula of variable $Q$ is given by
\begin{equation}
\begin{aligned} \label{eq:upwind_biased_stencils}
{\bar Q}_{i+\frac{1}{2}}^{(0)} &= \frac{1}{6}\left(2Q_{i-2} - 7Q_{i-1} + 11Q_{i} \right)\\
{\bar Q}_{i+\frac{1}{2}}^{(1)} &= \frac{1}{6}\left(-Q_{i-1} + 5Q_{i} + 2Q_{i+1} \right) \\
{\bar Q}_{i+\frac{1}{2}}^{(2)} &= \frac{1}{6}\left(2Q_{i} + 5Q_{i+1} - Q_{i+2} \right)
\end{aligned}
\end{equation}
where ${\bar Q}_{i+\frac{1}{2}}^{(k)}$ are approximated values at cell interfaces from different sub-stencils and $Q_i$ are the values at cell centers. In WENO literature, the variable $Q$ can either be fluxes \cite{Zhang2008, Liu2015}, conservative variables \cite{Nonomura2013}, primitive variables or characteristic variables \cite{johnsen2006implementation}. In this current work the conservative variables $(Q)$, which are also the fluxes $(E)$, are directly interpolated. The three third order upwind approximation polynomials in Eqn \ref{eq:upwind_biased_stencils}, are chosen dynamically by a nonlinear convex combination which adapts either to a higher order approximation in smooth regions of the solution, or to a lower-order spatial discretization that avoids interpolation across discontinuities and provides the necessary numerical dissipation for shock capturing. The fifth order WENO scheme can be expressed as
\begin{equation}
{\bar Q}_{i+\frac{1}{2}} = \sum\limits_{k=0}^{2} \omega_k {Q}_{i+\frac{1}{2}}^{(k)},
\end{equation}
where $\omega_k$ are the nonlinear weights which are given by,
\begin{equation} \label{eq:JS5_nonlinear_weights}
	\omega_k = \frac{\alpha_k}{\sum\limits_{k=0}^{2}\alpha_k}, \:
    \alpha_k = \frac{\gamma_k}{\left(\beta_k + \epsilon \right)^m}, \: k = 0, 1, 2,
\end{equation}
where $m$, $\gamma_k$ and $\beta_k$ are a positive integer, ideal linear weights, and smoothness indicators, respectively. $\epsilon=10^{-6}$ is a small constant to prevent division by zero. The non-linear weights of the convex combination are based on local smoothness indicators $\beta_k$, which measure the sum of the normalized squares of the scaled $L^2$ norms of all derivatives of the lower order polynomials. The basic weighting strategy is to assign small weights to those lower order polynomials whose underlying stencils contain discontinuities so that an essentially non-oscillatory solution is obtained. The traditional smoothness indicators for fifth order upwind interpolation are given by Jiang, and Shu \cite{jiang1995} denoted as WENO-JS.
\begin{equation} \label{eq:old_beta}
\beta_i = \sum_{l=1}^{k} \Delta x^{2l-1} \int_{x_{i-\frac 12}}^{x_{i+\frac 12}} \left( \frac{d^l}{dx^l} p_j(x) \right)^2 \, dx
\end{equation}
where \(k\) is the polynomial degree of \(p_j(x)\). Evaluating of each \(k\), we obtain the following equations
\begin{equation}
\begin{aligned} \label{eq:smoothness}
\beta_0 &= \frac{1}{4} \left(  Q_{i-2} - 4Q_{i-1} +3  Q_{i} \right)^2 + \frac{13}{12} \left(  Q_{i-2} - 2Q_{i-1} +  Q_{i} \right)^2\\
\beta_1 &= \frac{1}{4} \left(  Q_{i-1} - Q_{i+1} \right)^2  + \frac{13}{12} \left( Q_{i-1} - 2 \bar Q_{i} +  Q_{i+1} \right)^2 \\
\beta_2 &= \frac{1}{4} \left( 3Q_{i} - 4Q_{i+1} + Q_{i+2} \right)^2+ \frac{13}{12} \left( Q_{i} - 2  Q_{i+1} + Q_{i+2} \right)^2
\end{aligned}
\end{equation}
Borges et al. \cite{Borges2008} proposed a new approach, denoted as WENO-5Z in this paper, for the nonlinear weights obtained by WENO-JS as they are known to lose accuracy at critical points and are also excessively dissipative in smooth regions. The improved non-linear weights are as follows:
\begin{align}\label{eqn:om_z}
\omega^z_k=\dfrac{\alpha^z_k}{\sum_{k=0}^2\alpha^z_k},\quad \alpha^z_{k}=\gamma_k\left(1+\left(\dfrac{\tau_5}{\epsilon+\beta_k}\right)^p\right),
\end{align}
where the smoothness indicators $\beta_k$'s are the same as those given in Eqs. (\ref{eq:smoothness}), $\epsilon=10^{-40}$, and $\tau_5$ is the smoothness indicator of the large stencil given by,
 \begin{align}\label{tau_5}
\tau_5=|\beta_0-\beta_2|= \frac{13}{12} (Q_i^{''} Q_i^{'''} \Delta x^5) +  O(\Delta x^6)
\end{align}
The variable $p$ is used to tune the dispersive and dissipative properties of the scheme. It is reported by Borges et al. \cite{Borges2008} that the scheme becomes more dissipative when $p$ is increased. In this paper, \(p=2\) is employed in all test problems for diffusion and advection-diffusion equations and \(p=1\) for magnetized electron fluids. Through spectral and approximate dispersion relation (ADR) analysis, Jia et al. \cite{jia2015spectral} found that the anti-dissipation of the WENO-5Z scheme is less than that of the WENO-JS scheme. They also demonstrated that the WENO-5Z scheme is not only less dissipative and dispersive but also relatively more accurate and safer. These properties of the scheme were found to be useful for magnetized electron fluid simulations discussed later.

															\subsection{Linear and compact interpolation}
For a smooth function, the fifth order WENO scheme is theoretically equivalent to the optimal fifth order linear upwind scheme, denoted as U-5E in this paper, that is, the nonlinear weights, $\omega_k$, are equal to the ideal linear weights, $\gamma_k$. The fifth order interpolation formulas for left and right interfaces are given by,
\begin{equation}
{\bar Q}_{i+\frac{1}{2}} = \sum\limits_{k=0}^{2} \gamma_k { Q}_{i+\frac{1}{2}}^{(k)}, \: \gamma_0 = \frac{1}{10}, \:
\gamma_1 = \frac{6}{10}, \:
\gamma_2 = \frac{3}{10}  
\end{equation}

\begin{equation}
\begin{aligned} 
{\bar Q}_{i+\frac{1}{2}}^{(L)} &= \frac{1}{60} \left(2Q_{i-2} -13Q_{i-1} +47Q_{i}+27Q_{i+1}-3Q_{i+2}  \right), \\
{\bar Q}_{i+\frac{1}{2}}^{(R)} &= \frac{1}{60} \left(-3Q_{i-1} +27Q_{i} +47Q_{i+1} -13Q_{i+2} + 2Q_{i+3} \right).
\end{aligned}
\end{equation}

\noindent Similarly, third order extrapolation formulas, same as that of in Eq. (\ref{eq:upwind_biased_stencils}), denoted as U-3E in this paper, are given by,
\begin{equation}
\begin{aligned} 
{\bar Q}_{i+\frac{1}{2}}^{(L)} &= \frac{1}{6}\left(-Q_{i-1} + 5Q_{i} + 2Q_{i+1} \right) \\
{\bar Q}_{i+\frac{1}{2}}^{(R)} &= \frac{1}{6}\left(2Q_{i} + 5Q_{i+1} -Q_{i+2} \right).
\end{aligned}
\label{eqn:3linear}
\end{equation}
Finally, we have also implemented the compact upwind interpolation, denoted as U-5C in this paper, given by equations (\ref{eqn:upwind-compact}). Compact schemes are a family of interpolation schemes which are implicit in space and therefore requires an inversion of a tridiagonal matrix. They are characterized by high spectral resolution and have significantly lower dispersion errors compared to that of non-compact schemes. Implementation of the boundary conditions is same as that of the non-compact schemes.
\begin{subequations}
     \begin{alignat}{1}
\frac{1}{2} \bar Q^{L}_{i-\frac{1}{2}}+ \bar Q^{L}_{i+\frac{1}{2}} + \frac{1}{6}\bar  Q^{L}_{i+\frac{3}{2}}&= \frac{1}{18}  Q_{i-1}+  \frac{19}{18} Q_{i} + \frac{5}{9}  Q_{i+1} \label{eq:left}\\
\frac{1}{6} \bar Q^{R}_{i-\frac{1}{2}}+ \bar Q^{R}_{i+\frac{1}{2}} + \frac{1}{2} \bar Q^{R}_{i+\frac{3}{2}}&= \frac{5}{9}  Q_{i}+ \frac{19}{18} Q_{i+1} + \frac{1}{18}  Q_{i+2} \label{eq:CR}
     \end{alignat}
     \label{eqn:upwind-compact}
   \end{subequations}
\noindent Finally, the derivative $(\partial E/\partial x)$ is computed as follows for all the schemes:
\begin{equation}
  \Bigl(\frac{\partial E}{\partial x}\Bigr)_{i} =  \frac{1}{\Delta x} \Bigl[\hat{E}_{i + \frac{1}{2}} - \hat{E}_{i - \frac{1}{2}} \Bigr] \label{eqn:e2c}
\end{equation}   
In general, the integrals of the fluxes are discretized using a high-order Gaussian quadrature with suitable Gaussian integration points over the faces of the control volume\cite{Titarev2004, coralic2014finite} to achieve higher order accuracy in, third order or more, multidimensional finite volume method. In the present hyperbolic approach, the fluxes are no longer non-linear and are same as that of the conservative variables. {\color{black} Due to this, finite volume method can also obtain higher order accuracy by using point values and reconstruction of the fluxes at Gaussian quadrature points is not necessary. Therefore, we cannot distinguish the difference between cell-centered finite volume and finite difference schemes for the flux computation. In fact, we also computed the fluxes by WCNS in \cite{chamarthi} and the results are identical. In this paper, we focus only on the standard WENO schemes and not a comparison of WENO and WCNS.} The implementation of interpolation polynomials beyond $5^{th}$ order are straightforward and are not presented in this paper.

{\color{black}\subsection{Total variation diminishing schemes} Several TVD schemes are proposed in the literature for shock-capturing and preventing Gibbs oscillations. Two of the popular approaches are considered here. First one is the van Leer's standard TVD-MUSCL \cite{van1977towards} reconstruction for the left and right interfaces, ${\bar{Q}^L}_{i+\frac{1}{2}}$ and ${\bar{Q}^R}_{i+\frac{1}{2}}$, are given by, 
\begin{equation}\label{eq:muscl} 
\begin{aligned}
\bar{Q}^L_{i+\frac{1}{2}}&=Q_i+\frac{1}{4} \left( (1-\kappa) \phi(r^L) (Q_{i}-Q_{i-1}) + (1+\kappa)\phi\left(\frac{1}{r^L}\right) (Q_{i+1}-Q_{i})\right), \\
\bar{Q}^R_{i+\frac{1}{2}}&=Q_{i+1}-\frac{1}{4} \left( (1-\kappa) \phi(r^R) (Q_{i+2}-Q_{i+1}) + (1+\kappa)\phi\left(\frac{1}{r^R}\right) (Q_{i+1}-Q_{i})\right),
\end{aligned}
\end{equation}  
\noindent where $\kappa$ is a free parameter which is set to 1/3 for the third-order limiter and 
\begin{equation}\label{eq:ratio-2} 
\begin{aligned}
		&r^L_i= \frac{Q_{i+1}-Q_{i}}{Q_{i}-Q_{i-1}+\epsilon}, \\
		&r^R_i= \frac{Q_{i+1}-Q_{i}}{Q_{i+2}-Q_{i+1}+\epsilon},
\end{aligned}
\end{equation}
\noindent where $\epsilon$ is a small value $10^{-16}$. In this paper we considered minmod and van Leer limiters which are given by

\begin{equation}\label{eq:limiters} 
\begin{aligned}
&\phi_{minmod}= min(r,1),\\
		&\phi_{van Leer }= \frac{2r}{1+r}.
\end{aligned}
\end{equation}

Second approach is that of Cockburn and Shu \cite{cockburn1989tvb} which can be considered as Generalized MUSCL and is popular in discontinuous Galerkin methods. For $k=\frac{1}{3}$, Eq. (\ref{eq:muscl}) will reduce to third order linear interpolation formula, Eq. (\ref{eqn:3linear}), if limiters are not considered. Now we define,
\begin{equation}
\tilde{Q}_{i} = {Q}_{i+1/2}^{-} - {Q}_{i}, \quad  \tilde{Q}_{i+1} = {Q}_{i+1/2}^{+} - {Q}_{i+1}.
\end{equation}
Then, by using the definition of minmod limiter we can obtained a generalized limiter for 3 points
\begin{equation}\label{eq:minmodgeneral}
\tilde{Q}_{i}^{mod} =  \mbox{minmod}\left(\tilde{Q}_{i}, {{ Q}_i-{ Q}_{i-1}}, {{ Q}_{i+1}-{ Q}_{i}} \right),
\end{equation}
where the minmod function defined as 
\begin{equation}
\mbox{minmod}(a,b,c) = \left\{ \begin{array}{cl}
     sign(a)\mbox{min($|a|,|b|,|c|$)}  & \mbox{if $sign(a)=sign(b)=sign(c)$} \\
     0  & \mbox{otherwise}
                  \end{array}
                  \right. 
\end{equation}
}

									\subsection{Implementation of boundary conditions}
In this section, the implementation of boundary conditions for the all the equations is described. Two different approaches are employed for the numerical boundary conditions, weak boundary condition, and ghost-cells. In the hyperbolic approach the Neumann boundary conditions are also implemented as Dirichlet boundary condition through the gradient variables, and therefore we describe only Dirichlet boundary conditions. 
\subsubsection{Weak boundary formulation}
\begin{figure}[H]\label{fig:inflow}  
\centering
        		\includegraphics[width=0.6\textwidth]{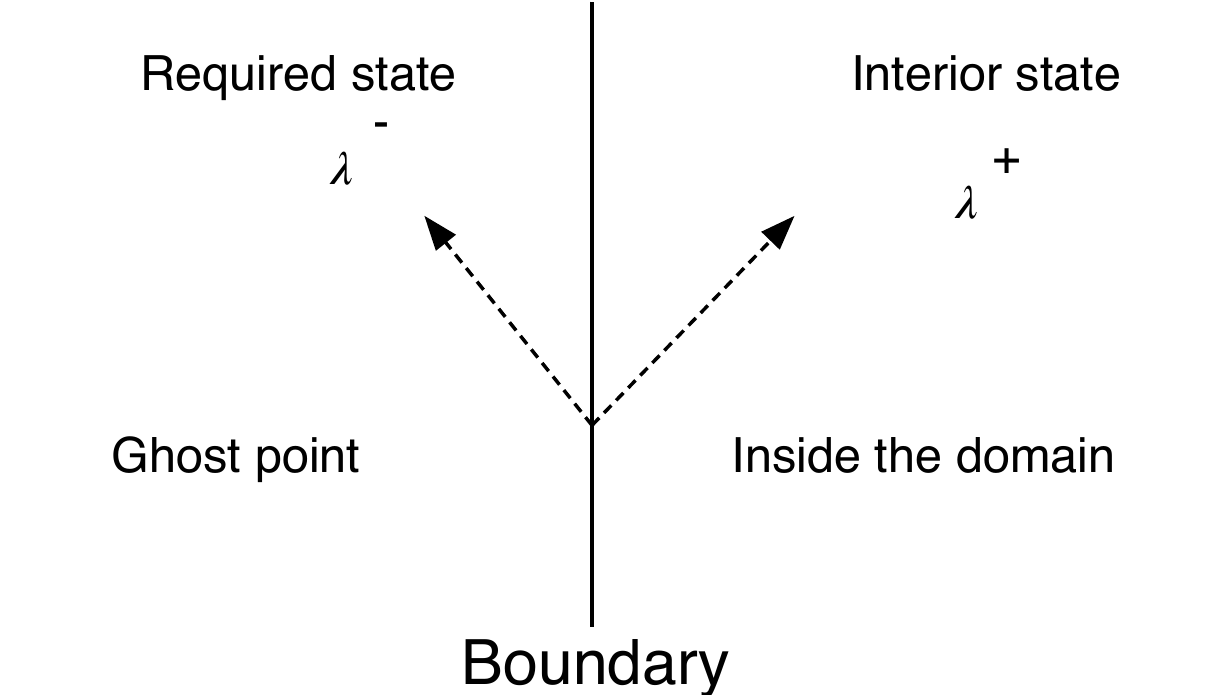}
        		\caption{Characteristic waves at the boundary for hyperbolic approach.}      
\end{figure}

The Dirichlet boundary condition $u$ = $u_{0}$ at, say, $x = 0$ can be implemented through the numerical flux by weak boundary condition as described by Nishikawa and Roe \cite{Nishikawa2016}. For a one-dimensional problem, the upwind flux for a is given by,
\begin{equation}
\begin{aligned}
E_{i+\frac{1}{2},j} &= \frac{1}{2}(\mathbf{E_L} + \mathbf{E_R}) - \frac{1}{2} |\mathbf {J_{dx}}|(\mathbf{Q_R}-\mathbf{Q_L}) \\
 &=  \frac{1}{2} \left( \begin{array}{ccc}
-(p_R+p_L) \\
-(u_R+u_L)\\
\end{array} \right) -\frac{1}{2} \left( \begin{array}{ccc}
1 & 0 \\
0  & 1\\
\end{array} \right)  \left( \begin{array}{ccc}
u_R-u_L \\
p_R-p_L\\
\end{array} \right) .
\end{aligned}
\end{equation}
\begin{equation}
\begin{aligned} 
E_{x=0} = E_\frac{1}{2} = \frac{1}{2}(E_L + E_R) - \frac{1}{2} |J_x|(Q_R-Q_L),
\end{aligned}
\end{equation}
\noindent where ($u_R$, $p_{R}$) are given by a higher order interpolation from the interior of the domain, and the left state ($u_L$, $p_{L}$) is specified by the boundary condition:
\begin{equation}
 (u_L, p_{L}) =  (u_0, p_{R}).
\end{equation}
Since the value of gradient variable is not known we set $p_{L}$ = $p_{R}$. This approach is consistent with the characteristic condition at $x=0$. We can specify only one condition in hyperbolic approach only one wave enters the domain as shown in Fig. \ref{fig:inflow}. For a Neumann boundary condition, say $p=p_{0}$, the value of $p$ is specified instead of $u$:
\begin{equation}
{\color{black} (u_L, p_{L}) =  (u_R, p_{0}).}
\end{equation}

\subsubsection{Lagrange type extrapolation for ghost cells}
In this approach, an additional "ghost cells" are introduced by extending the physical domain. Unlike the weak boundary condition, the Dirichlet boundary condition is employed at the cell interface as shown in {\color{black}Fig. \ref{ghost-2}}. The number of ghost-cells depends on the interior interpolation scheme and cell-interface to cell-center Lagrange extrapolation formula given by Eq. (\ref{lagrange}) is used to compute the values in ghost cells:
\begin{figure}[H]
\centering
        		\includegraphics[width=0.7\textwidth]{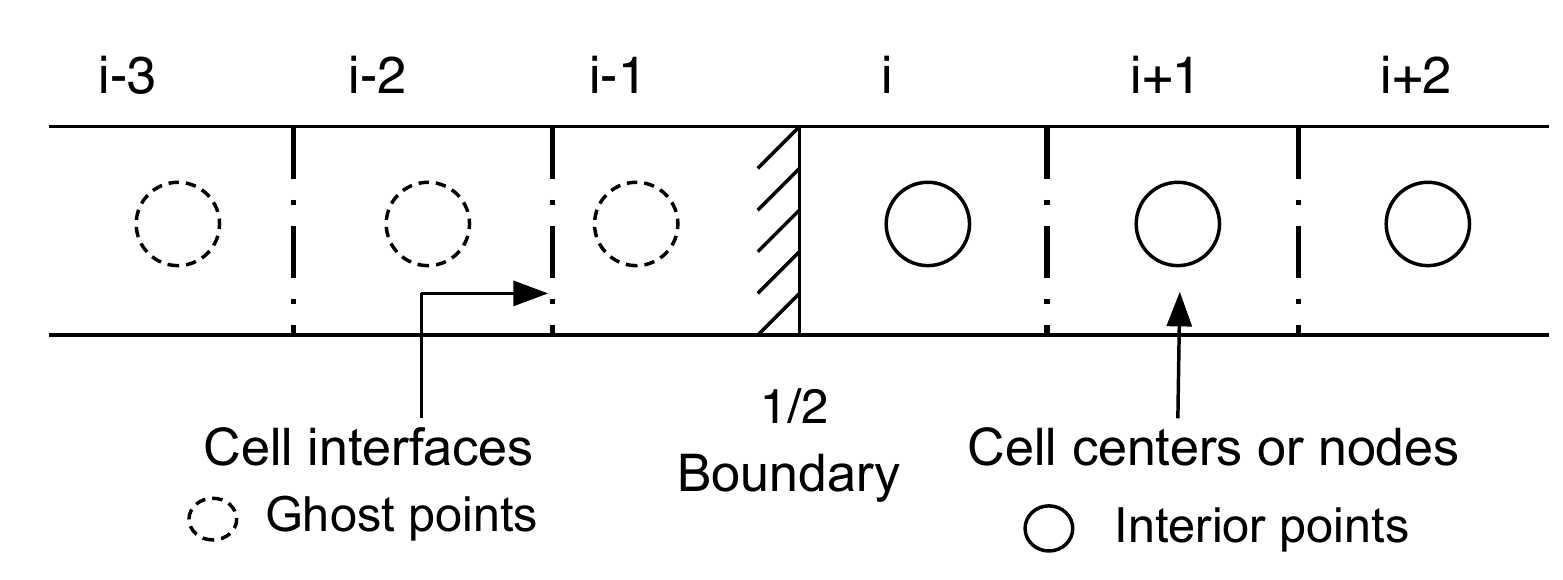}
        		\caption{Ghost cell approach showing Dirichlet boundary condition for $\phi$.}    
		\label{ghost-2}    
\end{figure}
\begin{equation}\label{lagrange}
u^{(r)}_{\frac{1}{2}}(x)=\sum _{j=0}^{k} u_{i-r+j}C_{rj}(x),
\end{equation}
where r is the order of the Lagrange polynomial and the constants $C_{rj}$ are obtained by,
\begin{equation}
C_{rj}(x)=\prod^{k} _{\begin{smallmatrix} l= 0\\l\neq j\end{smallmatrix}}{\frac {x-x_{i-r+l}}{x_{i-r+j}-x_{i-r+l}}}.
\end{equation}
%
\noindent For Dirichlet boundary condition, where the value of $u$ is prescribed at the cell interface, say $u_{\frac{1}{2}}$, one can extrapolate the ghost cell values by using third order accurate polynomial given by Eq. (\ref{ghost_3}).
\begin{equation}
\label{ghost_3}
u_0 = -2u_1 + \frac{1}{3}u_2 + \frac{8}{3} u_\frac{1}{2}.
\end{equation}
 For the gradient variable, $p$, the values in ghost cells are extrapolated by using the interior points by extrapolation, cell-center to cell-center \cite{Tan2010,morinishi1998fully}. Such extrapolation is also consistent with the outgoing characteristics at the boundary as shown in Fig. \ref{fig:inflow}. For example, At the boundary, say $x=0$, the values of ghost cell $p_{0}$ can be approximated by 3rd order extrapolation given by Eq. (\ref{extra}),  
\begin{equation}\label{extra}
p_0 = 3p_{1} - 3p_{2} +  p_{3}.
\end{equation} 
Lagrange extrapolation can also be approximated by a (s-k)th order Taylor expansion
\begin{equation}
u_j = \sum_{k=o}^{s-1}  \frac{(x_j -1)^k}{k!} u_R^{(k)},
\end{equation}
where $u_R^{(k)}$ is a (s-k)th order approximation of $\frac{\partial^k u}{\partial x^k}$ at x = 1. If $u(x)$ is smooth near the boundary, $u_R^{(k)}$ can be obtained by 
\begin{equation}
\begin{aligned} \label{eq:lag}
u_R^{(k)} & =  \frac{d^k P_{s-1}(x)}{d x^k}
\end{aligned}
\end{equation}
\noindent where $P_{s-1}(x)$ is a Lagrange polynomial of degree at most $s-1$.
\subsubsection{WENO type extrapolation for ghost points}
In situations where there are sharp gradients close to the boundary, the Lagrange extrapolation may result in severe oscillations. To overcome such difficulties a more robust WENO type extrapolation is also proposed by Tan and Shu \cite{Tan2010, Tan2011}. Analogous to the idea of WENO, three polynomials are constructed for the extrapolation given by,
\begin{equation}
\begin{aligned}
P_0(x)&= u_0,\\
P_1(x)&= \frac{({u_1}-{u_0})}{{\Delta x}}x+{u_0},\\
P_2(x)&={u_0}+\frac{-3 {u_0}+4 {u_1}-{u_2}}{2{\Delta x}}x+\frac{{u_0}-2 {u_1}+{u_2}}{2 {\Delta x}^2} x^2.
\end{aligned}
\end{equation}
Using the standard WENO procedure, the equation \ref{eq:lag} can be rewritten as,
\begin{equation}
u^k= \sum_{r=0}^{2} \omega_r \frac {d^k p_r(x)}{dx^k }, \quad {at\ x=\frac{\Delta x}{2}}
\end{equation} 
%
\noindent where $\omega_r$ are the typical nonlinear weights given by, $\omega_r = \frac {\alpha_r} {\sum_{r=0}^{2} \alpha_s}$ and $\alpha_r = \frac {d_r} {(\epsilon + \beta_r)^3}$, where the linear weights $d_r$ are chosen as 
\begin{equation}
d_0=\Delta x^2, d_1=\Delta x, d_2 = 1-\Delta x -\Delta x^2,
\end{equation}
\noindent and the smoothness indicators $\beta_r$ are obtained by,
\begin{equation}
\beta_j = \sum_{l=1}^{2} \Delta x^{2l-1} \int_{- \Delta x}^{0} \left( \frac{d^l}{dx^l} p_j(x) \right)^2 \, dx.
\end{equation} 
\noindent Readers can refer \cite{Tan2011} for the explicit expressions of smoothness indicators.  In this paper $3^{rd}$ order WENO extrapolation is considered for all the simulations.

\subsection{Time discretization}
After discretizing the spatial derivative, the set of ordinary differential equations obtained are,
\begin{equation}
\mathbf{Q_t} = \mathbf{Res(Q)}, 
\end{equation}
where the operator $Res(Q)=E'_x$ and  $E'_x$ is approximated by linear upwind interpolations or WENO interpolation. For time integration the following third order TVD Runge-Kutta method \cite{jiang1995} is used
\begin{eqnarray}  \label{rk}
\mathbf{Q^{(1)}} & = & \mathbf{Q^n} + \Delta t Res(\mathbf{Q^n}),  \nonumber \\
\mathbf{Q^{(2)}} & = & \frac{3}{4} \mathbf{Q^n} + \frac{1}{4} \mathbf{Q^{(1)}} + \frac{1}{4}\Delta t Res(\mathbf{Q^{(1)}}) ,\\
\mathbf{Q^{n+1}} & = & \frac{1}{3} \mathbf{Q^n} + \frac{2}{3} \mathbf{Q^{(2)}} + \frac{2}{3}\Delta t Res(\mathbf{Q^{(2)}}).  \nonumber
\end{eqnarray}
 A fourth order non-TVD Runge-Kutta scheme can also be employed for the computations. Numerical results obtained by Eq. (\ref{rk}) are only presented in this paper.

												\section{Numerical Results and Discussion}\label{sec-4}										
In this section, various test cases in one and two dimensions for diffusion equation, advection-diffusion equation, and electron fluid equations are tested by using the upwind schemes and the boundary conditions described in earlier section. All the simulations are carried out on a uniform Cartesian mesh, and TVD-RK is used for time integration for all the problems.
\subsection{Diffusion equation}
For the diffusion equation, the numerical solutions are also computed with the standard $2^{nd}$ order central scheme using {\color{black}successive over relaxation} for comparison. For the linear upwind schemes, depending on the order of the interior scheme the corresponding $r^{th}$ order extrapolation polynomial is used for the numerical boundary conditions. Only Lagrange extrapolation and weak boundary procedures are implemented for the hyperbolic form of diffusion equation as WENO extrapolation is found to be unstable.

\noindent \textbf{Example 4.1.1.}\label{sec:4.1.1} To investigate the implementation and accuracy of the numerical schemes the following one-dimensional diffusion equation including a source term is considered with the domain size of $x$ $\in$ [0,1].
	\begin{equation}
		\frac{\partial u}{\partial t}=\nu \frac{\partial^2 u}{\partial x^2}+A \cos\left(2\pi N\frac{x}{L}\right), 
		\label{app1-eq:Poissontest}
	\end{equation}
 where $\nu =1$ and Dirichlet boundary conditions, $u_{1}=2, u_{N}=1$, are considered. The exact {\color{black}steady state solution} for $u$ is given by:
	\begin{equation}
		u_e=-\left(\frac{L}{2\pi C}\right)^2A\cos\left(2\pi C\frac{x}{L}\right)+\frac{u_{N}-u_{1}}{L}x+A\left(\frac{L}{2\pi C}\right)^2+u_{1}.
	\end{equation}

\noindent where the constants and $A$ and $C$ are assumed to be 10.0 and 3.0 respectively. 

\begin{figure}[H]
\centering
\subfigure[Primary variable]{%
\includegraphics[width=0.48\textwidth]{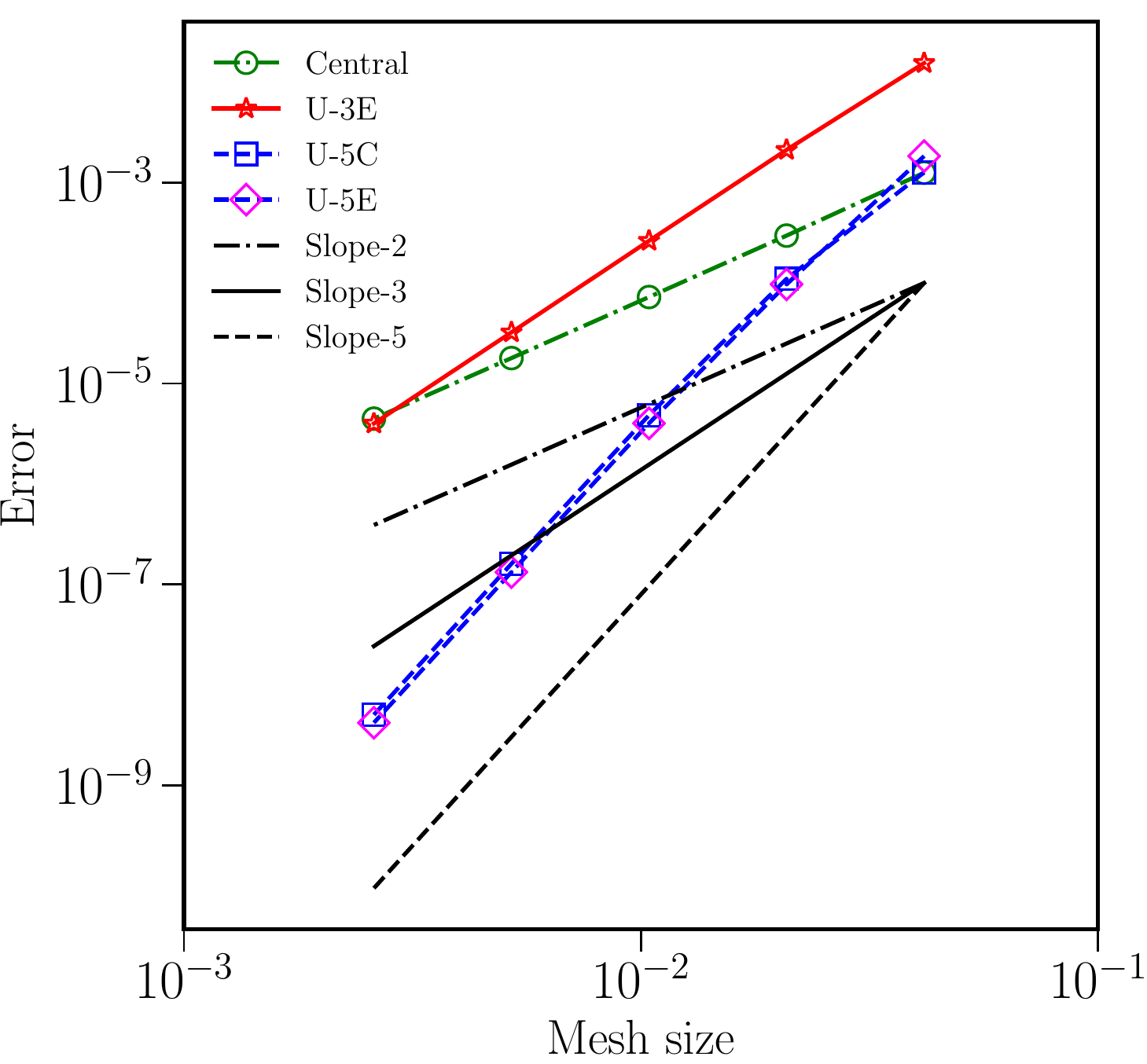}
\label{fig:primary-error}}
\subfigure[ Gradient variable]{%
\includegraphics[width=0.48\textwidth]{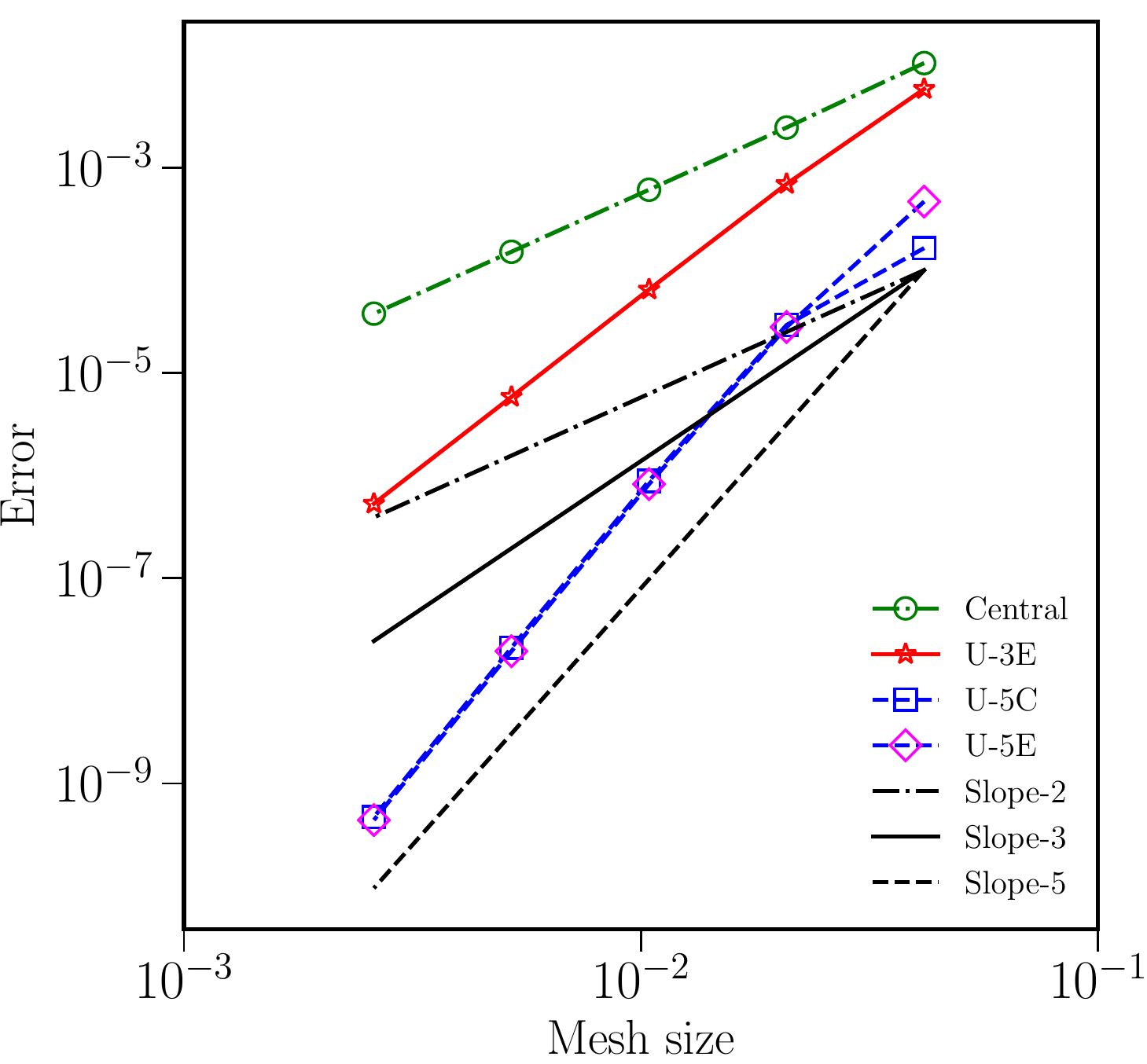}
\label{fig:gradient-error}}
\caption{{\color{black}$L_2$ convergence errors for one-dimensional test case using different schemes.}}
\end{figure}

\noindent The equivalent first order hyperbolic equation system can be written as follows:
\begin{equation}\label{eq:hyperpoi1}
\begin{aligned}
\frac {\partial u}{\partial t} - \frac {\partial p}{\partial x} =& -A cos(\frac{2 \pi Nx}{L}), \\
 \frac {\partial p}{\partial t} - \frac {\partial {\color{blue}u}}{\partial x} =& -p.
 \end{aligned}
\end{equation}
The conservative variables are ${\bf{Q}} = [\ u, \ p] $ and the flux vector is ${\mathbf{E}} = [-p, -\ u]$. The Jacobian matrix and the corresponding eigenvalues for the above system of equations is, 
\begin{equation}\label{eq:jacobian}
A(\underline{\mathbf{Q}}) = \frac{\partial {\mathbf{E}}}{\partial {\mathbf{Q}}} = 
\left[ \begin{array}{ccc}
\frac{\partial f_1}{\partial q_1} & \frac{\partial f_1}{\partial q_2} \\
\frac{\partial f_2}{\partial q_1} & \frac{\partial f_2}{\partial q_2} 
\end{array} \right] = \left[ \begin{array}{ccc}
0 & -1 \\
-1 & 0 \\
\end{array} \right] \quad; \lambda_1=-1,
\lambda_2=1.
\end{equation}
\noindent The right and left eigenvectors can be used to compute the absolute flux Jacobian
 \begin{equation}\label{eq:RAL}
 |{\bf J}_x| =R|\Lambda|L= \left[ \begin{array}{ccc}
1 & 0 \\
0  & 1 \\
\end{array} \right] .
\end{equation}
The simulations are conducted with grid refinements from $N$ = 24 to 384 by second-order central, and all the upwind schemes including TVD-MUSCL scheme. For upwind schemes, the numerical solution is computed by an explicit time-marching until the residuals are dropped below $10^{-12}$ in $L_1$ norm with a constant CFL $= 0.65$. Table \ref{table:1D_upwind-1} shows the $L_2$ error for the velocity and the order of accuracy for the U-3E, U-5E,U-5C and WENO-5Z schemes respectively. We can see that design order of accuracy is obtained for all the schemes.
%
\begin{table}[H]
  \centering
  \caption{$L_2$ errors and order of convergence of $u$, primary variable, by 3rd, 5th order explicit and compact and WENO schemes for 1D diffusion equation.}
\label{table:1D_upwind-1}
\footnotesize
    \begin{tabular}{| c | c | c | c | c | c | c | c | c|}
\hline
    Number & \multicolumn{2}{c|}{Upwind-3E} & \multicolumn{2}{c|}{Upwind-5E} & \multicolumn{2}{c|}{Upwind-5C}& \multicolumn{2}{c|}{WENO-5Z}  \\
    \cline{2-9}
    of points& error & order & error & order & error & order & error & order   \\
    \cline{1-9}
    24    & 1.54E-02 &       & 1.83E-03 &       & 1.26E-03 &       & 2.64E-03 &  \\
    \hline
    48    & 2.11E-03 & 2.87  & 9.70E-05 & 4.24  & 1.10E-04 & 3.52  & 7.70E-05 & 5.10 \\
    \hline
    96    & 2.63E-04 & 3.01  & 4.02E-06 & 4.59  & 4.84E-06 & 4.50  & 3.99E-06 & 4.27 \\
    \hline
    192   & 3.23E-05 & 3.02  & 1.33E-07 & 4.91  & 1.61E-07 & 4.91  & 1.33E-07 & 4.90 \\
    \hline
    384   & 4.00E-06 & 3.02  & 4.22E-09 & 4.98  & 5.07E-09 & 4.99  & 4.22E-09 & 4.98 \\
    \hline
    \end{tabular}%
  \label{tab:1D-poten}%
\end{table}%
Table \ref{tab:gradient} shows the $L_2$ error for the gradient variable, $p$, and the order of accuracy for the  U-3E, U-5E,U-5C and WENO-5Z schemes respectively. We can see that design order of accuracy is obtained for the gradient variable as well for all the schemes. $L_2$ error convergence results are shown in Fig. \ref{fig:primary-error} and Fig. \ref{fig:gradient-error} for the solution and the gradient variables respectively.
%
\begin{table}[H]
  \centering
  \caption{$L_2$ errors and order of convergence of gradient variable by 3rd, 5th order explicit and compact and WENO schemes one-dimensional diffusion problem, \hyperref[sec:4.1.1]{Example 4.1.1}.}
\footnotesize
    \begin{tabular}{| c | c | c | c | c | c | c | c | c|}
\hline
    Number & \multicolumn{2}{c|}{Upwind-3E} & \multicolumn{2}{c|}{Upwind-5E} & \multicolumn{2}{c|}{Upwind-5C}& \multicolumn{2}{c|}{WENO-5Z}  \\
    \cline{2-9}
     of points& error & order & error & order & error & order & error & order   \\
    \cline{1-9}
    24    & 5.82E-03 &       & 4.66E-04 &       & 1.64E-04 &       & 1.44E-03 &  \\
    \hline
    48    & 6.91E-04 & 3.07  & 2.79E-05 & 4.06  & 2.92E-05 & 2.48  & 2.73E-05 & 5.72 \\
    \hline
    96    & 6.48E-05 & 3.42  & 8.23E-07 & 5.08  & 8.86E-07 & 5.04  & 8.23E-07 & 5.05 \\
    \hline
    192   & 5.83E-06 & 3.47  & 1.94E-08 & 5.40  & 2.10E-08 & 5.40  & 1.94E-08 & 5.40 \\
    \hline
    384   & 5.31E-07 & 3.46  & 4.40E-10 & 5.47  & 4.71E-10 & 5.48  & 4.40E-10 & 5.46 \\
    \hline
    \end{tabular}%
  \label{tab:gradient}%
\end{table}%

\begin{figure}[H]
\centering
\subfigure[Primary variable]{%
\includegraphics[width=0.46\textwidth]{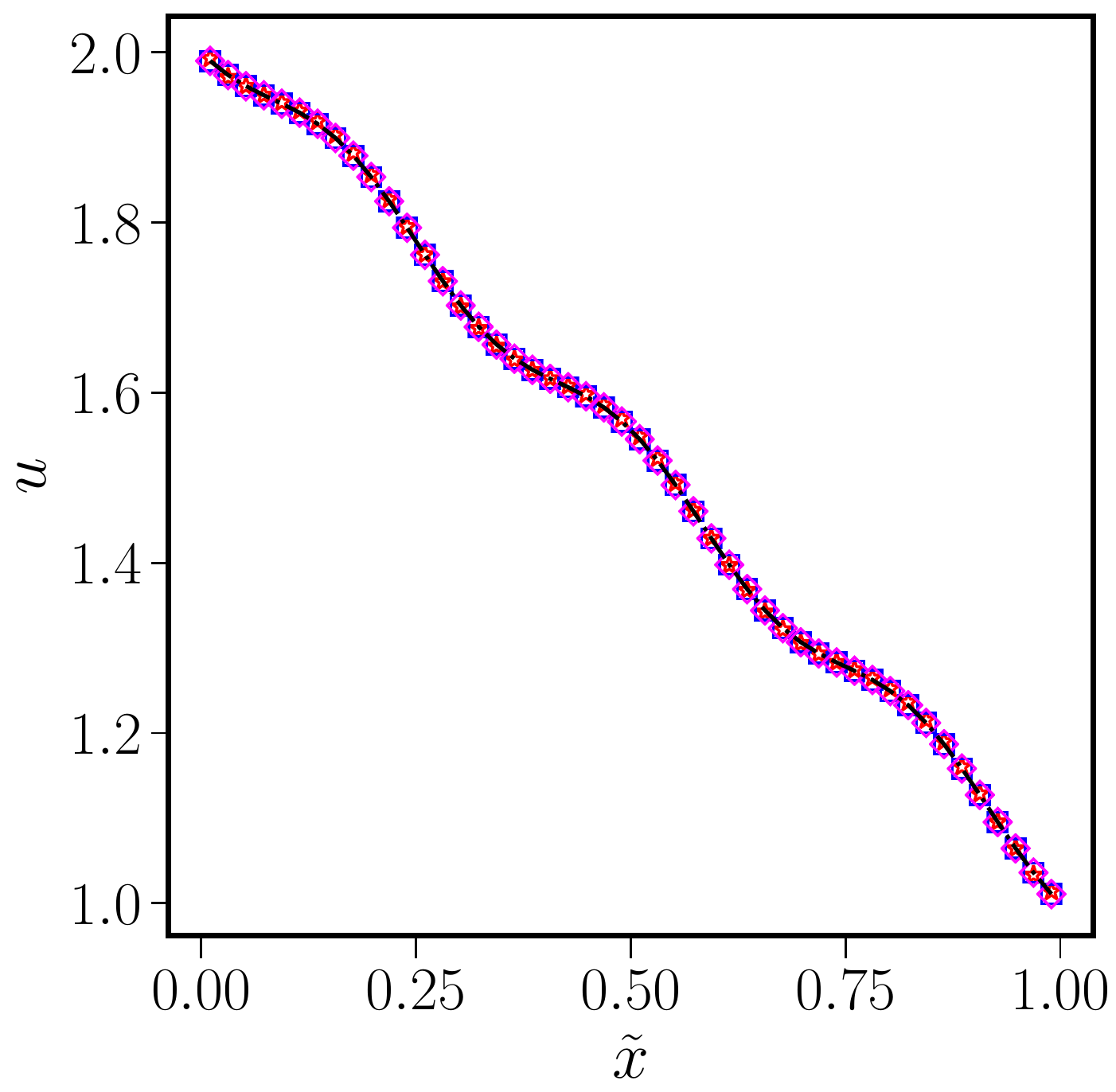}
\label{fig:primary-plot}}
\subfigure[Gradient variable]{%
\includegraphics[width=0.48\textwidth]{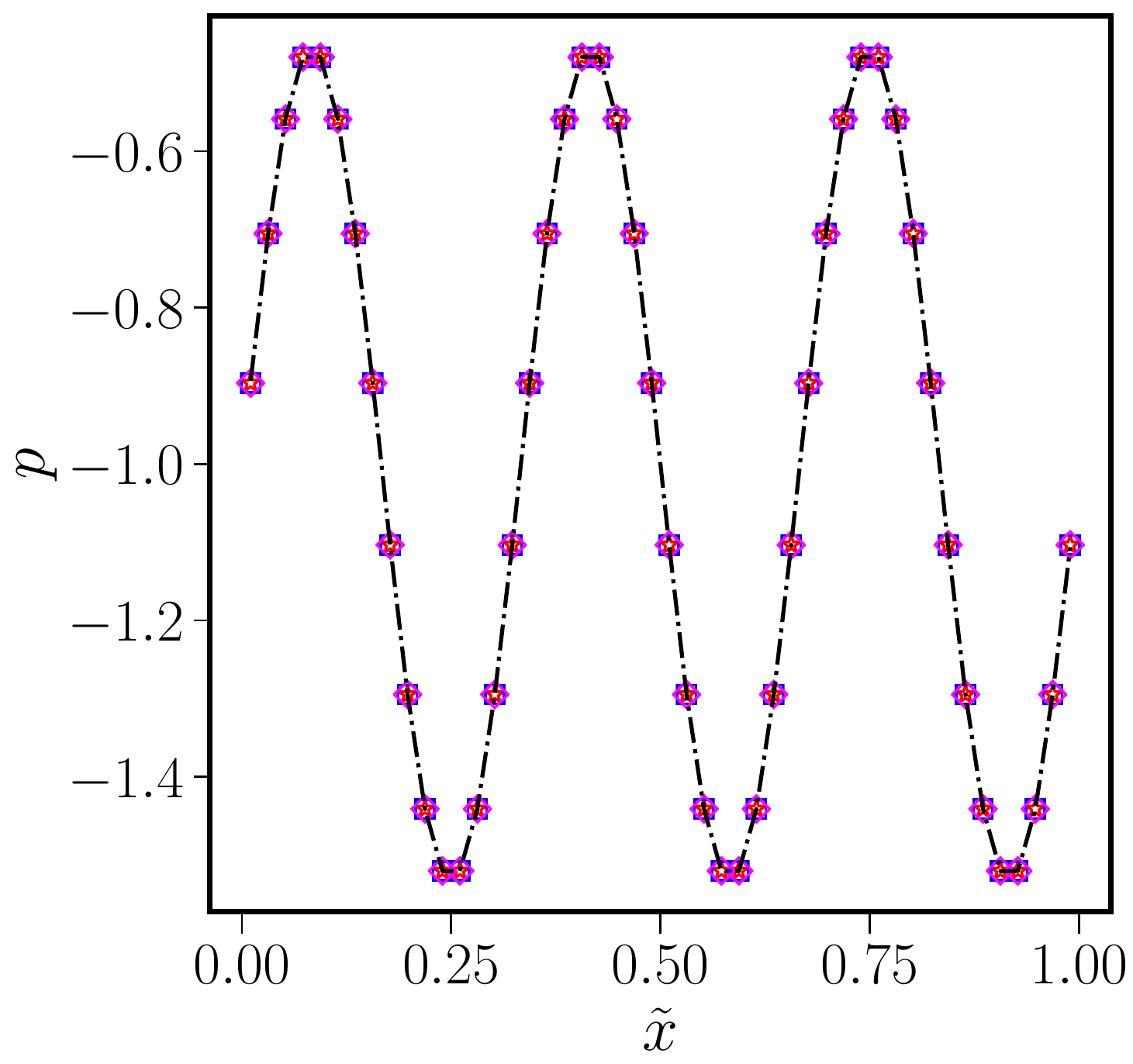}
\label{fig:gradient-plot}}
\subfigure[Effect of boundary conditions]{%
\includegraphics[width=0.46\textwidth]{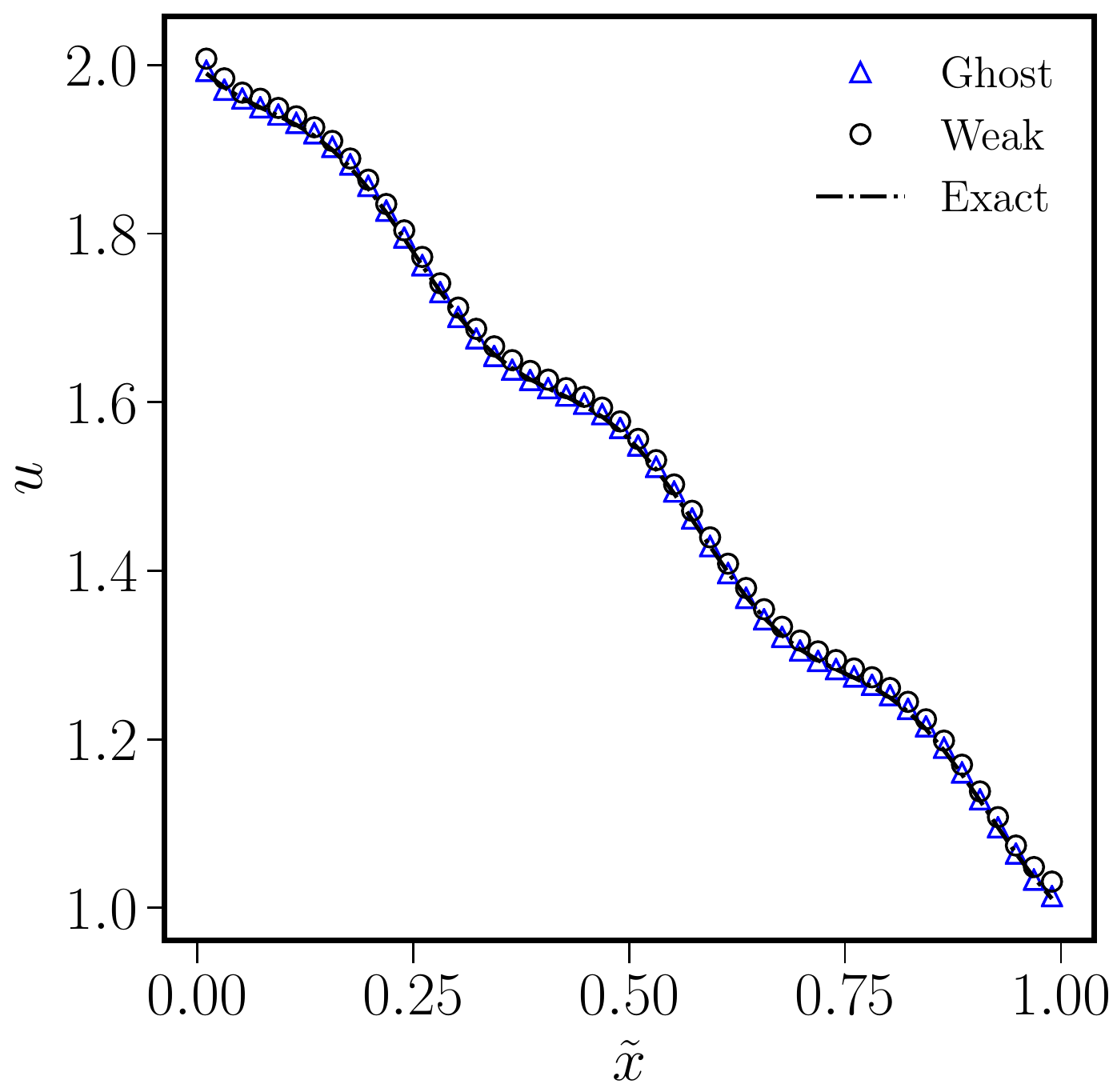}
\label{fig:bound}}
\caption{One-dimensional test case using different schemes and effect of boundary conditions.  {\color{black}In Fig. \ref{fig:primary-plot} and \ref{fig:gradient-plot}, Dashed line: analytical; red stars: U-3E; blue squares: U-5C; magenta diamonds: U-5E.} }
\label{fig:1D-case}
\end{figure}

Fig. \ref{fig:1D-case} shows the solutions contours of various schemes in comparison with the analytical solution for both primary and gradient variables. Fig.  \ref{fig:bound} shows the comparison of ghost cell approach and weak boundary implementation, and we can observe that ghost cell approach represents the solution more accurately on coarse meshes. 
Weak boundary implementation was unstable beyond $3^{rd}$ order accuracy whereas the ghost cell approach was found to be stable until 6th order accuracy. We note that the ghost cell approach may be complicated to implement on an unstructured mesh compared to the weak formulation and also the difference between these approaches is minimal on finer meshes.  Based on these observations only ghost cell approach has been employed for the all the test cases in the next subsections. WENO extrapolation is not considered for diffusion equation as it is found to be unnecessary.
%
%
%
%
%
%
%
%
%
%

{\color{black}In Fig. \ref{fig:lim_1}, we can observe the difference between the shock-capturing schemes, TVD-MUSCL and Generalized-MUSCL. TVD-MUSCL formulation has not contaminated the solution and it is reduced to the linear third order scheme, regardless of the limiter. On the other hand, the Generalized-MUSCL approach resulted in unnecessary oscillations and is only first order accurate, shown in Fig. \ref{fig:lim_acc}, and it may not be appropriate for the diffusion equation in hyperbolic form.
{\color{black}\begin{figure}[H]
\centering
\subfigure[{\color{black}Solution obtained by TVD schemes}]{\includegraphics[width=0.43\textwidth]{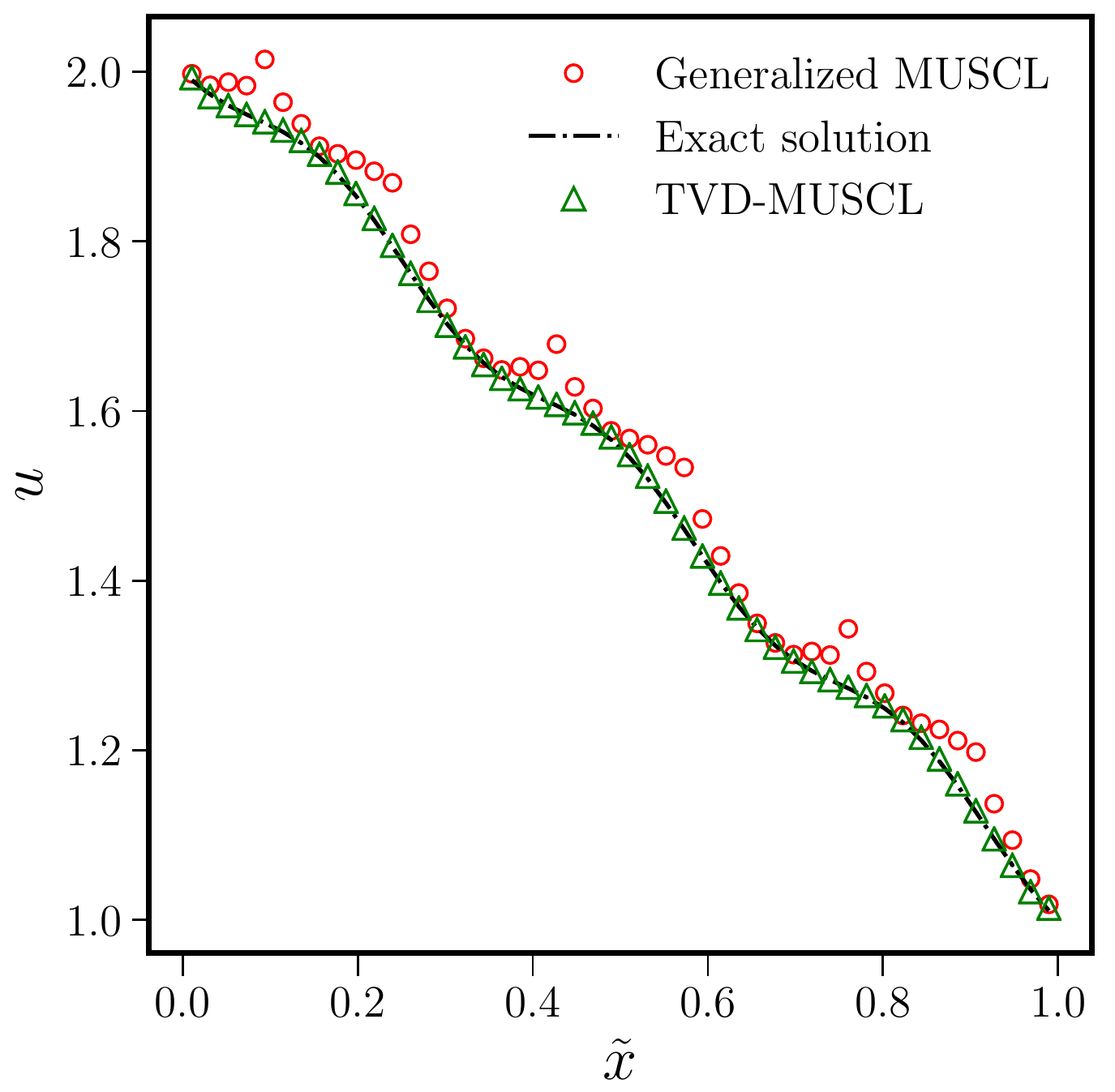}
\label{fig:lim_1}}
\subfigure[{\color{black}Accuracy of TVD schemes}]{\includegraphics[width=0.45\textwidth]{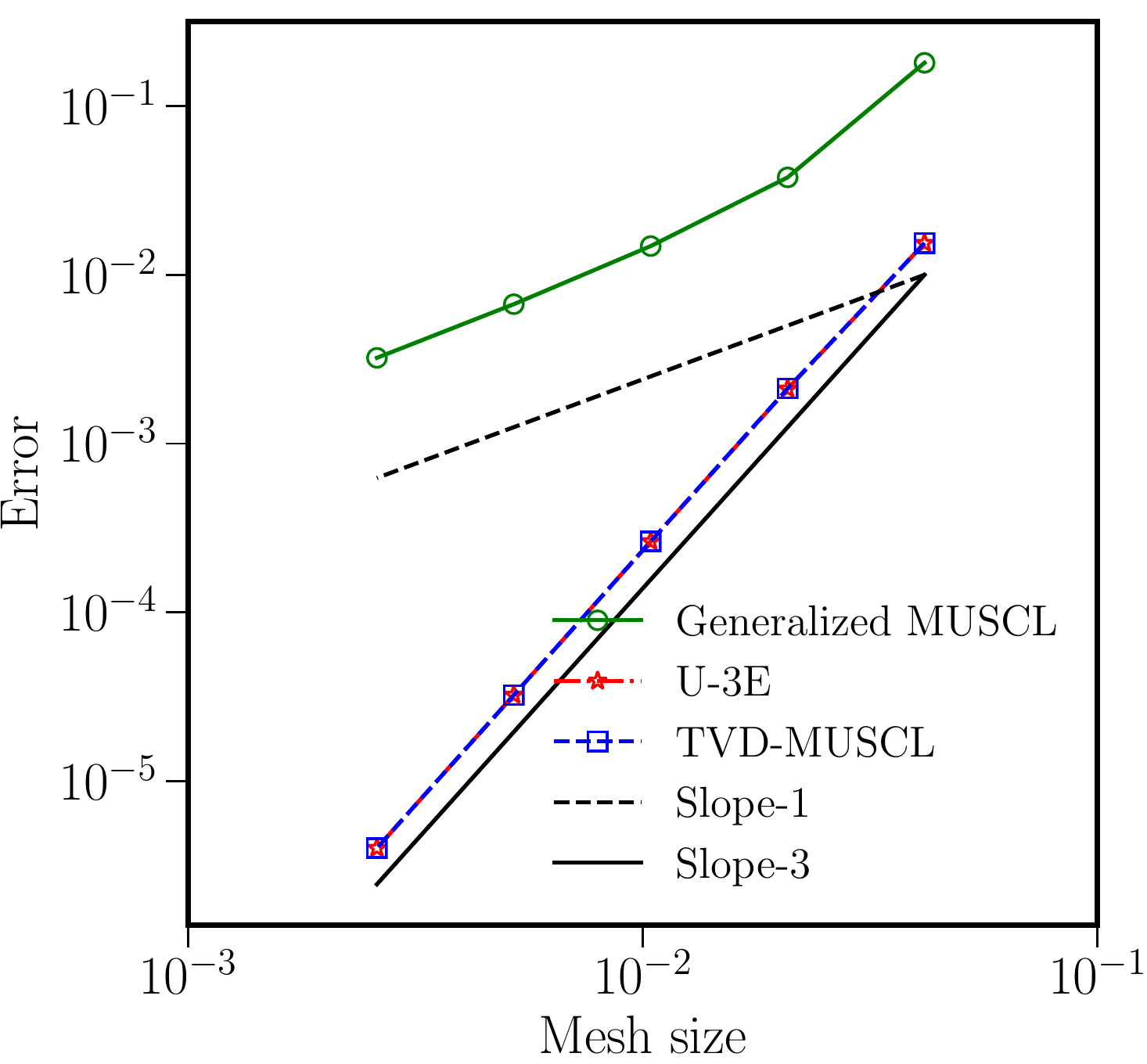}
\label{fig:lim_acc}}
  \caption{ {\color{black} \hyperref[sec:4.1.1]{Example 4.1.1} Effect of TVD schemes.}}
    \label{fig:shocked-tvd}
\end{figure}} 
Fig. \ref{fig:WENO_comp} shows the solution obtained by weighted essentially non-oscillatory schemes, WENO-JS, WENO-M and WENO-Z respectively. As expected, WENO-Z and WENO-M schemes have better accuracy than WENO-JS. WENO-M and WENO-Z  gave similar results but it is well known that WENO-M scheme is more computationally intensive. Based on this analysis WENO-Z is considered for all the simulations in following sections. Order of accuracy for WENO schemes is shown in Fig. \ref{fig:WENO_error}.
\begin{figure}[H]
\centering
\subfigure[{\color{black}Simulations by WENO schemes}]{\includegraphics[width=0.43\textwidth]{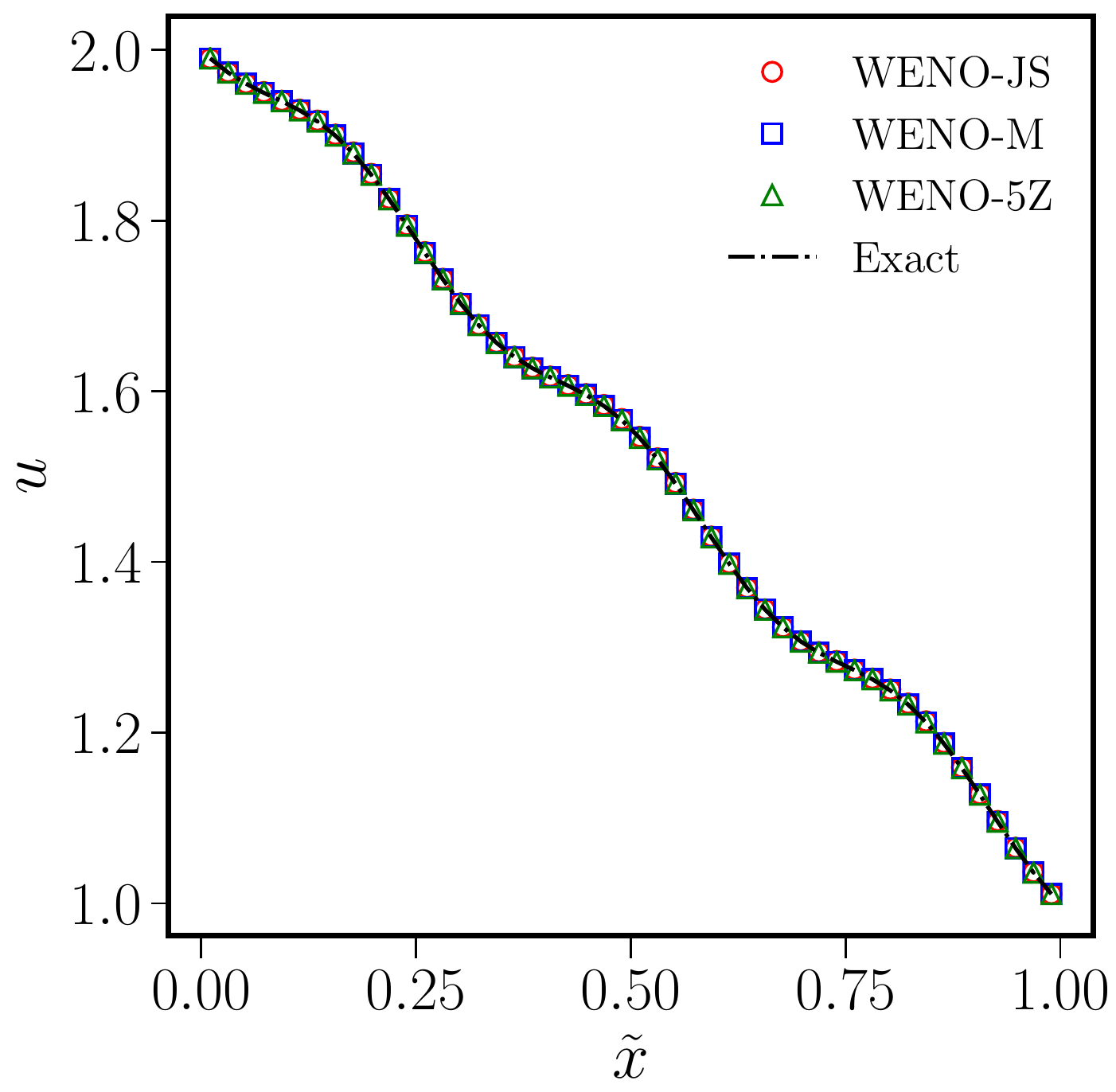}
\label{fig:WENO_comp}}
\subfigure[{\color{black}Accuracy of WENO schemes}]{\includegraphics[width=0.45\textwidth]{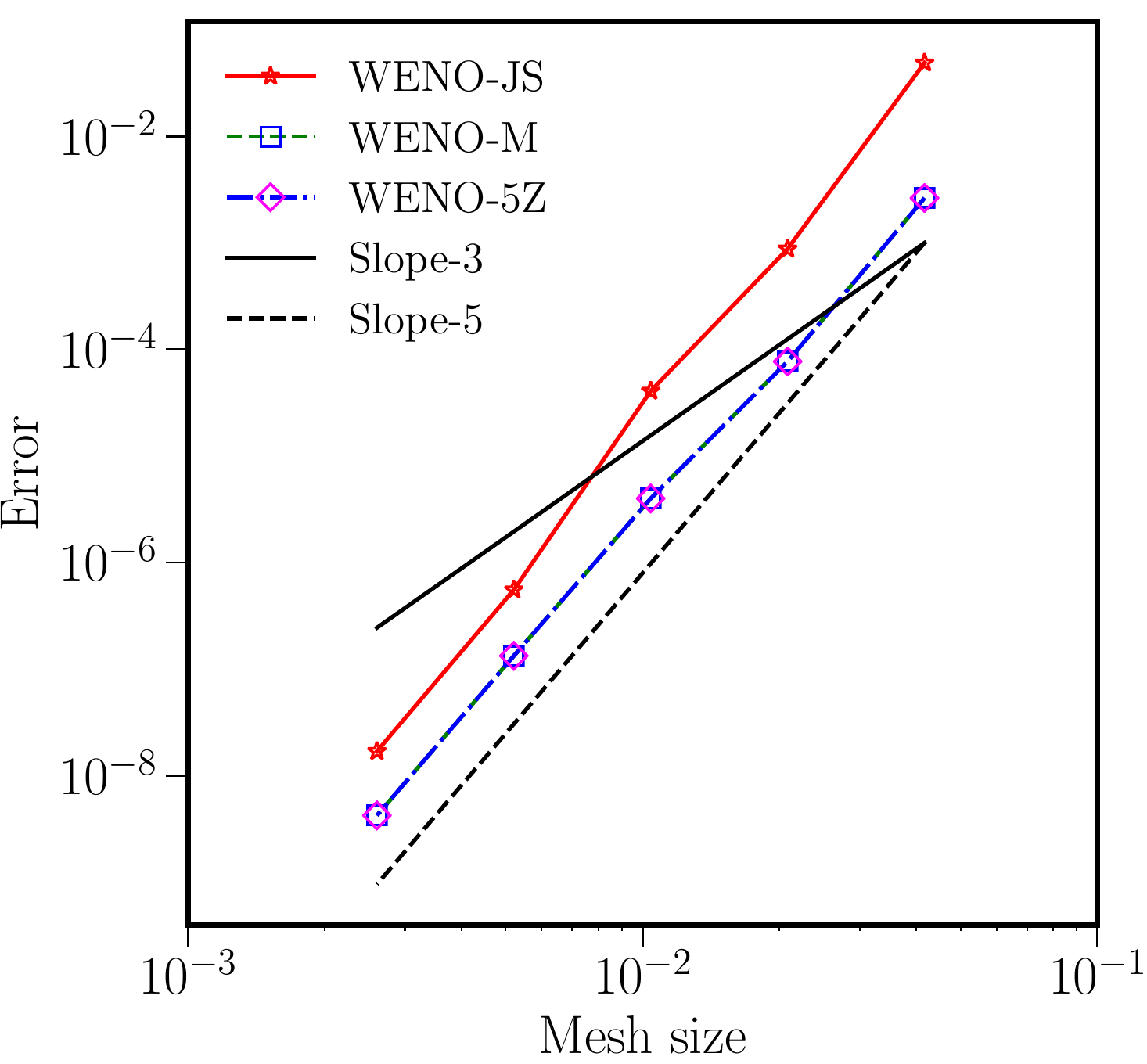}
\label{fig:WENO_error}}
  \caption{ {\color{black}\hyperref[sec:4.1.1]{Example 4.1.1} Effect of WENO schemes.}}
    \label{fig:shocked-WENO}
\end{figure} 
From this analysis, shock-capturing schemes can be considered unnecessary for diffusion equation in hyperbolic approach. TVD schemes contaminated the solution depending on the type of the scheme considered whereas the WENO scheme did not show any unnecessary oscillations for all the variations. It was also mentioned by Nishikawa \cite{Nishikawa2007} that no discontinuity capturing mechanism is required for the upwind formulation of the diffusion equation.} \\

\begin{figure}[H]
\centering
\subfigure[Exact solution]{%
\includegraphics[width=0.48\textwidth]{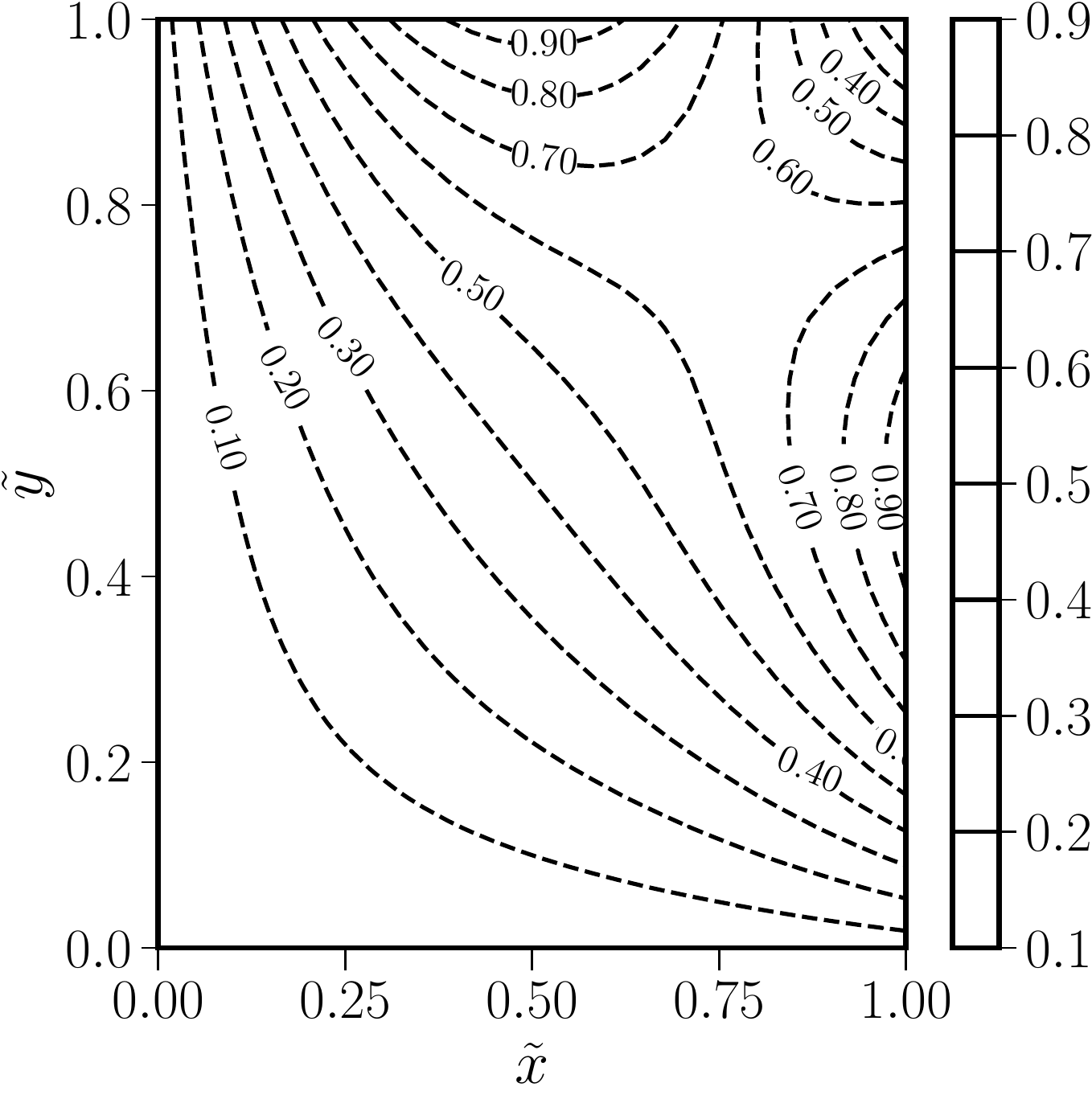}
\label{fig:exact-Nishi}}
\subfigure[Numerical solution]{%
\includegraphics[width=0.48\textwidth]{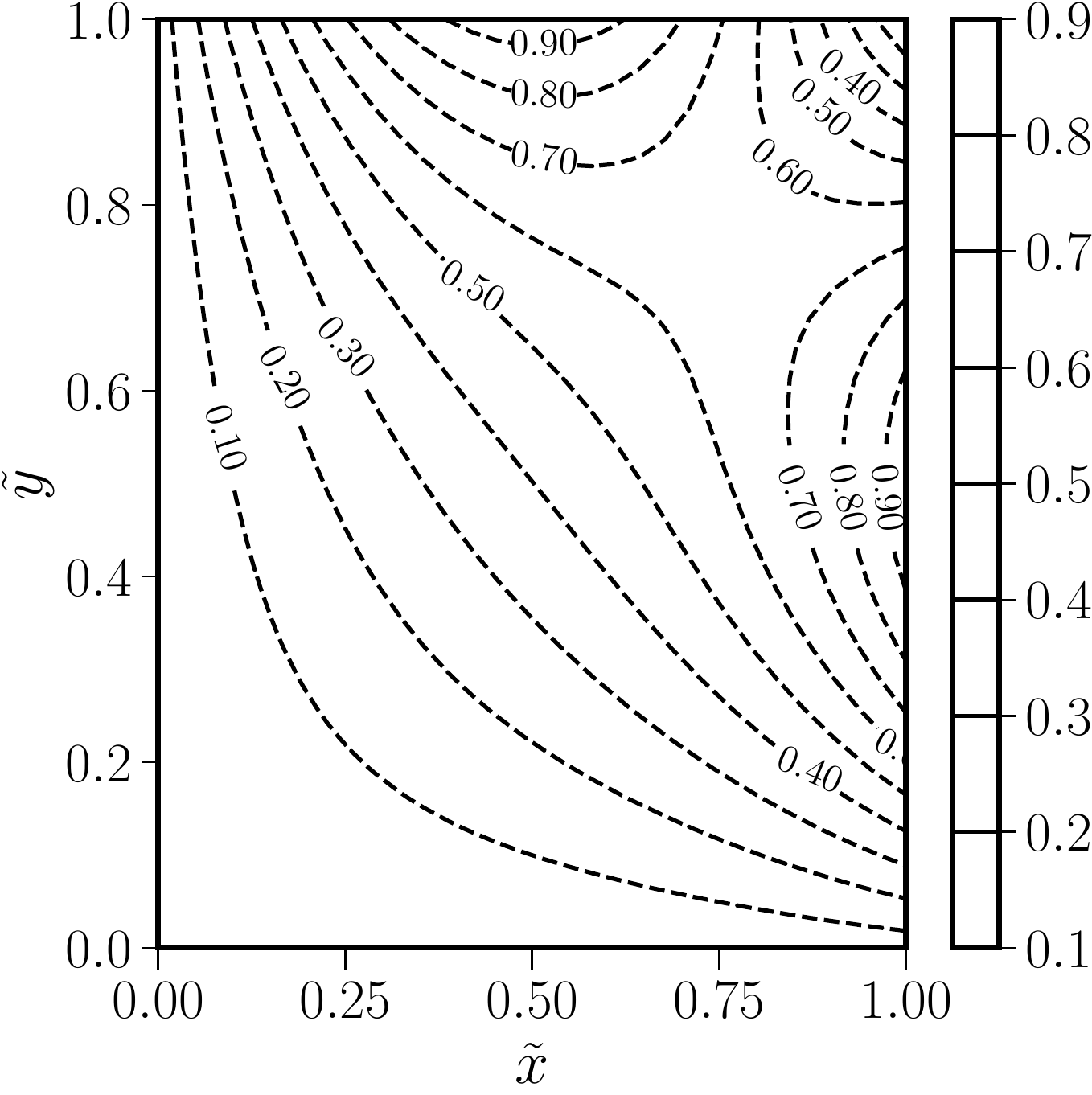}
\label{fig:3e-Nishi}}
\label{fig:2d-Nishi-1}
\caption{Comparison of analytical solution and by upwind scheme U-5E for diffusion equation,\hyperref[sec:4.1.1]{Example 4.1.1}, are shown in (a) and (b) respectively.}
\end{figure}

\begin{figure}[H]
\centering
\subfigure[Solution along geometric center line ]{%
\includegraphics[width=0.45\textwidth]{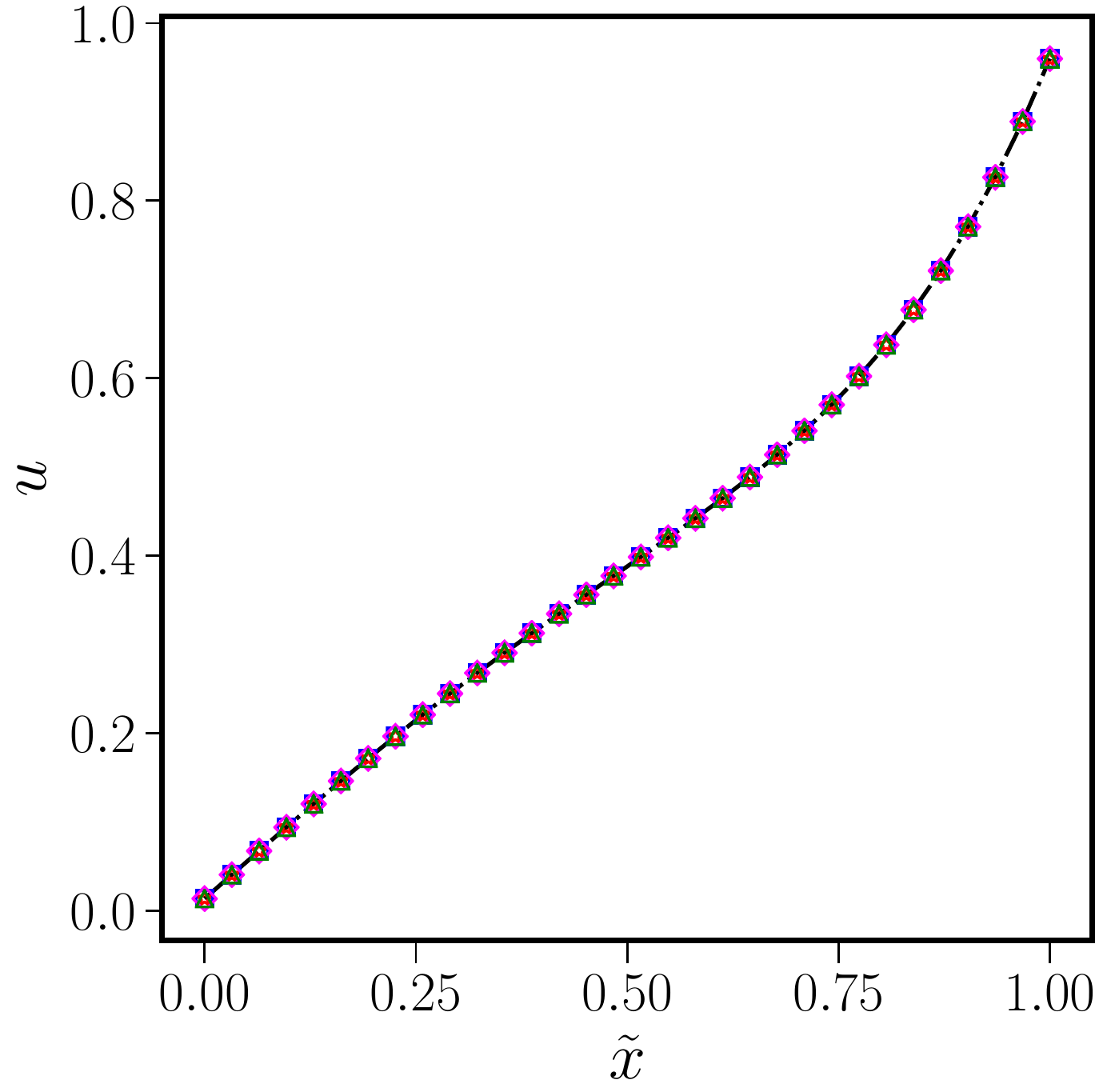}
\label{fig:2d-Nishi-center}}
\subfigure[Convergence of the $L_2$ error]{%
\includegraphics[width=0.48\textwidth]{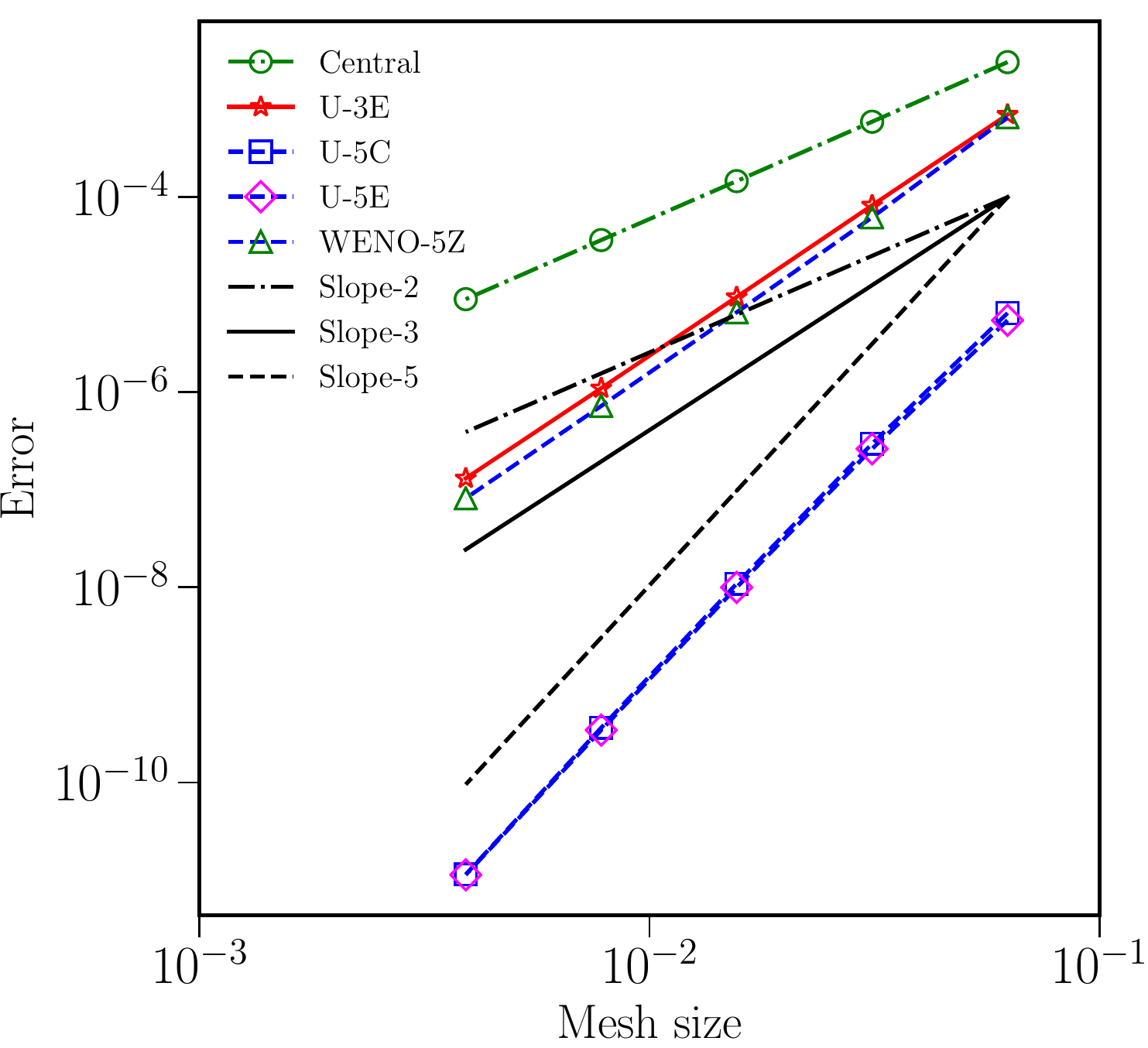}
\label{fig:2d-Nishi-error}}
\caption{{\color{black}Computed values at geometric center and $L_2$ convergence errors for various schemes are shown in (a) and (b) respectively. Dashed line: analytical; {\color{black}red stars}: U-3E; blue squares: U-5C; magenta diamonds: U-5E; green triangles: WENO-5Z.}}
\label{fig:2d-Nishi}
\end{figure}

\noindent \textbf{Example 4.1.2.}\label{sec:4.1.2} First test case is the following two-dimensional diffusion equation considered by Nishikawa \cite{Nishikawa2007} 
\begin{equation}\label{eq:diffu}
\begin{aligned}
\frac{\partial u}{\partial t} = \nu(\frac{\partial ^2 u}{\partial x^2} + \frac{\partial ^2 u}{\partial y^2}) 
\end{aligned}
\end{equation}
\noindent with a spatial domain of $[0,1] \times [0,1]$,  where $\nu$=1 and the following Dirichlet boundary conditions are considered:
\begin{equation}
    u= 
\begin{cases}
   0,&  x= 0\\
    \sin (\pi y),&  x = 1
\end{cases}
\quad 
 u = 
\begin{cases}
   0,&  y= 0\\
    \sin(\pi x),&  y = 1
\end{cases}  
\end{equation}
The exact steady state solution is given by,
\begin{equation}
u_{exact}(x,y) = \frac{\sinh(\pi x) \sin(\pi y) + \sinh(\pi y) \sin(\pi x)}{\sinh(\pi)}.
\end{equation}
The simulations are conducted with grid refinements from 16$\times$16 to 256 $\times$ 256 by second-order central, and all the upwind schemes. For upwind schemes, the numerical solution is computed by an explicit time-marching until the residuals are dropped below $10^{-12}$ in $L_1$ norm with a constant CFL $= 0.5$. The exact solution and numerical solution contours computed by the U-5E scheme are shown in Fig. \ref{fig:exact-Nishi} and Fig. \ref{fig:3e-Nishi}. Computed values of $u$ for various schemes along the geometric center line along the horizontal axis are shown in \ref{fig:2d-Nishi-center}. $L_2$ error convergence results are shown in Fig. \ref{fig:2d-Nishi-error} for the primary variable and we can observe that the design order of accuracy is obtained for all the schemes. 

\begin{figure}[H]
\centering
\subfigure[Solution along geometric center line]{%
\includegraphics[width=0.46\textwidth]{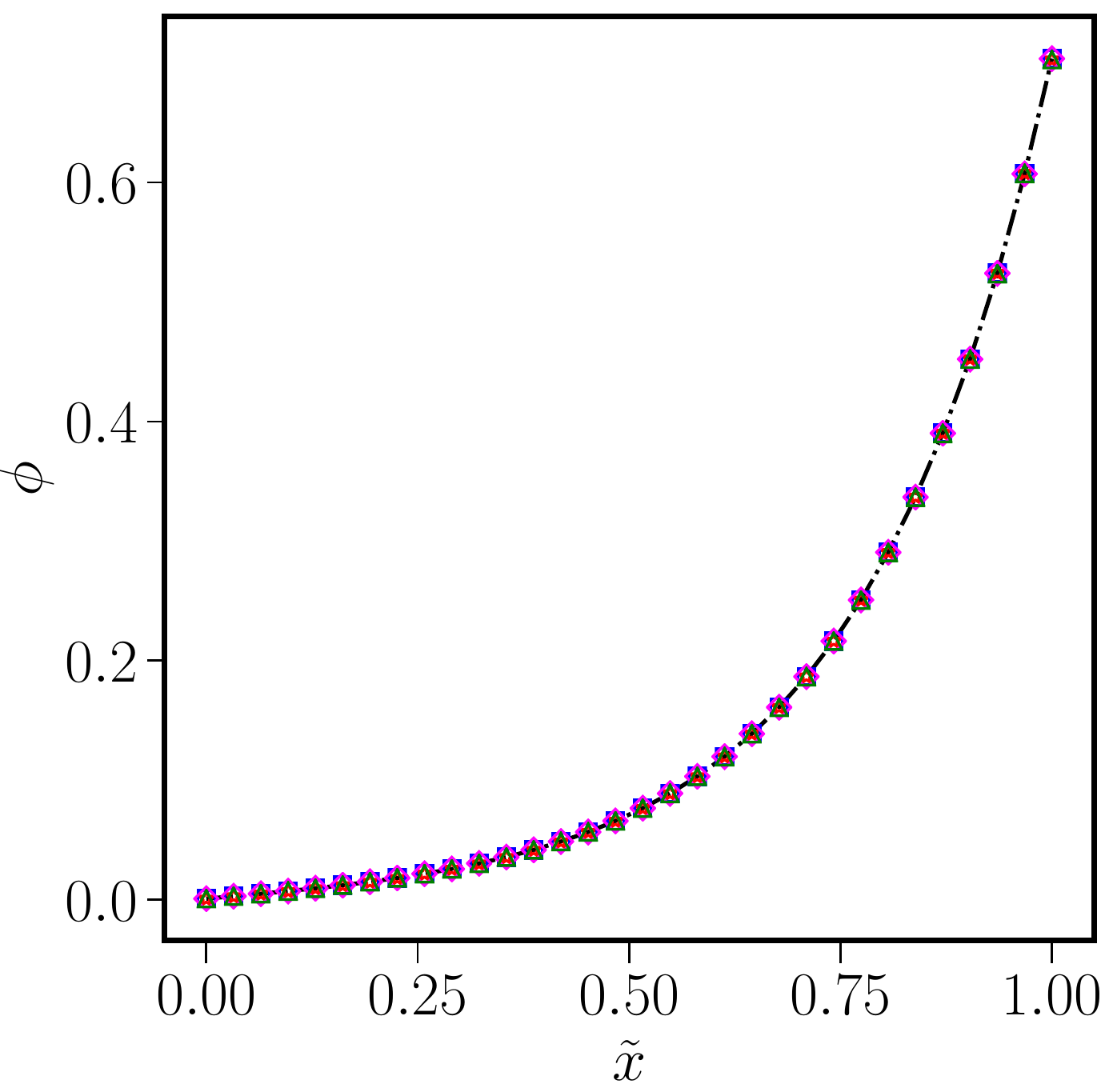}
\label{fig:2d-Barba_center}}
\subfigure[Convergence of the $L_2$ error]{%
\includegraphics[width=0.48\textwidth]{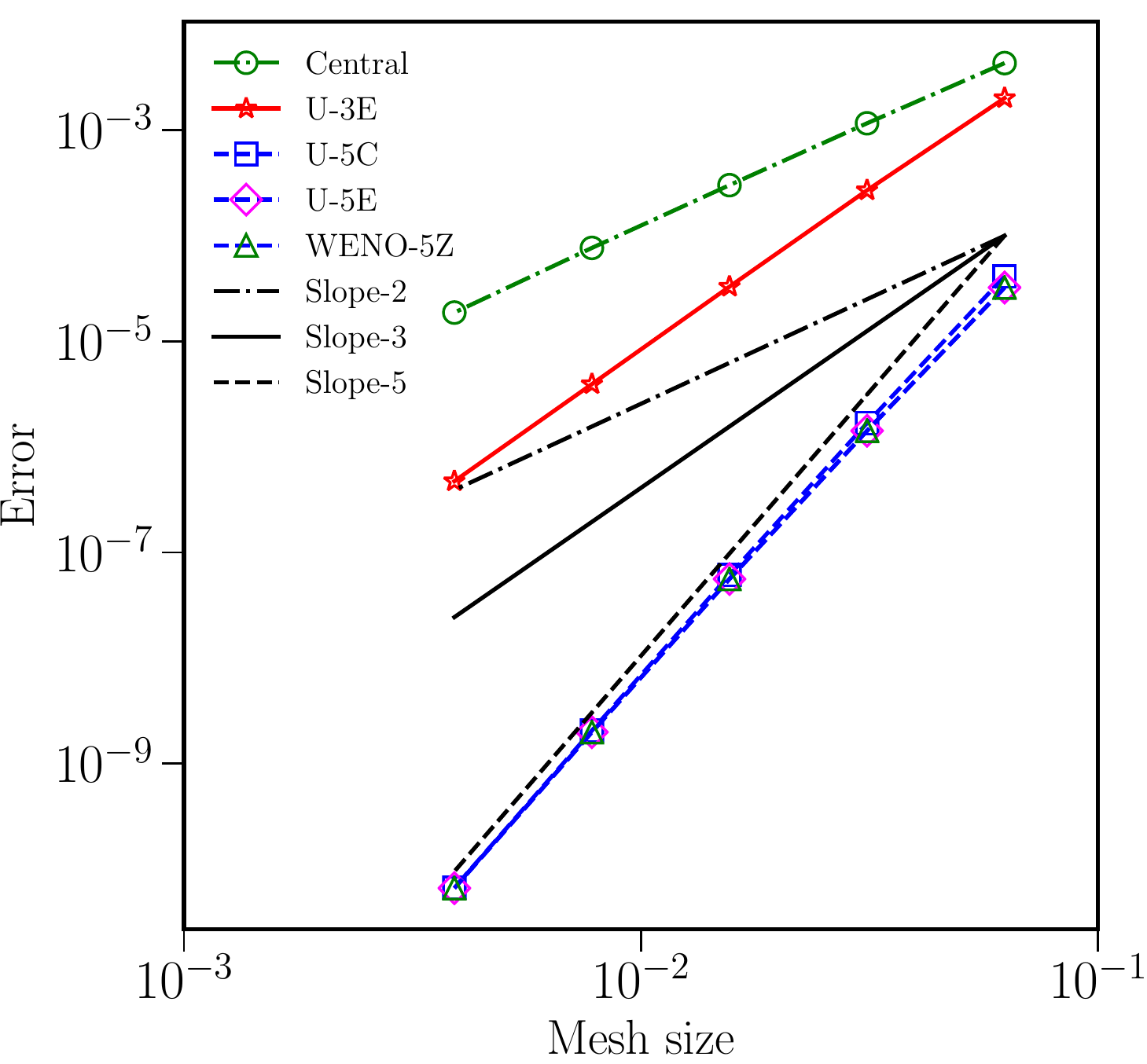}
\label{fig:2d-Barba_error}}
\caption{{\color{black}Computed values at geometric center and $L_2$ convergence errors for various schemes for Neumann boundary condition, \hyperref[sec:4.1.3]{Example 4.1.3}. Dashed line: analytical;  {\color{black}red stars}: U-3E; blue squares: U-5C; magenta diamonds: U-5E; green triangles: WENO-5Z.}}
\label{fig:2d-Barb}
\end{figure}

Table \ref{tab:2D-Nishi} shows the $L_2$ error for the primary variable, $u$, and the order of accuracy for the U-3E, U-5E,U-5C and WENO-5Z schemes respectively. We can see that design order of accuracy is obtained for the velocity for all the linear upwind schemes. For the WENO scheme, the 5th order boundary conditions are found to be unstable, and only 3rd order boundary conditions are considered. {\color{black} TVD-MUSCL has once again reduced to 3rd order linear scheme and is not discussed henceforth for diffusion problems.}
\begin{table}[H]
  \centering
  \caption{$L_2$ errors and order of convergence of primary variable, $u$, by 3rd order explicit, 5th order explicit and compact and WENO schemes one-dimensional diffusion problem, \hyperref[sec:4.1.1]{Example 4.1.1}.}
\footnotesize
    \begin{tabular}{| c | c | c | c | c | c | c | c | c|}
\hline
    Number & \multicolumn{2}{c|}{Upwind-3E} & \multicolumn{2}{c|}{Upwind-5E} & \multicolumn{2}{c|}{Upwind-5C}& \multicolumn{2}{c|}{WENO-5Z}  \\
    \cline{2-9}
     of points& error & order & error & order & error & order & error & order   \\
    \cline{1-9}
    16    & 7.01E-04 &       & 5.46E-06 &       & 6.48E-06 &       & 6.59E-04 &  \\
    \hline
    32    & 8.24E-05 & 3.09  & 2.63E-07 & 4.37  & 2.97E-07 & 4.45  & 6.20E-05 & 3.41 \\
        \hline
    64    & 9.47E-06 & 3.12  & 1.00E-08 & 4.72  & 1.09E-08 & 4.77  & 6.60E-06 & 3.23 \\
            \hline
    128   & 1.10E-06 & 3.10  & 3.44E-10 & 4.86  & 3.64E-10 & 4.90  & 7.23E-07 & 3.19 \\
                \hline
    256   & 1.31E-07 & 3.07  & 1.13E-11 & 4.93  & 1.14E-11 & 5.00  & 8.19E-08 & 3.14 \\
                    \hline
    \end{tabular}%
  \label{tab:2D-Nishi}%
\end{table}%
\noindent \textbf{Example 4.1.3}\label{sec:4.1.3} In this test case, the verification of Neumann boundary condition which is also implemented as Dirichlet boundary condition in the hyperbolic formulation for diffusion equation is considered. The following two-dimensional Laplace equation with the spatial domain of [0, 1] $\times$ [0, 1], Eq. (\ref{Laplace}), has been considered. 
\begin{equation}\label{Laplace}
  \frac{\partial ^2 \phi}{\partial x^2} + \frac{\partial ^2 \phi}{\partial y^2} = 0
\end{equation}
which has the following Neumann and Dirichlet boundary conditions
\begin{equation}
\begin{aligned}
    \phi &= 0,   x= 0\\
    \frac{\partial \phi}{\partial x} &=  0,  x = L
\end{aligned}
\quad 
 \phi = 
\begin{cases}
   0,&  y= 0\\
    \sin \left(  \frac{\frac{3}{2}\pi x}{L} \right),&  y = M,
\end{cases}  
\label{neu-1}
\end{equation}
\noindent where $L=1$ and $M=1$ and the analytical solution is given by Eq. (\ref{Laplace-an})
\begin{equation}\label{Laplace-an}
\phi_{exact}(x,y) = \frac{\sinh \left( \frac{\frac{3}{2} \pi y}{L}\right)}{\sinh \left(  \frac{\frac{3}{2} \pi M}{L}\right)} \sin \left( \frac{\frac{3}{2} \pi x}{L} \right)
\end{equation}
 The simulations are conducted with grid refinements from 16$\times$16 to 256 $\times$ 256 for the test case. For the upwind schemes, the numerical solution is again computed by an explicit time-marching until the residuals are dropped below $10^{-12}$ in $L_1$ norm with a constant CFL $= 0.5$. Computed values of $\phi$ for various schemes along the geometric center line along the horizontal axis are shown in Fig. \ref{fig:2d-Barb} for a 32 $\times$ 32 grid. Design accuracy is obtained for all the schemes, and the implementation of Neumann boundary condition is successfully verified through this test case. Unlike the previous test case, WENO scheme also shows 5th order accuracy. Table \ref{tab:2D-Neu} show the $L_2$ norms for all the schemes and the advantage of high-order methods over lower-order methods can be seen as they need less number of computational cells to get a solution with the same accuracy. This advantage enables high-order methods to use coarse meshes, in comparison with the lower-order methods.
\begin{table}[H]
  \centering
  \caption{$L_2$ errors and order of convergence of potential by 3rd order explicit, 5th order explicit and compact and WENO schemes for Neumann boundary test case.}
\footnotesize
    \begin{tabular}{| c | c | c | c | c | c | c | c | c|}
\hline
    Number & \multicolumn{2}{c|}{Upwind-3E} & \multicolumn{2}{c|}{Upwind-5E} & \multicolumn{2}{c|}{Upwind-5C}& \multicolumn{2}{c|}{WENO-5Z}  \\
    \cline{2-9}
    of points& error & order & error & order & error & order & error & order   \\
    \cline{1-9}
    $16^2$    & 2.01E-03 &       & 3.23E-05 &       & 4.11E-05 &       & 3.22E-05 &  \\
    \hline
    $32^2$    & 2.69E-04 & 2.90  & 1.42E-06 & 4.50  & 1.68E-06 & 4.61  & 1.42E-06 & 4.50 \\
    \hline
    $64^2$    & 3.29E-05 & 3.03  & 5.61E-08 & 4.67  & 6.12E-08 & 4.78  & 5.61E-08 & 4.67 \\
    \hline
    $128^2$   & 3.93E-06 & 3.06  & 1.99E-09 & 4.82  & 2.06E-09 & 4.89  & 1.99E-09 & 4.82 \\
    \hline
    $256^2$   & 4.73E-07 & 3.06  & 6.63E-11 & 4.91  & 6.68E-11 & 4.95  & 6.62E-11 & 4.91 \\
    \hline
    \end{tabular}%
  \label{tab:2D-Neu}%
\end{table}

\begin{figure}[H]
\centering
\subfigure[Solution along geometric center line]
{\includegraphics[width=0.48\textwidth]{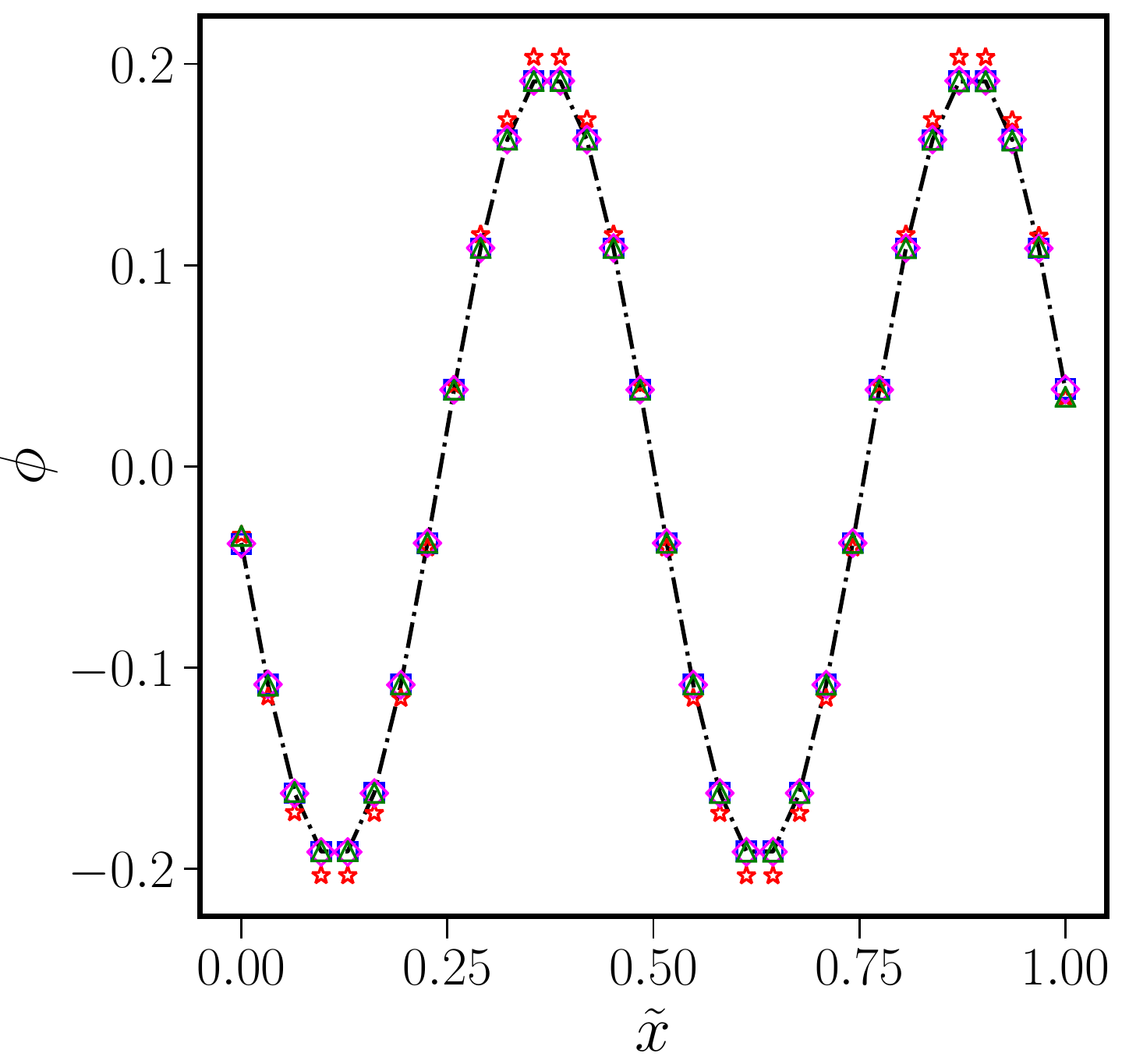}
\label{fig:s2d-Poi-center}}
\subfigure[Convergence of the $L_2$ error]
{\includegraphics[width=0.48\textwidth]{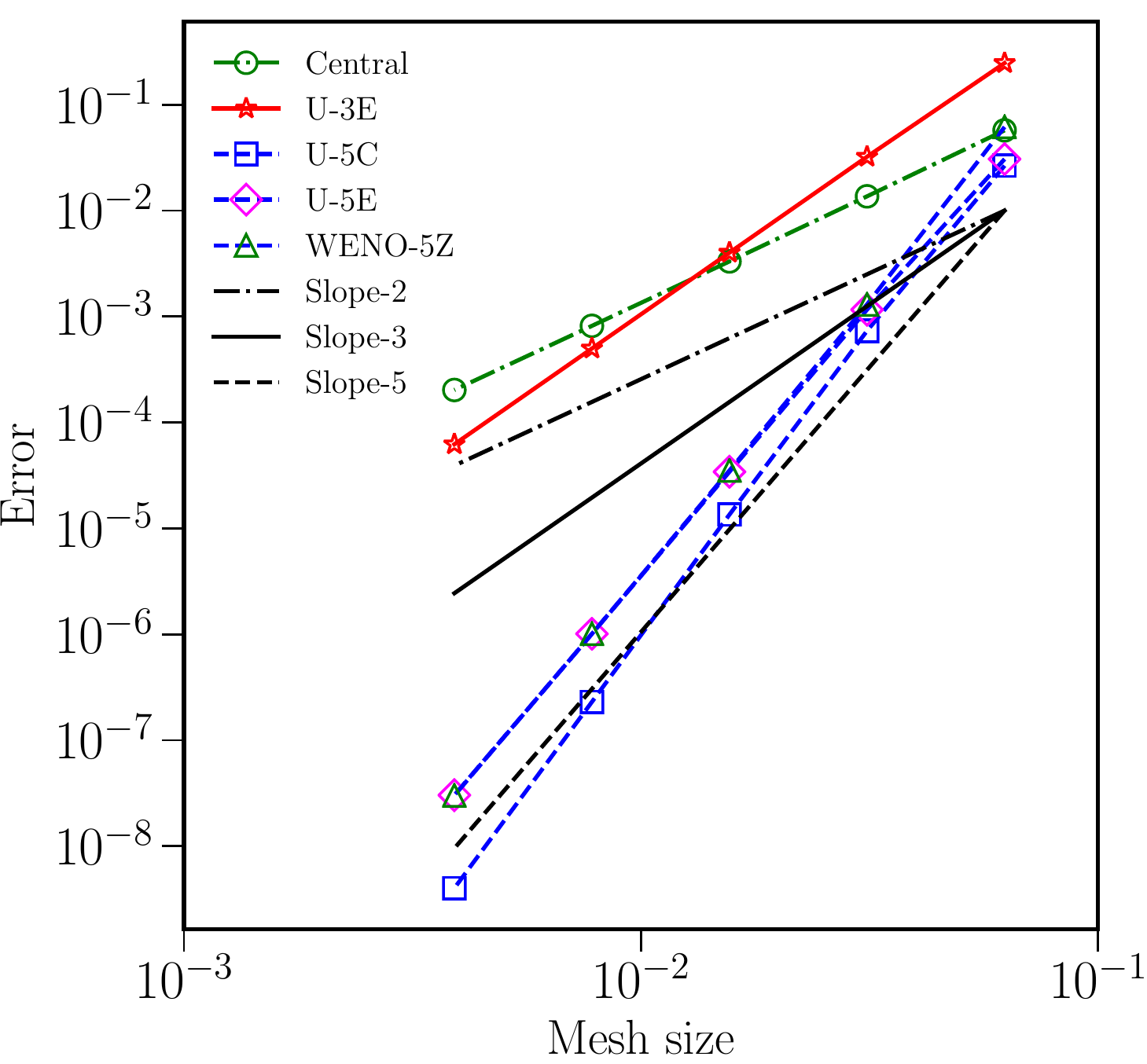}
\label{fig:2d-Poi-error}}
\caption{{\color{black}Computed values at geometric center and $L_2$ convergence errors for various schemes for Poisson equation, \hyperref[sec:4.1.4]{Example 4.1.4}, are shown here. Dashed line: analytical; {\color{black}red stars}: U-3E; blue squares: U-5C; magenta diamonds: U-5E; green triangles: WENO-5Z.}}
\end{figure}

\begin{figure}[H]
\centering
\subfigure[Exact solution]{%
\includegraphics[width=0.45\textwidth]{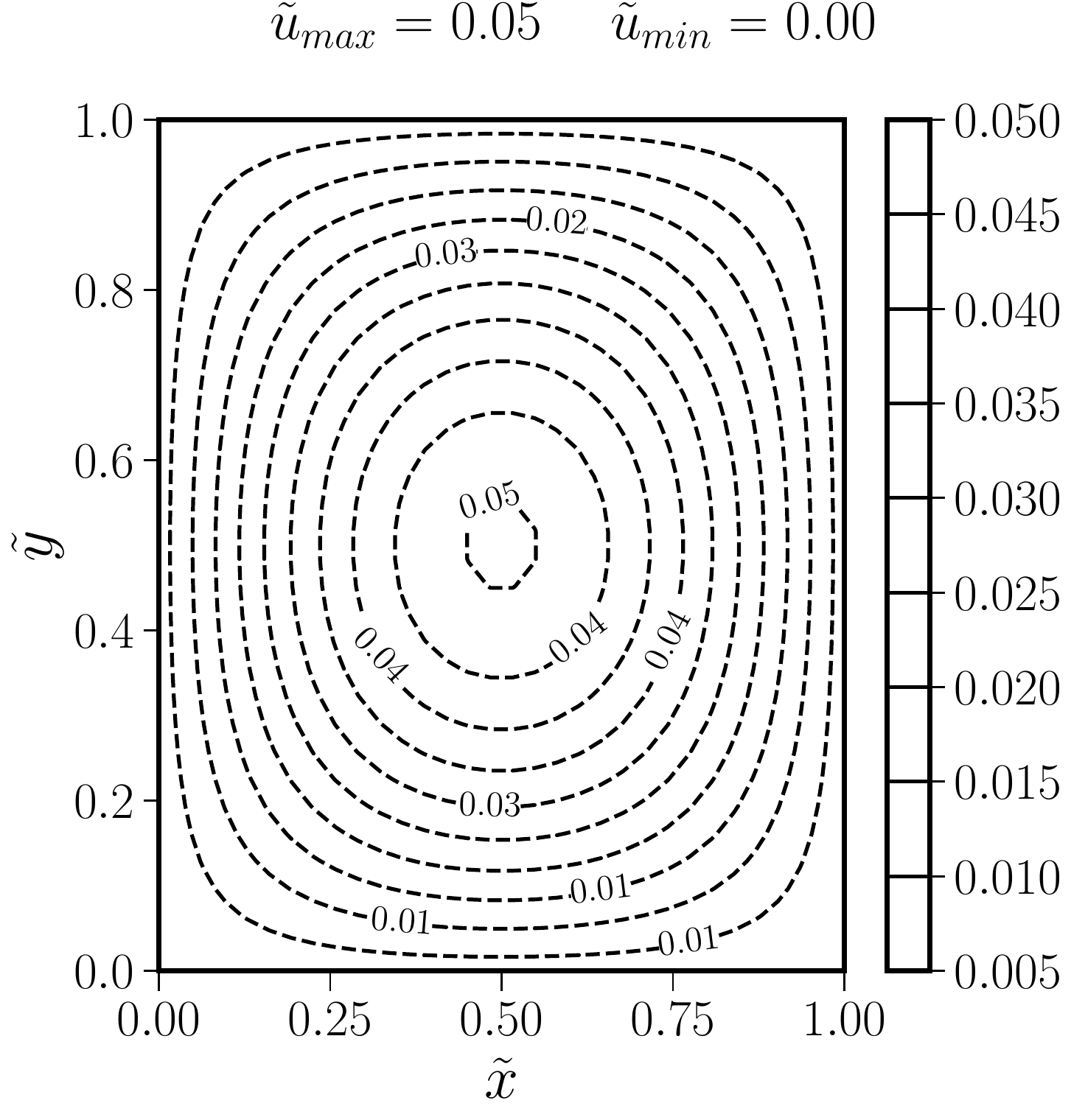}
\label{fig:exact-aniso-source}}
\subfigure[Numerical solution by U-5E]{%
\includegraphics[width=0.45\textwidth]{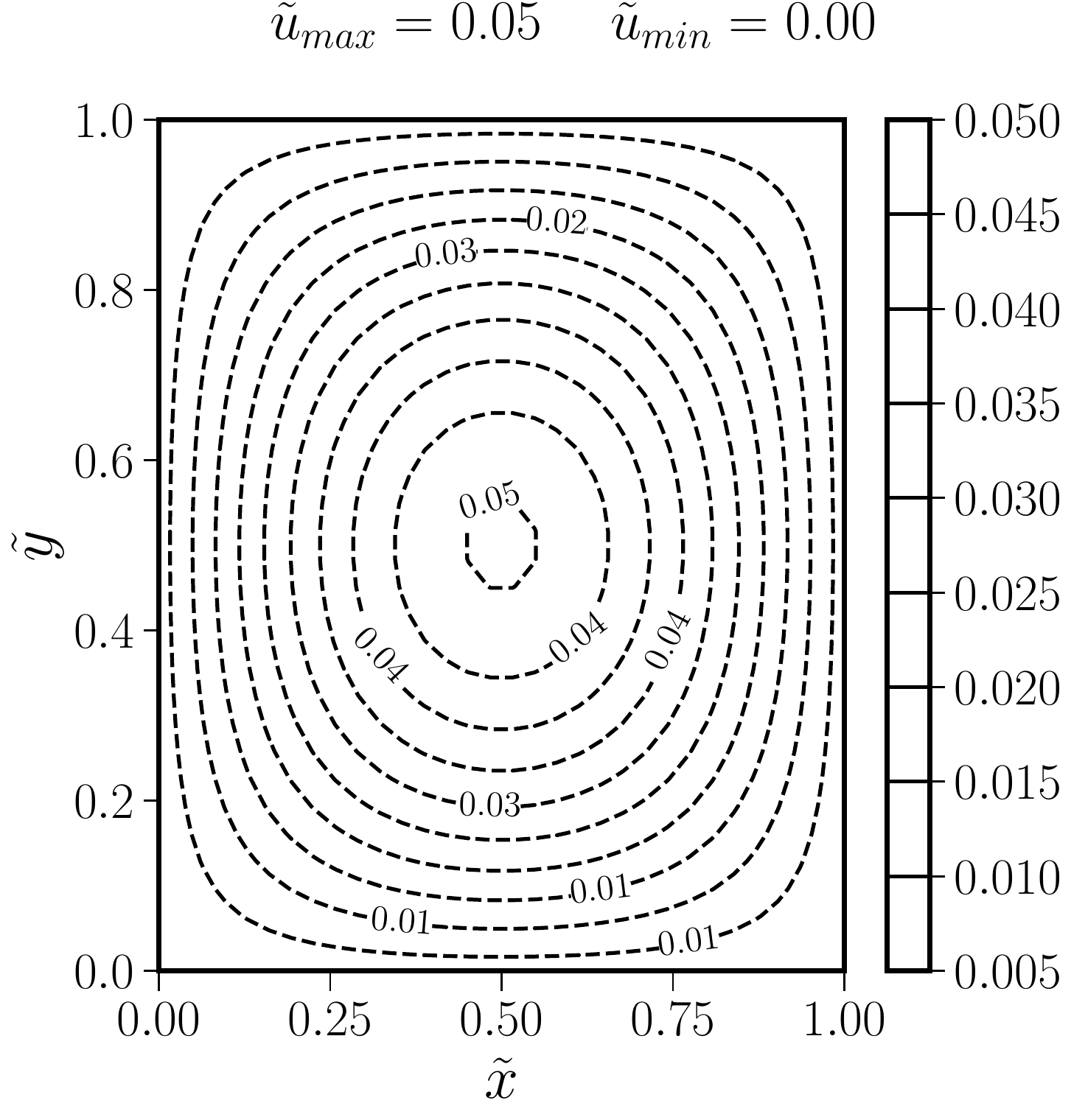}
\label{fig:3e-aniso-source}}
\caption{Solution contours for anisotropic diffusion by analytical solution and by upwind scheme U-5E.}
\end{figure}

\begin{figure}[H]
\centering
\subfigure[Solution along geometric center line]
{\includegraphics[width=0.45\textwidth]{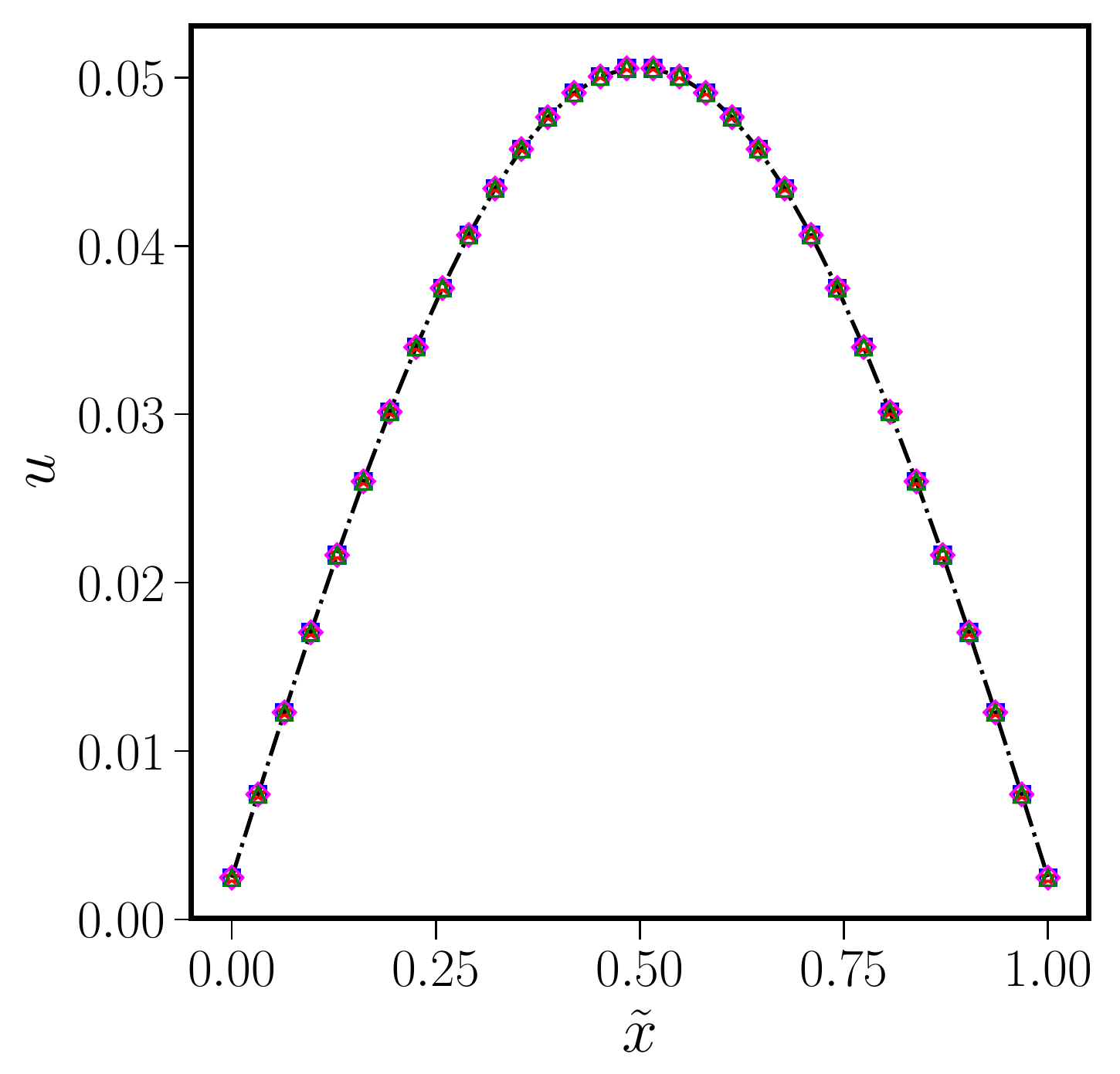}
\label{fig:s2d-aniso-center}}
\subfigure[Convergence of the $L_2$ error]
{\includegraphics[width=0.46\textwidth]{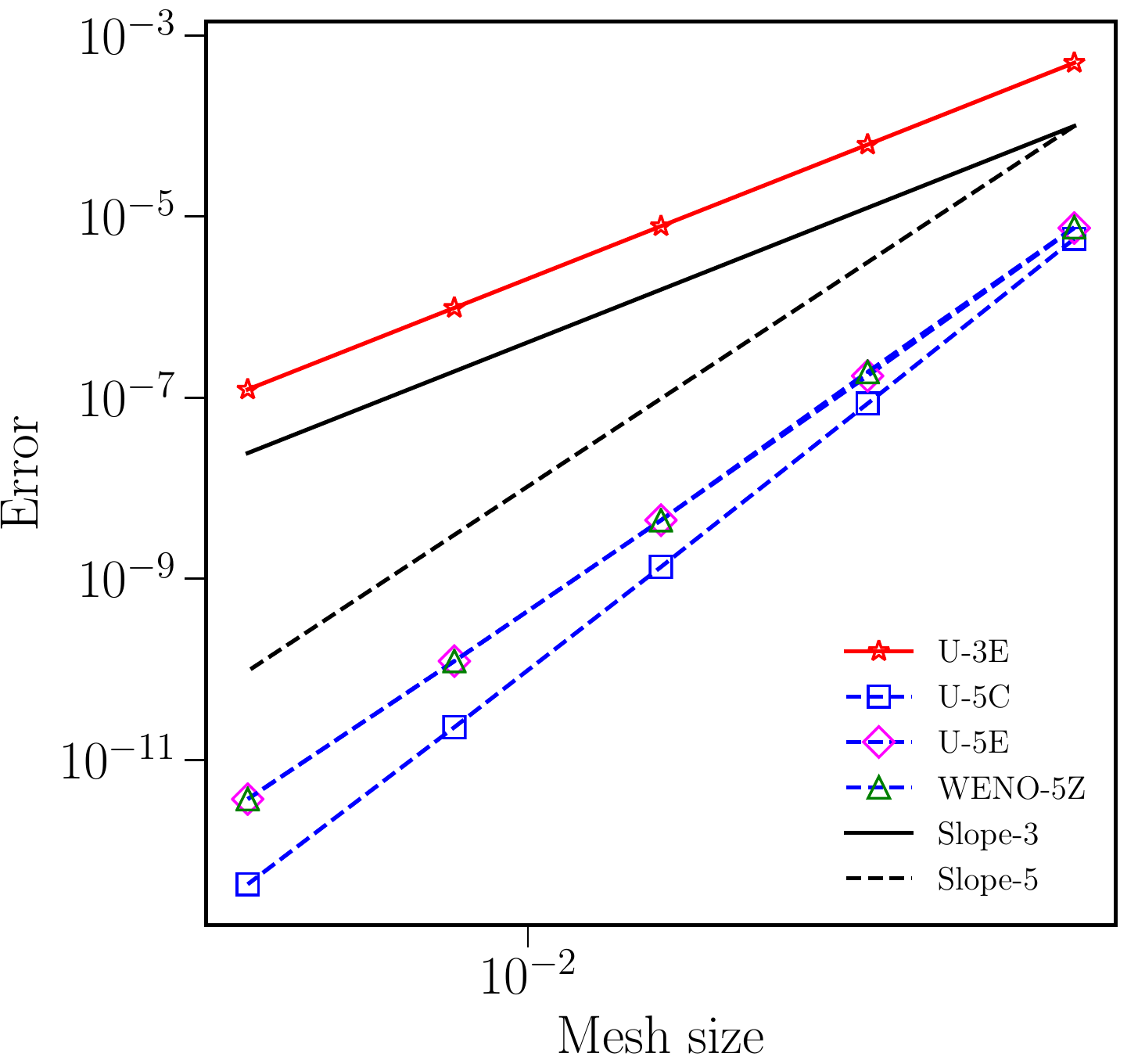}
\label{fig:2d-aniso-error}}
\caption{{\color{black}Computed values at geometric center and $L_2$ convergence errors for various schemes for anisotropic diffusion problem, \hyperref[sec:4.1.5]{Example 4.1.5}, are shown here. Dashed line: analytical; {\color{black}red stars}: U-3E; blue squares: U-5C; magenta diamonds: U-5E; green triangles: WENO-5Z.}}
\end{figure}

\noindent \textbf{Example 4.1.4.}\label{sec:4.1.4}
For the previous test cases and the given numerical grids, we could not distinguish the numerical results by different methods. Hence, we have only presented the numerical results of these schemes. In this example, we consider the following Poisson equation, i.e., with a source term, given by Eq. (\ref{Poisson})
\begin{equation}\label{Poisson}
\frac{\partial ^2 \phi}{\partial x^2} + \frac{\partial ^2 \phi}{\partial y^2} = 32 \pi^{2}\sin(4 \pi  x)\sin(4 \pi y)
\end{equation}
where the domain is [0, 1] $\times$ [0, 1] and $\phi = 0$ at all the boundaries. The exact solution for this test case is given by
 \begin{equation}
\phi_{exact}(x,y) = \sin(4\pi x)\sin(4\pi y)
\end{equation}
In this test case, the advantage of 5th order schemes can be observed. The simulations are conducted with grid refinements from 16$\times$16 to 256 $\times$ 256 for the test case. For the upwind schemes, the numerical solution is deemed to have reached a steady state when the residuals are dropped below $10^{-12}$ in $L_1$ norm and constant CFL $= 0.5$ is used. In Fig. \ref{fig:s2d-Poi-center} the computed results at geometric center on a 32 $\times$ 32 grid is shown and we can observe that the third order explicit scheme has deviated considerably from the exact solution in comparison with the other schemes. We can see from the $L_2$ norms shown in the Fig. \ref{fig:2d-Poi-error} that even though third order accuracy is obtained for the upwind scheme, U-3E, the solution is inferior to standard second order case on coarse meshes. Such results are also observed in edge based methods proposed by Nishikawa in [Ref. \cite{Nishikawa2014b}, Figs. 4 and 5]. The source term may have a significant effect on the solution accuracy for the upwind formulation. Table \ref{table:2D_Poi} shows the $L_2$ error for the primary variable, $\phi$, and the order of accuracy for the U-3E, U-5E, U-5C and WENO-5Z schemes respectively. Compact schemes show better accuracy for problems with source term with increasing grid size in comparison with the explicit scheme.
\begin{table}[H]
\footnotesize
  \centering
  \caption{$L_2$ errors and order of convergence for two-dimensional Poisson problem,\hyperref[sec:4.1.4]{Example 4.1.4}, by 3rd order explicit, 5th order explicit and compact and WENO schemes.}
    \begin{tabular}{| c | c | c | c | c | c | c | c | c|}
\hline
    Number & \multicolumn{2}{c|}{Upwind-3E} & \multicolumn{2}{c|}{Upwind-5E} & \multicolumn{2}{c|}{Upwind-5C}& \multicolumn{2}{c|}{WENO-5Z}  \\
    \cline{2-9}
    of points& error & order & error & order & error & order & error & order   \\
    \cline{1-9}
    $16^2$    & 2.466E-01 &       & 3.06E-02 &       & 2.65E-02 &       & 6.15E-02 &  \\
   \hline
    $32^2$    & 3.221E-02 & 2.94  & 1.16E-03 & 4.72  & 7.32E-04 & 5.18  & 1.30E-03 & 5.56 \\
   \hline
    $64^2$    & 4.006E-03 & 3.01  & 3.42E-05 & 5.09  & 1.35E-05 & 5.76  & 3.52E-05 & 5.21 \\
   \hline
    $128^2$   & 4.985E-04 & 3.01  & 1.01E-06 & 5.09  & 2.30E-07 & 5.87  & 1.01E-06 & 5.12 \\
   \hline
    $256^2$   & 6.218E-05 & 3.00  & 3.03E-08 & 5.05  & 4.01E-09 & 5.84  & 3.01E-08 & 5.07 \\
   \hline
    \end{tabular}%
    \label{table:2D_Poi}
\end{table}%
\noindent \textbf{Example 4.1.5.}\label{sec:4.1.5} In this test case the following anisotropic diffusion equation identified by Kuzmin et al. \cite{kuzmin}, with a spatial domain $(x,y) \in [0,1] \times [0,1]$,  has been considered. 
\begin{equation}\label{eq:diffu-ani}
\begin{aligned}
\frac{\partial u}{\partial t} = \nu_1 \frac{\partial ^2 u}{\partial x^2} + \nu_2 \frac{\partial ^2 u}{\partial y^2} + S,
\end{aligned}
\end{equation}
where the diffusion tensor and source term are given by  
 \begin{equation}\label{eq:tensor}
 \mathbf{D} = \left[ \begin{array}{ccc}
\nu_1 & 0 \\
0 & \nu_2 \\
\end{array} \right] =  \left[ \begin{array}{ccc}
100 & 0 \\
0 & 1 \\
\end{array} \right], \quad  S = 50.5 \sin(\pi x)\sin(\pi y).
\end{equation}
 The {\color{black}exact steady state solution} for this test case is given by, 
 \begin{equation}\label{eq:kuz}
 u_{exact}(x,y) =\frac {1}{2\pi^2}  \sin(\pi x)\sin(\pi y),
\end{equation}
, which is also imposed as Dirichlet boundary conditions. Again, grid refinement study is carried out for all the upwind schemes using explicit time stepping. The solution contours obtained on a 32 $\times$ 32 grid shown in Figs. \ref{fig:exact-aniso-source} and \ref{fig:3e-aniso-source} are indistinguishable and also unlike the slope limiting approach of Kuzmin et al. the current scheme does not pollute the smooth extrema and also satisfies the discrete maximum principle. Table \ref{tab:aniso} shows the design order of accuracy for the primary variable, $u$, for the U-3E, U-5E,U-5C and WENO-5Z schemes respectively. Once again, compact schemes show better accuracy for problems with increasing grid size in comparison with the explicit scheme. Computed values at geometric center and $L_2$ convergence errors for various schemes are shown in Figs. \ref{fig:s2d-aniso-center} and \ref{fig:2d-aniso-error} respectively. Convergence rates for anisotropic diffusion problem by using explicit time stepping are extremely slow and can be significantly improved by implicit time marching approach \cite{nishikawa2017effects}.

\begin{table}[H]
  \centering
\footnotesize
  \caption{$L_2$ errors and order of convergence by 3rd order explicit, 5th order explicit and compact and WENO schemes for anisotropic diffusion problem given in \hyperref[sec:4.1.5]{Example 4.1.5}.}
    \begin{tabular}{| c | c | c | c | c | c | c | c | c|}
\hline
    Number & \multicolumn{2}{c|}{Upwind-3E} & \multicolumn{2}{c|}{Upwind-5E} & \multicolumn{2}{c|}{Upwind-5C}& \multicolumn{2}{c|}{WENO-5Z}  \\
    \cline{2-9}
    of points& error & order & error & order & error & order & error & order   \\
    \cline{1-9}
    $16^2$    & 4.99E-04 &       & 7.47E-06 &       & 5.68E-06 &       & 7.61E-06 &  \\
    \hline
    $32^2$    & 6.23E-05 & 3.00  & 1.74E-07 & 5.42  & 8.61E-08 & 6.04  & 1.92E-07 & 5.31 \\
    \hline
    $64^2$    & 7.83E-06 & 2.99  & 4.43E-09 & 5.30  & 1.35E-09 & 5.99  & 4.42E-09 & 5.44 \\
    \hline
    $128^2$   & 9.81E-07 & 3.00  & 1.23E-10 & 5.17  & 2.29E-11 & 5.88  & 1.23E-10 & 5.17 \\
    \hline
    $256^2$   & 1.23E-07 & 3.00  & 3.68E-12 & 5.06  & 4.22E-13 & 5.77  & 3.68E-12 & 5.06 \\
    \hline
    \end{tabular}%
  \label{tab:aniso}%
\end{table}%
%
%

%
										\subsection{Advection-Diffusion equation}

For the advection-diffusion equation, the numerical solutions are also computed on a uniform cartesian mesh. For the linear upwind schemes, depending on the order of the interior interpolation the corresponding $r^{th}$ order extrapolation polynomial is used for the ghost boundary conditions. For WENO scheme both Lagrange and WENO extrapolations are considered for boundary conditions.

\noindent \textbf{Example 4.2.1.}\label{sec:4.2.1}
The following one dimensional-advection diffusion equation is considered with the spatial domain of [0,1] and $u(0) = 0$ and $u(1) = 1$ 
\begin{equation}
u_{t} + a u_x = \nu u_{xx} + s(x),
\end{equation}
where, $s(x) = \pi Re [ a\cos(\pi x) + \pi \nu sin(\pi x)$ and the Reynolds number, $Re$ = $\frac{a}{\nu}$. The {\color{black} exact steady state solution} is given by \cite{Nishikawa2010b}
\begin{equation}
u_{exact}(x) = \frac{exp(-Re)-exp(x Re - Re)} {exp(-Re)-1} + \frac{\sin(\pi x)}{Re}.
\end{equation}
The exact solution is a smooth sine curve for low Reynolds numbers, which are diffusion dominant, and develops a sharp gradient close to the boundary for high Reynolds numbers when advection becomes the dominant effect. The main objective of this test case to verify the design accuracy and implementation of the numerical schemes and also to show the advantage of WENO scheme in capturing the sharp gradients. 

\begin{figure}[H]
\centering
\subfigure[Linear schemes]{\includegraphics[width=0.44\textwidth]{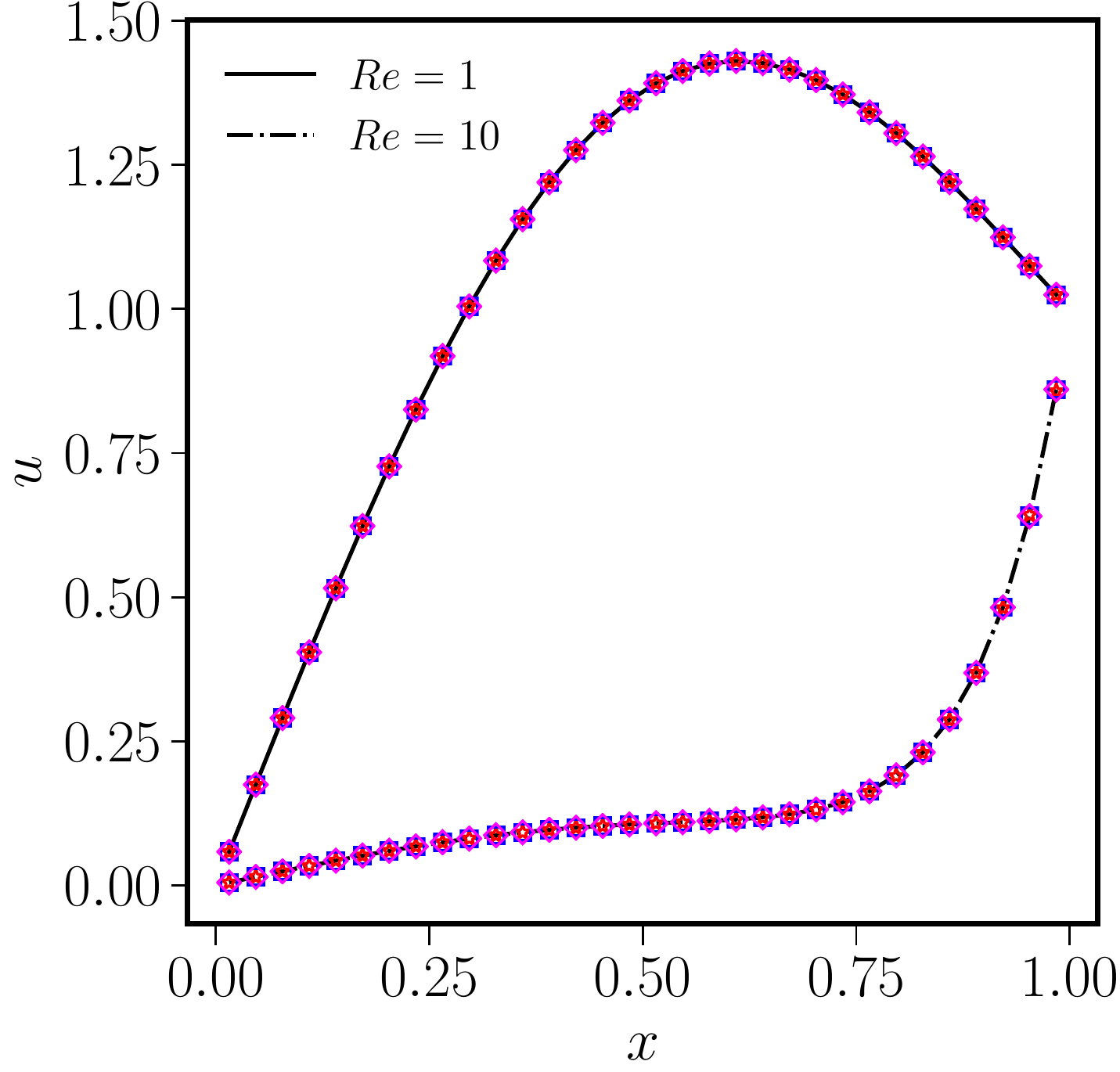}
\label{fig:Re-1-10}}
\subfigure[{\color{black}Shock-capturing schemes}]{\includegraphics[width=0.44\textwidth]{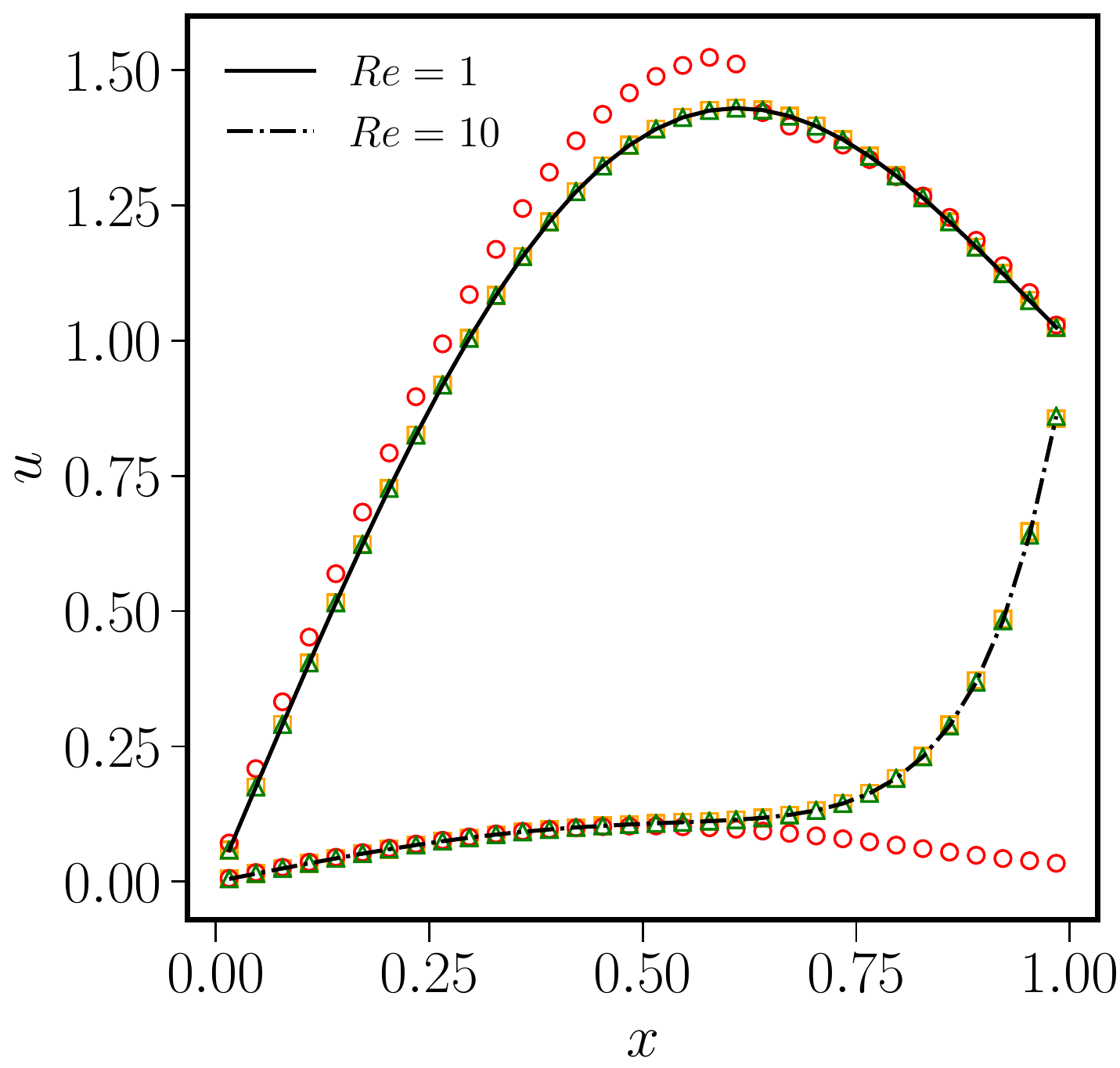}
\label{fig:Re-TVD}}
\caption{{\color{black}Figure 15: Numerical solutions obtained by using different schemes for $Re$ = 1 and 10 for \hyperref[sec:4.2.1]{Example 4.2.1}. Red circles: Generalized-MUSCL; {\color{black}red stars:} U-3E; {\color{black}orange squares: TVD-MUSCL; blue squares: U-5C; magenta diamonds: U-5E; green triangles: WENO-5Z.}}}
\label{fig:Re-ana}
\end{figure}

\begin{figure}[H]
\centering
\subfigure[$Re=0.1$]{\includegraphics[width=0.3\textwidth]{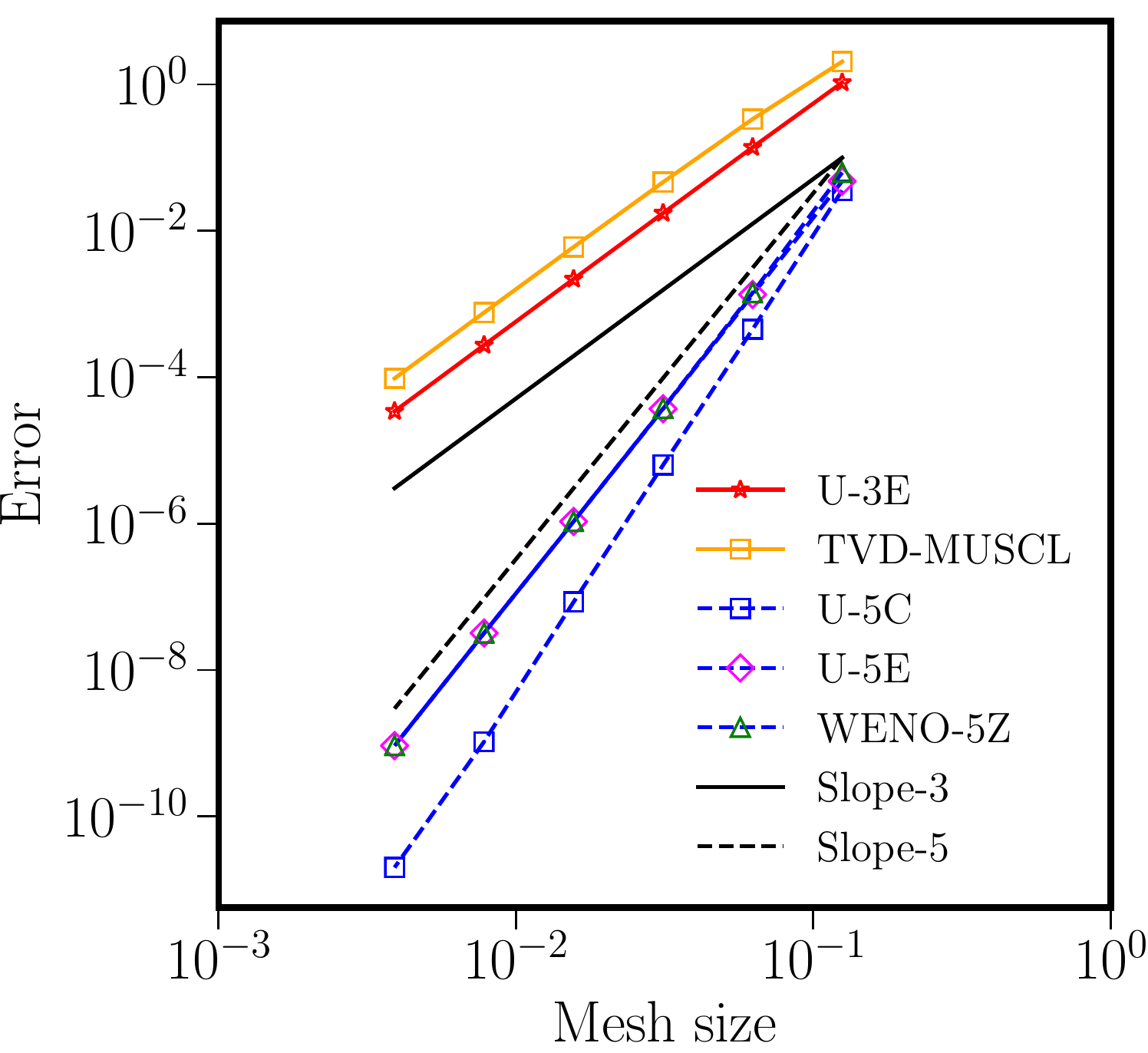}
\label{fig:Re-0.1-error}}
\subfigure[$Re=1$]{\includegraphics[width=0.3\textwidth]{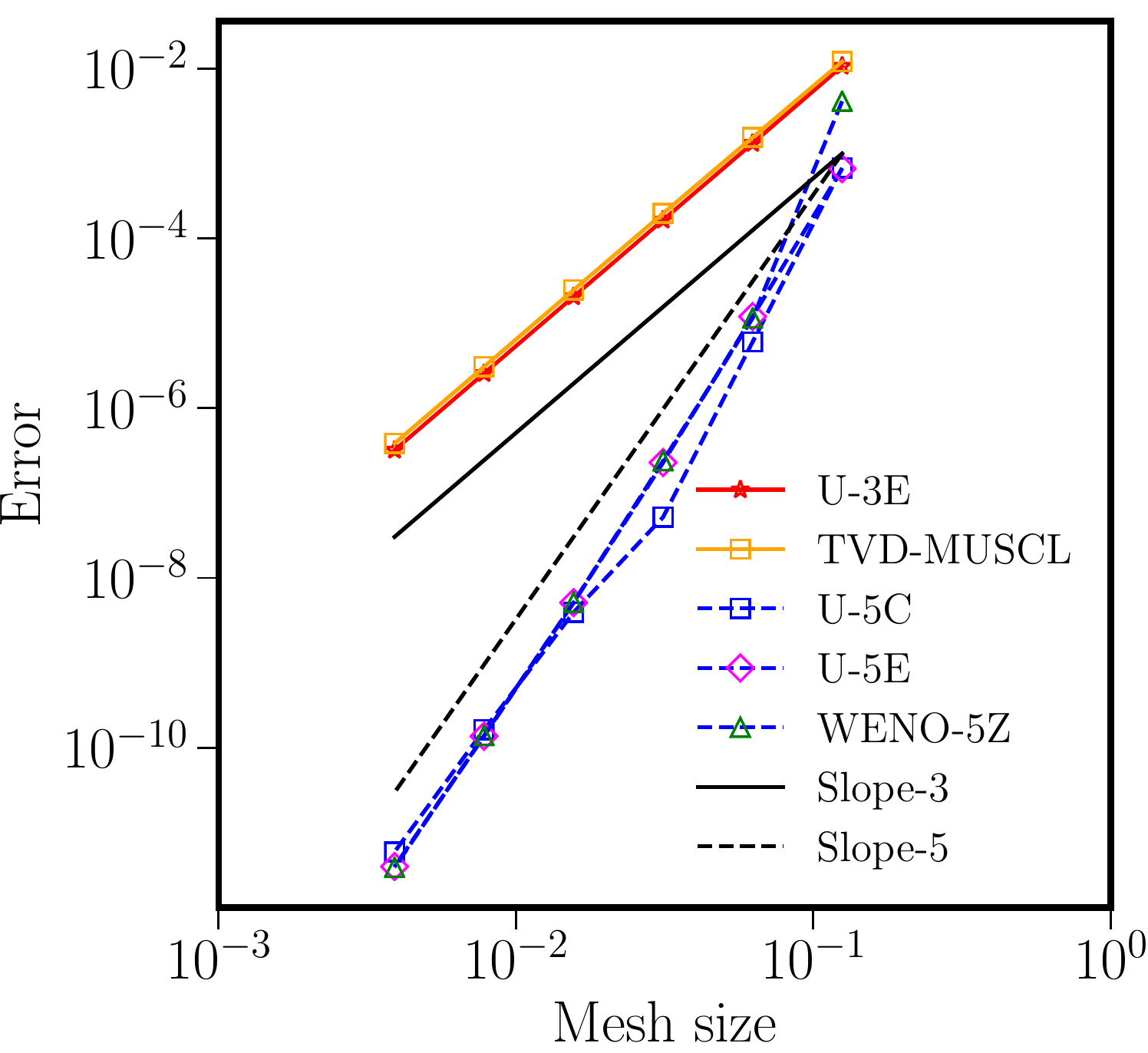}
\label{fig:Re-1-error}}
\subfigure[$Re=10$]{\includegraphics[width=0.3\textwidth]{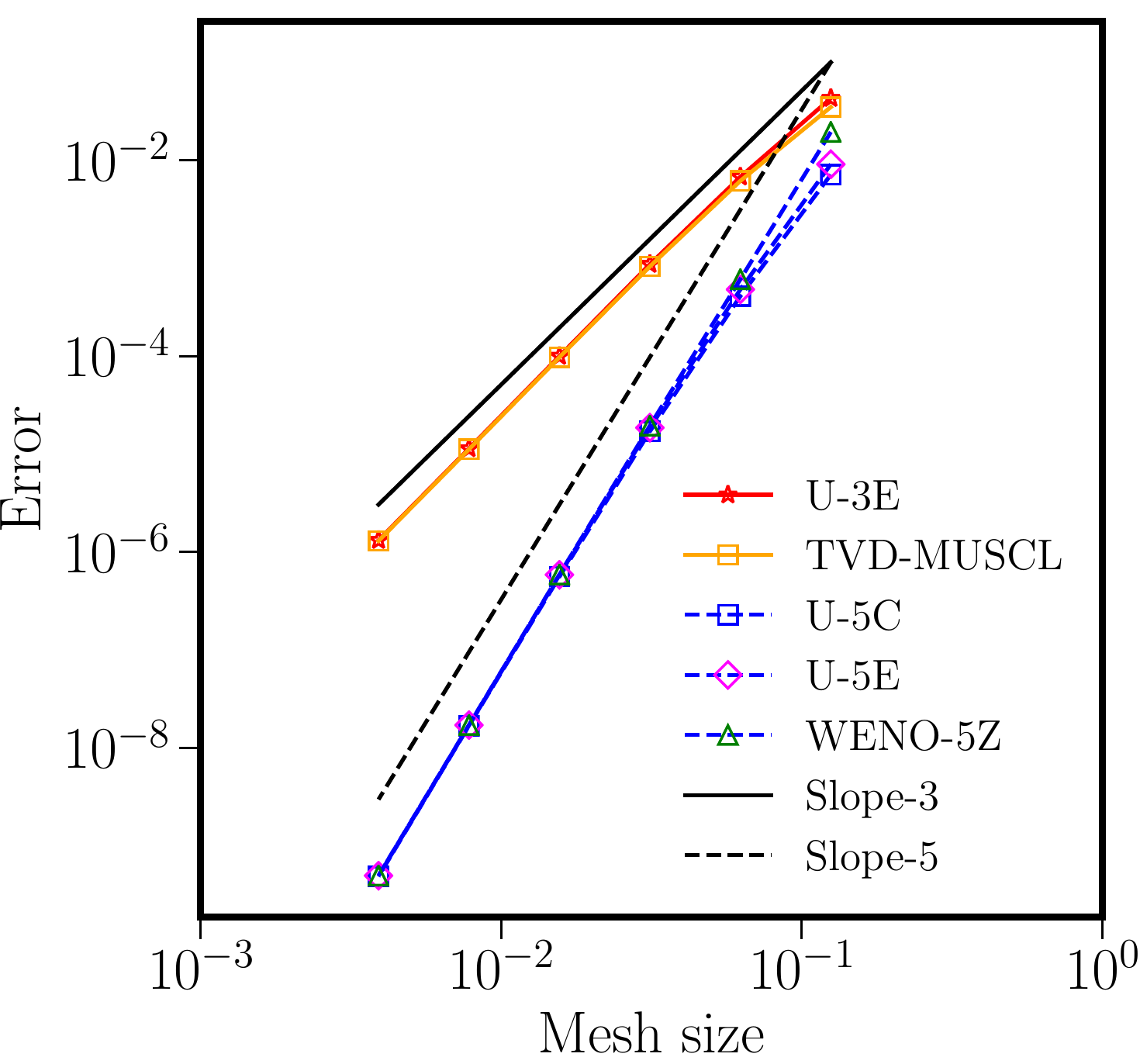}
\label{fig:Re-10-error}}
\caption{{\color{black}Convergence of the $L_2$ error for upwind schemes for $Re$ = 0.1, 1 and 10 for \hyperref[sec:4.2.1]{Example 4.2.1}. }}
\label{fig:1D-Re}
\end{figure}

\begin{figure}[H]
\centering
\subfigure[Global profile of all schemes]{\includegraphics[width=0.46\textwidth]{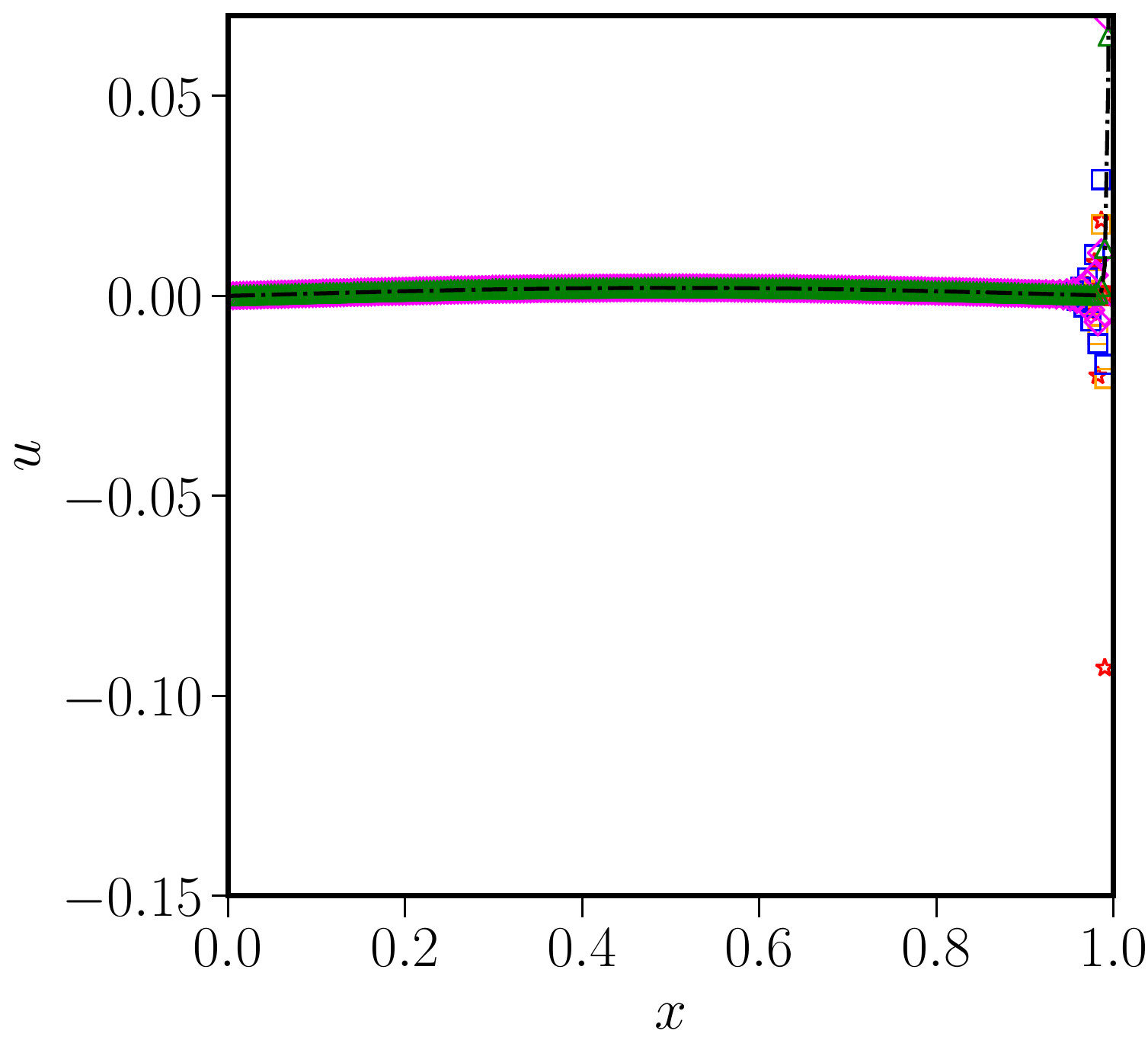}
\label{fig:global}}
\subfigure[Local profile of all schemes]{\includegraphics[width=0.46\textwidth]{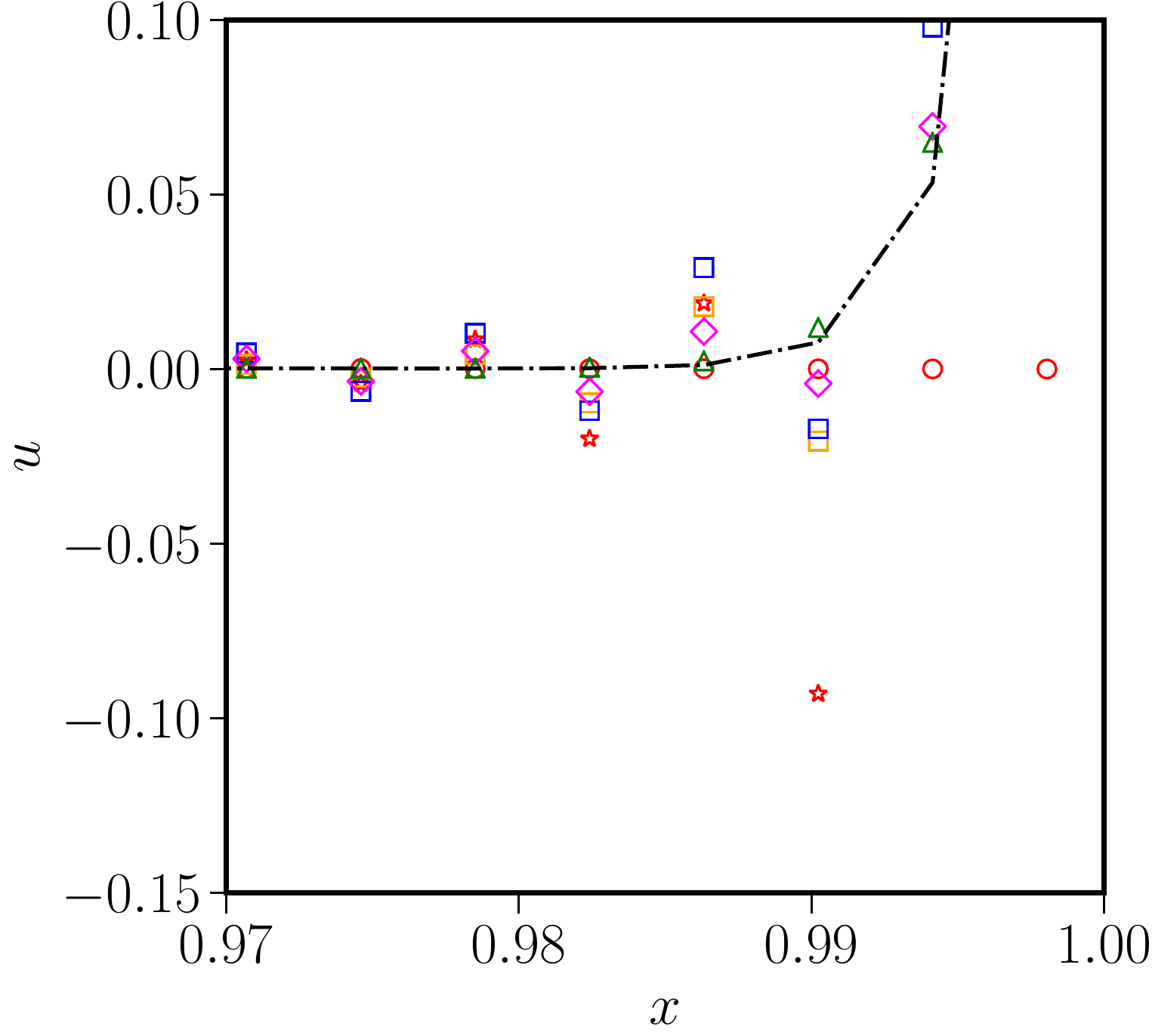}
\label{fig:local-all}}
\subfigure[Local profile third order schemes]{\includegraphics[width=0.46\textwidth]{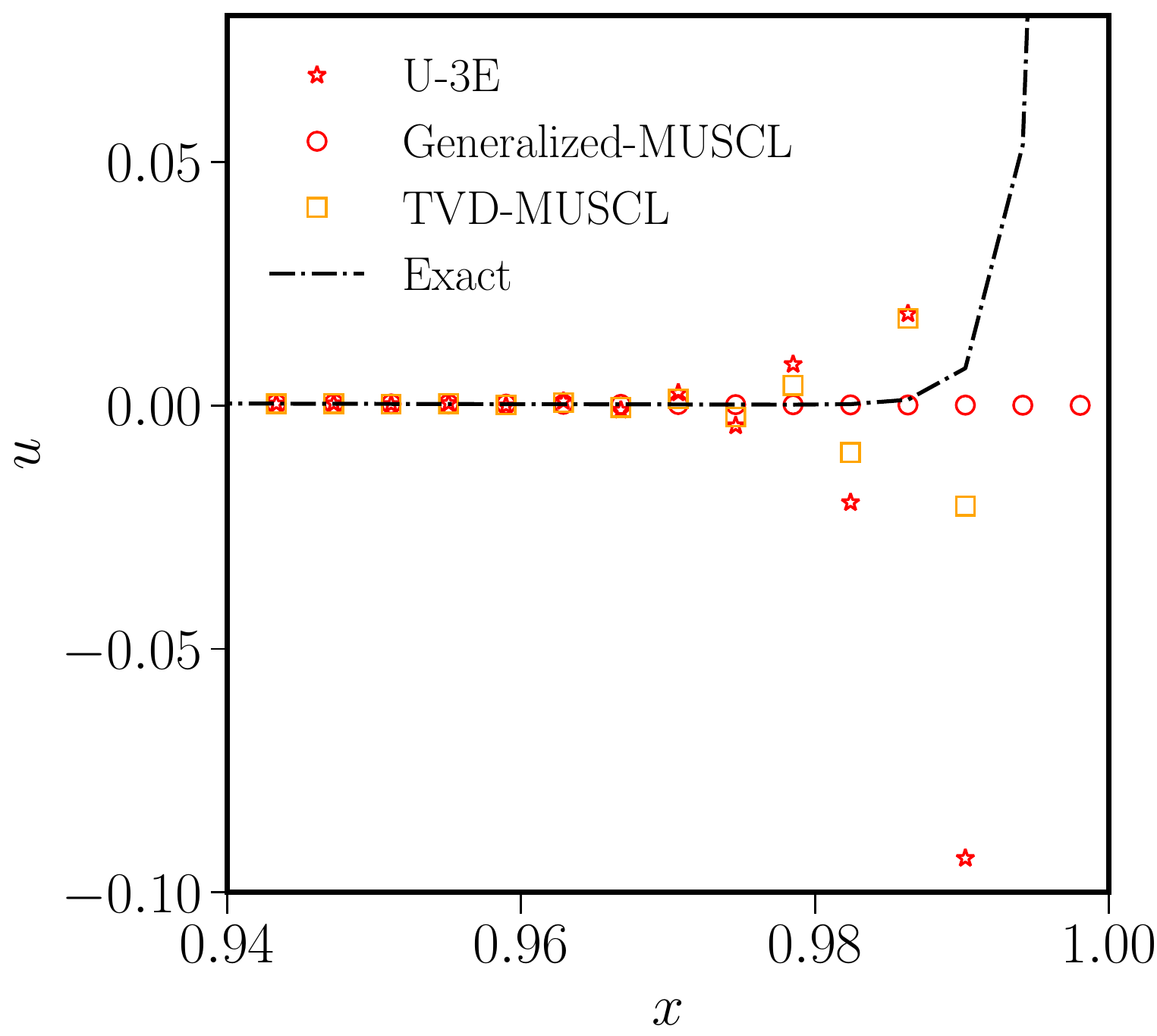}
\label{fig:local-3}}
\subfigure[Local profile fifth order schemes]{\includegraphics[width=0.46\textwidth]{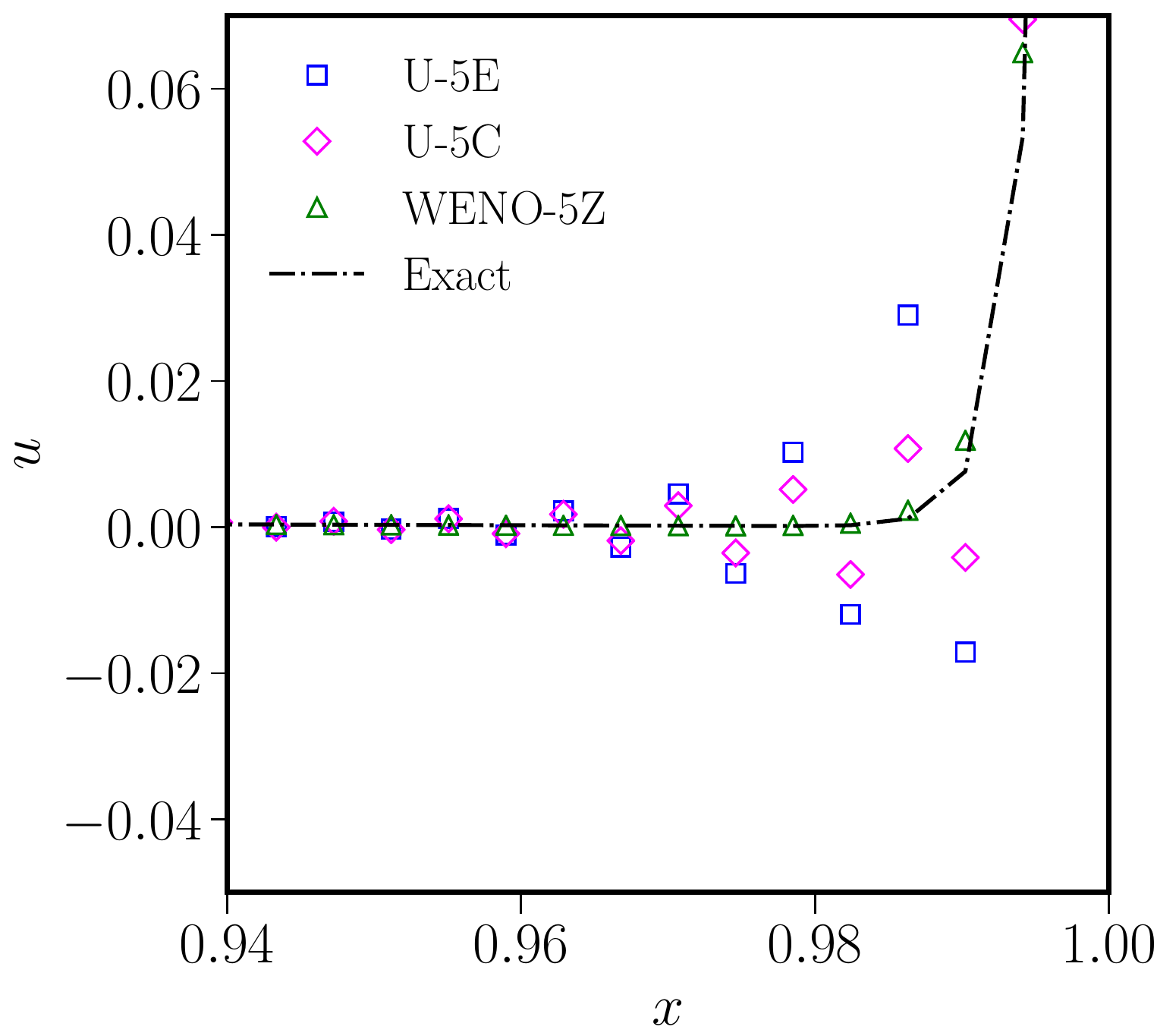}
\label{fig:local-5}}
\caption{{\color{black}Numerical and exact (dotted line) solutions obtained by various schemes for $Re= 500$ on a grid of 256 points.}}
\label{fig:re-500}
\end{figure}

The simulations are conducted on a uniform grid with grid refinements from $N$ = 8 to 256 by all the upwind schemes. The numerical solution is computed by an explicit time-marching until the residuals are dropped below $10^{-10}$ in $L_1$ norm. A constant value of CFL is considered depending on the order of the scheme. Simulations are carried out for three Reynolds numbers: $Re= 0.1, 1 ,10$. We can see that the design order of accuracy is obtained for all the schemes for all the Reynolds numbers in Fig. \ref{fig:1D-Re}. The 5th order compact scheme is one order of magnitude more than the design accuracy for $Re=0.1$. Fig. \ref{fig:Re-ana} shows the solutions contours of various schemes in comparison with the analytical solution for the primary variable, $u$. {\color{black} From Fig. \ref{fig:Re-TVD}, it is obvious to notice that the Generalized MUSCL scheme can pollute the smooth profile whereas TVD-MUSCL reduces to linear 3rd order scheme.} For Re = 1 there were unnecessary oscillations in the profile, and for the Reynolds number of 10, even though there are no oscillations, the sharp gradient has been completely cut-off. Numerical results obtained by linear upwind schemes were, in fact, better than the results obtained by {\color{black} Generalized MUSCL scheme.}
To obtain the solution without oscillations in the case of high-Re, extremely fine grids are required to meet the well-known requirement on the mesh Reynolds-number, i.e., $Re < 2$, as explained by Nishikawa \cite{Nishikawa2010b}. Computations are carried out by Nishikawa on substantially coarser grids by grid stretching to meet the criteria. Mazaheri and Nishikawa \cite{mazaheri2014N} had merely refined the grid to satisfy the mesh Reynolds-number requirement by using a highly efficient implicit solver. In this test case, we incorporated the non-oscillatory schemes, WENO, to capture the sharp gradients without oscillations on uniform meshes which might not satisfy the mesh Reynolds-number requirement. Simulations carried out for the Reynolds number of 500 on a uniform grid size of 256 has been shown in Fig. \ref{fig:re-500}. {\color{black}Generalized MUSCL scheme did not show any oscillations but completely cut-off the sharp gradient and thereby failing to capture it. On the other hand, TVD-MUSCL has reduced oscillations compared to that of linear 3rd order scheme. Oscillations are also observed in both 5th order schemes. However, the amplitude of the oscillations is much smaller for the compact scheme than the explicit scheme as expected. Finally, sharp gradient developed at $x=1$ is easily captured by WENO scheme without spurious oscillations as shown in Fig. \ref{fig:local-5}.} Also, it is important to note that the linear schemes can also give oscillation free solutions on dense grids by satisfying the mesh-Reynolds number criteria. These important inferences were found to be useful for the simulation of magnetized electrons which can develop sharp gradients with increasing strength of magnetic confinement.
\noindent \textbf{Example 4.2.2.}\label{sec:4.2.2}
The two-dimensional advection-diffusion equation is solved by split flux approach for two test cases. 
\begin{equation}\label{eqn:adv-2d}
u_{t} + a u_x + b u_y = \nu (u_{xx}+u_{yy}), \text {in domain (x,y) $\in$ [0,1] $\times$ [0,1]} 
\end{equation}

\begin{enumerate}

\item The first test case is an advection-dominated problem considered by Nishikawa and Liu \cite{Nishikawa2018} which has the following exact solution,
 \begin{equation}
u_{exact}(x,y) = \cos(2 \pi \eta) \exp \left(\frac{-8\pi^2\nu\xi}{1+\sqrt{1+16\pi^2\nu^2}}\right),
\label{eqn:adv-only}
\end{equation}
where $\xi$ = ax + by, $\eta$ = bx - ay and the viscosity, $\nu =0.01$. The advection vector $(a,b)$ is given by $(1.0, 0.0)$. This exact solution is given by Eq. (\ref{eqn:adv-only}) is very smooth and therefore can be used to verify the accuracy of numerical schemes. 

\item For the second test case the following exact solution is considered, 
 \begin{equation}\label{eqn:adv-diff}
u_{exact}(x,y) = -\cos( \pi \eta) \exp \left(\frac{0.5 \xi(1-\sqrt{1+4\pi^2\nu^2})}{\nu}\right),
\end{equation}
where the viscosity coefficient is $\nu = 0.1$, and the advection vector $(a,b)$ is given by $(7.0, 4.0)$ respectively. This test case was originally proposed by Nishikawa and Roe in \cite{Nishi_fluctuation}, where both advective and the diffusive terms are equally important which may exist in the middle of the boundary layer.
\end{enumerate}
Computations are carried out with grid refinements from 8$\times$8 to 128 $\times$ 128 for both the test cases. For all the numerical schemes, the simulations are carried out until the residuals are dropped below $10^{-14}$ in $L_1$ norm and CFL is dependent on the numerical scheme used, which is in between from 0.3 to 0.7. Both WENO extrapolation denoted as WENO-5Z-W, and Lagrange extrapolation, WENO-5Z-L, are implemented for the numerical boundary conditions for the WENO scheme, which are 3rd order accurate.

\begin{figure}[H]
\footnotesize
\centering
\subfigure[Case-1]{%
\includegraphics[width=0.45\textwidth]{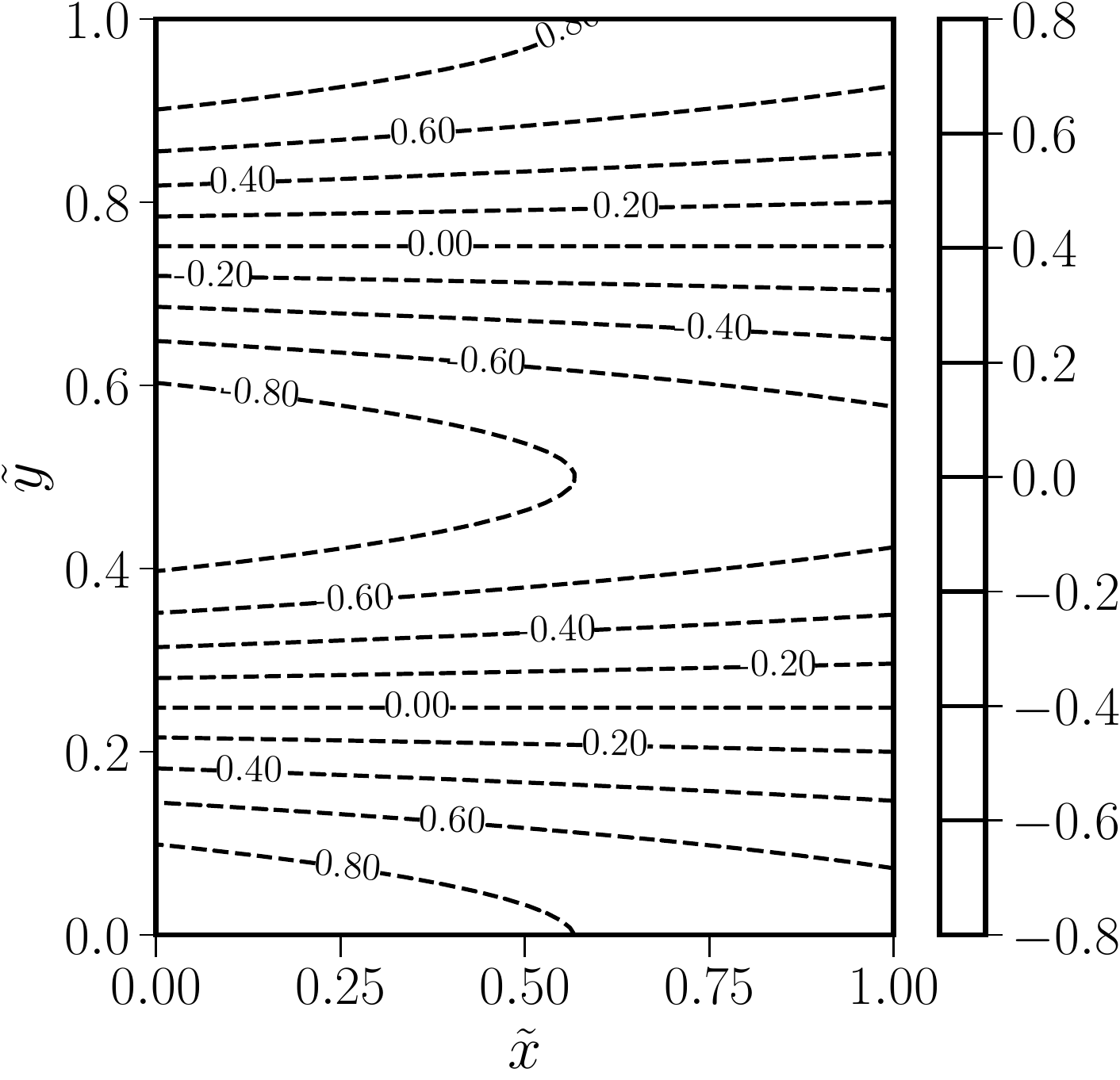}
\label{fig:5e-case-1}}
\subfigure[Case-2]{%
\includegraphics[width=0.45\textwidth]{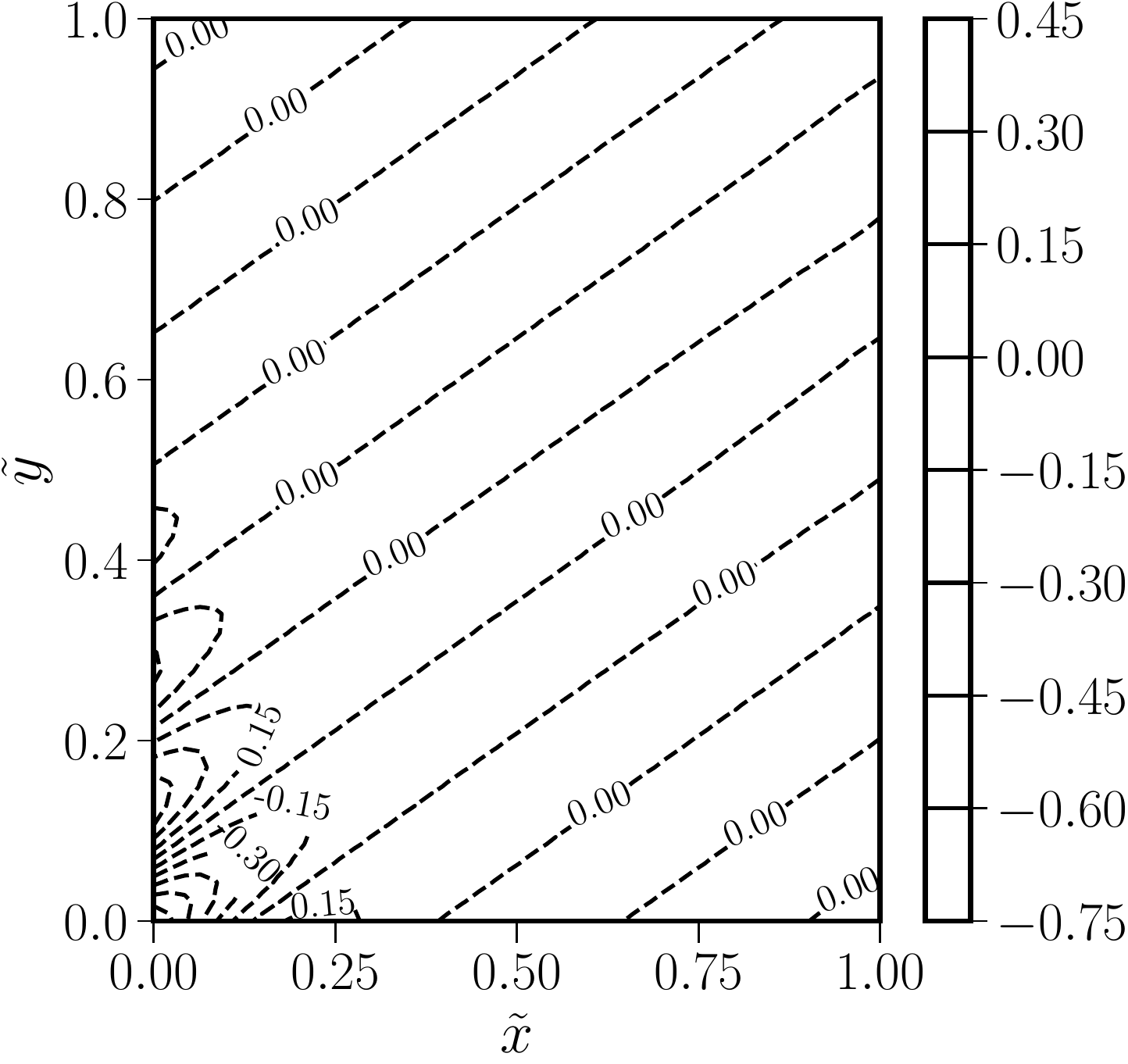}
\label{fig:5e-case-2}}
\caption{Comparison of velocity contours, on a grid size of 48 $\times$ 48, for test conditions given by \hyperref[sec:4.2.2]{Example 4.2.2}.}
\label{fig:2d-cases}
\end{figure}

\begin{figure}[H]
\footnotesize
\centering
\subfigure[Advection dominated]{\includegraphics[width=0.45\textwidth]{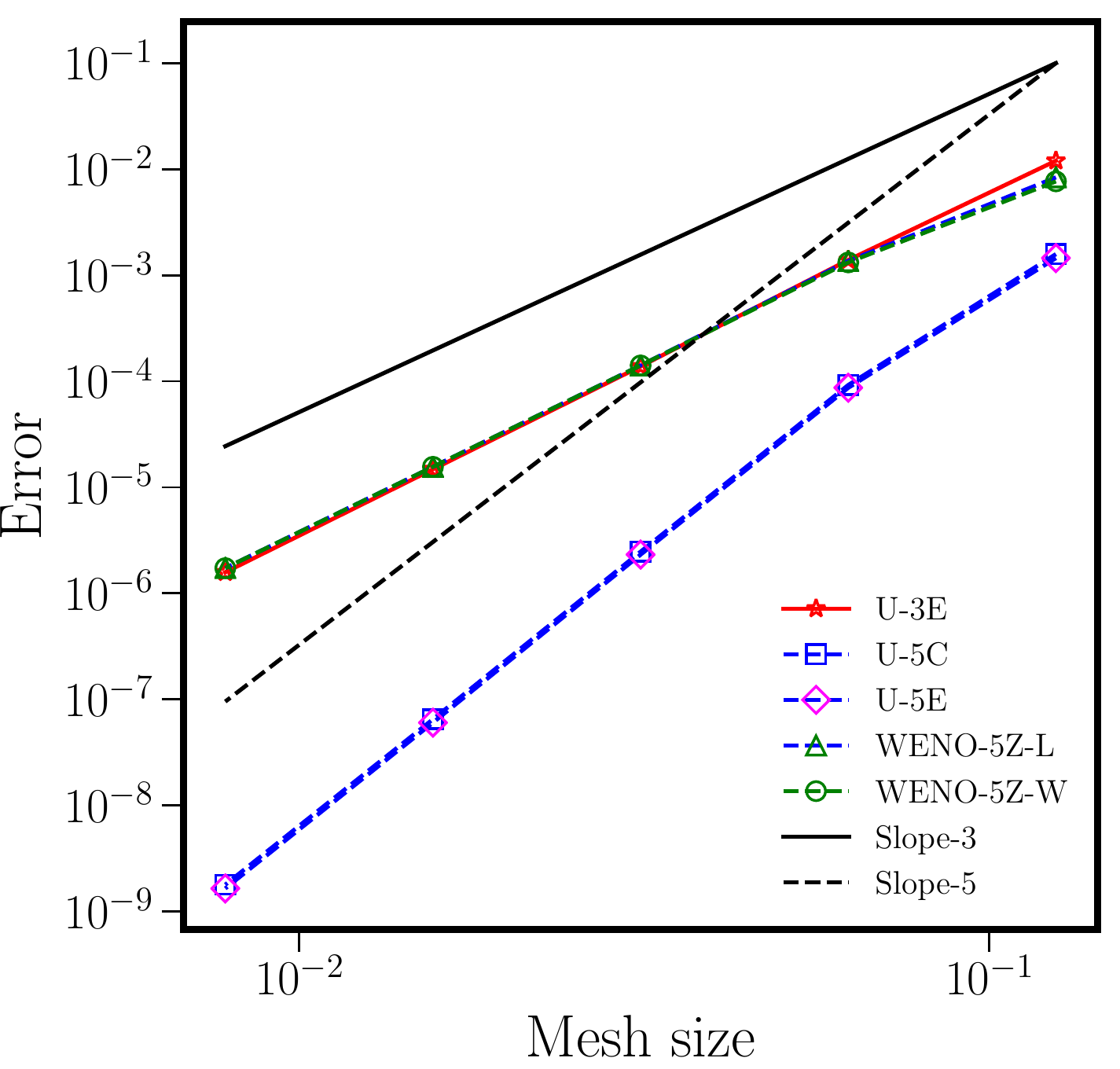}
\label{fig:case-1-error}}
\subfigure[Both Advection-Diffusion dominant]{\includegraphics[width=0.45\textwidth]{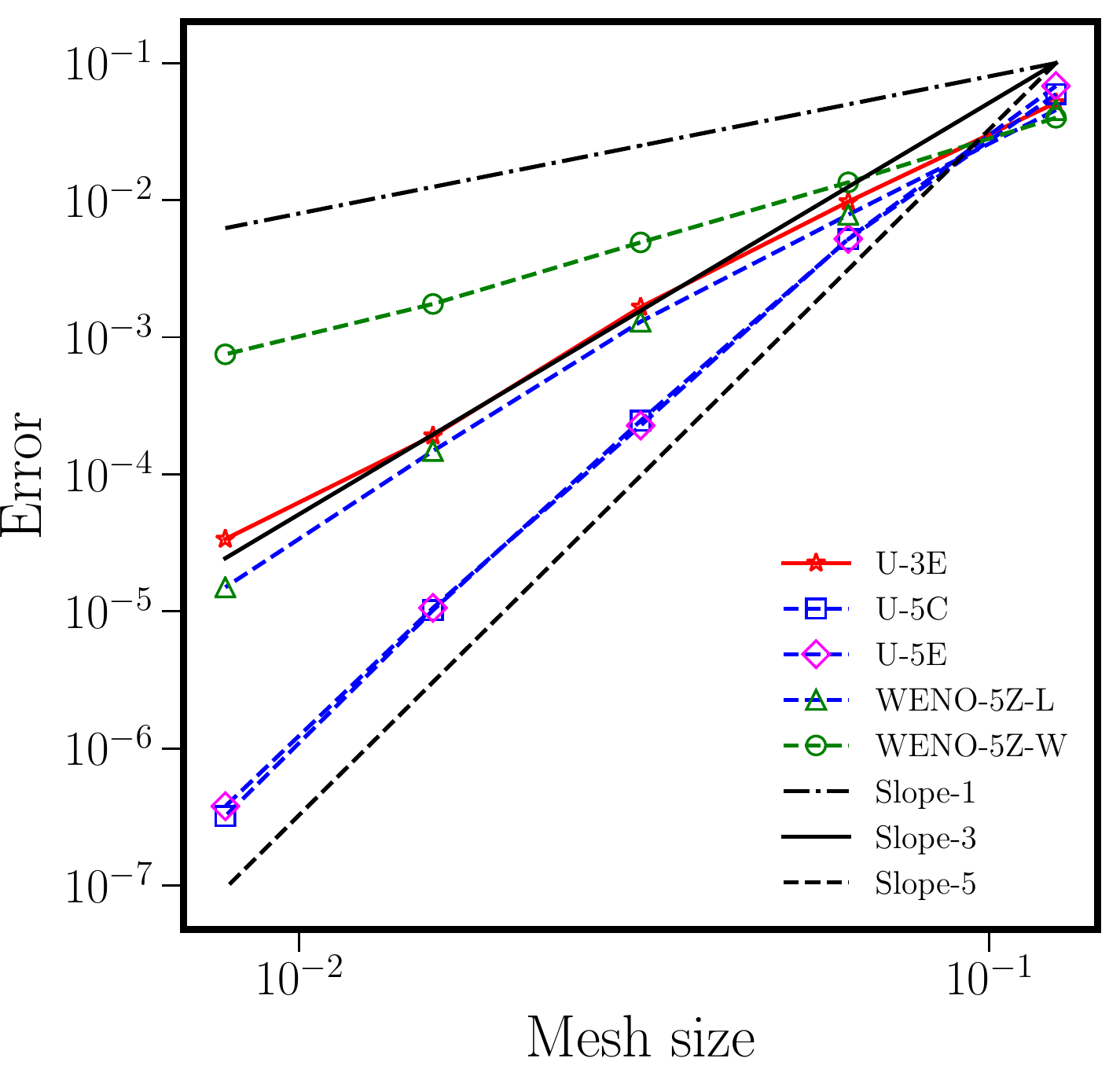}
\label{fig:case-2-error}}
\caption{Convergence of the $L_2$ error for all the upwind schemes for both the test conditions in \hyperref[sec:4.2.2]{Example 4.2.2}.}
\label{fig:2d-cases-l2}
\end{figure}

Figs. \ref{fig:5e-case-1} and \ref{fig:5e-case-2} show the velocity contours obtained by the upwind scheme, U-5E, for both the test cases on a 48 $\times$ 48 grid. Design accuracy is obtained for all the linear upwind schemes, shown in Fig. \ref{fig:case-1-error} and the implementation of boundary conditions for advection dominated problem are verified through this test case. First the first test condition, WENO scheme is third accurate for both WENO and Lagrange type extrapolations which is unlike the one-dimensional test case where the WENO scheme was fifth order accurate.  %
%
%
%
The second test case was found to be more challenging than the advection-dominant flow. Design accuracy is obtained for all the linear upwind schemes, shown in Fig. \ref{fig:case-2-error}. For the WENO scheme, the WENO extrapolation technique for the boundary conditions was observed to be first order accurate whereas the Lagrange extrapolation is 3rd order accurate similar to the advection-dominated problem. Table \ref{tab:cases} show the $L_2$ norms for both the test cases for both WENO implementations.
\begin{table}[H]
  \centering
\footnotesize
  \caption{$L_2$ errors and order of convergence for \hyperref[sec:4.2.2]{Example 4.2.2} by using WENO scheme along with WENO and Lagrange extrapolation techniques for boundary conditions.}
    \begin{tabular}{| c | c | c | c | c || c | c | c | c|}
    \hline
     Number & \multicolumn{4}{c||}{Advection} & \multicolumn{4}{c|}{Advection-Diffusion}  \\
    \cline{2-9}
     of points & \multicolumn{2}{c|}{WENO-5Z-L} & \multicolumn{2}{c||}{WENO-5Z-W} & \multicolumn{2}{c|}{WENO-5Z-L}& \multicolumn{2}{c|}{WENO-5Z-W}  \\
    \cline{2-9} 
    & error & order & error & order & error & order & error & order   \\
    \cline{1-9}
     $8^2$     & 8.36E-03 &       & 7.67E-03 &       & 4.55E-02 &       & 3.99E-02 &  \\
     \hline
    $16^2$    & 1.36E-03 & 2.62  & 1.32E-03 & 2.53  & 7.83E-03 & 2.54  & 1.35E-02 & 1.56 \\
     \hline
    $32^2$    & 1.41E-04 & 3.27  & 1.41E-04 & 3.23  & 1.30E-03 & 2.59  & 4.92E-03 & 1.45 \\
     \hline
    $64^2$    & 1.56E-05 & 3.18  & 1.56E-05 & 3.18  & 1.48E-04 & 3.14  & 1.75E-03 & 1.49 \\
     \hline
    $128^2$   & 1.73E-06 & 3.17  & 1.73E-06 & 3.17  & 1.50E-05 & 3.30  & 7.51E-04 & 1.22 \\
     \hline
    \end{tabular}%
  \label{tab:cases}%
\end{table}%
%
			
															\noindent \textbf{Example 4.2.3.}\label{sec:4.2.3}
In this test case, we solve the steady state problem of the two-dimensional advection-diffusion equation, with boundary layers along x = 1 and y = 1, considered by Chou and Shu \cite{chou2007high} which has the following exact solution given by 
\begin{equation}
u_{exact}(x,y) = e^{\frac{x-1}{\nu} + \frac{y-1}{\nu}}.
\end{equation}
The viscosity coefficient $\nu$ is taken as 0.05, and the exact solution is imposed on the boundaries. Linear upwind schemes are not considered for this problem as they will produce spurious oscillations. The numerical results are shown in Table \ref{tab:adv-w-2} indicate that the third order accuracy is obtained for WENO-5Z-L whereas the accuracy of WENO-5Z-W is reduced to first order. This test case demonstrates that the WENO scheme can resolve the boundary layers by using first-order hyperbolic approach. 
\begin{table}[H]
  \centering
  \caption{$L_2$ errors and order of convergence for \hyperref[sec:4.2.3]{Example 4.2.3} by using WENO scheme along with WENO and Lagrange extrapolation techniques for boundary conditions.}
\footnotesize
    \begin{tabular}{| c | c | c | c | c |}
\hline
    Number & \multicolumn{2}{c|}{WENO-5Z-W}& \multicolumn{2}{c|}{WENO-5Z-L}  \\
    \cline{2-5}
    of points & error & order & error & order   \\
    \cline{1-5}
        $20^2$      & 2.21E-03 &   & 1.91E-03 &  \\
        \hline
    $40^2$      & 1.07E-03 & 1.04  & 2.00E-04 & 3.26 \\
    \hline
    $80^2$      & 4.60E-04 & 1.22  & 2.34E-05 & 3.09 \\
        \hline
    $160^2$      & 1.99E-04 & 1.21  & 2.59E-06 & 3.18 \\
        \hline
    $320^2$     & 9.06E-05 & 1.14  & 2.68E-07 & 3.27 \\
        \hline
    \end{tabular}%
  \label{tab:adv-w-2}%
\end{table}%

										\subsection{Magnetized electron fluid simulation}
										
Magnetized electron fluid simulations can have all the features that are discussed in earlier sections like sharp gradients in the flow field, strongly diffusion dominated flow with anisotropic diffusion and also have the convection aspect of the flow. A key distinction between our present methodology and the earlier efforts based on first order upwind and third order accurate monotonicity preserving TVD discretization is in the implementation of high-order finite-volume WENO discretization. The gain in accuracy and reduction in numerical oscillations are significantly more pronounced for the WENO formulations when compared with TVD discretization. Before we discuss the results of the magnetized electron fluids we briefly summarized the inferences from the diffusion and advection-equations to better understand the issues faced:
\begin{itemize}
\item Both diffusion and advection-diffusion equations, linear upwind schemes are sufficient for problems with smooth solutions.
\item {\color{black}WENO schemes are more appropriate for capturing the sharp gradients than TVD type scheme.}
\item Ghost cell approach is preferable over the weak boundary. WENO extrapolation can reduce to first order in some instances and yet provide robust results.
\end{itemize}

\subsubsection{Steady state results and analysis}
\begin{figure}[H]
\centering
{\includegraphics[width=0.70\textwidth]{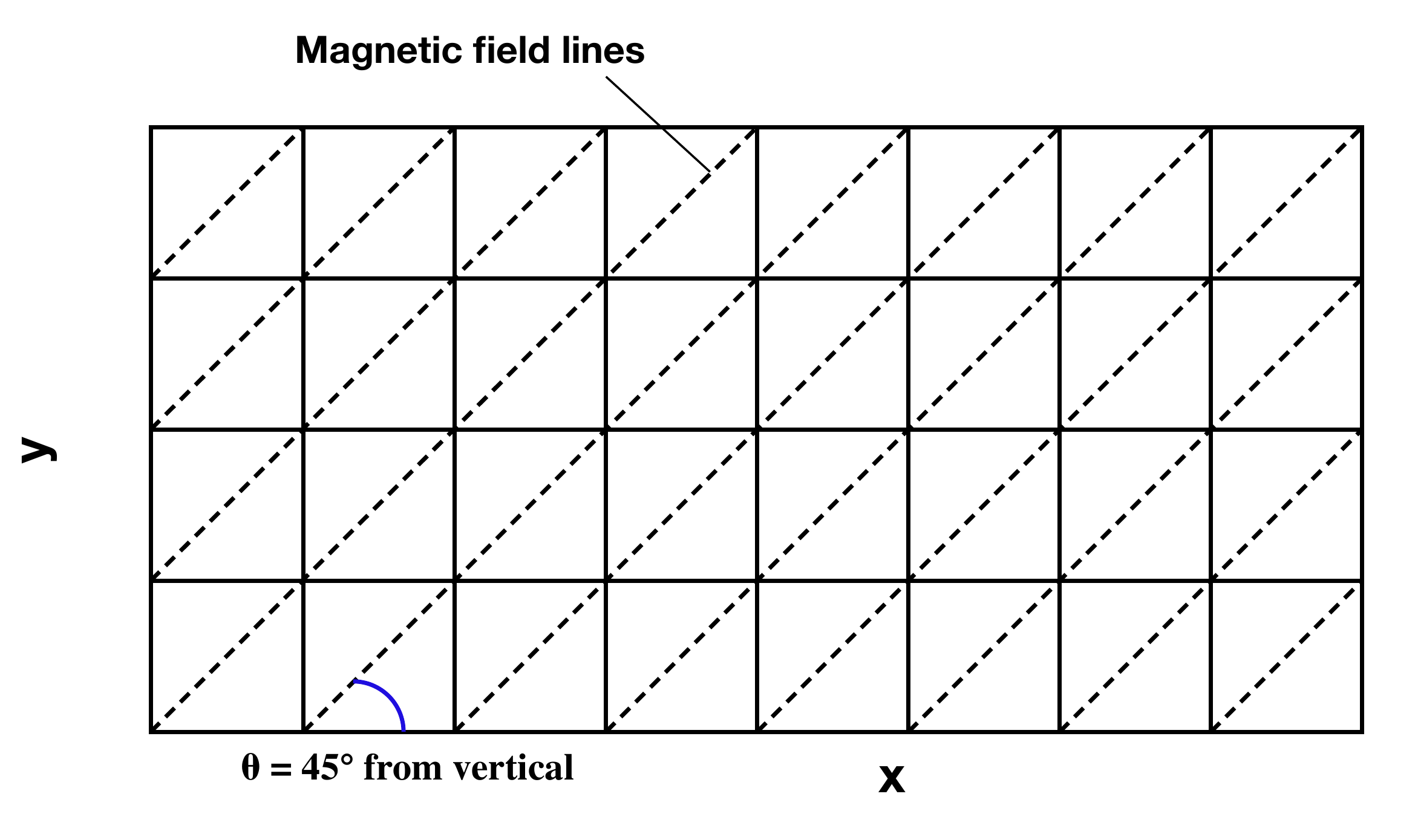}}
\caption{Sketch of the magnetic field lines for  $45^{\circ}$ angle.}
\label{fig:45}
\end{figure}

We consider the test calculation considered by Kawashima et al. \cite{Kawashima2015} with uniformly angled Magnetic lines of force at $45^{\circ}$ from the vertical, shown in Fig. \ref{fig:45}. For a Cartesian mesh, the condition of 45$^{\circ}$ magnetic lines of force can give significant false diffusion \cite{patankar1980}. False diffusion will occur due to the oblique flow direction and non-zero gradient in the direction normal to the flow. As a consequence of false diffusion, nonphysical local extrema can occur in regions in which the gradients of the solution are steep and not aligned with the orientation of grid, and the discretization method will be unable to capture them properly. Results were obtained for three different values of magnetic confinements, ${\mu}_{||}/{\mu}_{\perp}$=100, 500 and 1000, which are uniform throughout the domain. Dirichlet boundary conditions for the non-dimensional space potential are defined at the left and right side boundaries and zero-flux conditions, which are also Dirichlet, are used for the top and bottom boundaries, shown in Eq. (\ref{dirvel}). 
\begin{equation}
    \tilde\phi= 
\begin{cases}
   1,&  x= 0\\
    0,&  x = 200
\end{cases}
\quad 
\tilde u_y = 
\begin{cases}
   0,&  y= 0\\
    0,&  y = 100
\end{cases}  
\label{dirvel}
\end{equation}
\noindent Unlike the test cases of diffusion and advection-diffusion where the numerical schemes are compared with the analytical solutions, the linear and WENO upwind schemes are compared with the Magnetic Field Aligned Mesh (MFAM). MFAM is described by Mikellides et al. \cite{mikemfam} and Kawashima et al. \cite{Kawashima2015}, which eliminates the false diffusion by aligning the mesh with the magnetic field. The simulations are conducted for both MFAM and the upwind schemes with grid refinements of $NX, NY$ = [96,48] to [768,384] for all the values of ${\mu}_{||}/{\mu}_{\perp}$. For the upwind methods, the numerical solution is computed by explicit time marching with a constant CFL of $0.32$. Regardless of the interior scheme, 3rd order accurate boundary conditions are considered for all the upwind schemes for stable computation. The steady-state solution is deemed to be reached if the $L_1$ norms are below $10^{-12}$. As far the MFAM approach, successive over-relaxation is used to numerically relax the solution until the $L_2$ norms are below $10^{-10}$ by second-order central discretization.

\begin{figure}[H]
  \begin{sideways}
  \begin{minipage}{17.5cm}
\subfigure[${\mu}_{||}/{\mu}_{\perp}$=100]{\includegraphics[width=0.32\textwidth]{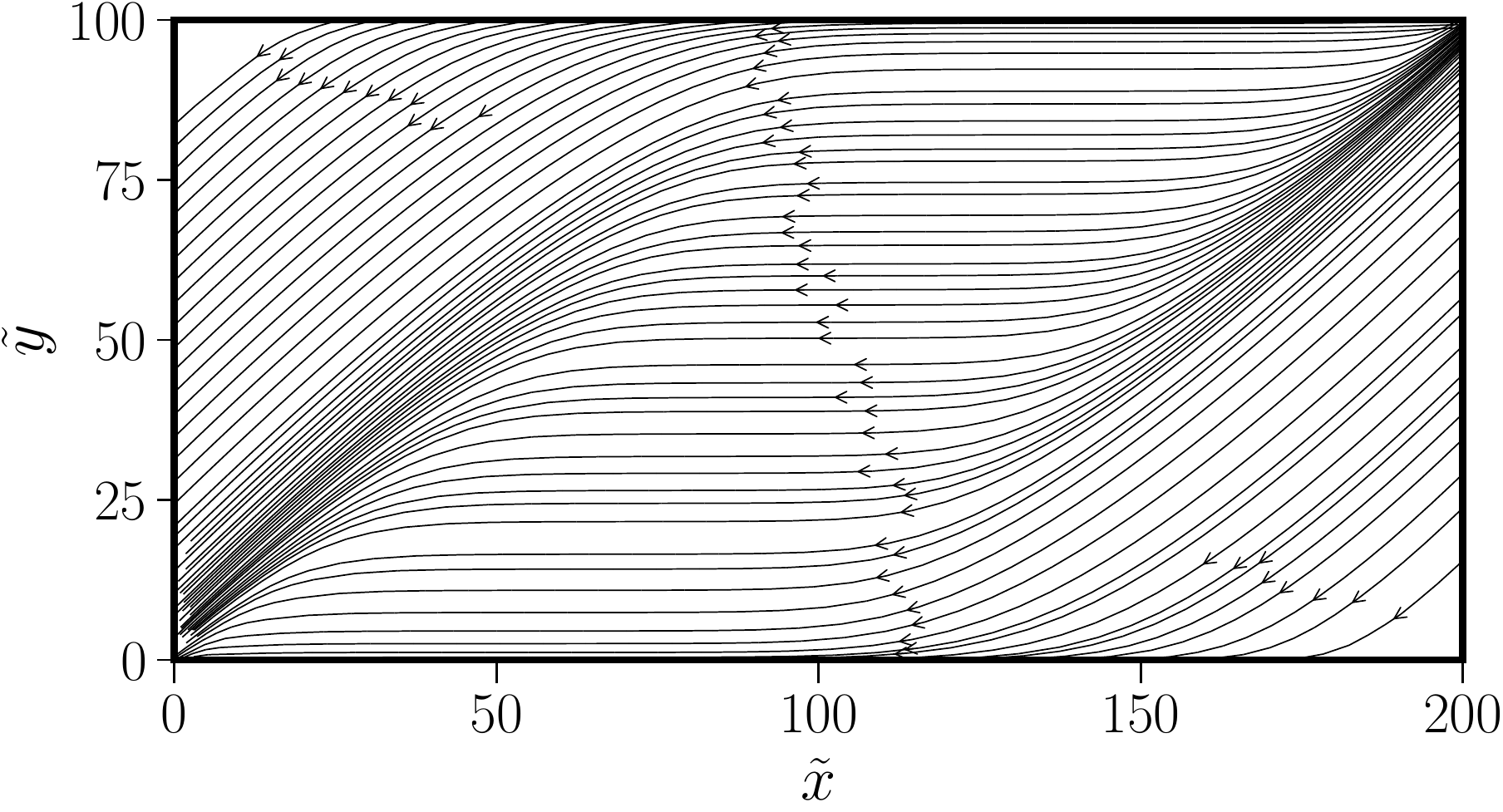}\hspace{0.01em}
\label{fig:3E-100-S}}
\subfigure[${\mu}_{||}/{\mu}_{\perp}$=500]{\includegraphics[width=0.32\textwidth]{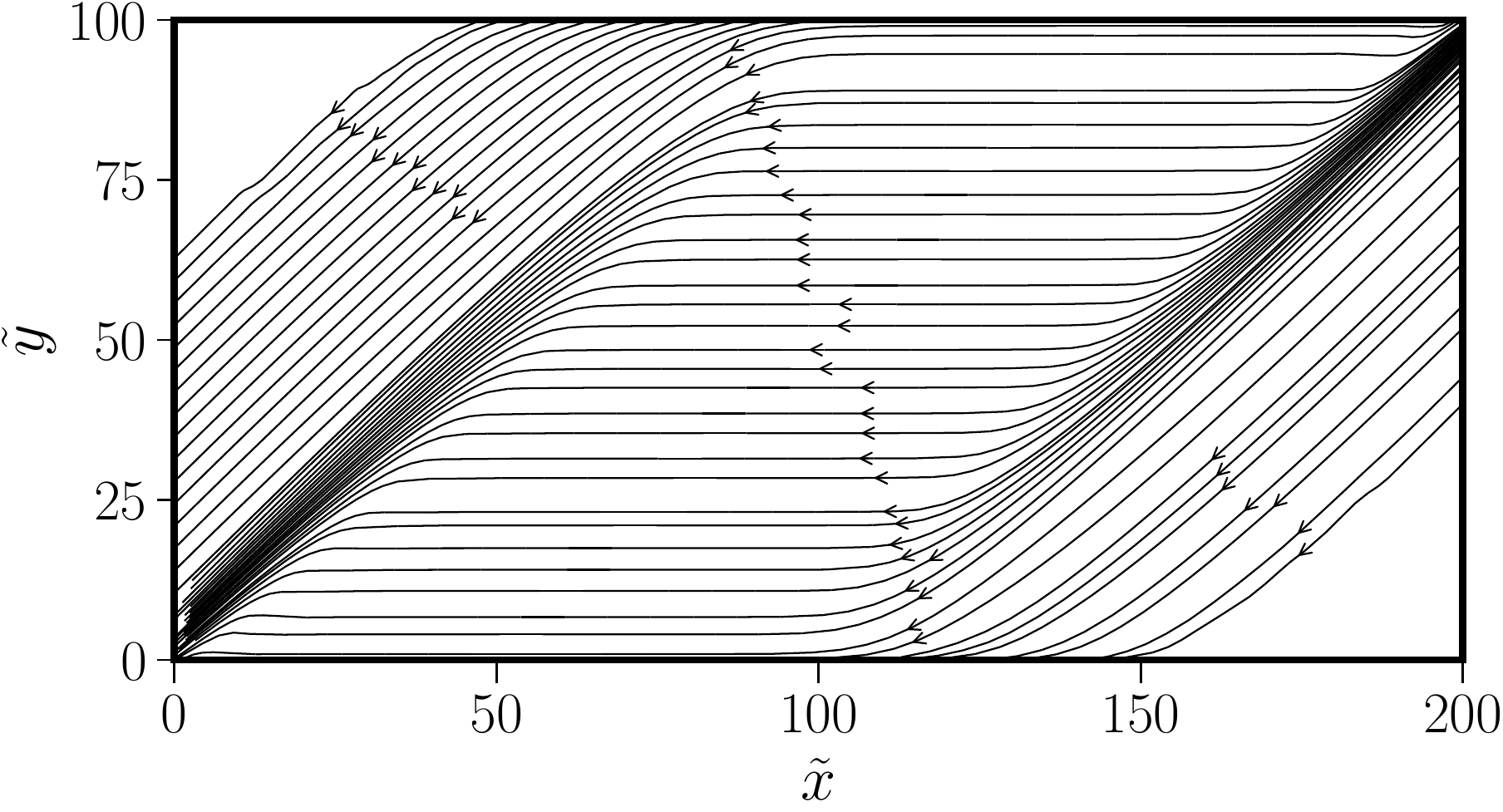}\hspace{0.01em}
\label{fig:3E-500-S}}
\subfigure[${\mu}_{||}/{\mu}_{\perp}$=1000]{\includegraphics[width=0.32\textwidth]{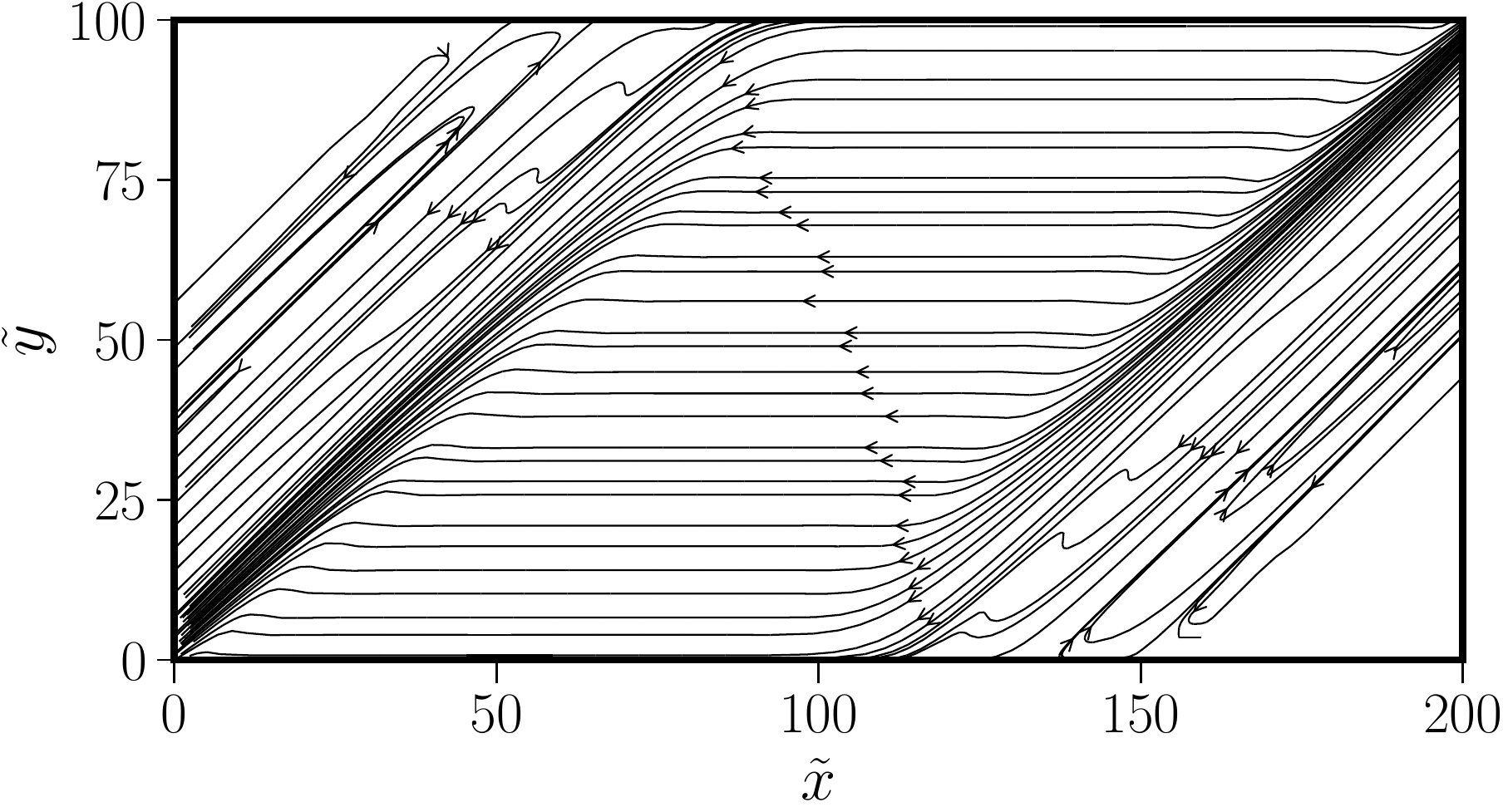}
\label{fig:3E-1000-S}}
\subfigure[${\mu}_{||}/{\mu}_{\perp}$=100]{\includegraphics[width=0.32\textwidth]{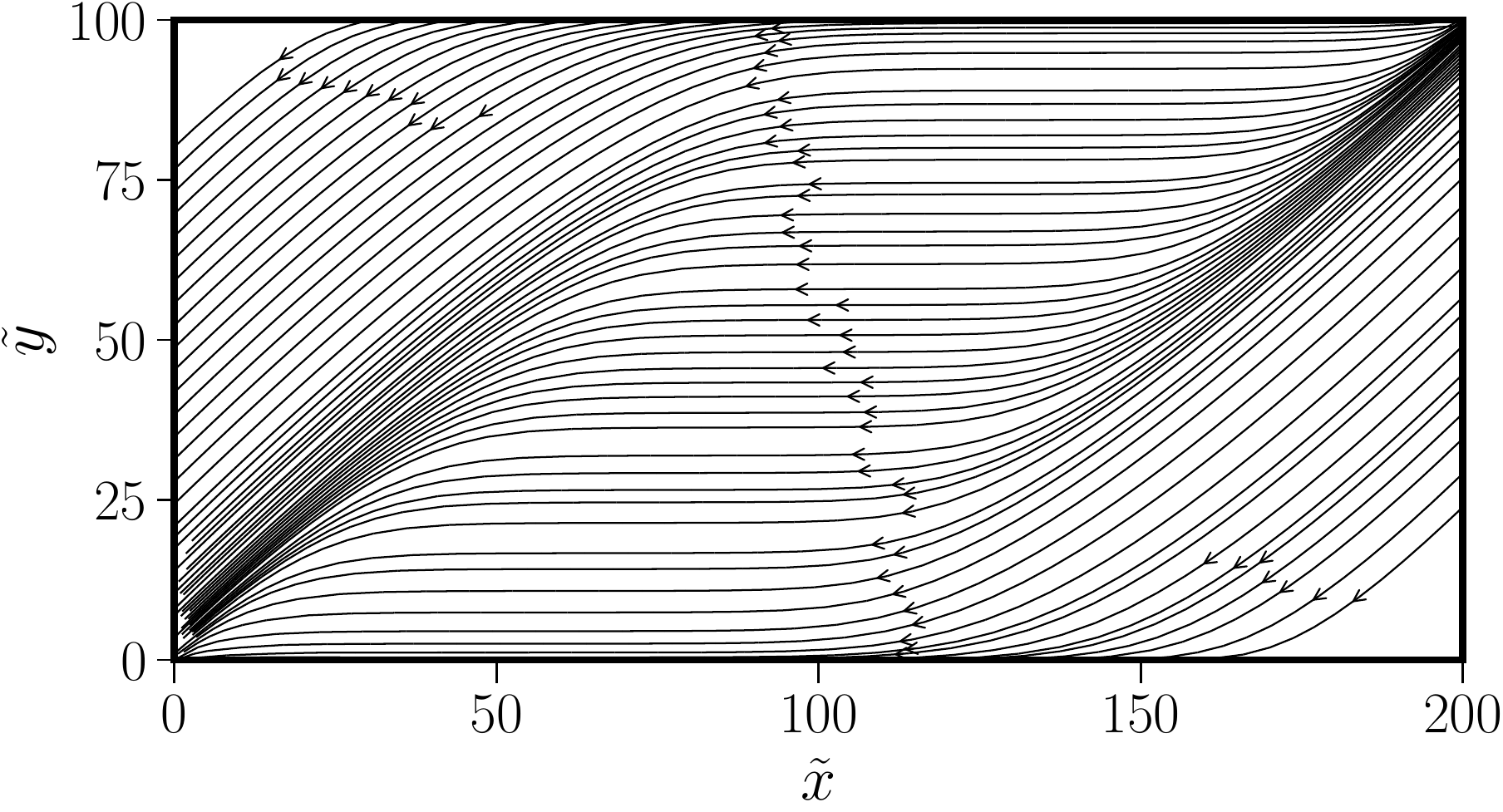}
\label{fig:5C-100-S}}
\subfigure[${\mu}_{||}/{\mu}_{\perp}$=500]{\includegraphics[width=0.32\textwidth]{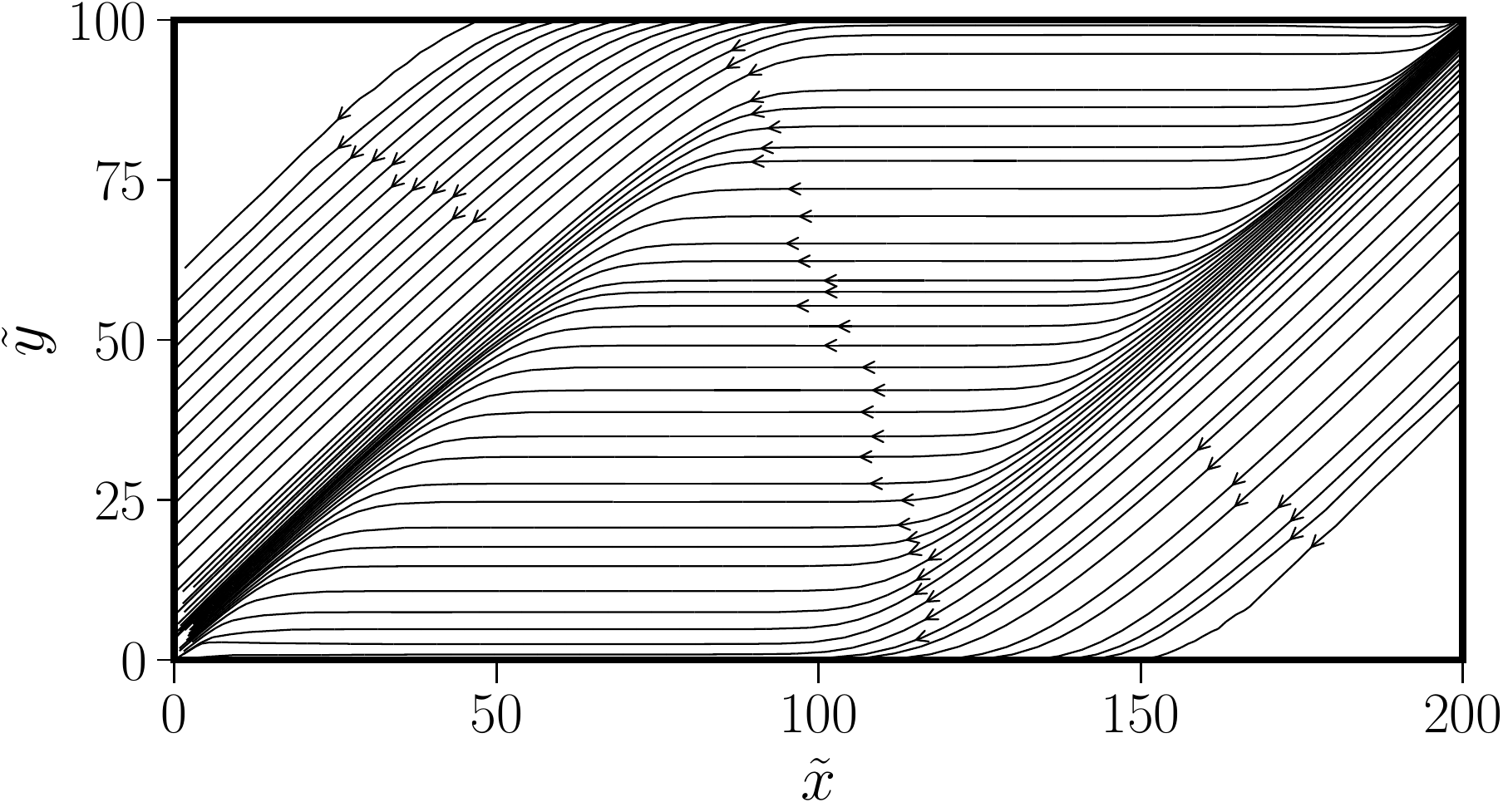}
\label{fig:5C-500-S}}
\subfigure[${\mu}_{||}/{\mu}_{\perp}$=1000]{\includegraphics[width=0.32\textwidth]{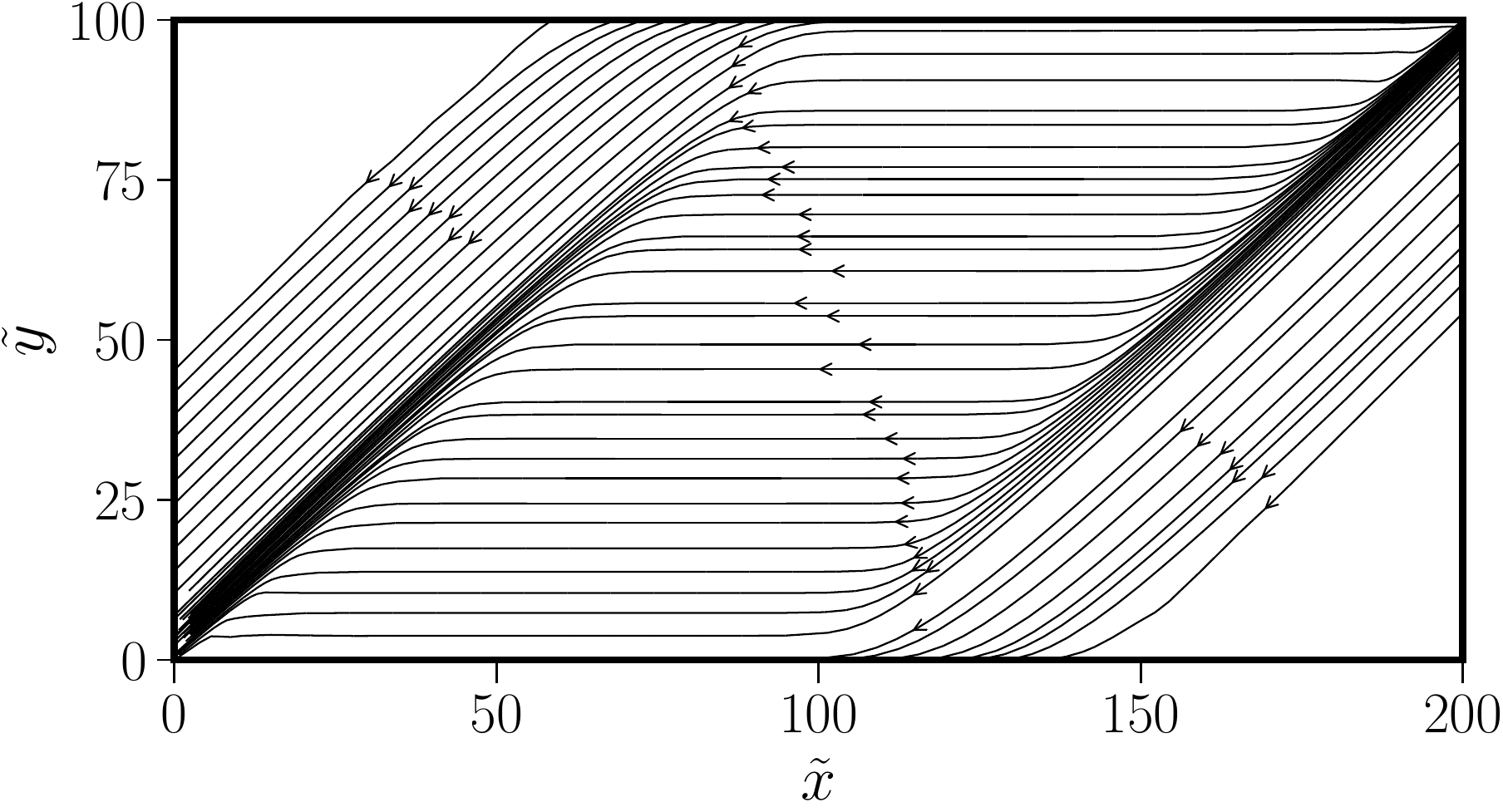}
\label{fig:5C-1000-S}}
\subfigure[${\mu}_{||}/{\mu}_{\perp}$=100]{\includegraphics[width=0.32\textwidth]{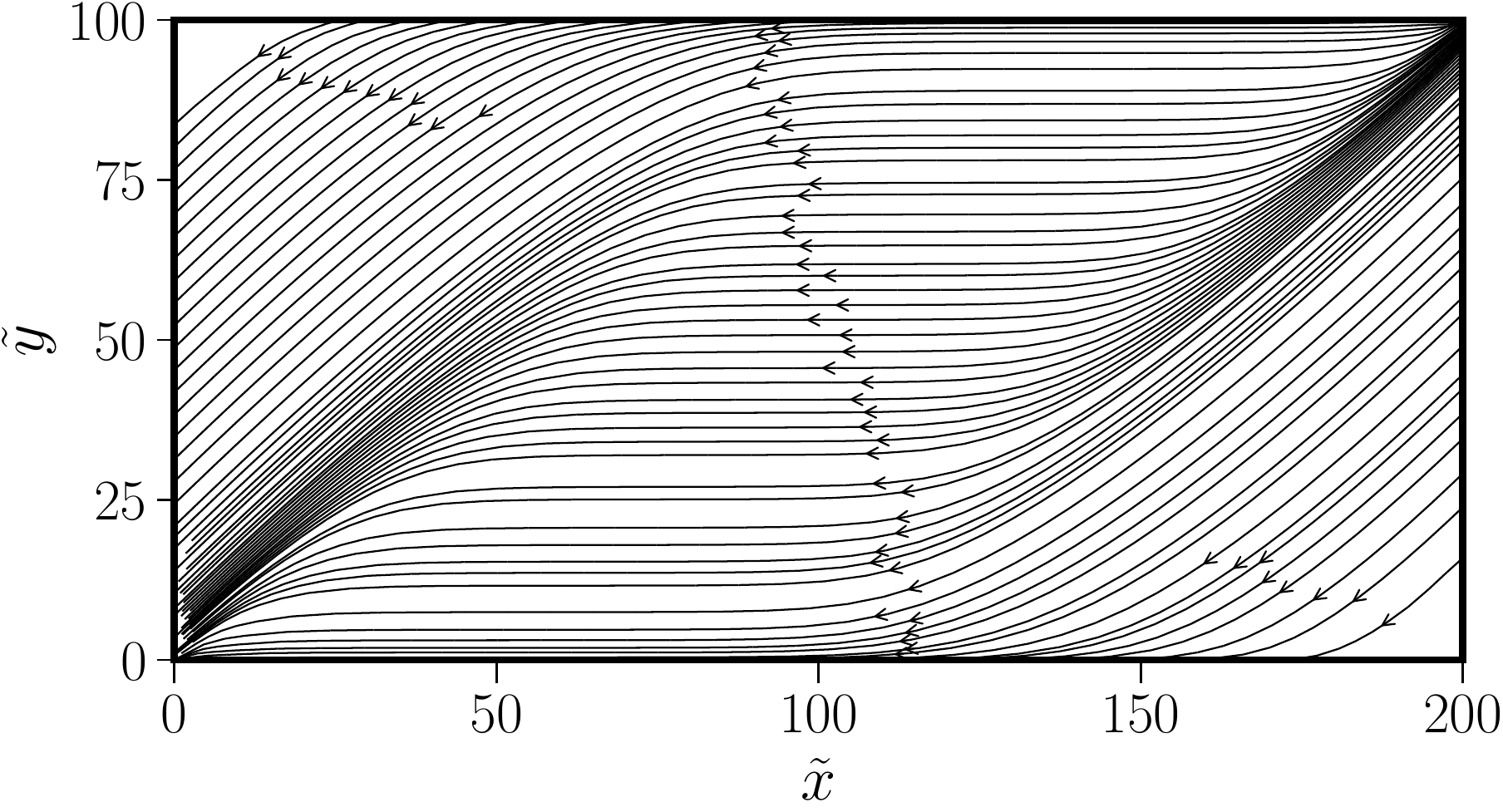}
\label{fig:5E-100-S}}
\subfigure[${\mu}_{||}/{\mu}_{\perp}$=500]{\includegraphics[width=0.32\textwidth]{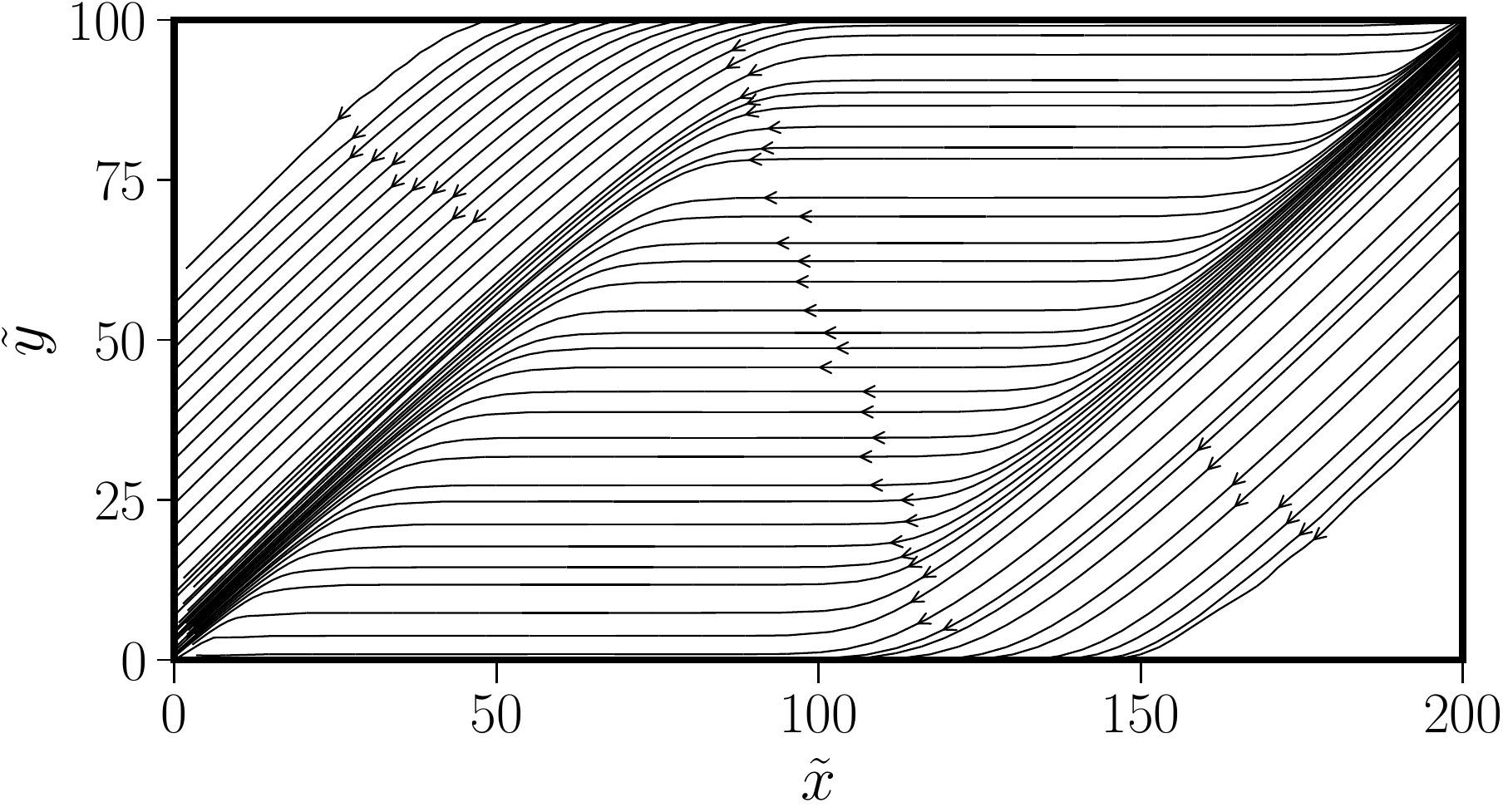}
\label{fig:5E-500-S}}
\subfigure[${\mu}_{||}/{\mu}_{\perp}$=1000]{\includegraphics[width=0.32\textwidth]{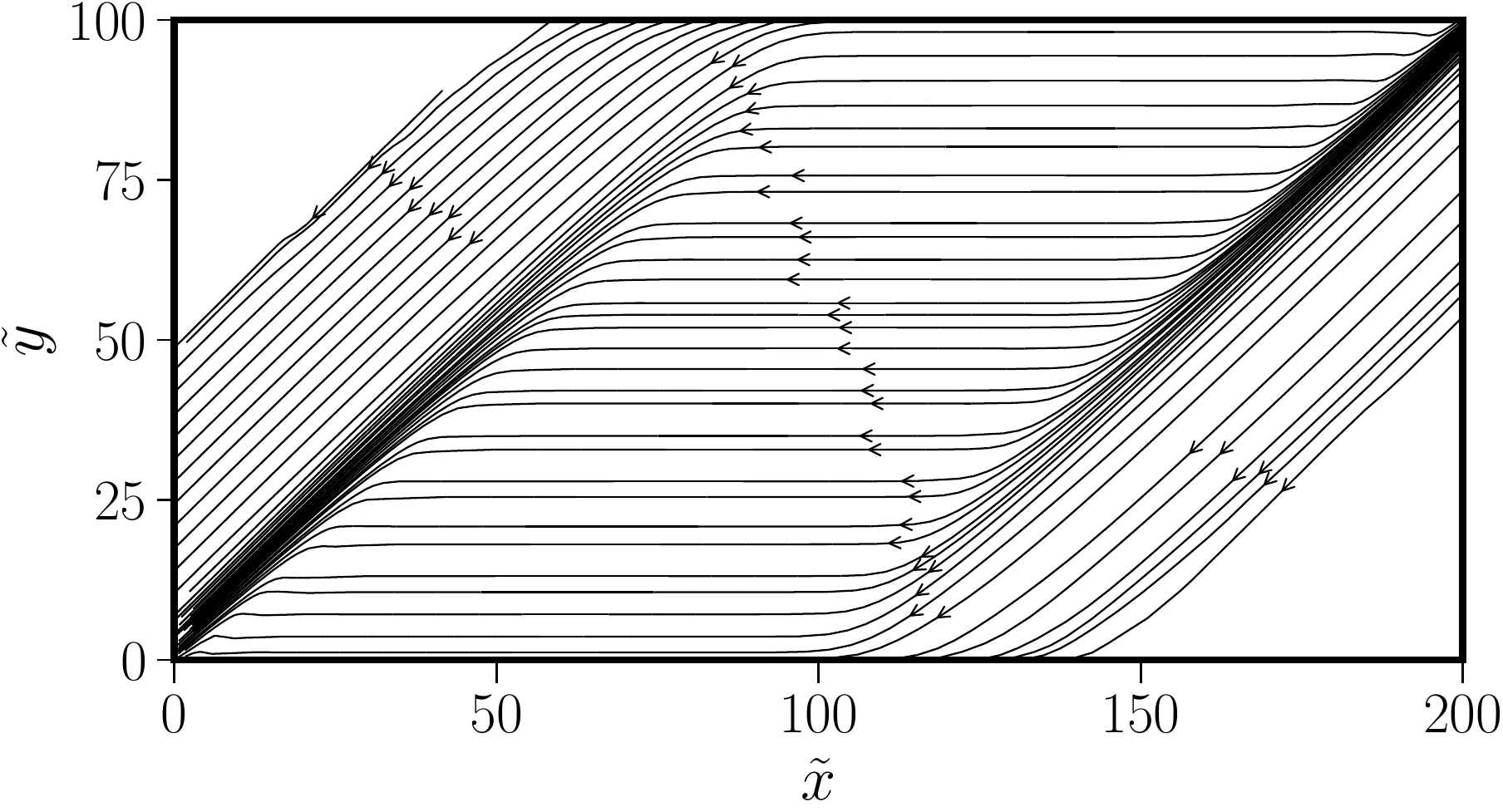}
\label{fig:5E-1000-S}}
  \centering
\caption{{\color{black}Comparison of velocity streamlines with increasing strength of magnetic confinement on a grid of 192 $\times$ 96. Top row: Upwind-3E;  middle row: Upwind-5C; bottom row: Upwind-5E. For interpretation of legend in this figure, the reader is referred to the web version of this article.}}
\label{fig:var-stream-1}
\end{minipage}
  \end{sideways}
\end{figure}

\begin{figure}[H]
  \begin{sideways}
  \begin{minipage}{17.5cm}
\subfigure[${\mu}_{||}/{\mu}_{\perp}$=100]{\includegraphics[width=0.32\textwidth]{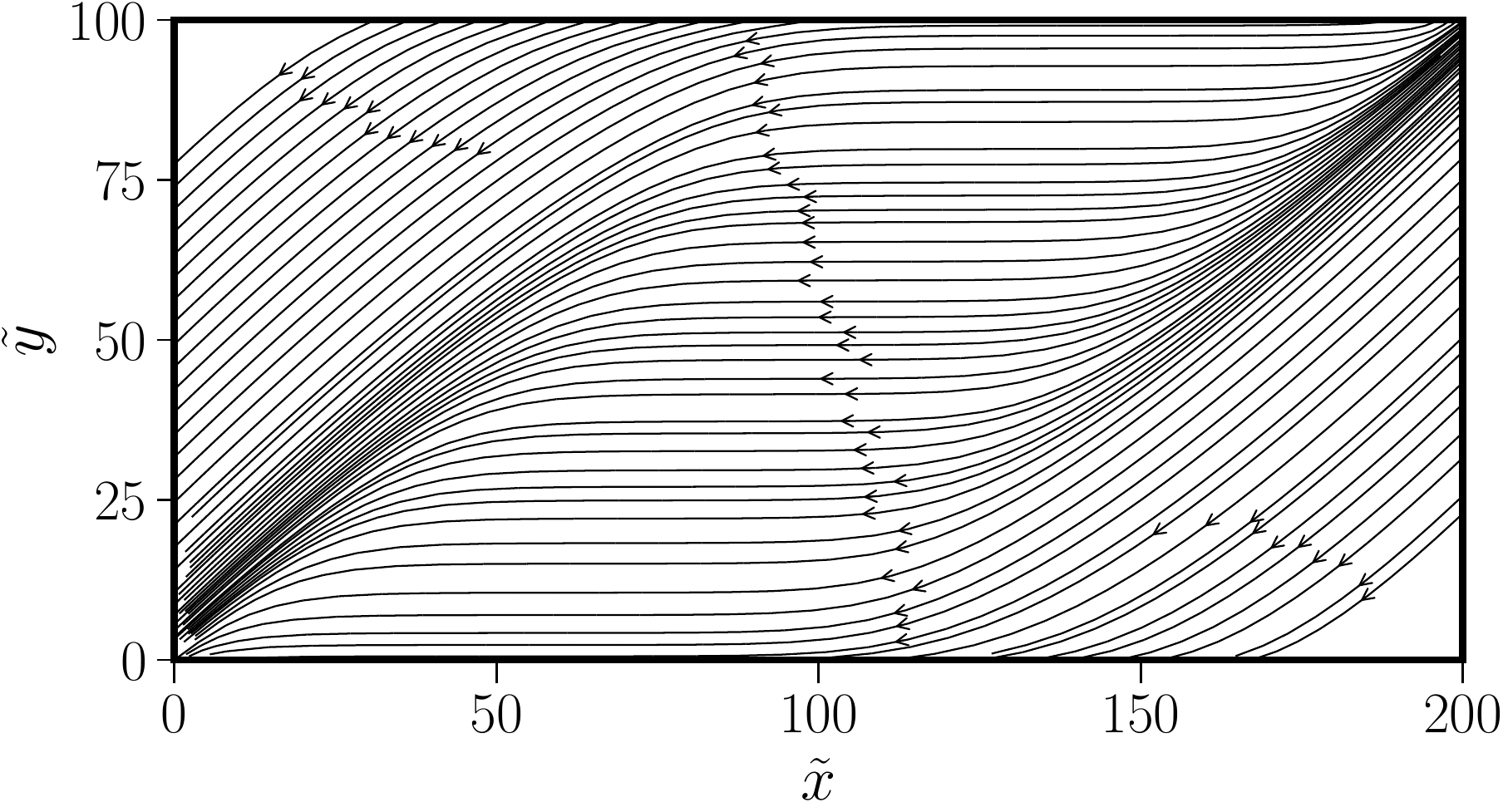}
\label{fig:MFAM-100-S1}}
\subfigure[${\mu}_{||}/{\mu}_{\perp}$=500]{\includegraphics[width=0.32\textwidth]{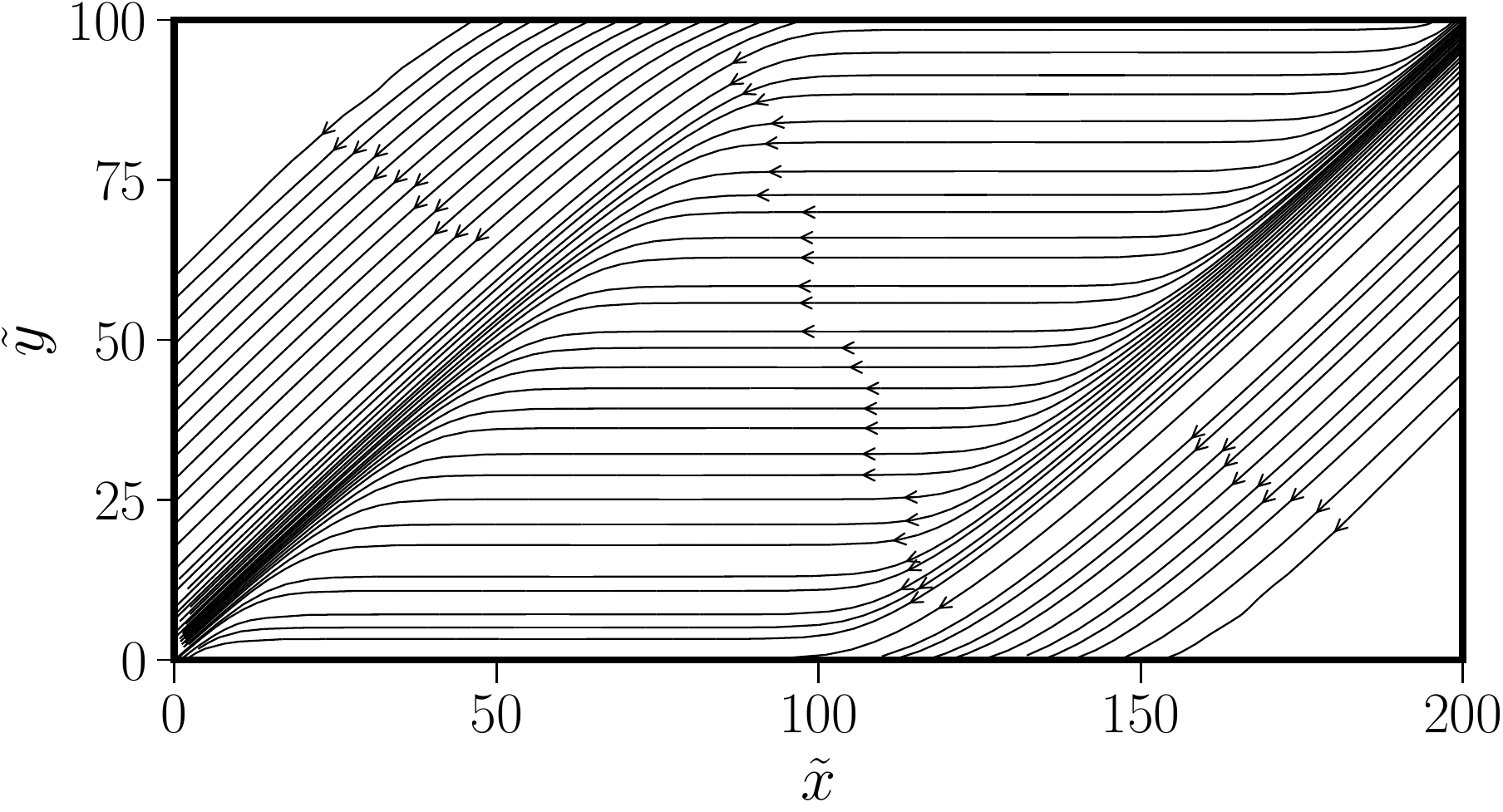}
\label{fig:MFAM-500-S1}}
\subfigure[${\mu}_{||}/{\mu}_{\perp}$=1000]{\includegraphics[width=0.32\textwidth]{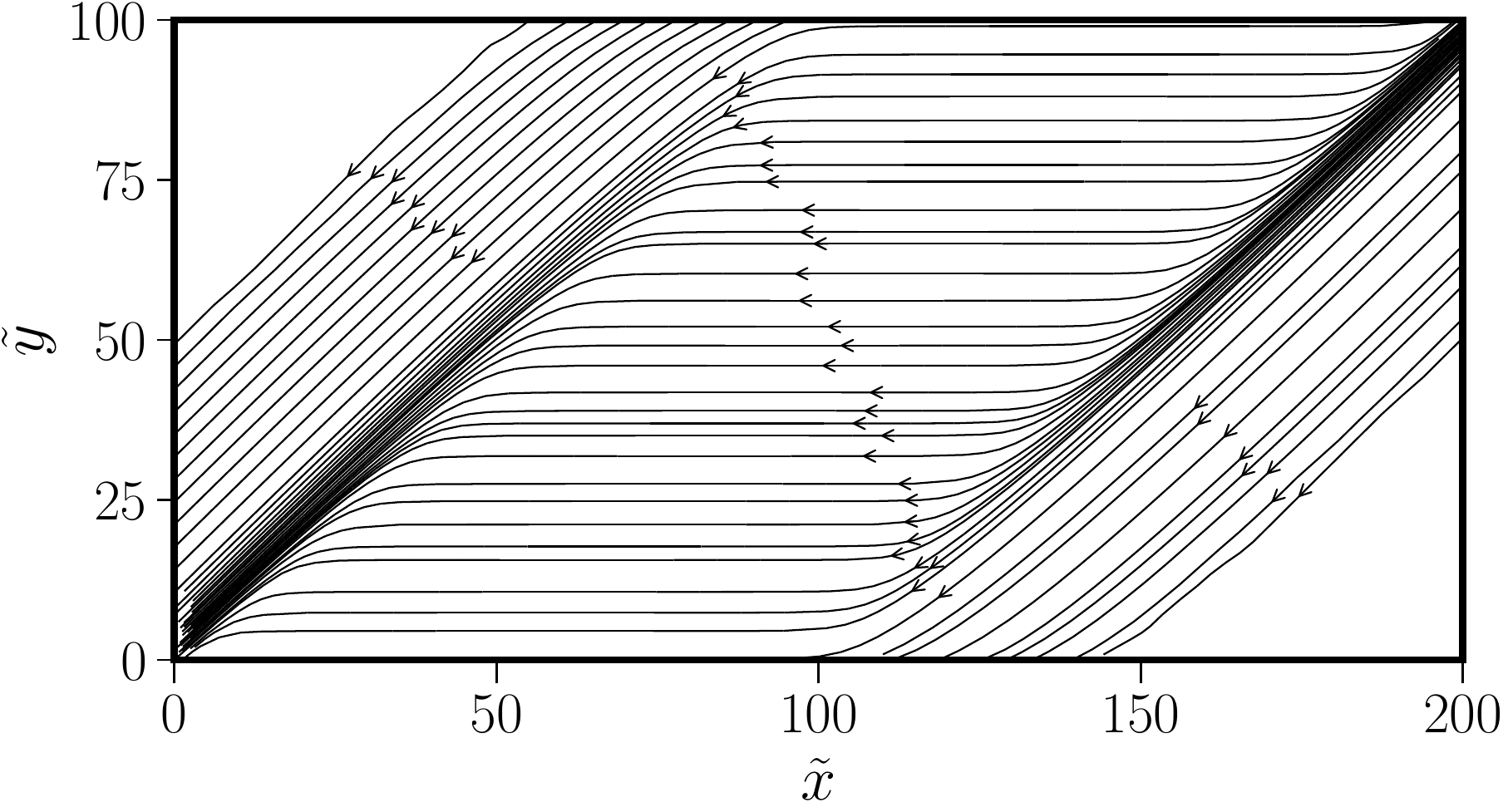}
\label{fig:MFAM-1000-S1}}
\subfigure[${\mu}_{||}/{\mu}_{\perp}$=100]{\includegraphics[width=0.32\textwidth]{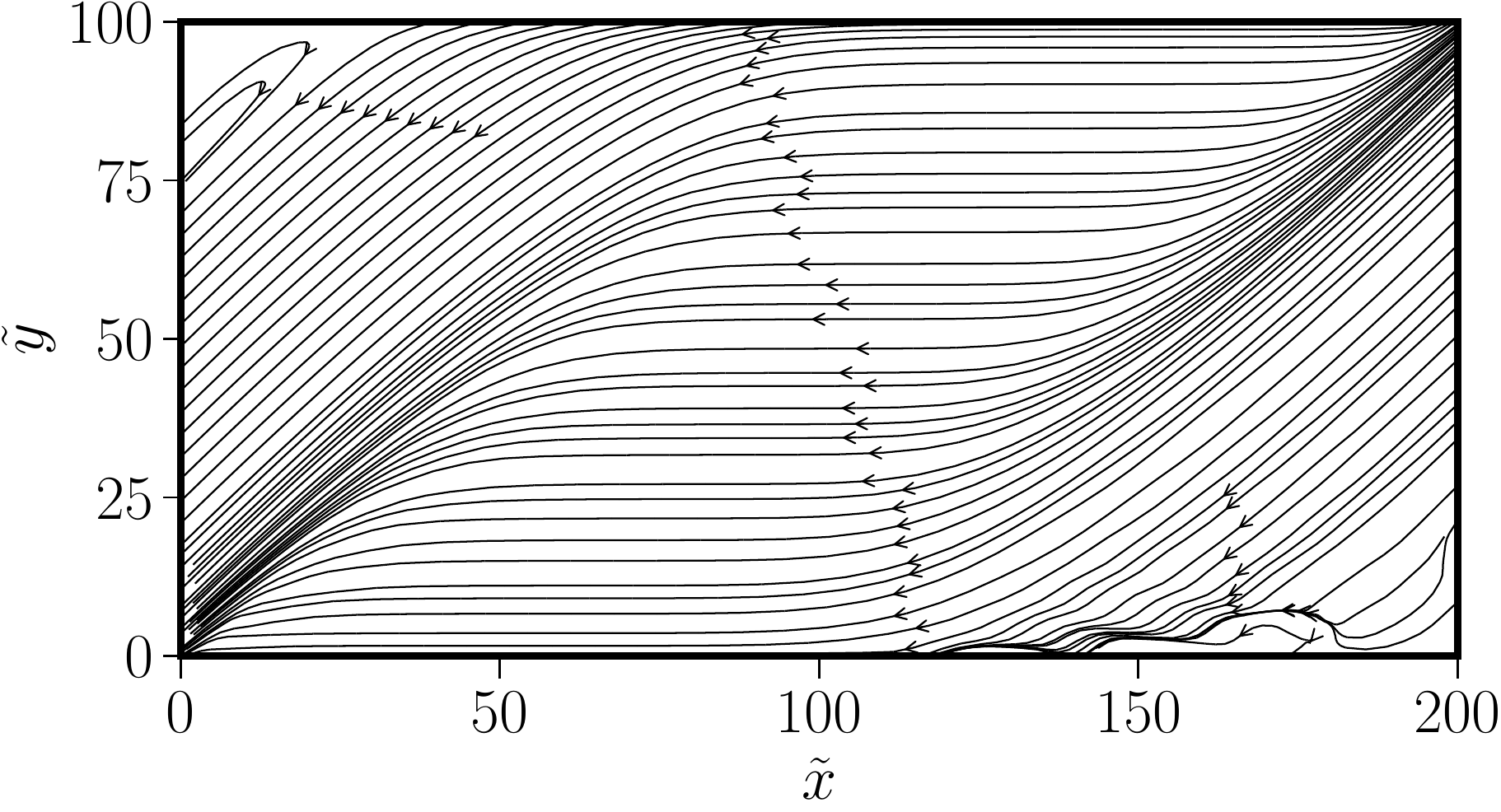}
\label{fig:L-100-S}}
\subfigure[${\mu}_{||}/{\mu}_{\perp}$=500]{\includegraphics[width=0.32\textwidth]{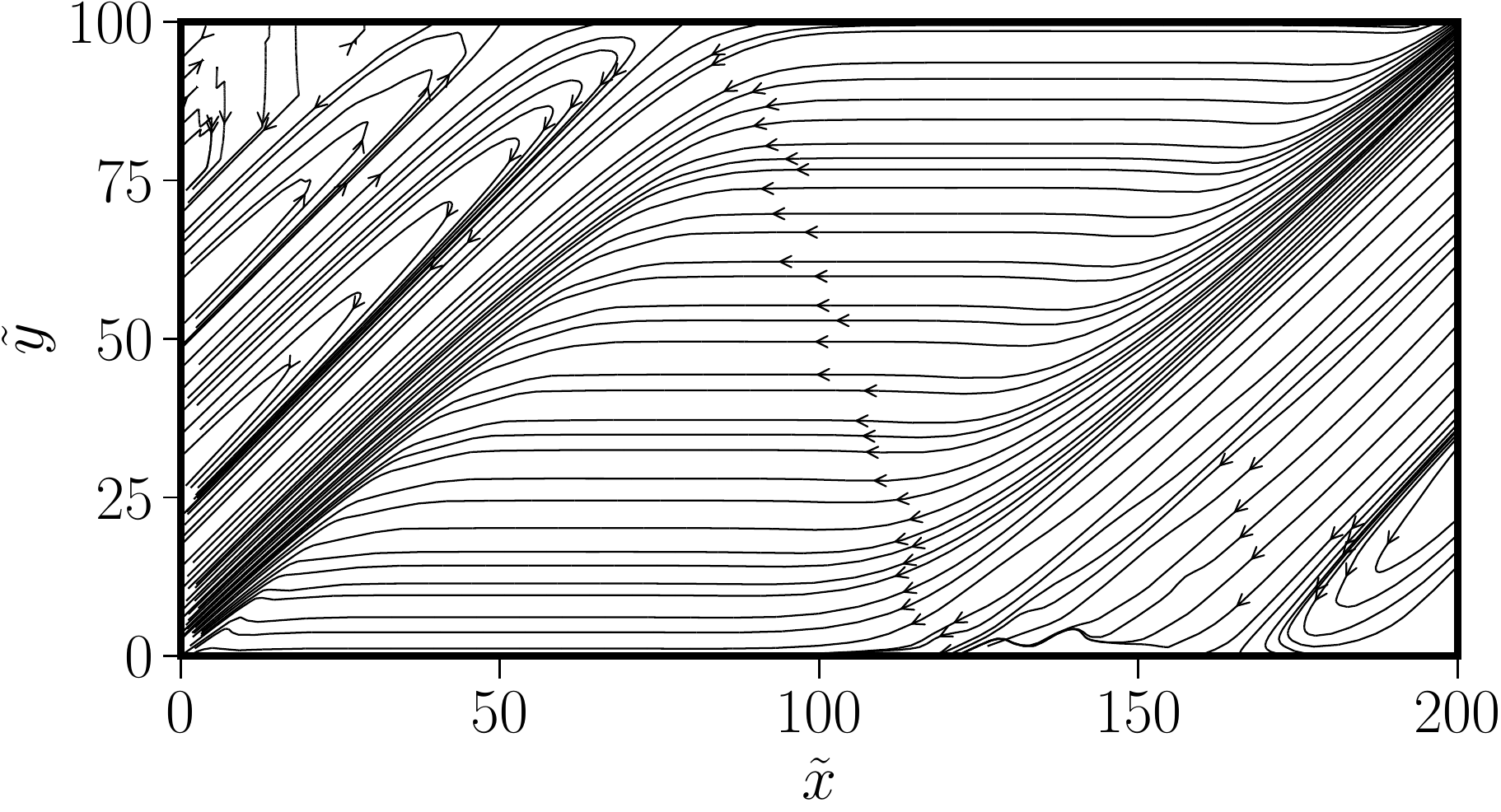}
\label{fig:L-500-S}}
\subfigure[${\mu}_{||}/{\mu}_{\perp}$=1000]{\includegraphics[width=0.32\textwidth]{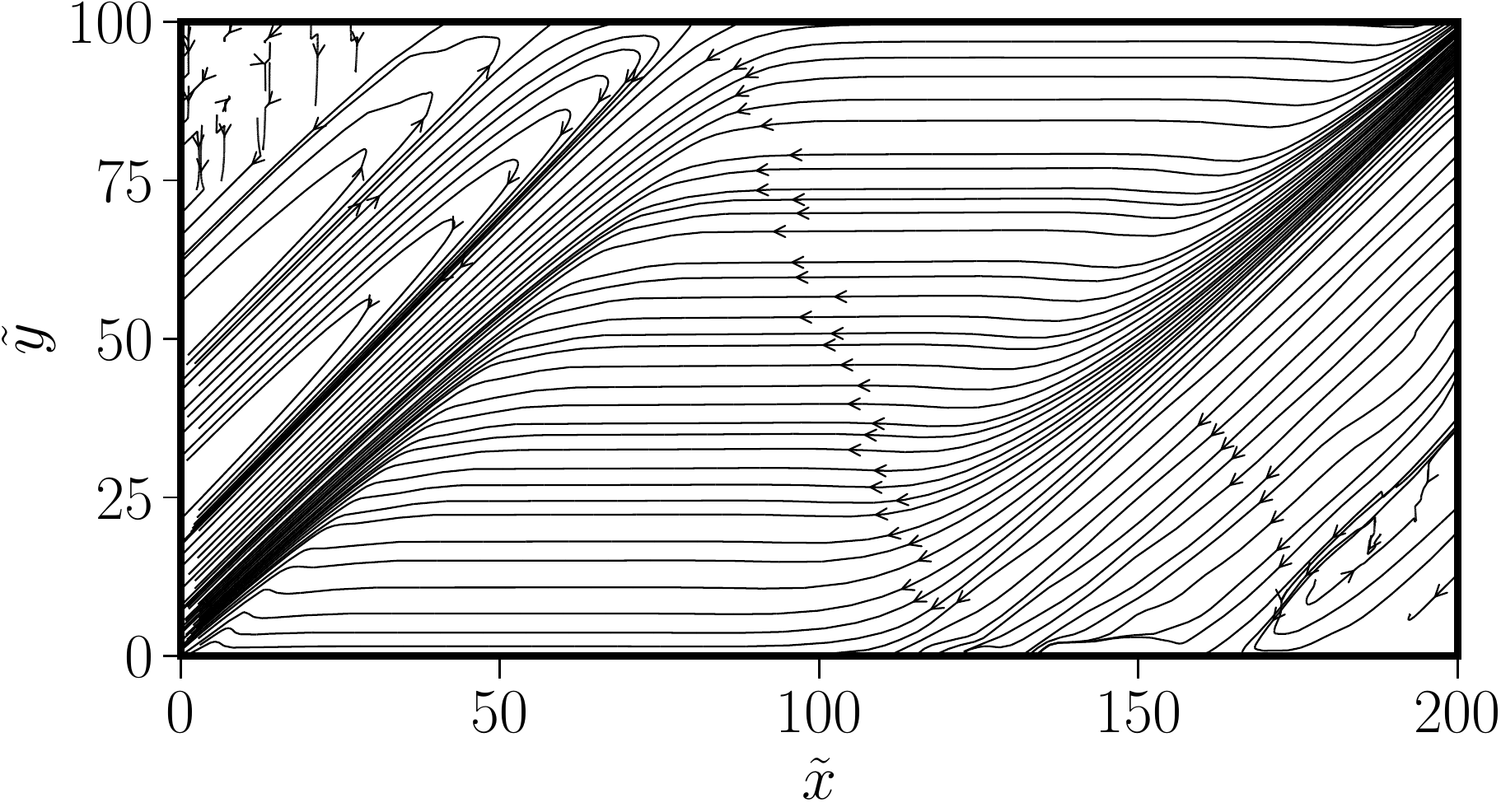}
\label{fig:L-1000-S}}
\subfigure[${\mu}_{||}/{\mu}_{\perp}$=100]{\includegraphics[width=0.32\textwidth]{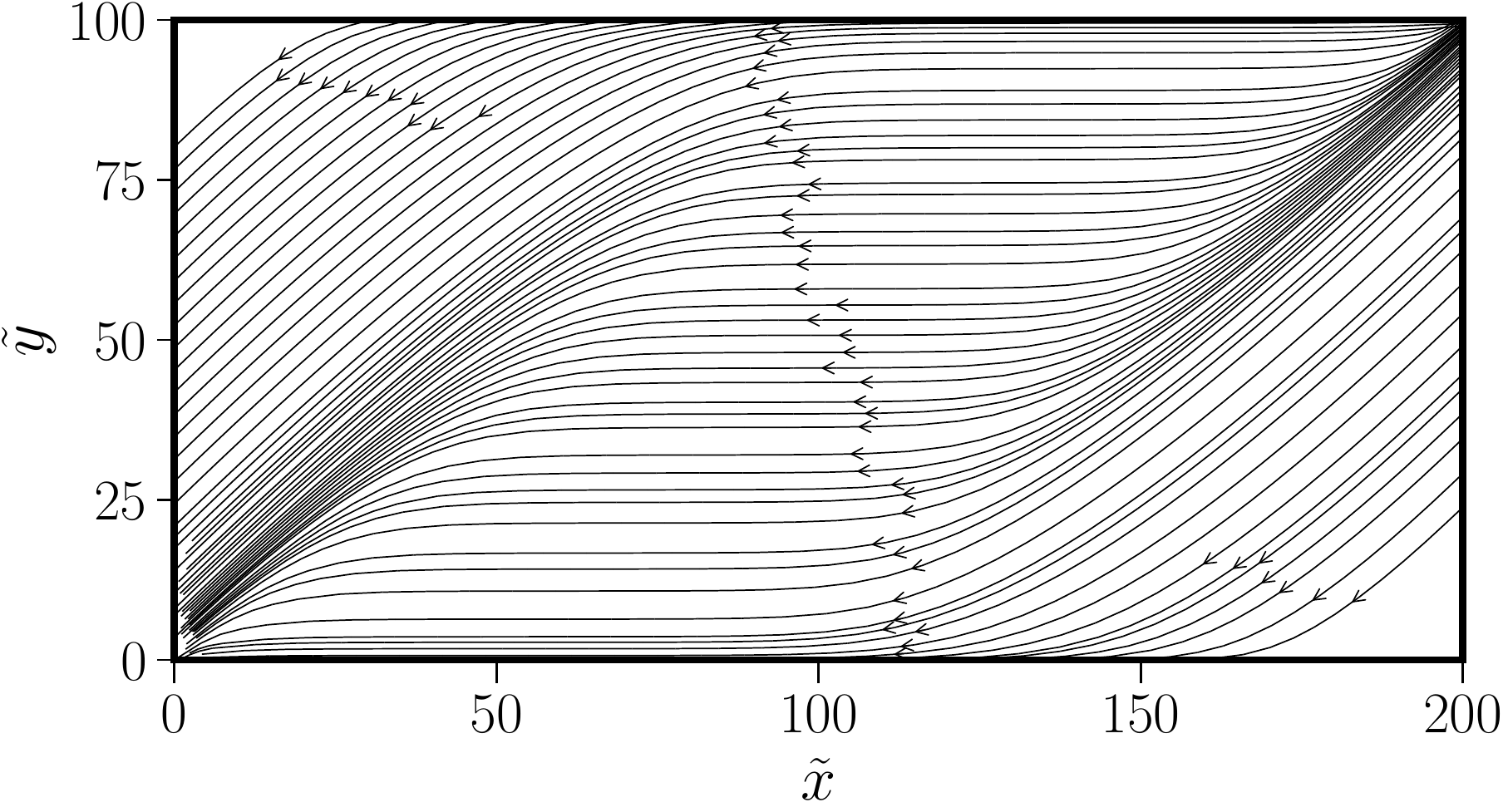}
\label{fig:WENO-100-S}}
\subfigure[${\mu}_{||}/{\mu}_{\perp}$=500]{\includegraphics[width=0.32\textwidth]{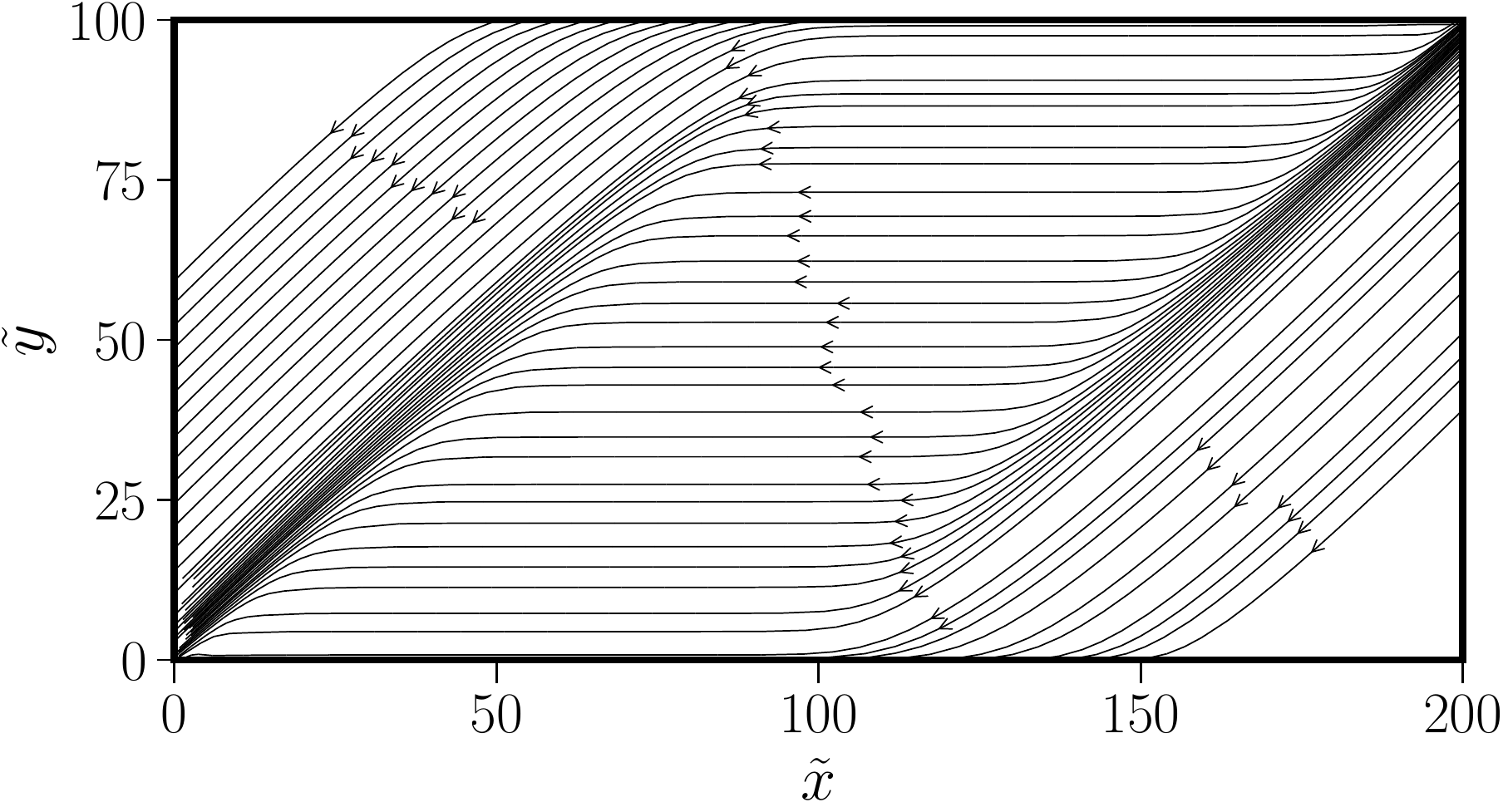}
\label{fig:WENO-500-S}}
\subfigure[${\mu}_{||}/{\mu}_{\perp}$=1000]{\includegraphics[width=0.32\textwidth]{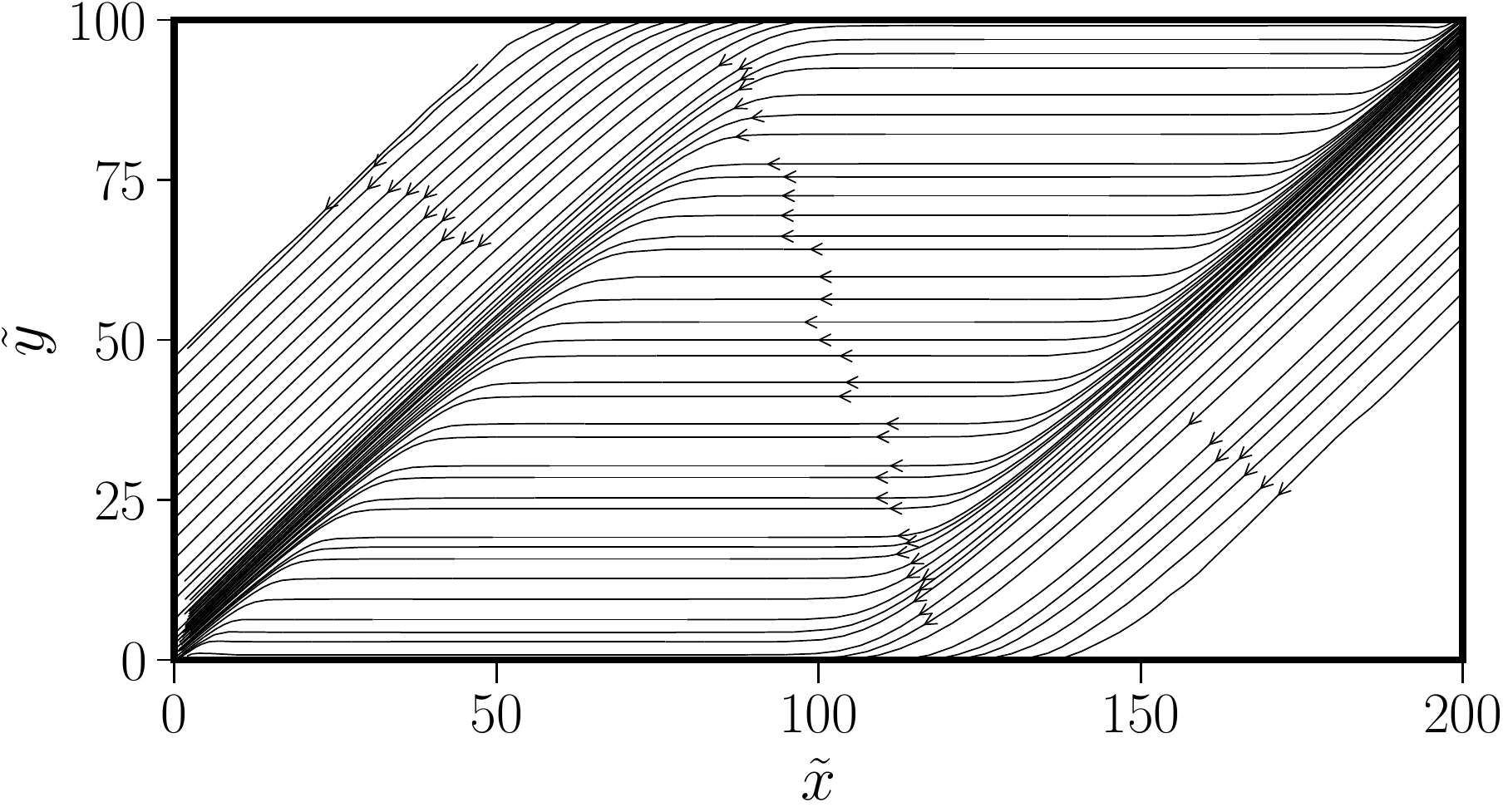}
\label{fig:WENO-1000-S}}
\caption{{\color{black}Comparison of velocity streamlines with increasing strength of magnetic confinement on a grid of 192 $\times$ 96. Top row: MFAM; Middle row: Generalized-MUSCL; bottom row: WENO-5Z-L. For interpretation of legend in this figure, the reader is referred to the web version of this article.}}
\label{fig:var-stream-2}
\end{minipage}
  \end{sideways}
\end{figure}

\begin{figure}[H]
  \begin{sideways}
  \begin{minipage}{17.5cm}
\centering
\subfigure[${\mu}_{||}/{\mu}_{\perp}$=100]{\includegraphics[width=0.31\textwidth]{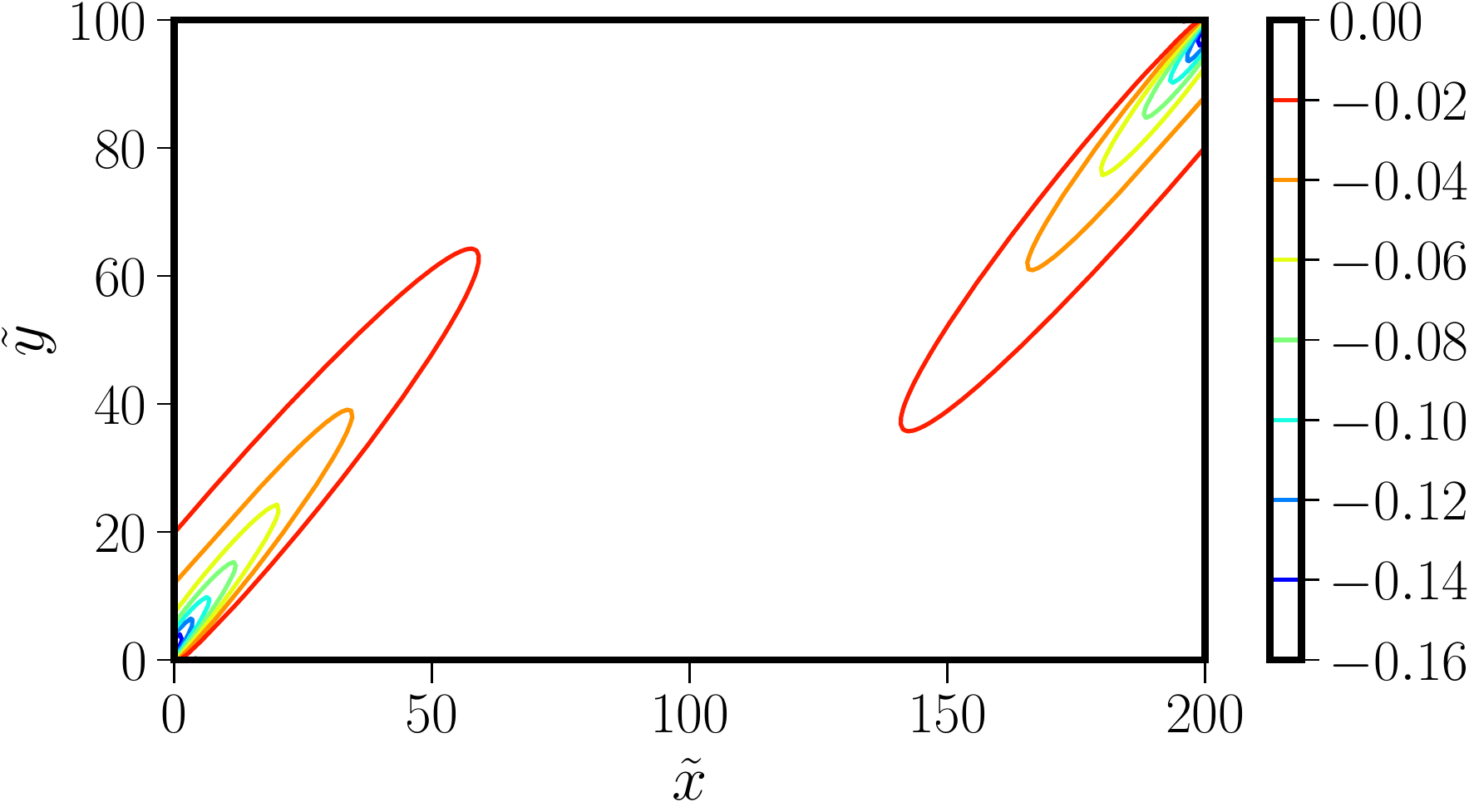}
\label{fig:3E-100-V}}
\subfigure[${\mu}_{||}/{\mu}_{\perp}$=500]{\includegraphics[width=0.31\textwidth]{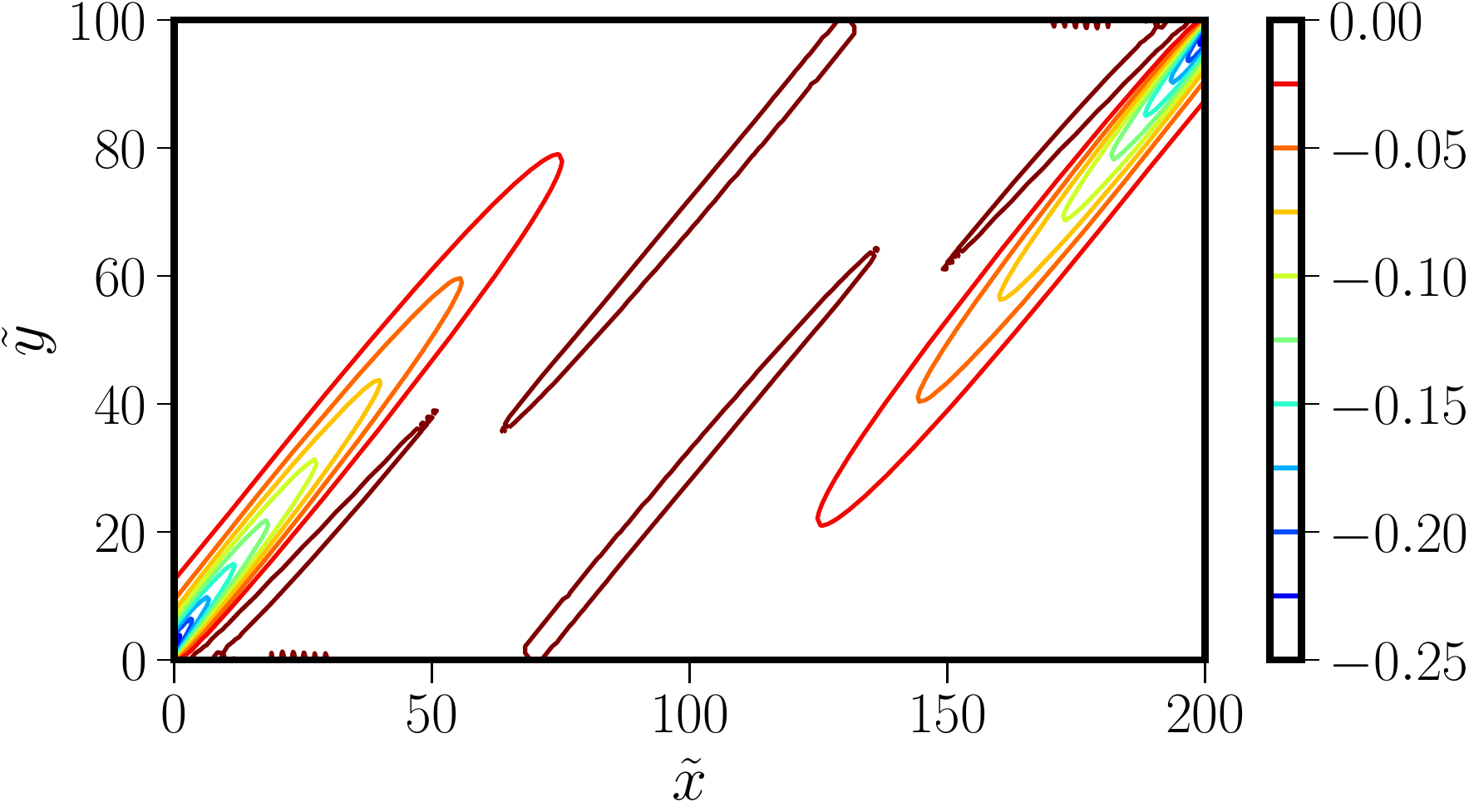}
\label{fig:3E-500-V}}
\subfigure[${\mu}_{||}/{\mu}_{\perp}$=1000]{\includegraphics[width=0.31\textwidth]{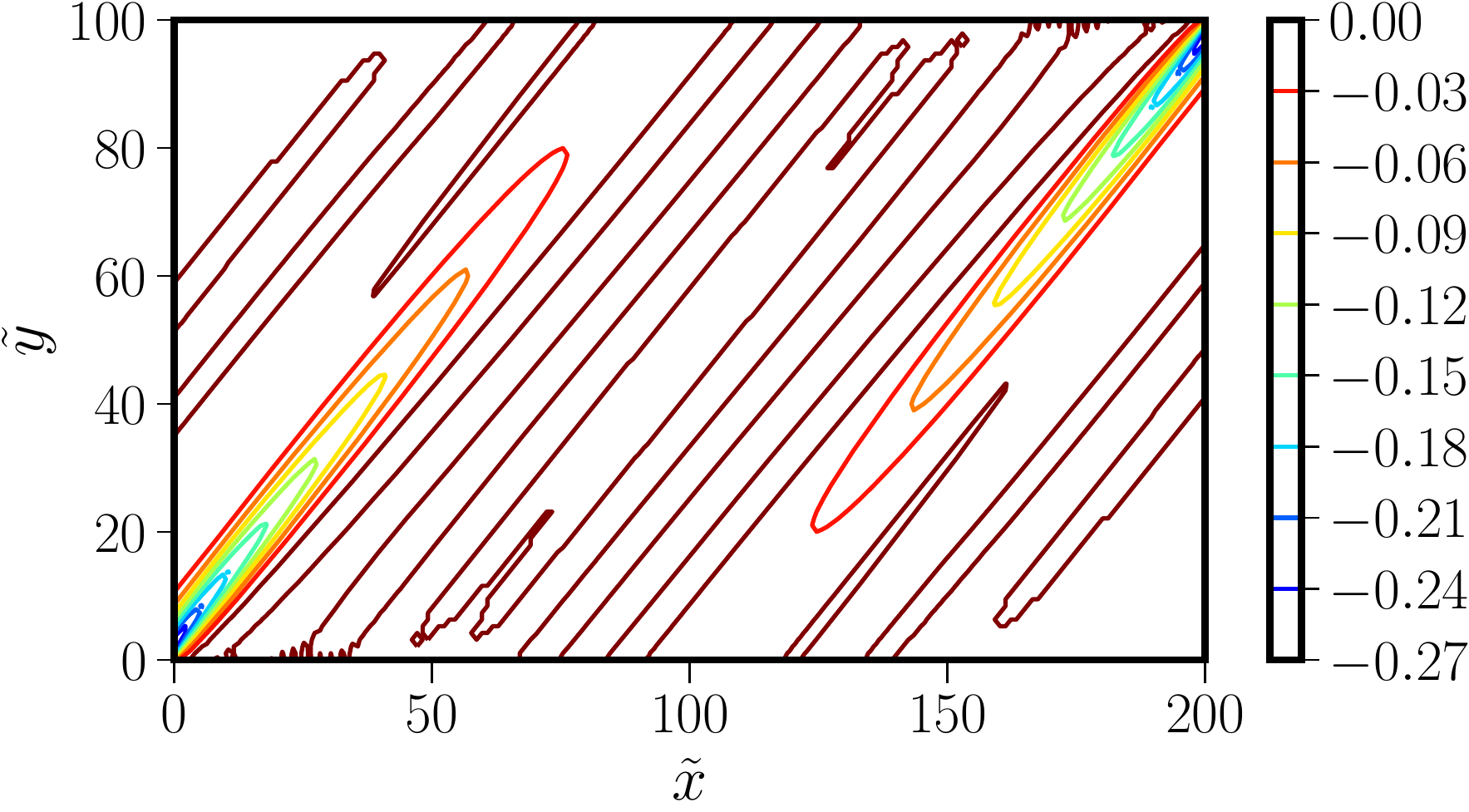}
\label{fig:3E-1000-V}}
\subfigure[${\mu}_{||}/{\mu}_{\perp}$=100]{\includegraphics[width=0.31\textwidth]{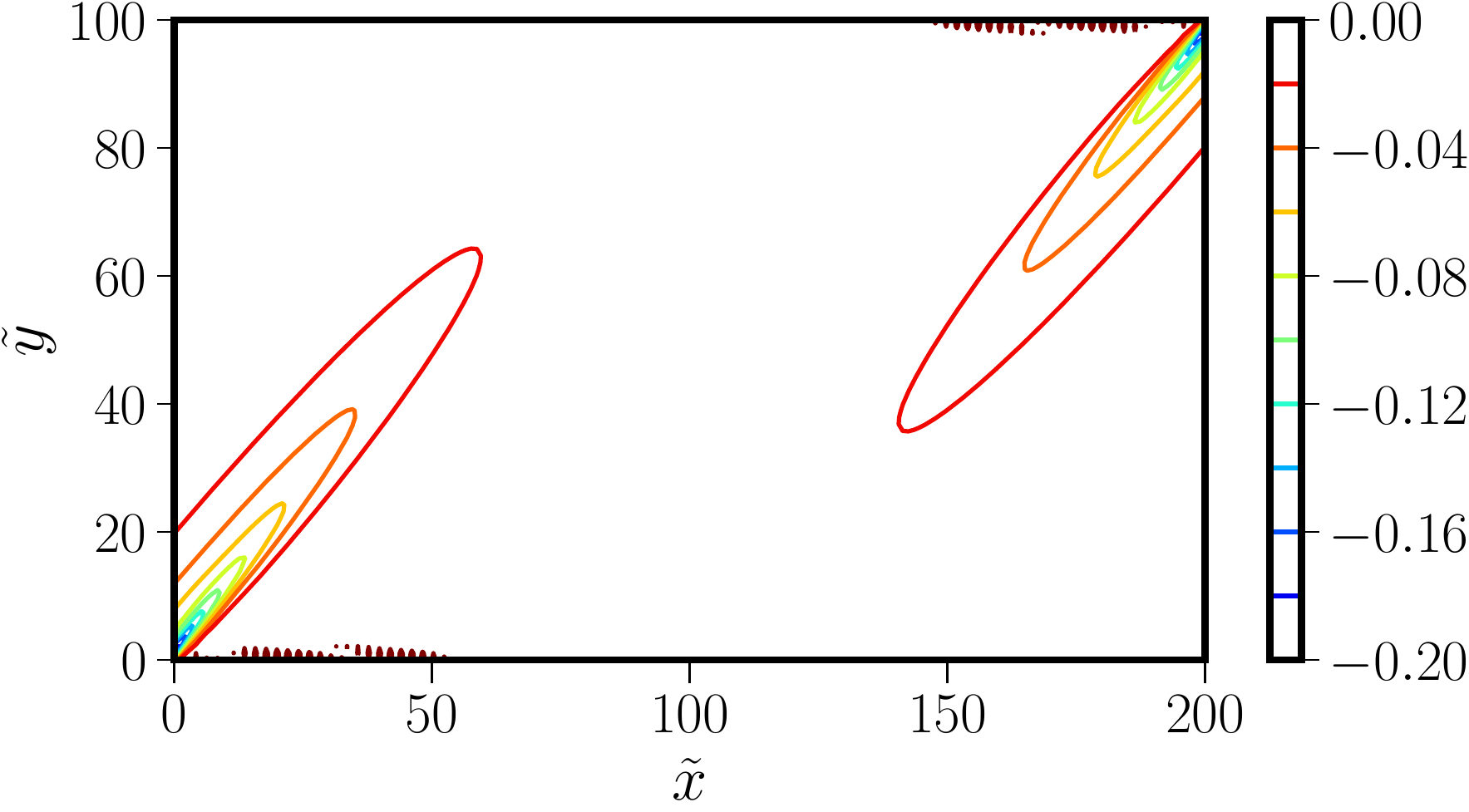}
\label{fig:5C-100-V1}}
\subfigure[${\mu}_{||}/{\mu}_{\perp}$=500]{\includegraphics[width=0.31\textwidth]{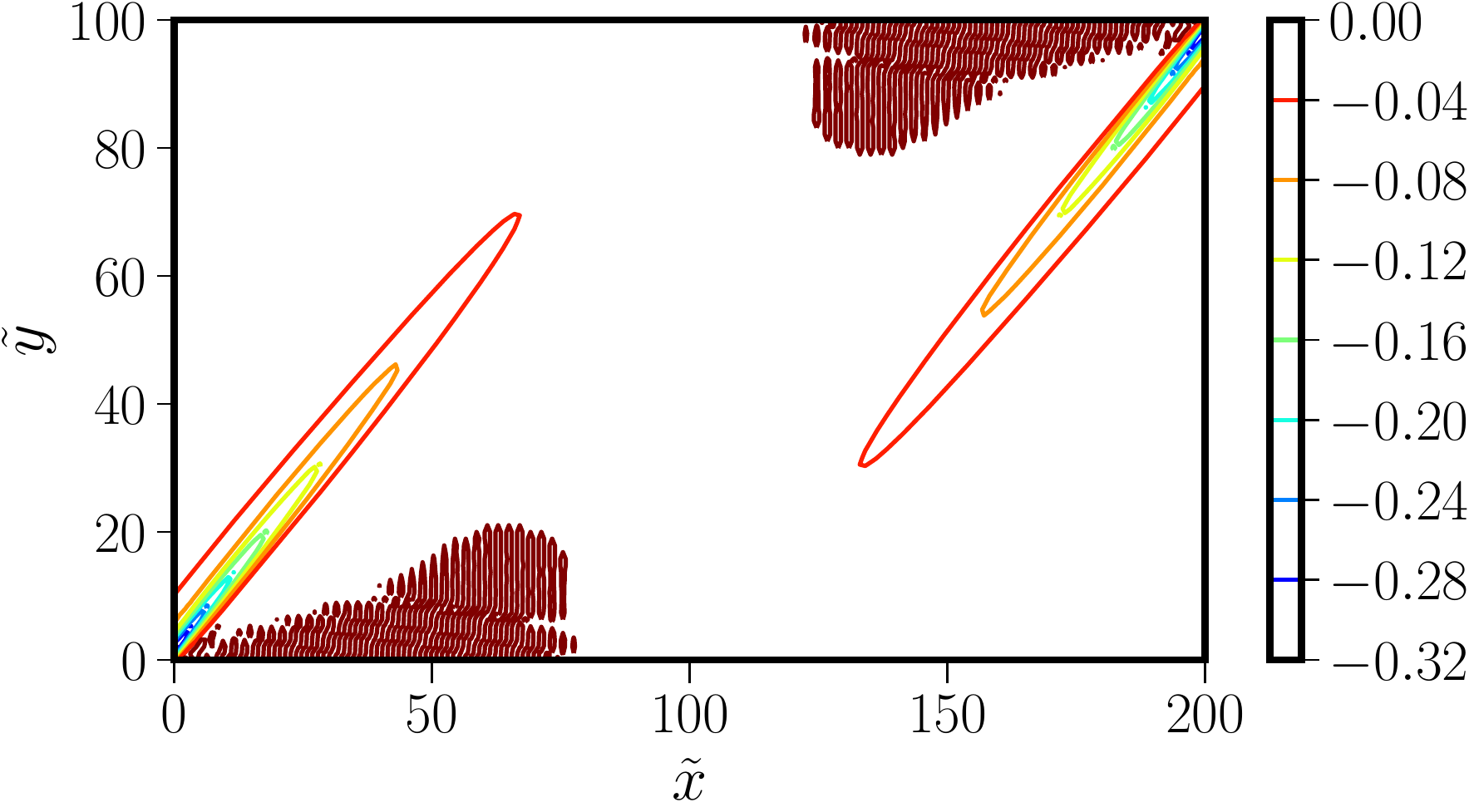}
\label{fig:5C-500-V1}}
\subfigure[${\mu}_{||}/{\mu}_{\perp}$=1000]{\includegraphics[width=0.31\textwidth]{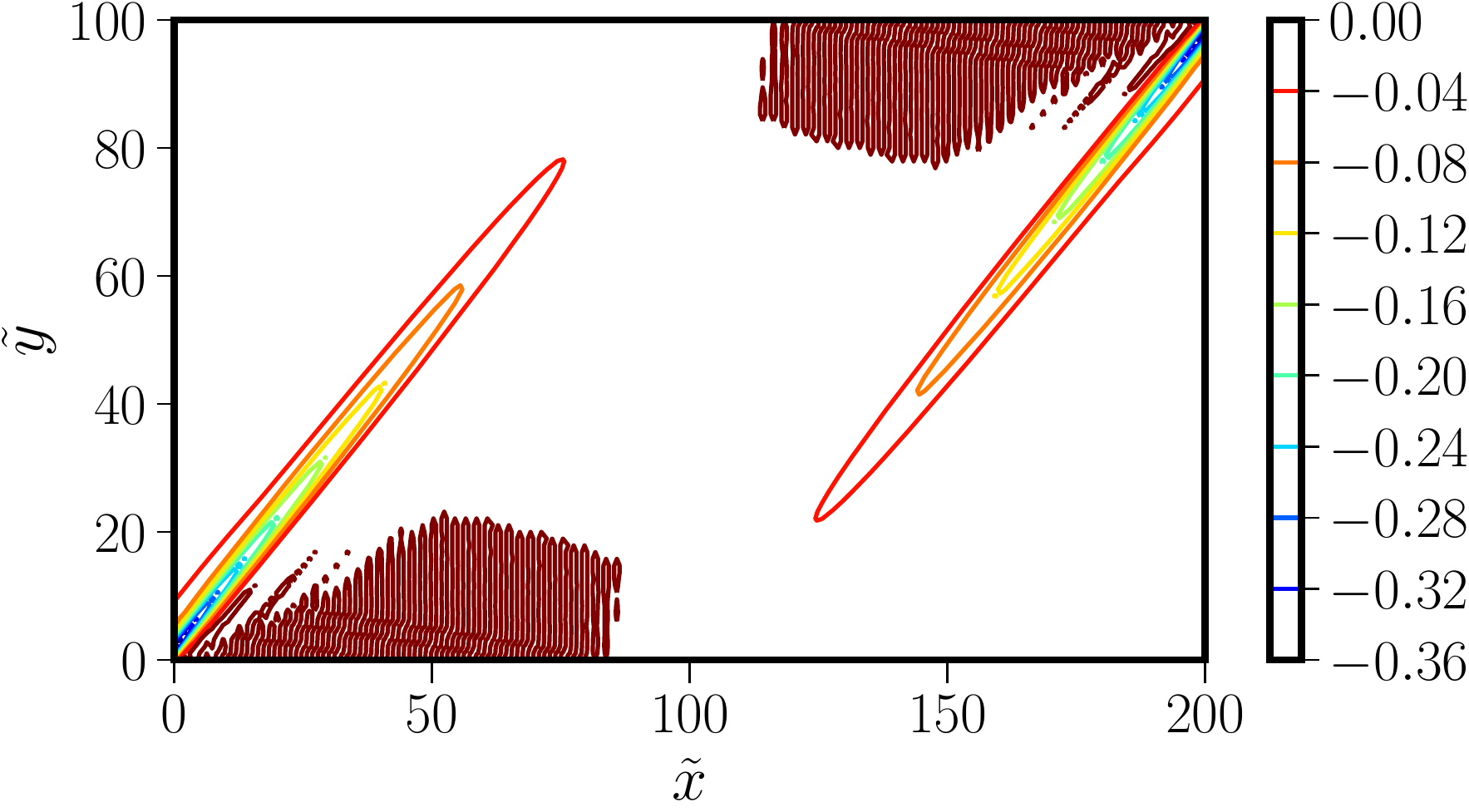}
\label{fig:5C-1000-V1}}
\subfigure[${\mu}_{||}/{\mu}_{\perp}$=100]{\includegraphics[width=0.31\textwidth]{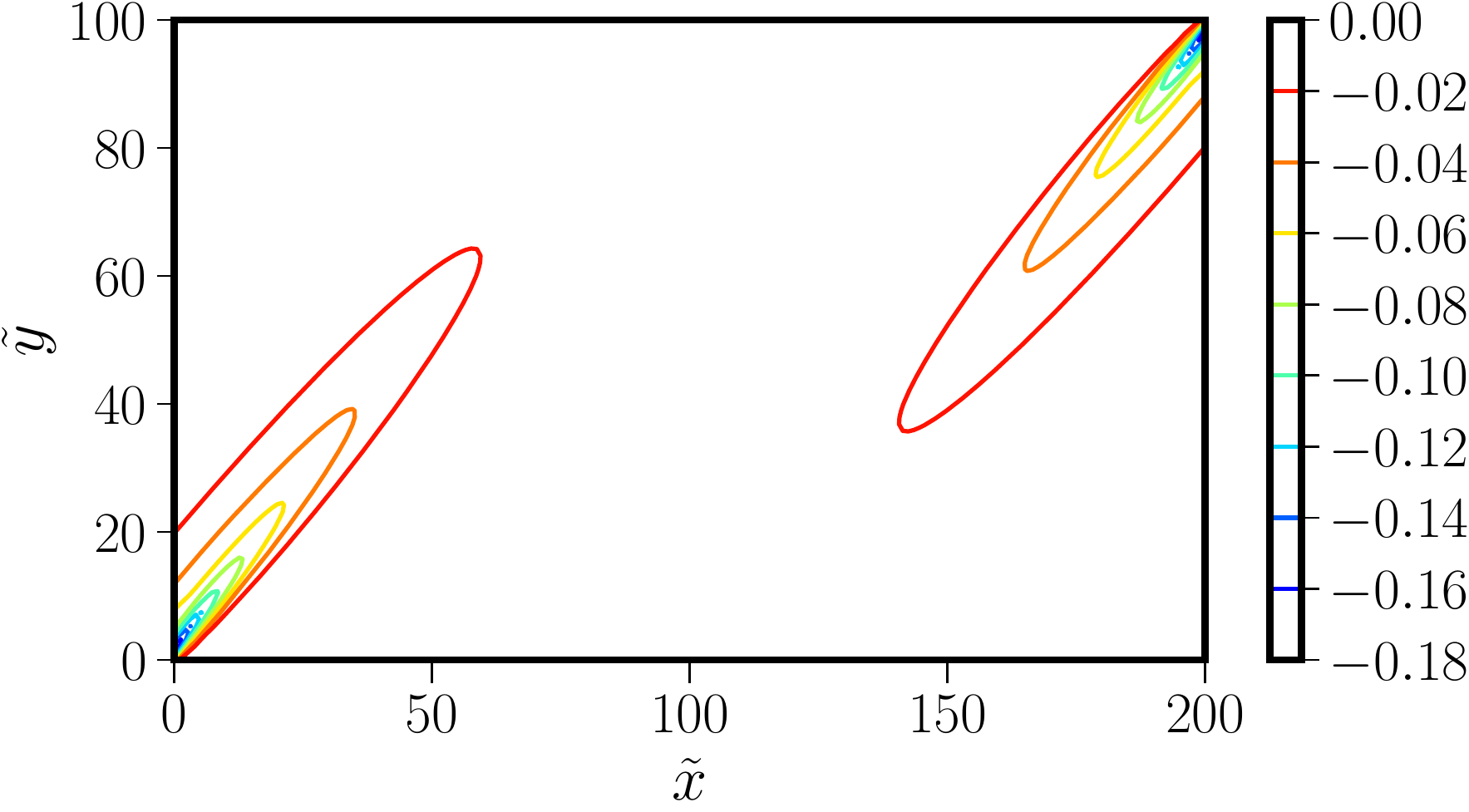}
\label{fig:500-100-V}}
\subfigure[${\mu}_{||}/{\mu}_{\perp}$=500]{\includegraphics[width=0.31\textwidth]{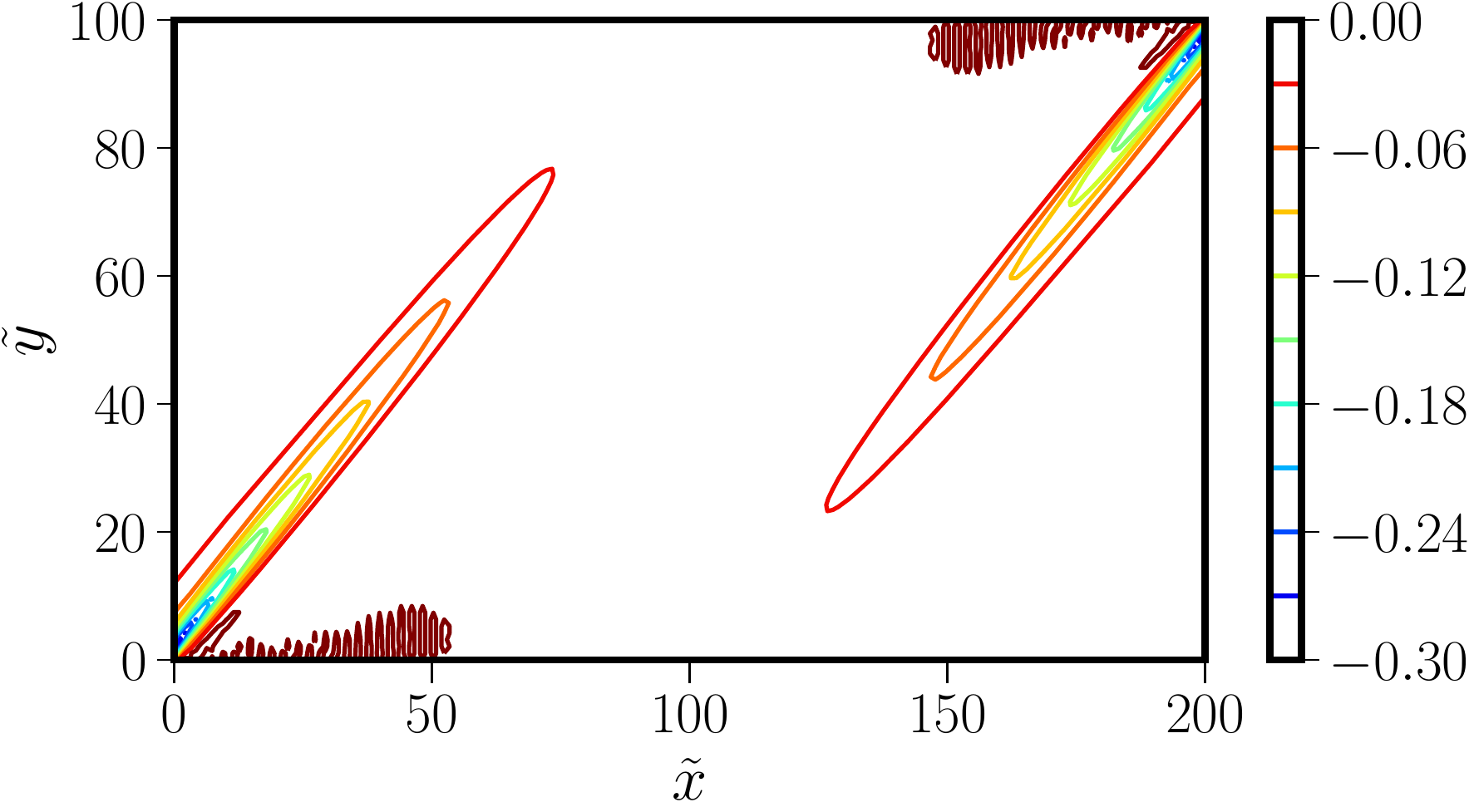}
\label{fig:5E-500-V}}
\subfigure[${\mu}_{||}/{\mu}_{\perp}$=1000]{\includegraphics[width=0.31\textwidth]{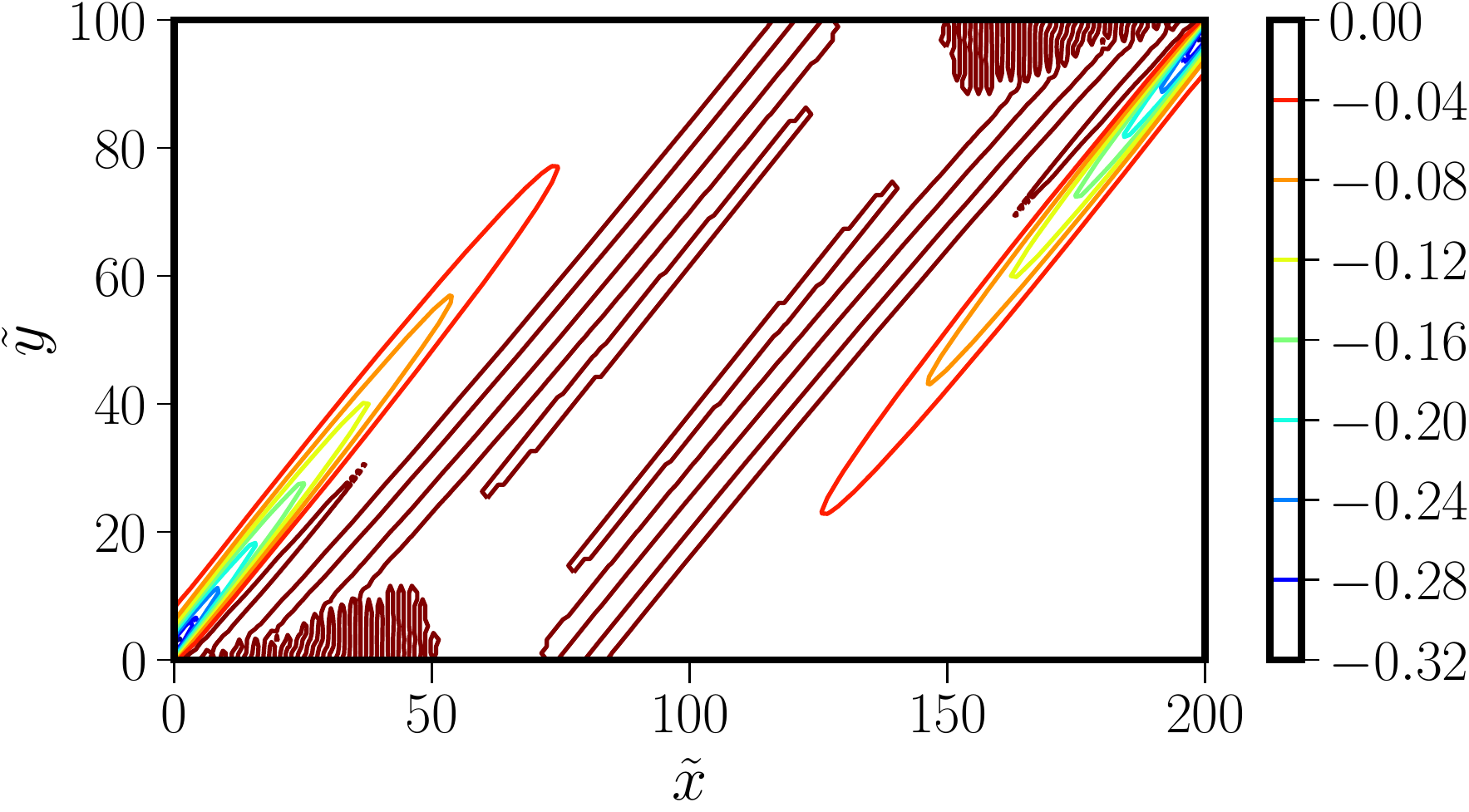}
\label{fig:5E-1000-V}}
\caption{Comparison of normalized velocity in y-direction with increasing ${\mu}_{||}/{\mu}_{\perp}$on a 192 $\times$ 96 grid. Top row: Upwind-3E;  middle row: Upwind-5C; bottom row: Upwind-5E.}
\label{fig:var-V-1}
\end{minipage}
  \end{sideways}
  \end{figure}

\begin{figure}[H]
  \begin{sideways}
  \begin{minipage}{17.5cm}
\centering
\subfigure[${\mu}_{||}/{\mu}_{\perp}$=100]{\includegraphics[width=0.31\textwidth]{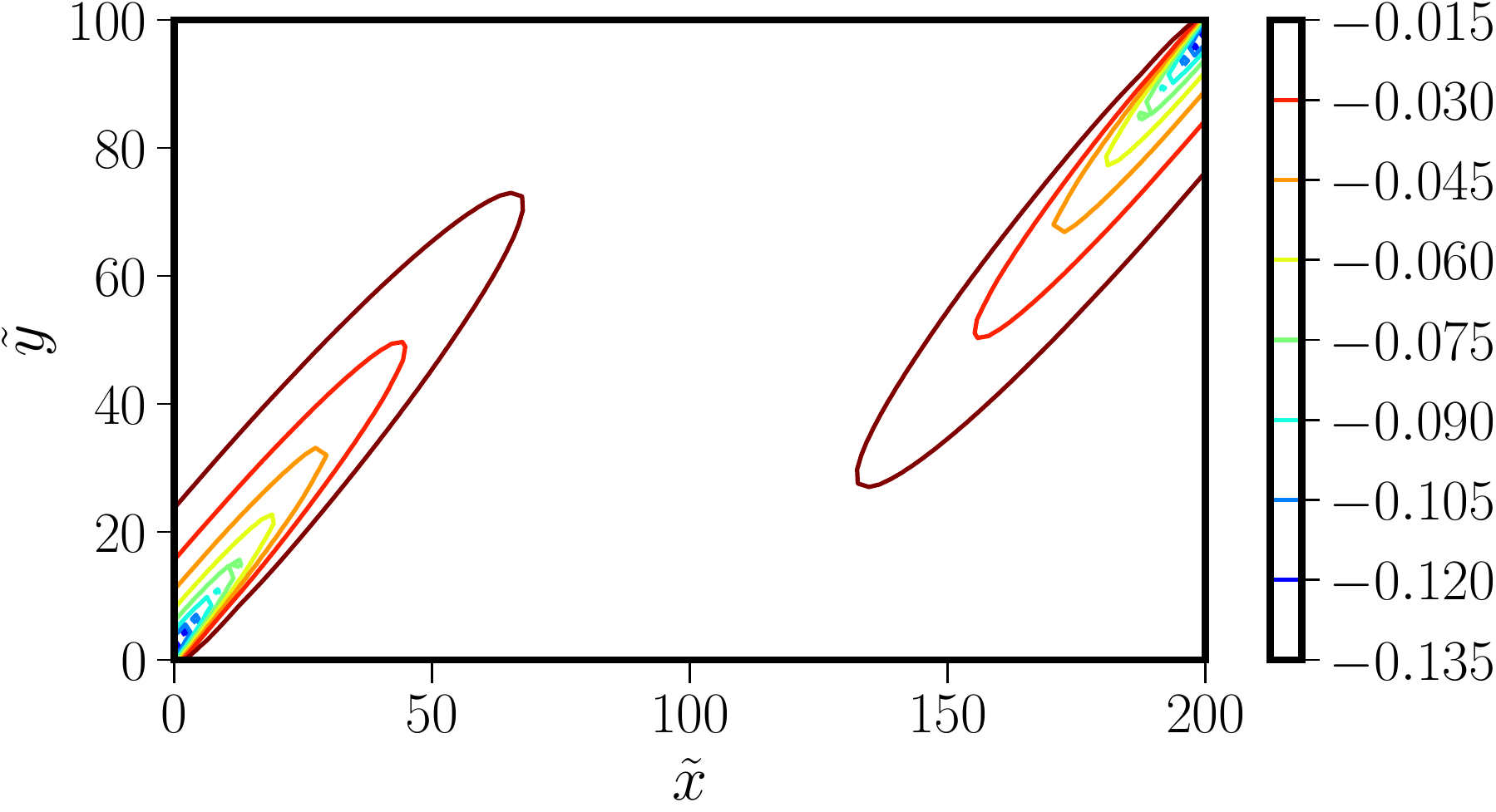}
\label{fig:MFAM-100-V}}
\subfigure[${\mu}_{||}/{\mu}_{\perp}$=500]{\includegraphics[width=0.31\textwidth]{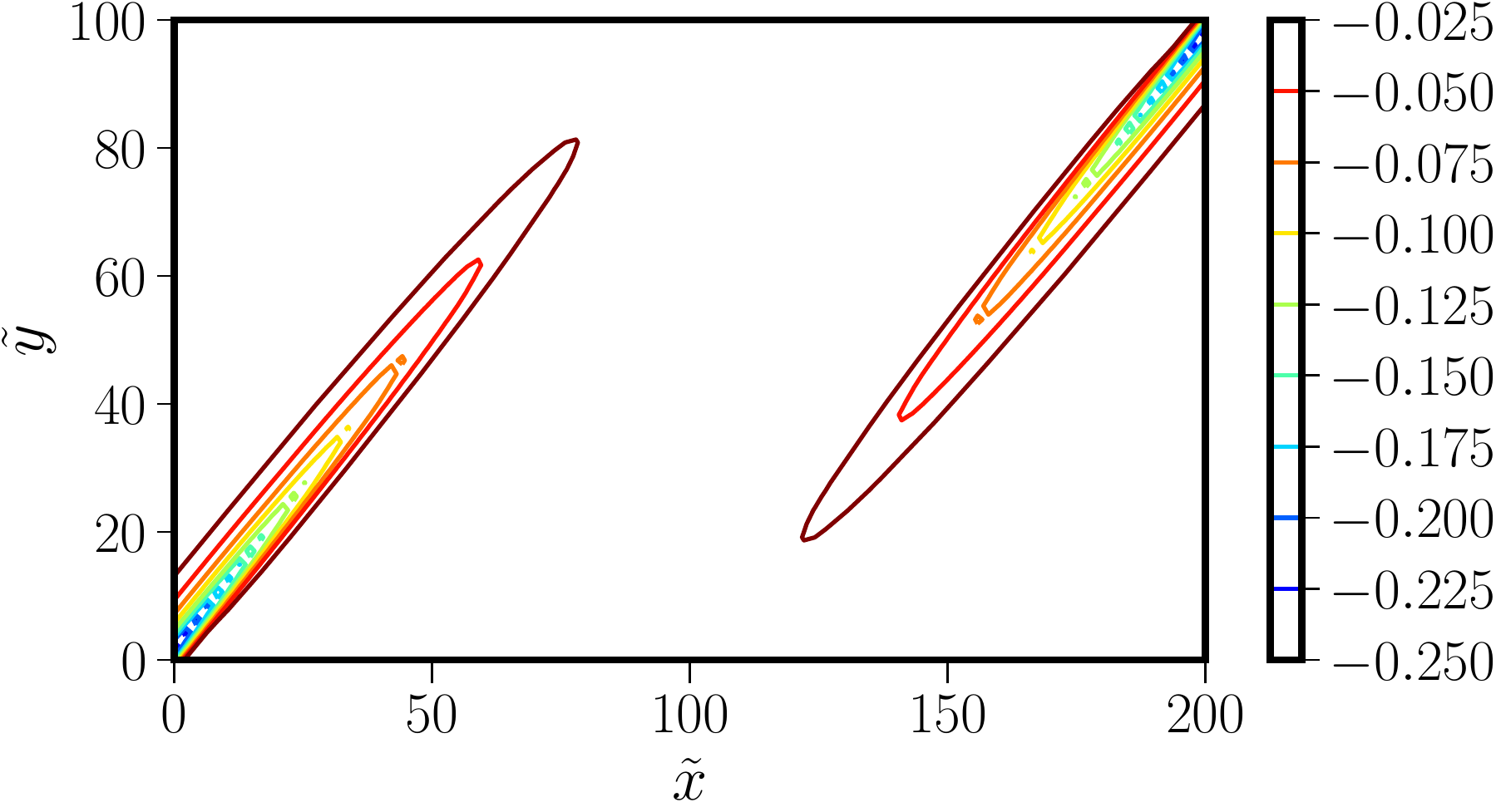}
\label{fig:MFAM-500-V}}
\subfigure[${\mu}_{||}/{\mu}_{\perp}$=1000]{\includegraphics[width=0.31\textwidth]{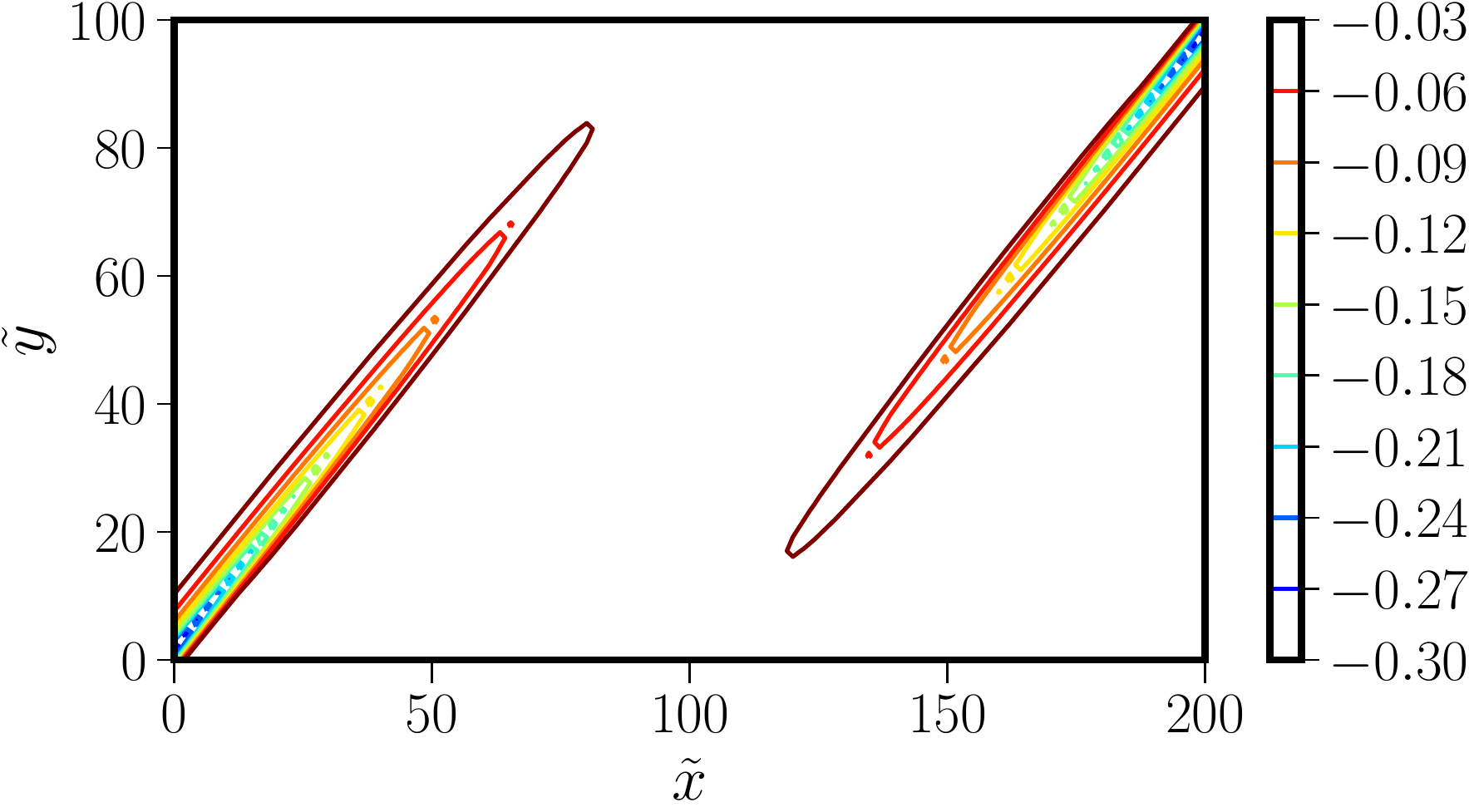}
\label{fig:MFAM-1000-V}}
\subfigure[${\mu}_{||}/{\mu}_{\perp}$=100]{\includegraphics[width=0.31\textwidth]{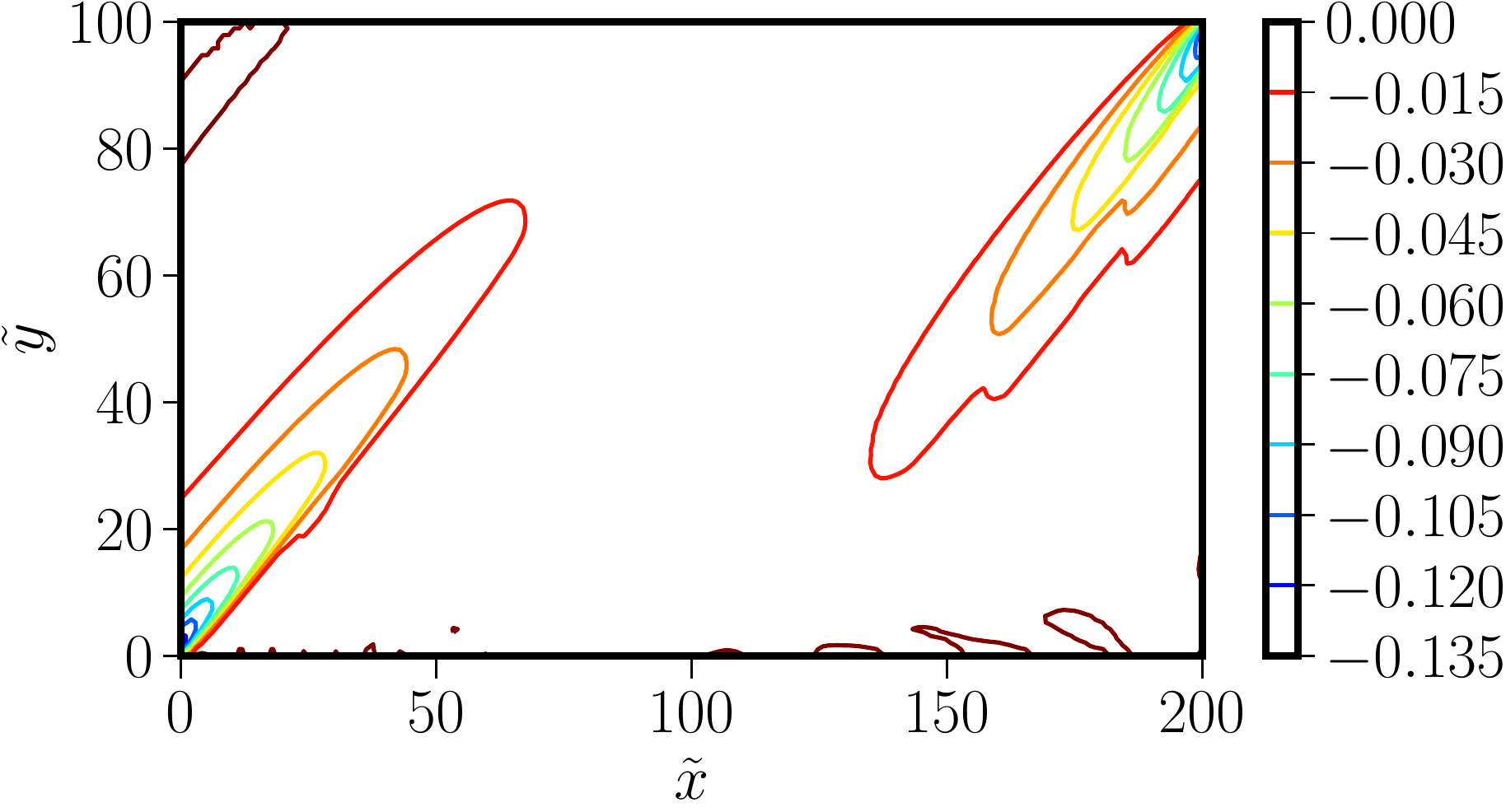}
\label{fig:L-100-V}}
\subfigure[${\mu}_{||}/{\mu}_{\perp}$=500]{\includegraphics[width=0.31\textwidth]{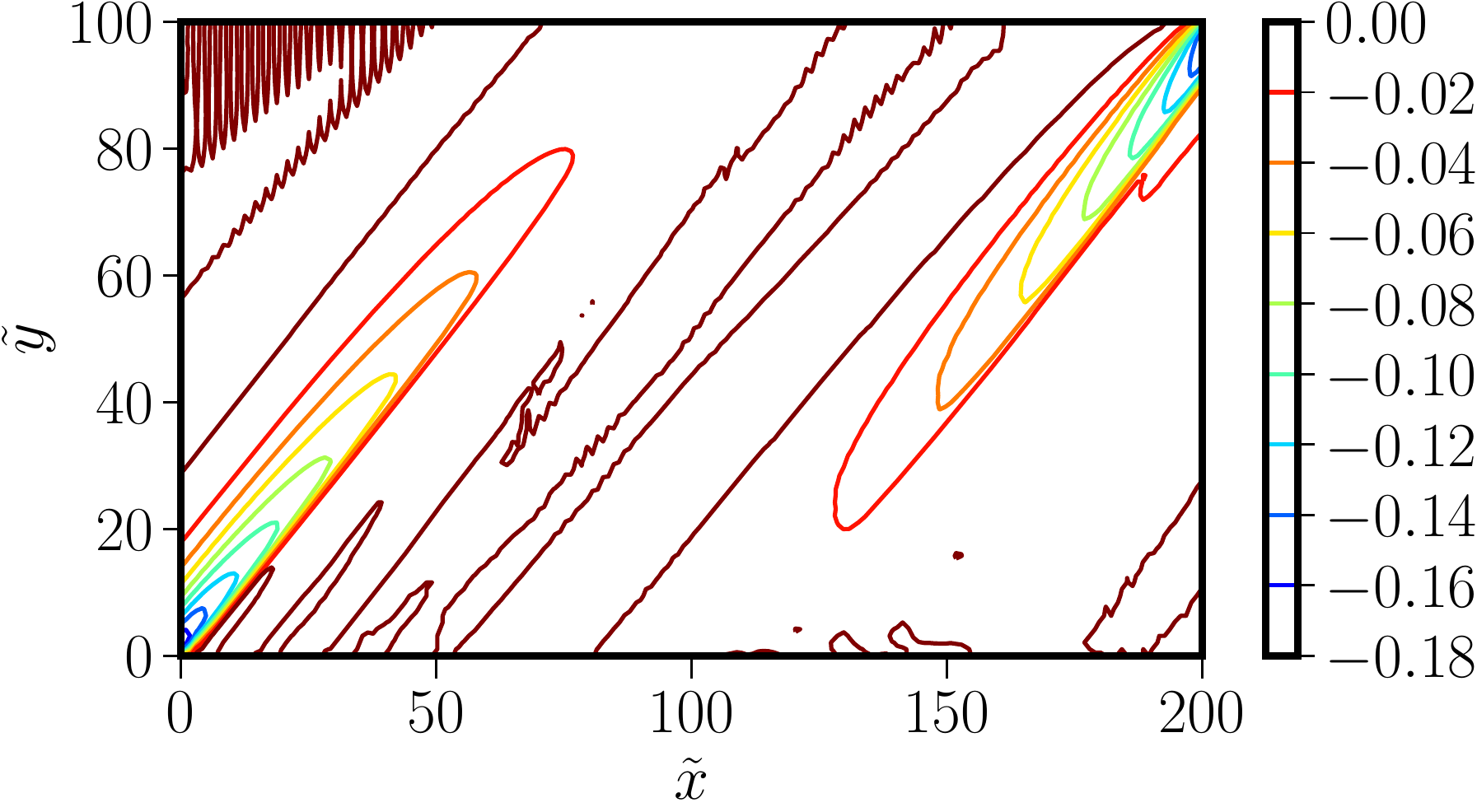}
\label{fig:L-500-V}}
\subfigure[${\mu}_{||}/{\mu}_{\perp}$=1000]{\includegraphics[width=0.31\textwidth]{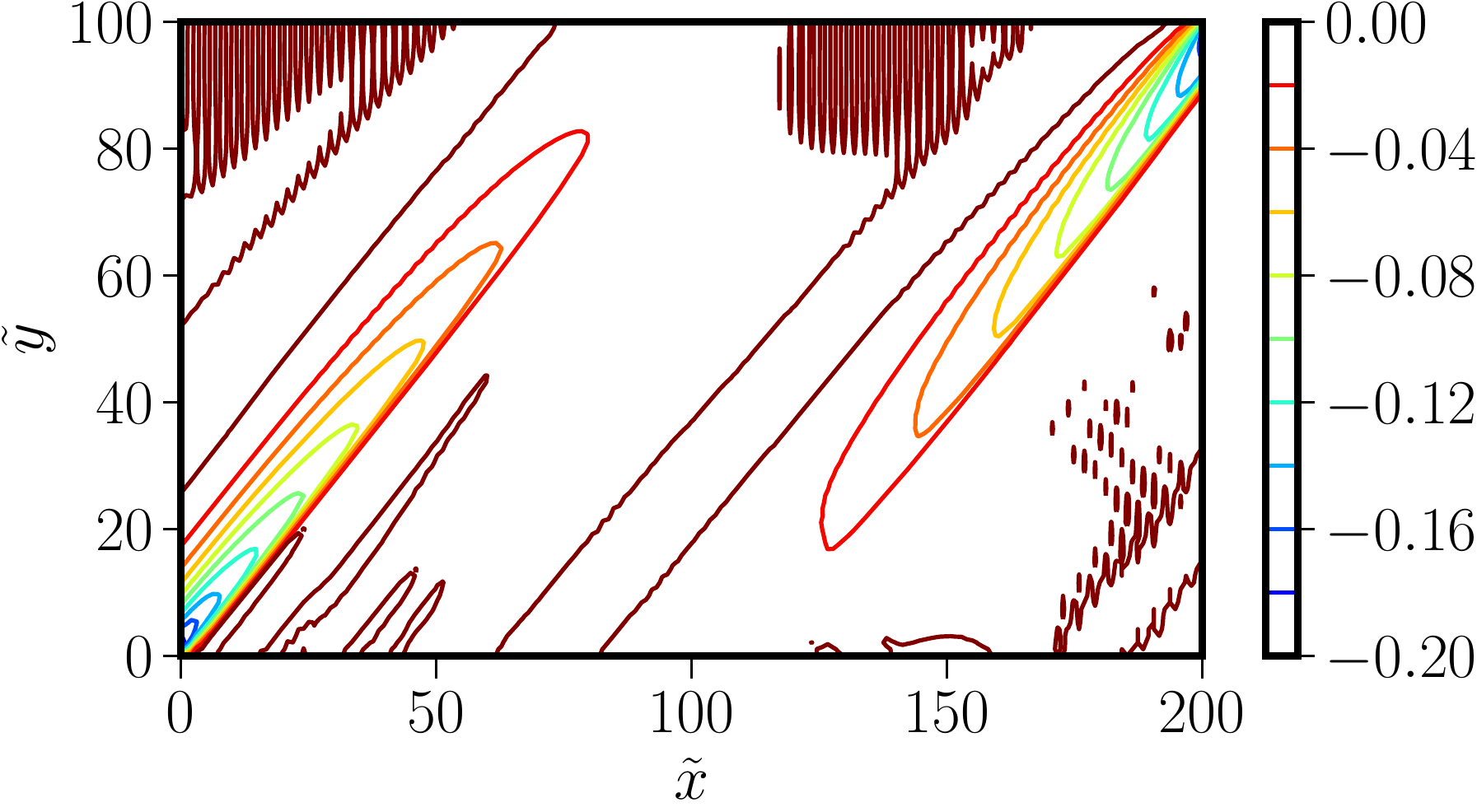}
\label{fig:L-1000-V}}
\subfigure[${\mu}_{||}/{\mu}_{\perp}$=100]{\includegraphics[width=0.31\textwidth]{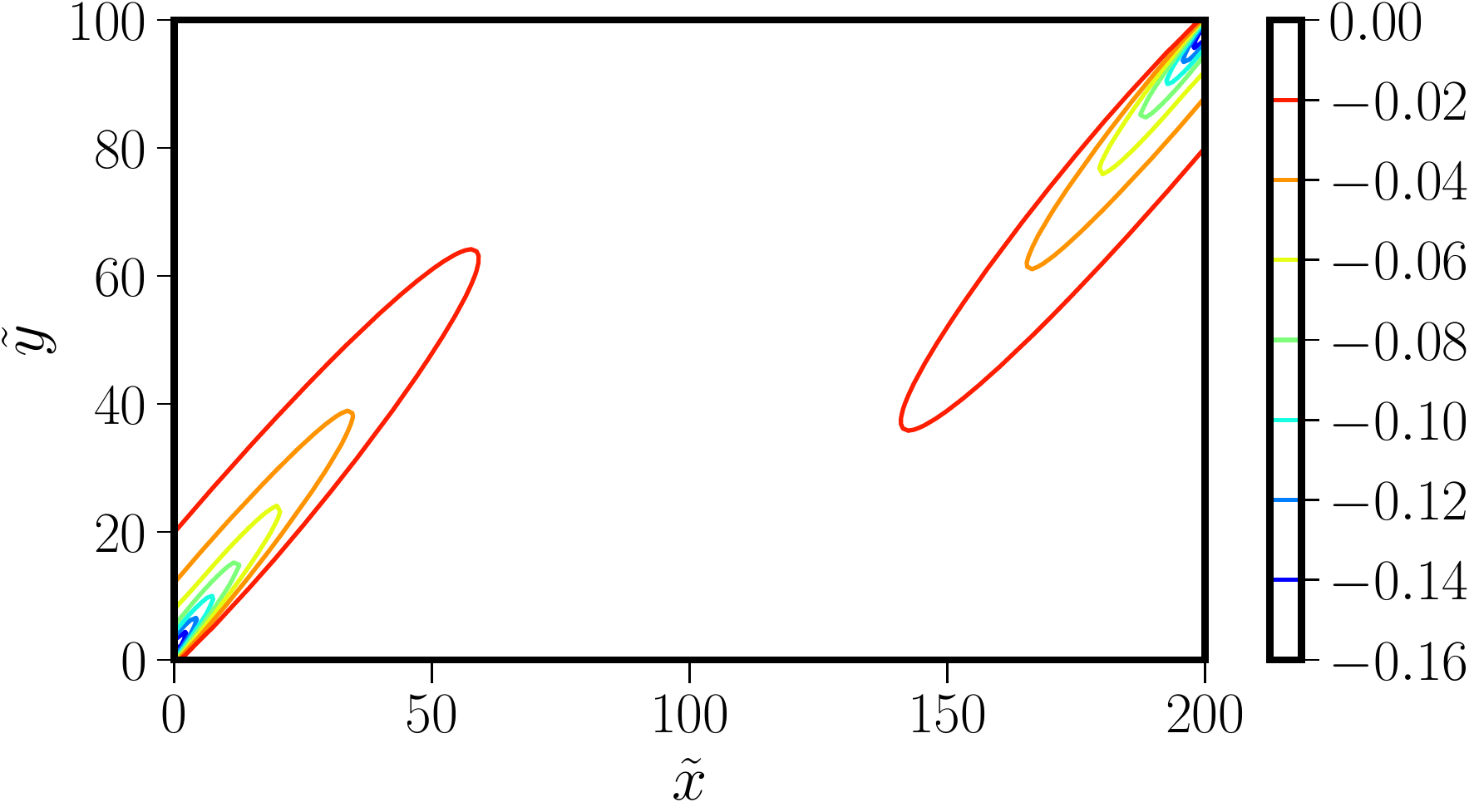}
\label{fig:WENO-100-V}}
\subfigure[${\mu}_{||}/{\mu}_{\perp}$=500]{\includegraphics[width=0.31\textwidth]{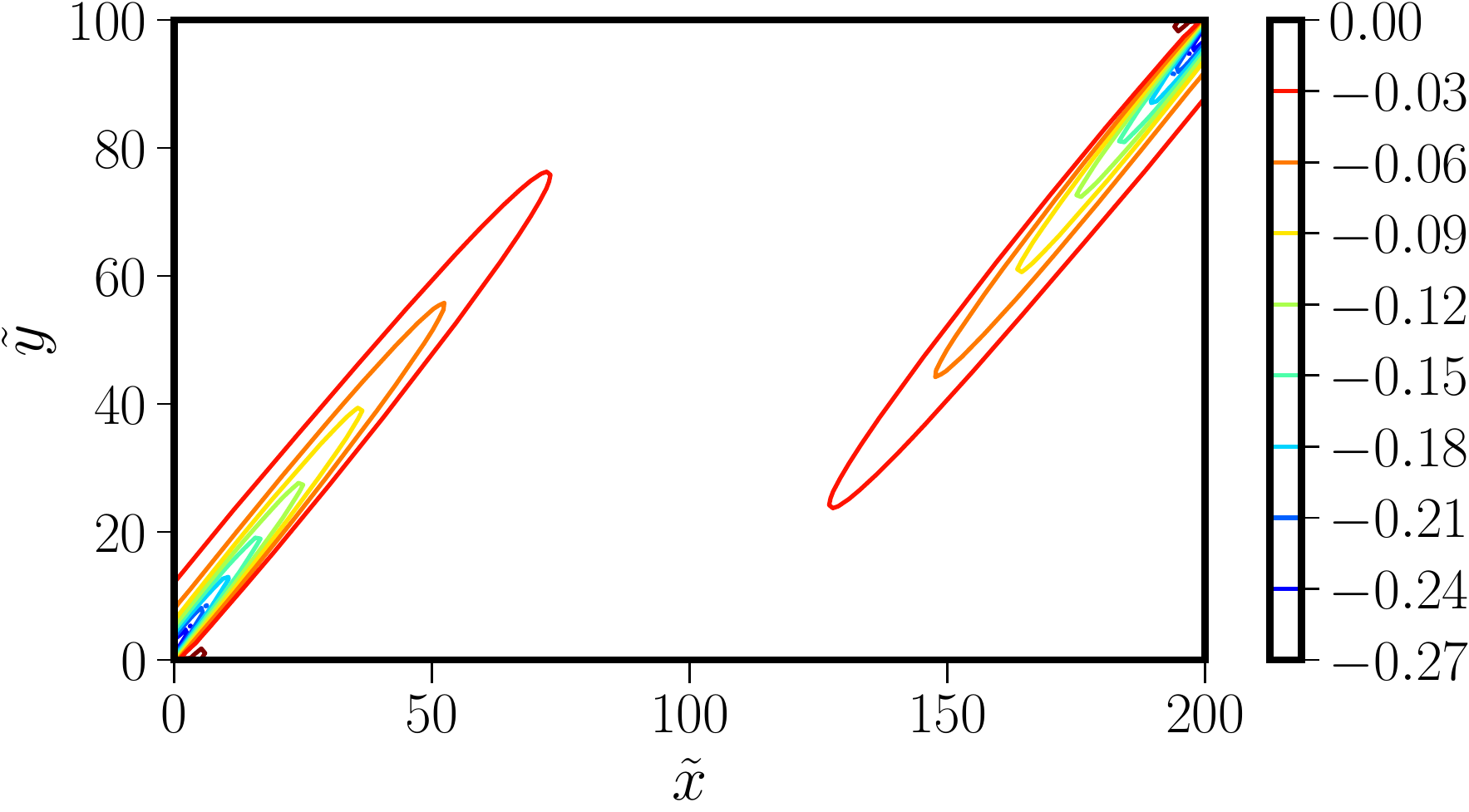}
\label{fig:WENO-500-V}}
\subfigure[${\mu}_{||}/{\mu}_{\perp}$=1000]{\includegraphics[width=0.31\textwidth]{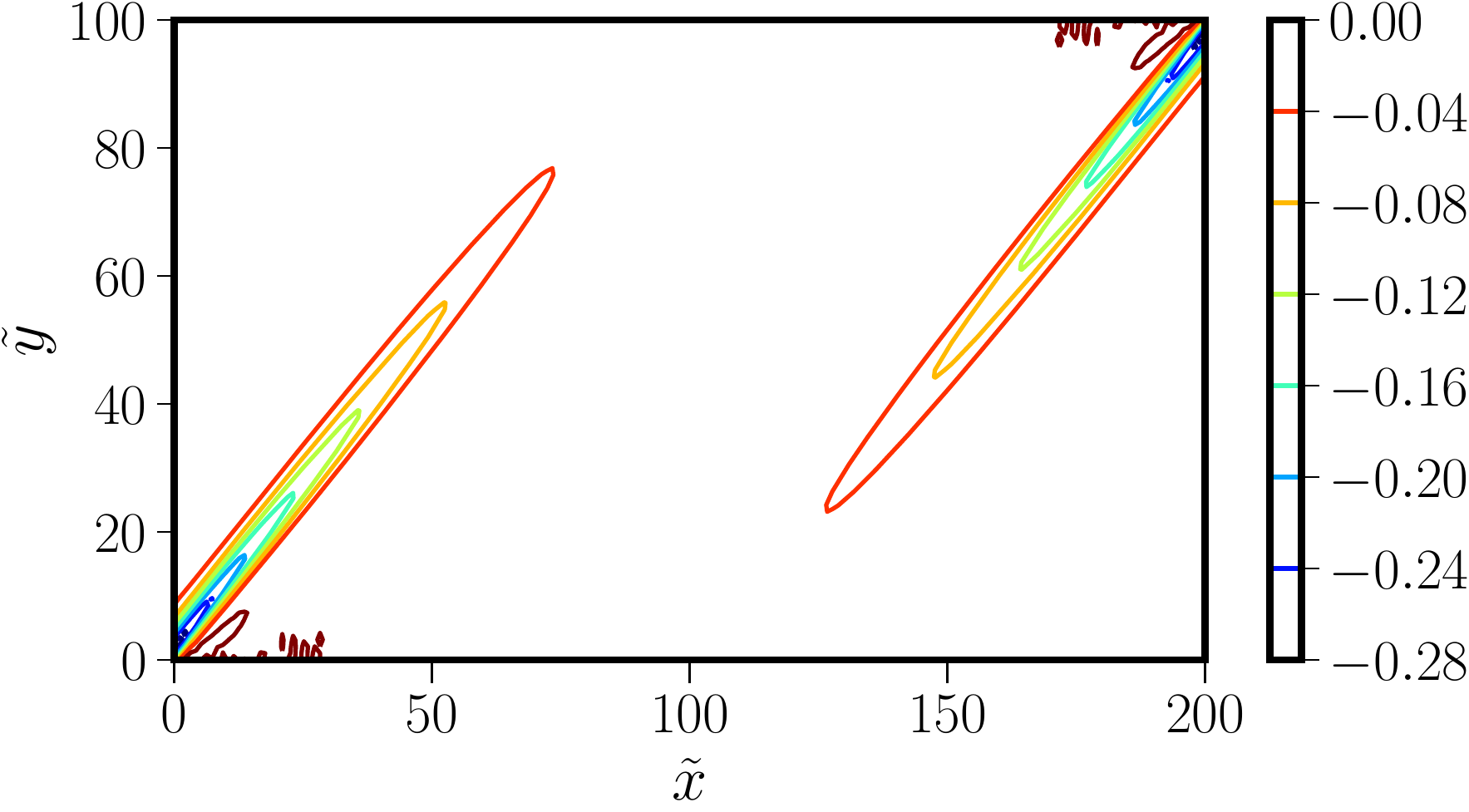}
\label{fig:WENO-1000-V}}
\caption{Comparison of  normalized velocity in y-direction with increasing ${\mu}_{||}/{\mu}_{\perp}$on a 192 $\times$ 96 grid. Top row: MFAM; Middle row: Generalized-MUSCL; bottom row: WENO-5Z-L.}
\label{fig:var-V-2}
\end{minipage}
  \end{sideways}
  \end{figure}

Figs. \ref{fig:var-stream-1} and \ref{fig:var-stream-2} show the velocity streamlines computed by using different schemes for mobility ratios ${\mu}_{||}/{\mu}_{\perp}$=10, 500 and 1000. Velocity streamlines computed by MFAM are shown in Figs. \ref{fig:MFAM-100-S1}, \ref{fig:MFAM-500-S1} and \ref{fig:MFAM-1000-S1} respectively and we can see that as ${\mu}_{||}/{\mu}_{\perp}$ is increasing the plasma is more confined to the center. For ${\mu}_{||}/{\mu}_{\perp} =100$ all the linear schemes and WENO has similar results compared to that of the MFAM whereas the {\color{black} Generalized MUSCL approach} is showing unphysical streamlines even for such small anisotropies. From the Figs. \ref{fig:3E-1000-S} and \ref{fig:L-1000-S}, we can see that there are significant unphysical vortices in Upwind-3E for ${\mu}_{||}/{\mu}_{\perp} =1000$. {\color{black} Results obtained by TVD-MUSCL are same as that of Upwind-3E scheme, similar to the results observed for isotropic diffusion problems}. However, the fifth order schemes, Upwind-5C, Upwind-5E, and WENO-5Z-L, were able to reduce the unphysical oscillations significantly and the difference is between them is very small. The reason for the unphysical oscillations can be understood from the velocity contour plots shown in Figs. \ref{fig:var-V-1} and \ref{fig:var-V-2}. 
\\
%
%
%
%
\newpage
We see from a series of Figs. $\ref{fig:3E-100-V}-\ref{fig:WENO-1000-V}$, that there can be serious numerical oscillations in the solutions of the flow field with increasingly strong magnetic confinement. The main reason is not only the anisotropic diffusion itself but also because of the sharp gradients that are not aligned with the grid. In Fig. \ref{fig:MFAM-1000-V}, MFAM approach do not show such behavior through an alignment of field lines with the grid and thereby captures the sharp gradient without numerical oscillations. If the angle between the field lines and the mesh is aligned, even the central scheme could also be easily implemented. As the gradient becomes more and more skewed, it diverges much more from the grid lines, thereby generating larger oscillations, as seen in Fig. \ref{fig:3E-1000-V}. WENO-5Z was able to capture the gradients without significant oscillations as the disturbances in the y-velocity are minimal when the solution has reached the steady state. Steady state solutions obtained for linear upwind schemes indicate that the disturbances behind the strong gradient have polluted the entire domain but similar to the boundary layer problem the solutions may be reduced with finer grids.

The oscillations due to sharp gradients can be drastically reduced on finer meshes by WENO-5Z. Even though MFAM approach did not show any spurious oscillations, such alignment may not be practical in many simulations with multiple sharp gradients. Furthermore, the boundary conditions and coding can be challenging to implement. Based on these simulations we can say that the non-oscillatory approach of WENO can be a reasonable and viable alternative.
\newpage 
%
%

\begin{figure}[H]
\centering
\subfigure[Upwind-3E ]{\includegraphics[width=0.47\textwidth]{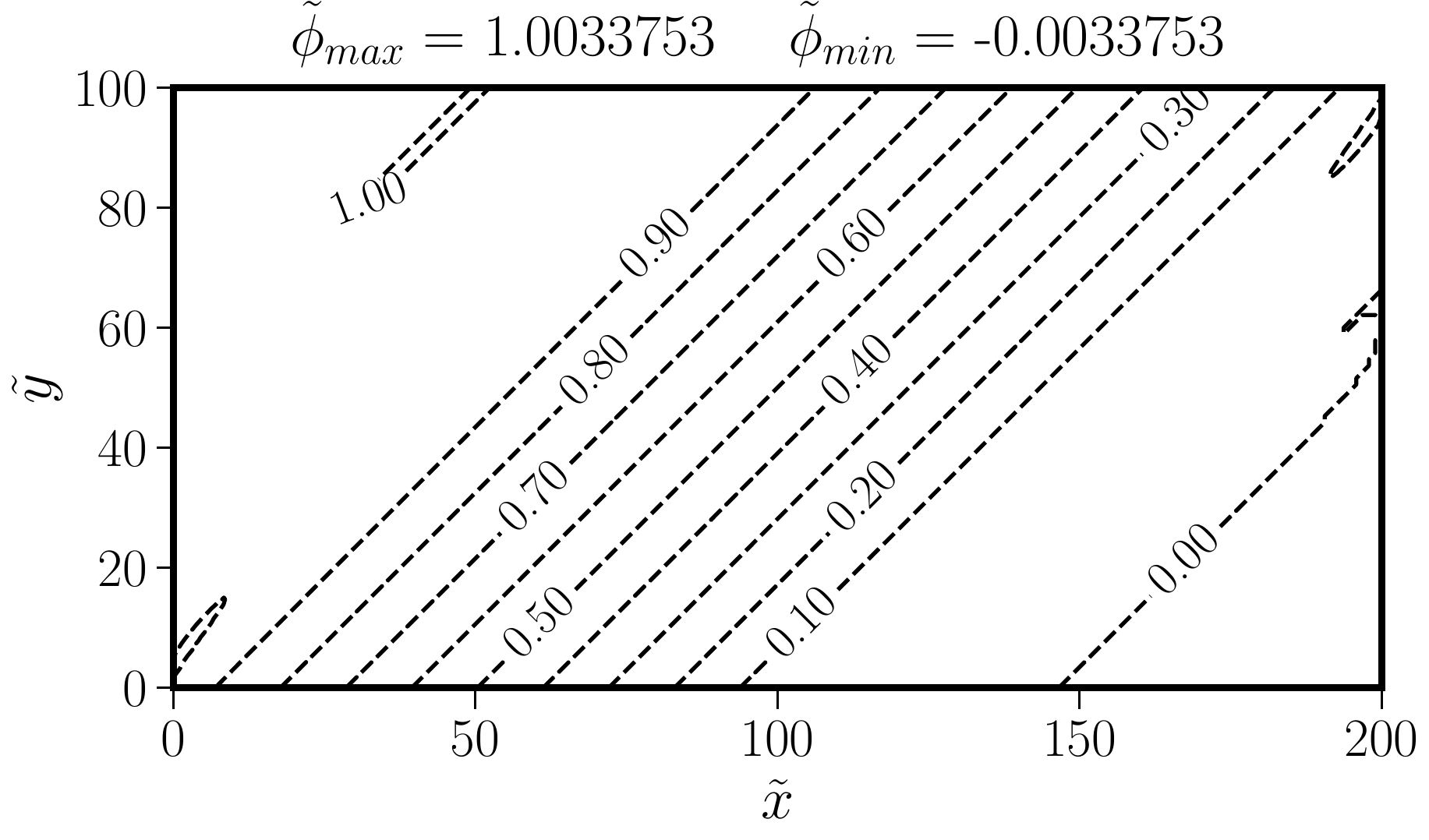}
\label{fig:3E-P1}}
\subfigure[Upwind-5E]{\includegraphics[width=0.47\textwidth]{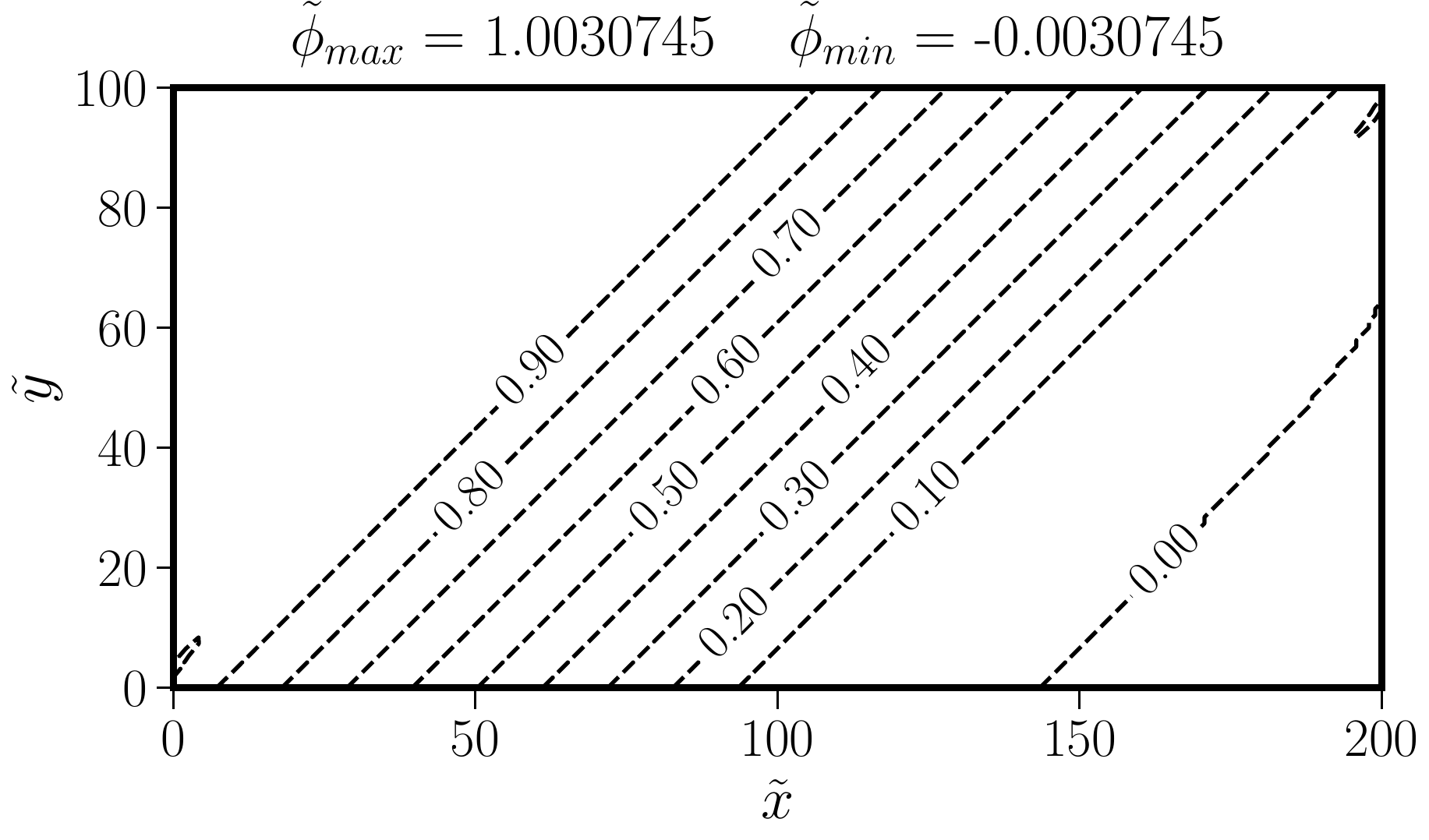}
\label{fig:5E-P1}}
\subfigure[Upwind-5C]{\includegraphics[width=0.47\textwidth]{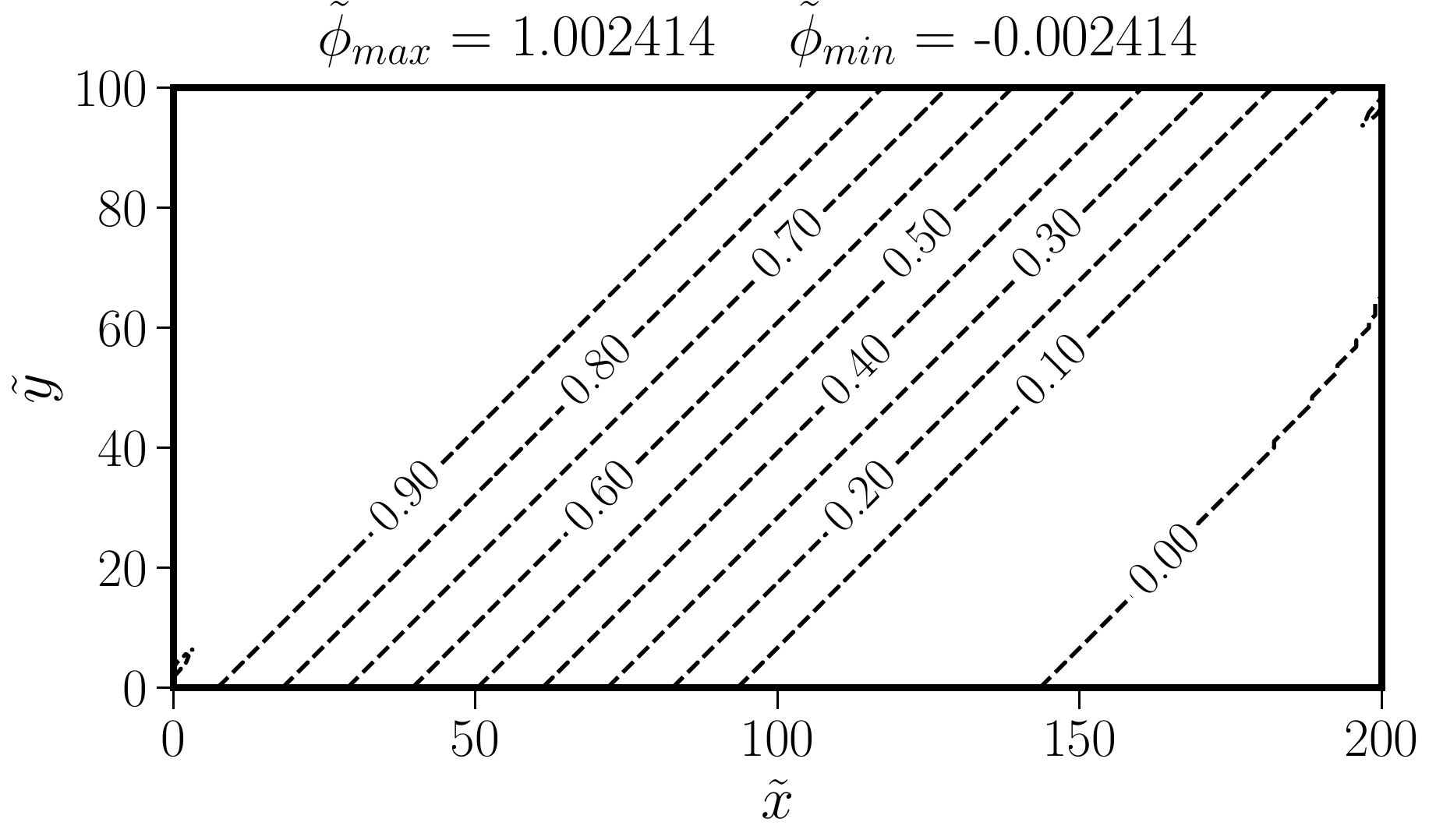}
\label{fig:5C-P1}}
\subfigure[{\color{black}Generalized-MUSCL}]{\includegraphics[width=0.47\textwidth]{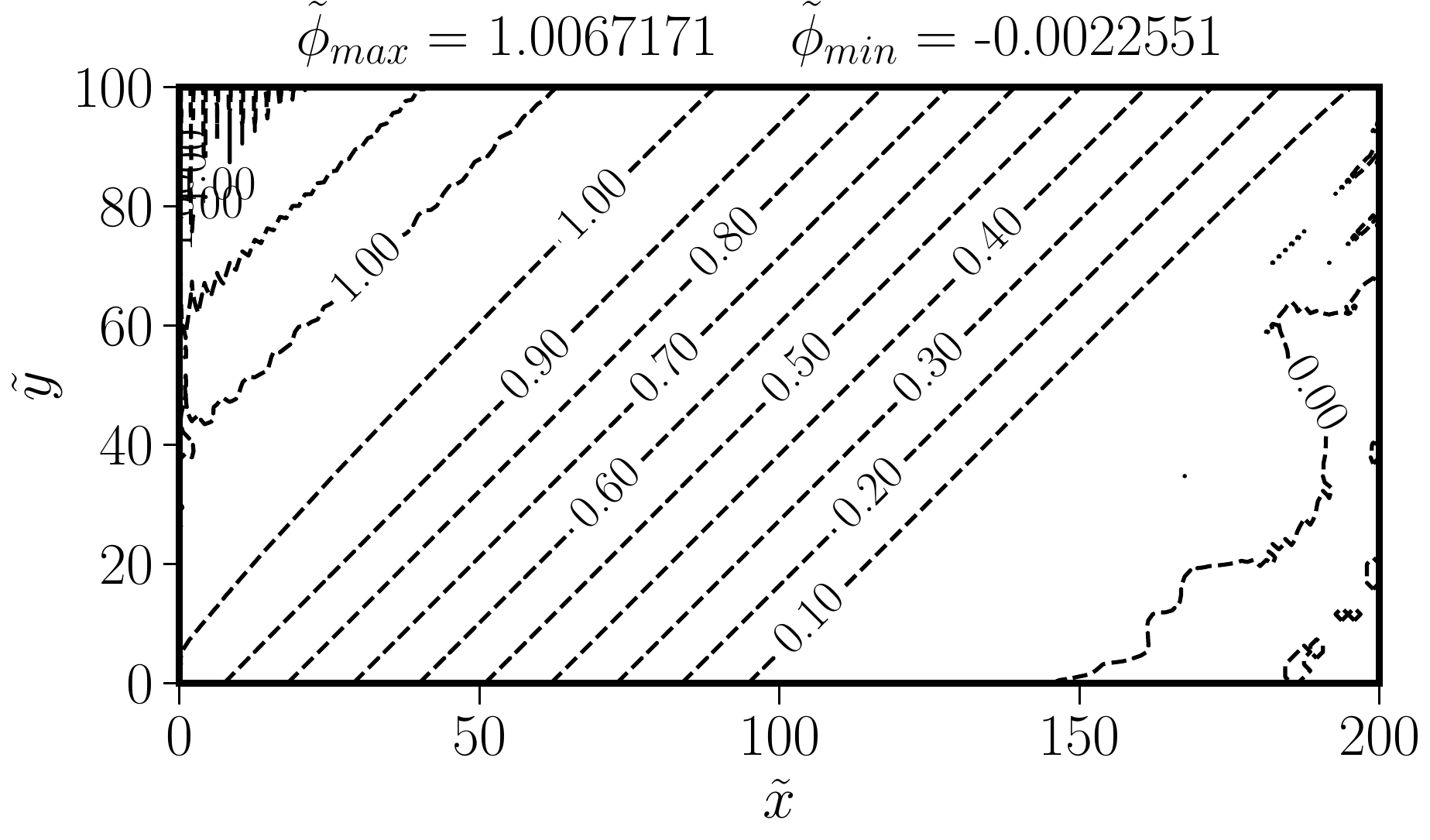}
\label{fig:L-P1}}
\caption{Distribution of dimensionless space potential, calculated by using various schemes on a grid of 192 $\times$ 96 for ${\mu}_{||}/{\mu}_{\perp}$=1000.}
\label{fig:var-poten-1000}
\end{figure}

\begin{figure}[H]
\centering
\includegraphics[width=0.45\textwidth]{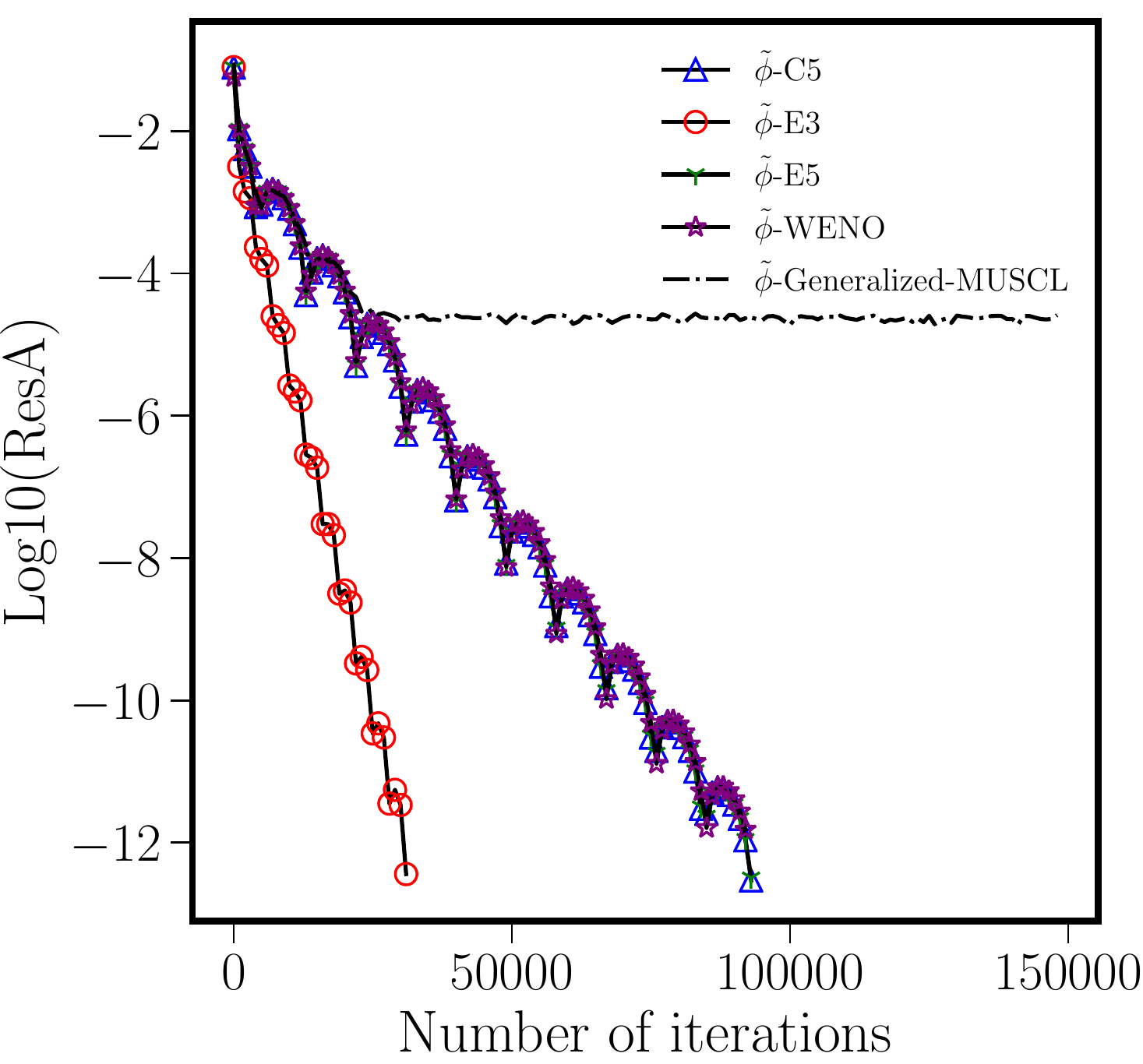}
\caption{Convergence histories by all the upwind schemes on a grid size of 192 $\times$ 96 for ${\mu}_{||}/{\mu}_{\perp}$=1000 and $\theta$ =45$^{\circ}$.}
\label{fig:residuals}
\end{figure}

%
%
Similar patterns are also observed in potential distributions shown in Fig. \ref{fig:var-poten-1000}. The maximum principle is violated as a consequence of these spurious oscillations as we can see the values of space potential are more than the theoretical bounds defined by the boundary conditions. Although the values are relatively smaller, it should be noted that the values are non-dimensionalized and the actual error can be much higher. On coarse grids, the entire domain can be polluted with unphysical extrema. The grid size of 192 $\times$ 96 is chosen in Fig. \ref{fig:var-poten-1000} is to reiterate the fact that the spurious oscillations can be reduced by the linear upwind schemes whereas they remain the same for the {\color{black}Generalized-MUSCL}. Spurious oscillations in potential are comparatively smaller in the compact scheme than that of explicit schemes. 
As explained earlier, the steady state is considered to be reached when the $L_1$ norms attain negligibly small values in the order of 10$^{-12}$, that are defined by Equation (\ref{eq:Dnorm})
   \begin{equation}
      L_{1}=\sqrt{\frac{1}{N_{\rm cell}}\sum^{N_{\rm cell}}\left({|Q^{n+1}-Q^{n}|}\right)},
      \label{eq:Dnorm}
   \end{equation}
 \noindent where $Q$ denotes each of the conservative variables $(\tilde{\phi}, \tilde{u}_{\rm x}, \tilde{u}_{\rm y})$. The convergence history of space potential for various schemes is shown in Fig. \ref{fig:residuals} when the test problem is calculated on a 192 $\times$ 96 grid. The residuals are monotonically decreasing and reached the limit as mentioned earlier for all the linear upwind schemes and WENO scheme. For the Generalized MUSCL scheme, the residual does not converge beyond four orders of magnitude.
\subsubsection{Effect of boundary conditions}
Simulations carried out by WENO-5Z along with Lagrange extrapolation, WENO-5Z-L, for ghost cells are free of spurious values in space potential if ${\mu}_{||}/{\mu}_{\perp} \leq 500$.  But with increasingly higher anisotropies small overshoots and undershoots, similar to those that are observed in linear upwind schemes, are also found in the cells next to the boundary. By keeping the objective of the uniform grid,  WENO extrapolation method, WENO-5Z-W, is used for computing the values in ghost-cells.

\begin{figure}[H]
\centering
\subfigure[WENO-5Z-L]{\includegraphics[width=0.7\textwidth]{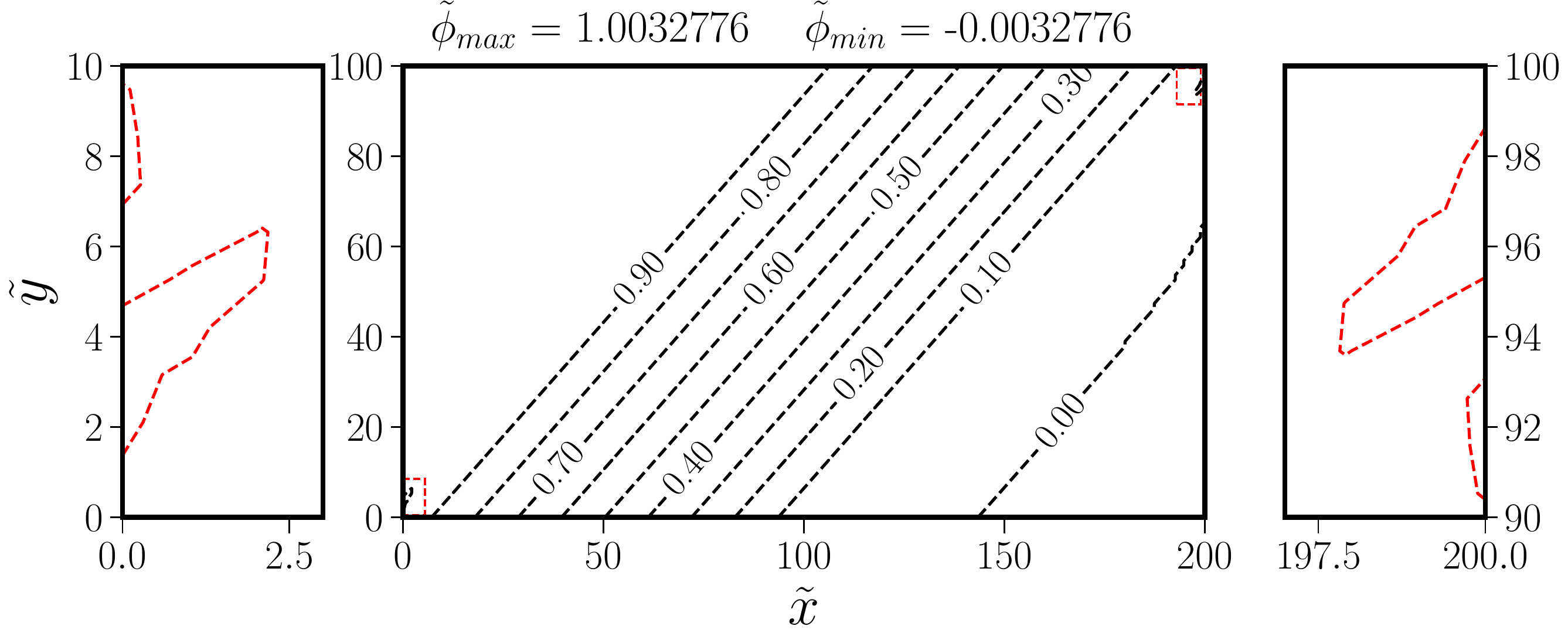}
\label{fig:WENO-P_L}}
\subfigure[MFAM]{\includegraphics[width=0.48\textwidth]{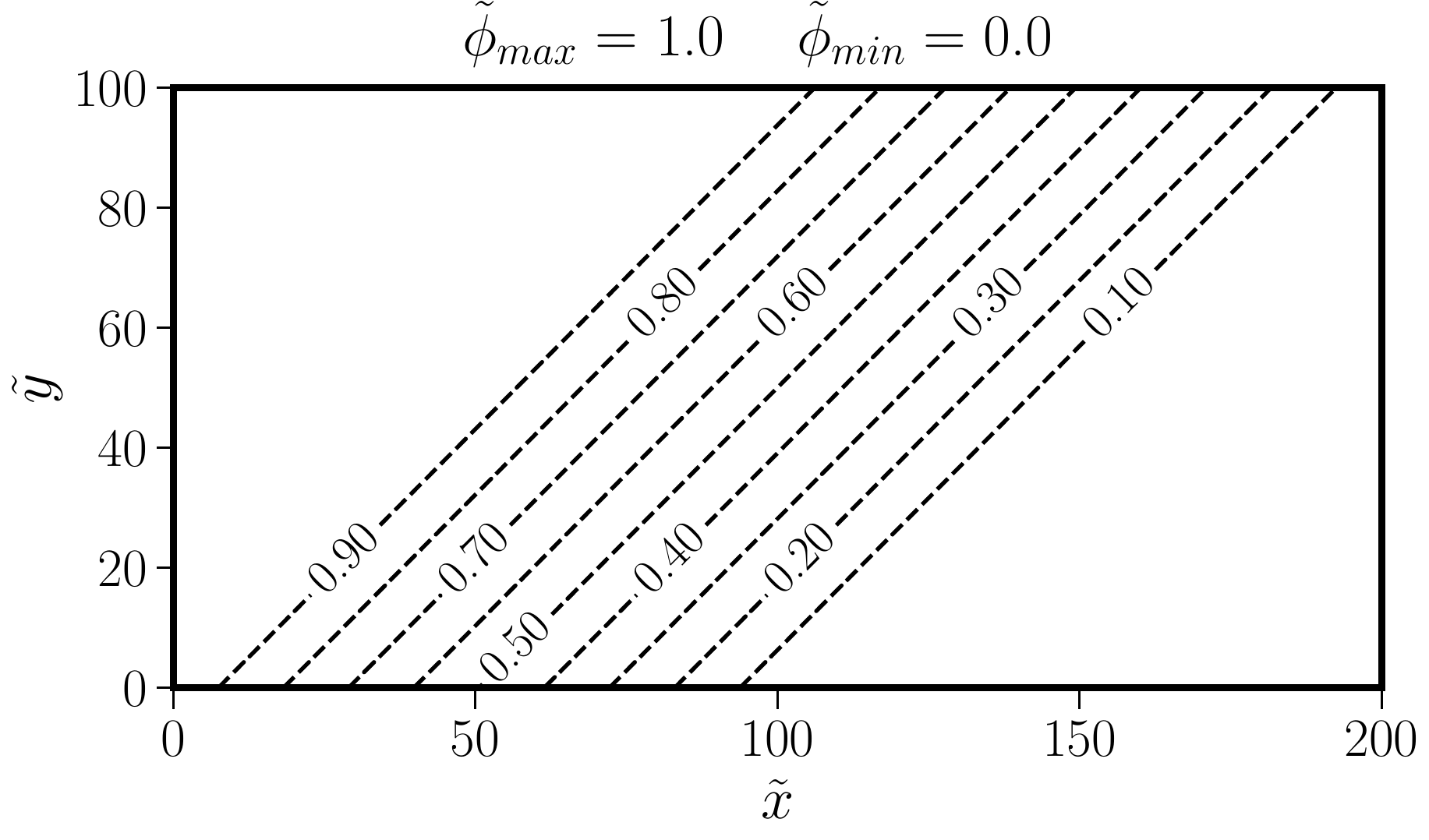}
\label{fig:MFAM-P_L}}
\subfigure[WENO-5Z-W]{\includegraphics[width=0.47\textwidth]{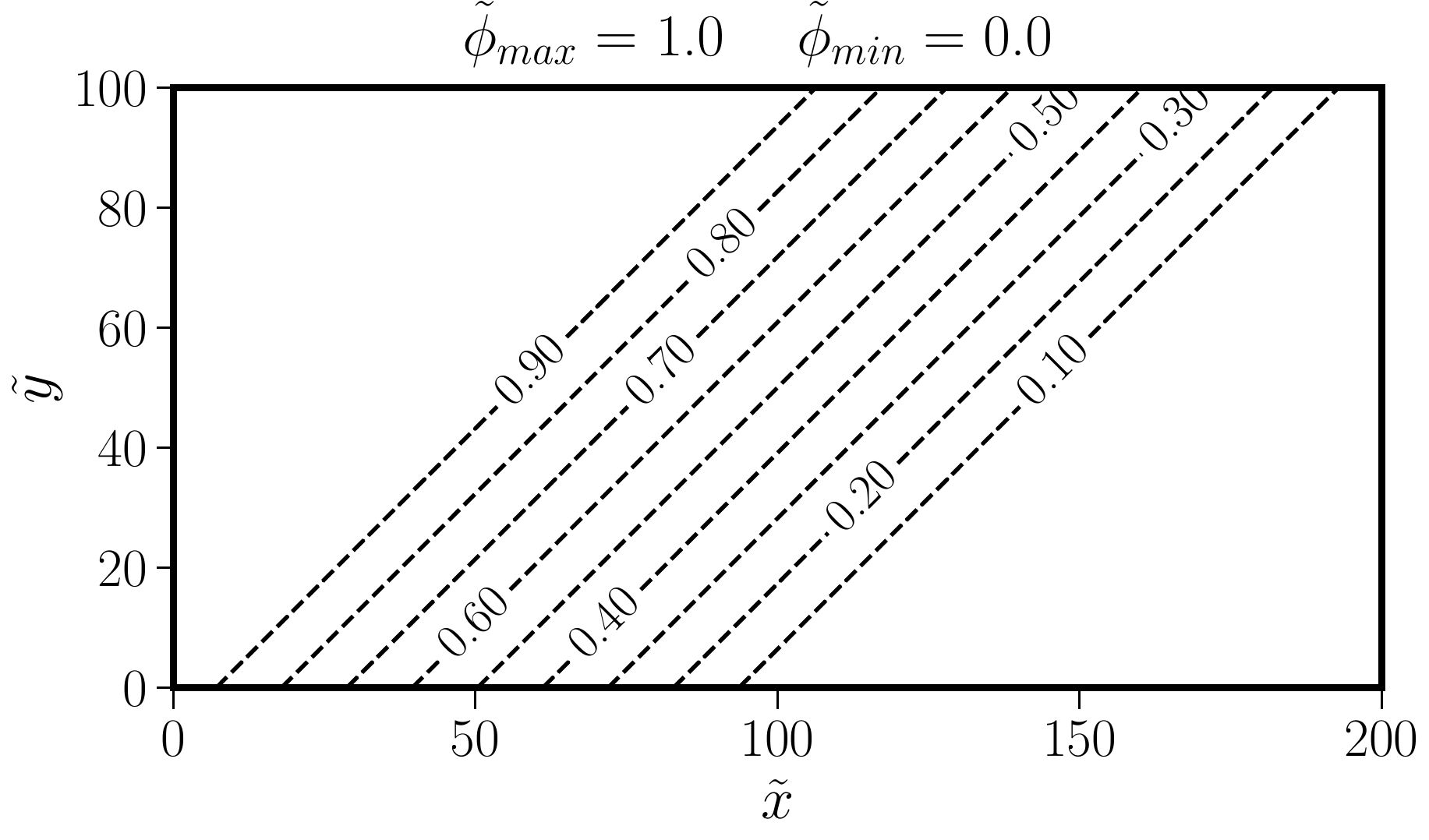}
\label{fig:WENO-P_W}}
\caption{Distribution of dimensionless space potential computed by WENO extrapolation and Lagrange extrapolation in comparison with MFAM for ${\mu}_{||}/{\mu}_{\perp}$=1000 on a grid of 96 $\times$ 96. Dotted regions are enlarged and shown in Fig. \ref{fig:WENO-P_L}. }
\label{fig:WENO-poten}
\end{figure}

The WENO-5Z-W scheme was able to adaptively reduce to lower accurate order boundaries and thereby prevented the unphysical oscillations in the flow field, as shown in Fig. \ref{fig:WENO-poten}. WENO extrapolation approach was also tested for smaller anisotropies, and the results indicated that the performances are similar to that of WENO-5Z-L and no unphysical extrema are generated. The minimum grid size required for WENO-5Z-W scheme to prevent unphysical extrema in space potential distribution is 16 $\times$ 16 for ${\mu}_{||}/{\mu}_{\perp}$ $\le500$ and  96 $\times$ 96 if ${\mu}_{||}/{\mu}_{\perp}$ $>500$. 

Test calculations are also carried out for the cases when the magnetic field lines are aligned at an angle $60^{\circ}$ from the vertical and even for complicated shapes constructed using the magnetic stream function. These results further confirm the fact that the WENO scheme has better capabilities in capturing the sharp gradients without spurious oscillations even for complex magnetic field shapes.
 %
\begin{figure}[H]
\centering
\subfigure[Space Potential for 60$^{\circ}$ B-field]{\includegraphics[width=0.47\textwidth]{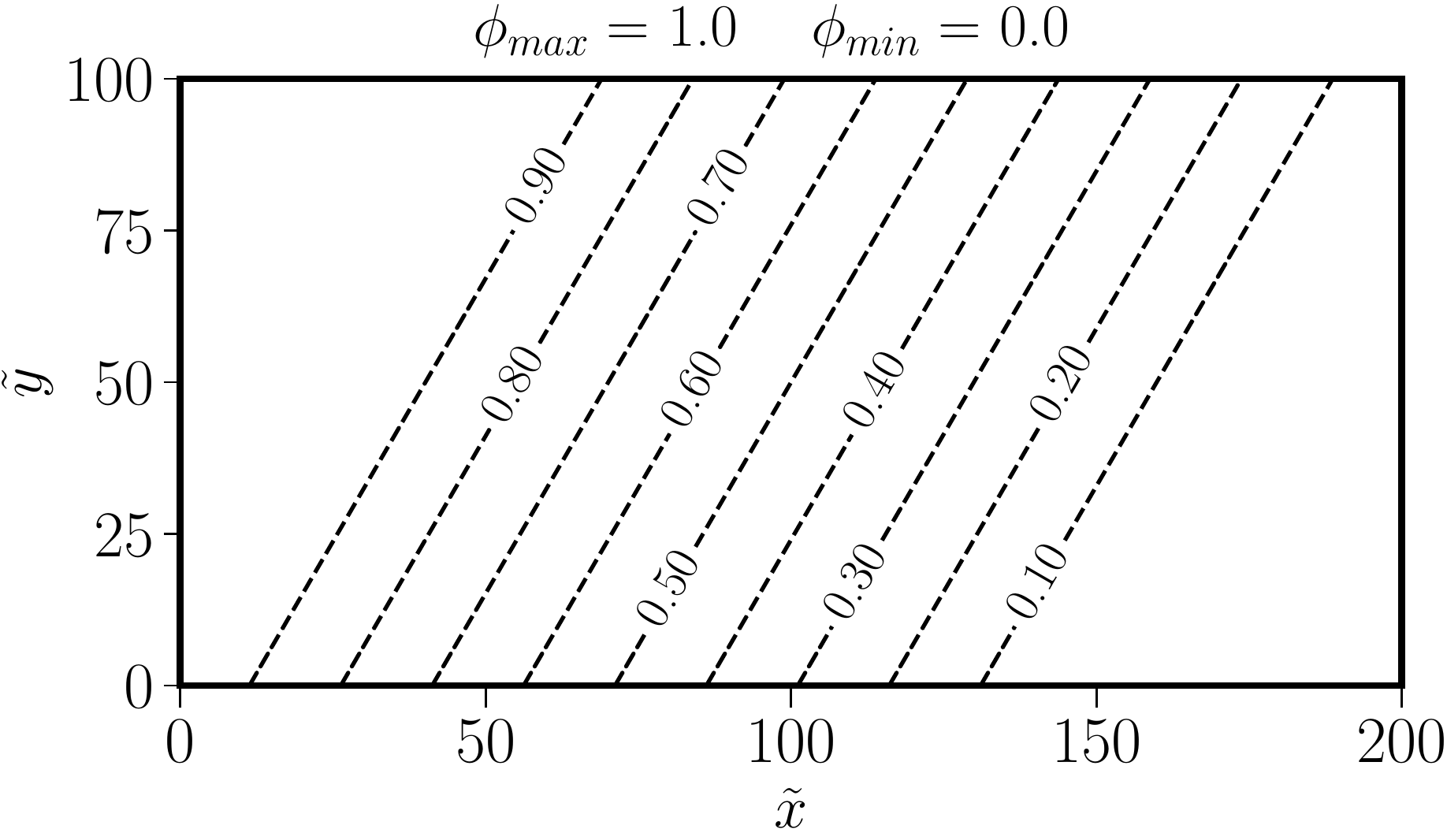}
\label{fig:WENO-field-60}}
\subfigure[Streamlines for 60$^{\circ}$ B-field]{\includegraphics[width=0.47\textwidth]{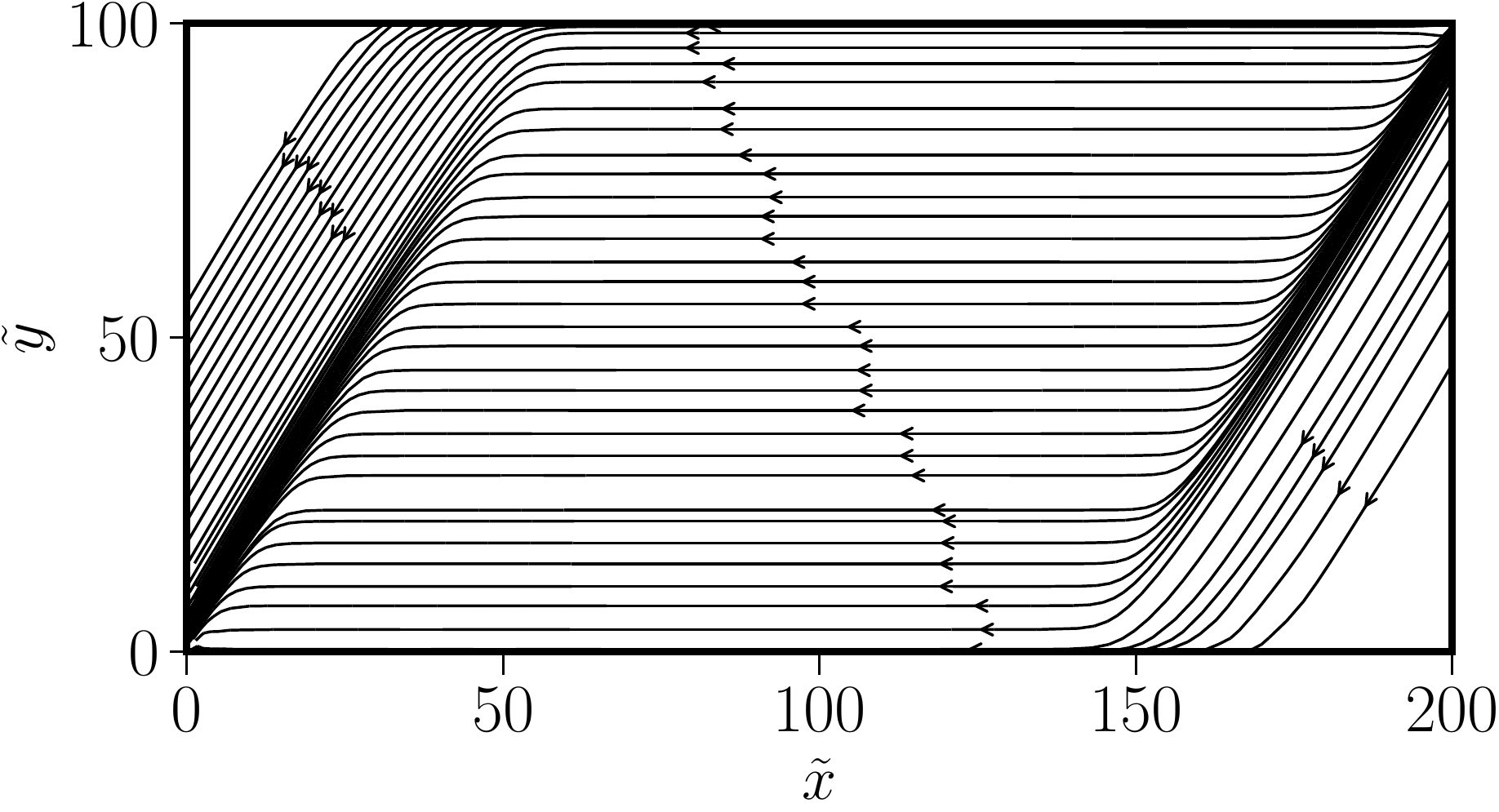}
\label{fig:WENO-field-stream-60}}
\subfigure[Space Potential for curved B-field]{\includegraphics[width=0.47\textwidth]{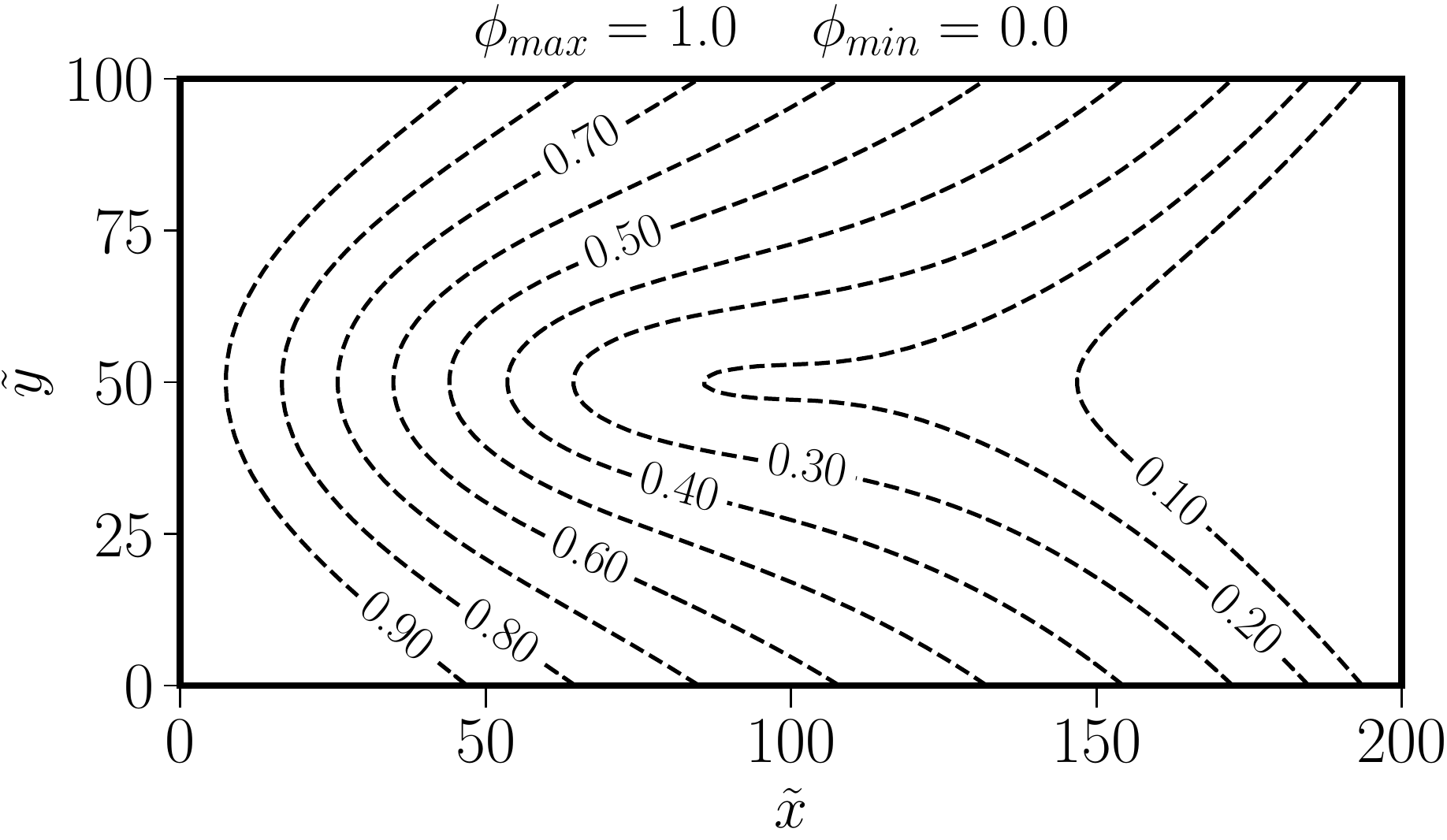}
\label{fig:WENO-field-P}}
\subfigure[Streamlines for curved B-field]{\includegraphics[width=0.47\textwidth]{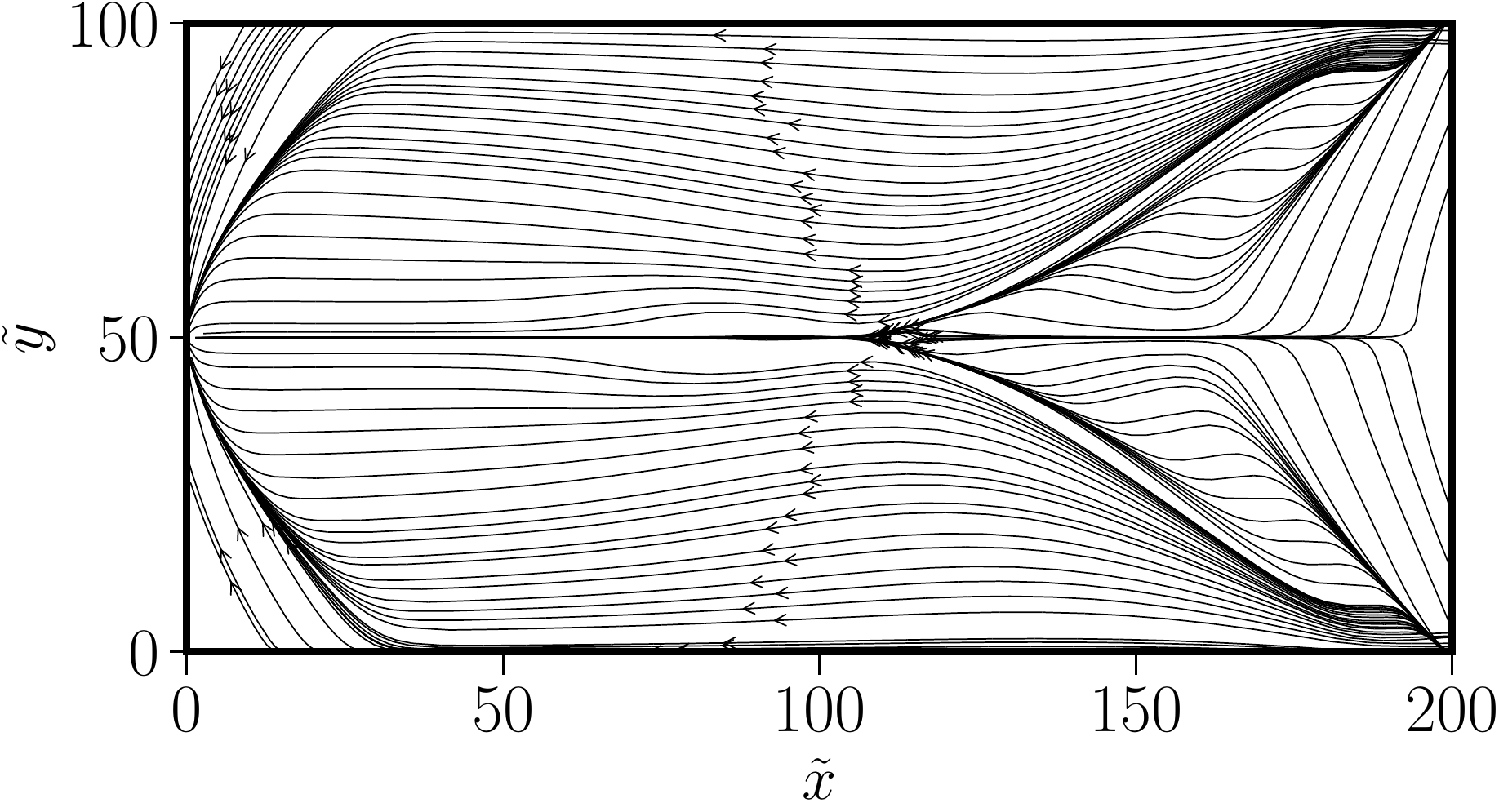}
\label{fig:WENO-field-stream}}
\caption{Distribution of dimensionless space potential and streamlines for curved magnetic field shape and a constant angle of 60$^{\circ}$, computed by WENO-5Z-W on a grid of 96 $\times$ 96 for ${\mu}_{||}/{\mu}_{\perp}$=1000.}
\label{fig:field-poten}
\end{figure}
\subsubsection{Mesh convergence and accuracy for electron fluids}
The computational accuracy of anisotropic diffusion by HES approach is compared with that of MFAM by using the transverse electron flux which is defined as the total electron flux from right to left boundary given by Eq. (\ref{eq:transverse}),
      \begin{equation}
      \Gamma_{\rm e}= \int_{\Omega_L}^{}  \left(
      \tilde{u}_{\rm e,x}\right)d\tilde{y},
      \label{eq:transverse}
   \end{equation}
  where $\Omega_L$ is the left boundary of the calculation field.
The grid convergence of the calculated electron transverse flux is evaluated only for WENO-5Z-W and MFAM schemes as for the other schemes the solution is polluted due to spurious oscillations and might not be appropriate. Table \ref{table:trans-2} shows the difference between computed transverse electron fluxes by both the approaches. The error reduces from 1.12$\%$ to 0$\%$ on mesh refinement.
\begin{table}[H]
\footnotesize
\centering
\caption{Error in transverse electron flux  for ${\mu}_{||}/{\mu}_{\perp}$=1000 and $\theta$ =45$^{\circ}$. }
\label{table:trans-2}
    \begin{tabular}{| c | c | c | c | c |  }
    \hline
    Number & \multicolumn{2}{c|}{WENO-5Z-W} & \multicolumn{2}{c|}{MFAM} \\
    \cline{2-5}
        of points&$\Gamma_{\rm e}$ & Error $\%$ & $\Gamma_{\rm e}$ & Error $\%$    \\
    \cline{1-5}
    96$\times$48    &  -0.01861 &1.12 &-0.01878  & 0.21    \\
    \hline
    192$\times$96    & -0.01878 & 0.21 &-0.01879  & 0.16   \\
    \hline
    384$\times$192   & -0.01881 & 0.05  &-0.01882  & 0.0    \\
    \hline
      768$\times$384   & -0.01882 &0.0 &-0.01882  & 0.0   \\
    \hline
\end{tabular}
\end{table}
\noindent Another criteria considered is the relative numerical error, $|l_2|$ error, which is defined as
\begin{equation}
|l_2| = \frac{1}{n} \left( \sum\limits \left( \frac{Q_{WENO}-Q_{MFAM}}{Q_{MFAM}}\right)^2 \right)^{0.5},
\end{equation}
where Q is each of the conservative variable. The relative errors are calculated in comparison with the field aligned mesh results. The Table \ref{table:trans-1} show the relative $l_2$ errors computed for field aligned mesh and WENO-5Z-W and the results show that the solution seems to improve with finer grid resolution.
\begin{table}[H]
\footnotesize
  \centering
  \caption{Relative $l_2$ error  for MFAM and WENO-5Z-W for ${\mu}_{||}/{\mu}_{\perp}$=1000.} 
\label{table:trans-1}
    \begin{tabular}{| c | c | c | c | c | }
      \hline
         Number of points & \multicolumn{1}{c|}{$\tilde \phi$} & \multicolumn{1}{c|}{$\tilde  u_x$} & \multicolumn{1}{c|}{$ \tilde u_y$} \\
         \hline
    96$\times$48 & 3.28E-03 & 1.01E-02 & 1.03E-02 \\
   \hline
    192$\times$96 & 2.62E-03 & 9.92E-03 & 1.02E-02 \\
     \hline
    384$\times$192 & 1.73E-03 & 8.71E-03 & 8.92E-03 \\
       \hline
    768$\times$384 &   9.82E-04    & 8.12E-03       & 8.35E-03 \\
      \hline
    \end{tabular}%
\end{table}%
		
										\section{Conclusions}\label{sec-5}

Higher order linear and non-linear schemes (WENO) are proposed for the simulation of magnetized electrons in quasi-neutral plasmas by the hyperbolic method. We first implemented the approach for diffusion equation, which has similar upwind structure, in hyperbolic form to verify the design accuracy and the implementation of boundary conditions. Then the schemes are implemented to simulate advection-diffusion equation to capture the sharp gradients in boundary layer type problems without spurious oscillations. Finally, the schemes are applied to anisotropic diffusion electron fluid equations to reduce the spurious oscillations due to sharp gradients significantly. The current methodology can be extended to the energy equation for the complete simulation of electron flow in magnetized plasmas. Implicit time stepping can substantially accelerate the steady-state convergence of all the problems considered here and will be discussed in future work. The critical findings of the paper are summarized as follows

\begin{enumerate}
\item High-order and high-resolution methods are implemented successfully for the diffusion equation in the hyperbolic form on uniform meshes. Design order of accuracy is obtained for all the schemes for all the test problems considered. 
\item {\color{black}Shock-capturing schemes are found to be unnecessary for simple diffusion equation, and through various test cases, the inapplicability of certain TVD schemes is explained. Even though the WENO scheme is a shock-capturing scheme the steady-state solutions are not contaminated and are similar to the linear upwind schemes.}
\item Ghost cell approach is found to be more accurate and stable than the weak boundary implementation, especially on coarse meshes. Linear upwind schemes are consistently stable with the corresponding higher order boundary conditions whereas WENO schemes were stable only with $3^{rd}$ order boundary conditions due to their inherent non-linearity.
\item Weighted essentially non-oscillatory schemes are implemented to capture the sharp gradients without spurious oscillations for boundary layer type problem for advection-diffusion equation. For smooth solutions design order of accuracy is obtained for all the higher order methods. 
\item The difference between $5^{th}$ order explicit and compact schemes is indistinguishable for diffusion equation. As for the advection-diffusion equation, the compact schemes are slightly more accurate than explicit schemes. The advantage of compact schemes may be more pronounced if the current approach is extended to time-dependent problems and also to hyperbolic compressible and incompressible Navier-Stokes equations proposed by Nishikawa \cite{nishikawa2017hyperbolic} and are currently being studied. 
\item Significant improvement is achieved by utilizing higher order linear upwind methods and WENO schemes for the simulation of magnetized electron fluids as the results are much closer to the field-aligned mesh. {\color{black} For small anisotropies, linear upwind schemes as well as TVD-MUSCL can be appropriate.}
\item Similar to the boundary layer problem WENO approach is found to be more suitable and robust to reduce the spurious oscillations associated with the sharp gradients with increasing anisotropic diffusion in electron fluid equations. Boundary conditions based on WENO extrapolation are found more appropriate to prevent unphysical extrema for anisotropies higher than 500.
\item For very high anisotropic diffusion problems, ${\mu}_{||}/{\mu}_{\perp}$ of the order $10^4-10^9$ that can be seen in practical applications like tokamaks and space propulsion devices, all the schemes would result in spurious oscillations on coarse meshes, say 96 $\times$ 96, and reducing such oscillations is the subject of our future work.
\end{enumerate}


\bibliographystyle{elsarticle-num} 
\bibliography{JCP_Sainath_completed.bbl}




\end{document}